\documentclass[twoside]{article}
\usepackage{amsmath,amsfonts,amssymb,wasysym,stmaryrd,manfnt}
\usepackage{qic,epsfig}
\usepackage{cite}
\usepackage{color,colortbl}
\usepackage{graphicx,booktabs}
\usepackage{array,latexsym}
\usepackage{textcomp,booktabs}
\usepackage[usenames,dvipsnames]{xcolor}
\usepackage{graphicx}
\usepackage{booktabs}
\usepackage{cases}
\usepackage{tikz}
\definecolor{classic}{cmyk}{1.0,0.6,0.1,0.35}
\definecolor{classic1}{cmyk}{1.0,0,1.0,0.55}
\definecolor{classic2}{cmyk}{0.8,0.5,0.5,0.15}

\renewcommand{\theequation}{\arabic{section}.\arabic{equation}}

\newtheorem{thm}{Theorem}
\newtheorem{cor}{Corollary}

\newtheorem{defn}{Definition}

\renewenvironment{proof}{\par\noindent{\bf Proof.}}{$\quad\Box$\par}
\newcommand{\ket}[1]{| #1 \rangle}
\newcommand{\bra}[1]{\langle #1 |}

\begin{document}
\setlength{\textheight}{8.0truein}    

\runninghead{Quotient Algebra Partition and Cartan Decomposition
for $su(N)$ III}
            {Zheng-Yao Su and Ming-Chung Tsai}

\normalsize\textlineskip \thispagestyle{empty}
\setcounter{page}{1}

\vspace*{0.88truein}

\alphfootnote

\fpage{1}

\centerline{\bf Quotient Algebra Partition and Cartan
Decomposition for $su(N)$ III} \vspace*{0.035truein}
\centerline{\footnotesize Zheng-Yao Su\footnote{Email:
zsu@nchc.narl.org.tw
}\hspace{.15cm} and Ming-Chung Tsai}\centerline{\footnotesize\it
National Center for High-Performance
 Computing,}
 \centerline{\footnotesize\it National Applied Research Laboratories,
 Taiwan, R.O.C.}

\vspace*{0.21truein}

\abstracts{
 In the $3$rd episode of the serial exposition~\cite{Su,SuTsai1,SuTsai3},
 quotient algebra partitions of rank zero earlier introduced undergo further partitions
 generated by bi-subalgebras of higher ranks.
 The {\em refined} versions of
 quotient algebra partitions admit not only Cartan decompositions of type {\bf AI}
 but also decompositions of types {\bf AII} and {\bf AIII},
 resorting to systematic applications of the operation {\em tri-addition}.
 Details of quotient algebra partitions of higher ranks
 are extensively examined in this longest episode of the serial.
 Furthermore,
 the computational universality is attained taking advantage of a special form of
 transformation called the {\em $s$-rotation}.
 The structure of quotient algebra partition is preserved
 under  mappings composed of spinor-to-spinor $s$-rotations.
 }{}{}

 \vspace{3pt} \vspace*{1pt}\textlineskip

 \section{Introduction}\label{secintro}
 \renewcommand{\theequation}{\arabic{section}.\arabic{equation}}
\setcounter{equation}{0} \noindent
 In this episode of the $3$rd,
 the Quotient Algebra Partition (QAP) considered in~\cite{Su,SuTsai1} undergoes further partitions.
 It will be clear that a bi-subalgebra {\em of rank} $r$ $\mathfrak{B}^{[r]}$ of a Cartan subalgebra
 in $su(2^p)$ is isomorphic to an $(r+p)$-th maximal subgroup of $Z^p_2$.
 Analogous to a partition of $Z^p_2$ produced by one of its subgroups,
 an arbitrary $\mathfrak{B}^{[r]}$ can generate a partition of $su(2^p)$
 that consists of a number $2^{p+r}$ of cosets,
 where each of the cosets serves as a conjugate-pair constituted of two abelian subspaces.
 More importantly, the operation {\em tri-addition} is
 introduced to this structure, by which these subspaces form an abelian group.
 Of great significance is the assertion that the Lie algebra $su(N)$
  admits a Cartan decomposition $su(N)=\mathfrak{t}\oplus\mathfrak{p}$ iff
  the subalgebra $\mathfrak{t}$ is a proper maximal subgroup of a QAP and
  $\mathfrak{p}$ is the coset of $\mathfrak{t}$ in this partition with the tri-addition.
  As an immediate consequence,
 all the three types of Cartan decompositions are exhaustively acquirable
 simply applying the tri-addition algorithmically to the partitions.
 Specifically, the type of a Cartan decomposition $\mathfrak{t}\oplus\mathfrak{p}$
 is determined by the trait of the maximal abelian subalgebra in $\mathfrak{p}$,
 that is,
 the decomposition $\mathfrak{t}\oplus\mathfrak{p}$
 is a Cartan decomposition of type {\bf AI} if a Cartan subalgebra $\mathfrak{C}$
 is recovered in the subspace $\mathfrak{p}$,
 or a decomposition of type {\bf AII}
 if the maximal abelian subalgebra in $\mathfrak{p}$
 is a proper maximal bi-subalgebra $\mathfrak{B}$ of $\mathfrak{C}$,
 or a type {\bf AIII} if the maximal abelian
 subalgebra in $\mathfrak{p}$ is a complement $\mathfrak{B}^{c}=\mathfrak{C}-\mathfrak{B}$.

 In addition, the computational universality is achieved in a composition of a set of $1$-qubit rotations
 of arbitrary angles and spinor-to-spinor $s$-rotations, {\em cf.} Appendix~B in~\cite{Su}.
 Similarly, a QAP preserving transformation is realized through spinor-to-spinor
 mappings formed in $s$-rotations.
 It is deduced that every quotient algebra partition is acquirable from the partition generated by the {\em intrinsic} bi-subalgebra.

 \section{$r$-th Maximal Bi-Subalgebra}\label{secr-thmaxB}
 \renewcommand{\theequation}{\arabic{section}.\arabic{equation}}
\setcounter{equation}{0} \noindent
 A concept equivalent to a subgroup of $Z^{2p}_2$ is introduced in~\cite{SuTsai1} that
 a set of spinor generators in $su(2^p)$
 forms a bi-subalgebra ${\cal B}$ of $su(2^p)$ iff
 $\forall\hspace{2pt} {\cal S}^{\zeta}_{\alpha},{\cal S}^{\eta}_{\beta}\in {\cal B}$,
 ${\cal S}^{\zeta+\eta}_{\alpha+\beta}\in{\cal B}$.
 Since constructing quotient algebras is the major interest of the serial, 
 the attention is confined to abelian members only and
 for details of non-abelian instances are
 referred to~\cite{SuTsai3}.
 Thereupon, a bi-subalgebra $\mathfrak{B}$ is maximal in a Cartan subalgebra $\mathfrak{C}$
 { if} ${\cal S}^{\zeta+\eta}_{\alpha+\beta}\in\mathfrak{B}$,
 $\forall\hspace{2pt}{\cal S}^{\zeta}_{\alpha},{\cal S}^{\eta}_{\beta}\in\mathfrak{B}^c=\mathfrak{C}-\mathfrak{B}$.
 The Cartan subalgebra $\mathfrak{C}$ is regarded the $0$th maximal bi-subalgebra of itself
 and has many the proper maximal, {\em i.e.}, $1$st maximal bi-subalgebras.
 Unless else specified, the set of maximal bi-subalgebras of $\mathfrak{C}$ will comprise the
 $0$th and $1$st ones.
 Although equally applicable to those nonabelian, an $r$-th maximal bi-subalgebra 
 is recursively defined as follows for $1\leq r\leq p$.
\vspace{6pt}
\begin{defn}\label{defr-thmaxB}
 Denoted as $\mathfrak{B}^{[r]}$, an $r$-th maximal bi-subalgebra
 of a Cartan subalgebra $\mathfrak{C}$ is a proper maximal
 bi-subalgebra of an $(r-1)$-th maximal bi-subalgebra $\mathfrak{B}^{[r-1]}\subset\mathfrak{C}$.
\end{defn}
\vspace{6pt}

 An $r$-th maximal bi-subalgebra 
 is in practice produced from a bisection of an $(r-1)$-th maximal bi-subalgebra 
 by fulfilling the condition of a bi-subalgebra, specifically
 its binary-partitioning and phase strings 
 respectively forming an additive subgroup of $Z^p_2$.
 A Cartan subalgebra $\mathfrak{C}\subset su(2^p)$ thus has from $1$st 
 through $p$-th maximal bi-subalgebras, where the final one contains solely the identity operator,
 {\em i.e.}, $\mathfrak{B}^{[p]}=\{{\cal S}^{\bf 0}_{\hspace{.01cm}{\bf 0}}\}$.
 Similar to the bisection
 $\mathfrak{B}^{[r-1]}=\mathfrak{B}^{[r]}\cup(\mathfrak{B}^{[r-1]}-\mathfrak{B}^{[r]})$,
 a Cartan subalgebra $\mathfrak{C}$ is recursively partitioned by
 $\mathfrak{B}^{[r]}\subset\mathfrak{C}$
 into a total number $2^r$ of cosets.
\vspace{6pt}
\begin{lemma}\label{lemcosetparinC}
 An $r$-th maximal bi-subalgebra $\mathfrak{B}^{[r]}\equiv\mathfrak{B}^{[r,\mathbf{0}]}$ of a Cartan subalgebra
 $\mathfrak{C}$ can generate a partition of $\mathfrak{C}$
 consisting of $2^r$ cosets $\mathfrak{B}^{[r,i]}$, that is,
 $\mathfrak{C}=\bigcup_{i\in{Z^r_2}}\mathfrak{B}^{[r,i]}$ 
 following the condition
 $\forall\hspace{2pt}{\cal S}^{\zeta}_{\alpha}\in\mathfrak{B}^{[r,i]}, 
 \hspace{2pt}{\cal S}^{\eta}_{\beta}\in\mathfrak{B}^{[r,j]}$, \hspace{0pt}
 $\exists !\hspace{3pt}\mathfrak{B}^{[r,l]}$,  such that
 ${\cal S}^{\zeta+\eta}_{\alpha+\beta}\in\mathfrak{B}^{[r,l]}$
 and $i+j=l$, $\forall\hspace{2pt}i,j,l\in{Z^r_2}$.
\end{lemma}
\vspace{3pt}
\begin{proof}
 The relation of bit-wise addition held among indices of cosets is attributed
 to the isomorphism between $\mathfrak{C}$ and $Z^p_2$, {\em cf.}
 Theorems~1 and~2 in~\cite{SuTsai1}.
 In this partition, the bi-subalgebra $\mathfrak{B}^{[r]}$ is the equivalence of
 the group identity, in other words, whose index is
 the all-zero string $\mathbf{0}\in Z^r_2$.
 For the other labellings, let an arbitrary generating set 
 $\{i_1,i_2,i_4\cdots,i_{2^j},\cdots,i_{2^{r-1}}\}$  of $Z^r_2$ be chosen
 that is composed of a number $r$ of independent $r$-digit strings.
 Else from $\mathfrak{B}^{[r,\mathbf{0}]}$, any two
 cosets can be labelled with the ``$1$st" two strings $i_1$ and $i_2$,
 which decide the ``$3$rd" coset labelled by $i_3=i_1+i_2$ following the recipe
 due to the isomorphism of $Z^p_2$ and $\mathfrak{C}$, {\em i.e.},
 ${\cal S}^{\zeta+\eta}_{\alpha+\beta}\in\mathfrak{B}^{[r,i_3]}$ for every pair
 ${\cal S}^{\zeta}_{\alpha}\in\mathfrak{B}^{[r,i_1]}$ and
 ${\cal S}^{\eta}_{\beta}\in\mathfrak{B}^{[r,i_2]}$.
 Next, any coset other from these labelled ones can be taken as the ``$4$th" member and
 be assigned the labelling string $i_4$. Then the labellings of
 the ``$5$th", ``$6$th" and ``$7$th" cosets are similarly determined. Continue this procedure until all
 cosets are labelled.
 Accordingly, the cosets bear the relation of bit-wise addition in their associated indices.
\end{proof}
\vspace{6pt}
 Labelling cosets by distinct binary strings and relating these indices by
 the bit-wise addition is generally seen in partitions of additive groups.
 As an example shown in Fig.~\ref{figsu8QArank2inC}, the $2$nd maximal bi-subalgebra
 $\mathfrak{B}^{[2,\hspace{2pt}00]}=\{{\cal S}^{000}_{000},\hspace{2pt}{\cal S}^{010}_{010}\}$
 of $\mathfrak{C}_{[\mathbf{0}]}$ in $su(8)$ generates a partition consisting of
 four cosets: besides $\mathfrak{B}^{[2,\hspace{2pt}00]}$ itself,
 $\mathfrak{B}^{[2,\hspace{2pt}01]}=\{{\cal S}^{001}_{000},\hspace{2pt}{\cal S}^{011}_{010}\}$,
 $\mathfrak{B}^{[2,\hspace{1pt}10]}=\{{\cal S}^{100}_{100},\hspace{2pt}{\cal S}^{110}_{110}\}$,
 and
 $\mathfrak{B}^{[2,\hspace{1pt}11]}=\{{\cal S}^{101}_{100},\hspace{2pt}{\cal S}^{111}_{110}\}$.
 The purpose of partitioning a Cartan subalgebra $\mathfrak{C}$ is to generate the corresponding
 partition of the quotient algebra given by $\mathfrak{C}$. The discussion in
 Section~\ref{secParQA} will be centered at this partition.

 Resulting from the isomorphism between a Cartan subalgebra $\mathfrak{C}\subset su(2^p)$
 and $Z^p_2$, {\em cf.} Theorems~$1$ and $2$ in~\cite{SuTsai1},
 some assertions regarding $r$-th maximal bi-subalgebras are noted.
\vspace{6pt}
\begin{lemma}\label{lemnum1}
 A bi-subalgebra $\mathfrak{B}$ of a Cartan subalgebra $\mathfrak{C}\subset su(2^p)$
 is an $r$-th maximal bi-subalgebra in $\mathfrak{C}$ if and only if
 it consists of a number $2^{p-r}$ of spinor generators.
\end{lemma}
\vspace{6pt}
\begin{proof}
 Since any two Cartan subalgebras of $su(2^p)$ are isomorphic via a
 conjugate transformation, {\em cf.} Theorem~1 in~\cite{SuTsai1} once again,
  it is convenient and without loss of
 generality to consider $\mathfrak{C}_{[\mathbf{0}]}$ only.
 The intrinsic Cartan subalgebra $\mathfrak{C}_{[\mathbf{0}]}
 =\{{\cal S}^{\nu}_{\mathbf{0}}:\forall\hspace{2pt}\nu\in Z^p_2\}$ exclusively consists
 of spinor generators with the all-zero string $\mathbf{0}$ of the binary partitioning.
 Then, it is easily recognized that the set of phase strings appearing in an $r$-th maximal bi-subalgebra
 of $\mathfrak{C}_{[\mathbf{0}]}$ is spanned by $p-r$ independent generators and
 forms an $r$-th maximal subgroup of $Z^p_2$, which affirms the lemma. 
\end{proof}
 Likewise, numbers of bi-subalgebras in a given Cartan subalgebra can be
 obtained by counting subgroups of $Z^p_2$.
\vspace{6pt}
\begin{lemma}\label{lemnum2}
 A Cartan subalgebra $\mathfrak{C}\subset su(2^p)$ has a total number
 $\prod^{p-r}_{i=1}\frac{2^{p-i+1}-1}{2^i-1}$
 of $r$-th maximal bi-subalgebras $\mathfrak{B}^{[r]}$.
\end{lemma}
\vspace{6pt}
\vspace{6pt}
\begin{lemma}\label{lemnum3}
 For an $r$-th maximal bi-subalgebra $\mathfrak{B}^{[r]}$ of a
 Cartan subalgebra $\mathfrak{C}\subset su(2^p)$, there exist a total number
 $\prod^{r-k}_{i=1}\frac{2^{r-i+1}-1}{2^i-1}$ of
 $k$-th maximal bi-subalgebras $\mathfrak{B}^{[k]}$ of $\mathfrak{C}$ 
 being a superset of
 $\mathfrak{B}^{[r]}$, i.e., $\mathfrak{B}^{[r]}\subset\mathfrak{B}^{[k]}$,
 $1\leq k<r\leq p$.
\end{lemma}
\vspace{6pt} \vspace{6pt}
\begin{lemma}\label{lemnumCartanBr}
 For an $r$-th maximal bi-subalgebra $\mathfrak{B}^{[r]}$
 of a Cartan subalgebra $\mathfrak{C}\subset{su(2^p)}$, $0\leq r\leq p$,
 there exist a total number $\prod^{r}_{s=1}(2^s+1)$ of Cartan
 subalgebras being a superset of
 $\mathfrak{B}^{[r]}$.
\end{lemma}
\vspace{2pt}
\begin{proof}
 Due to the fact that the Cartan subalgebra $\mathfrak{C}$ is a $p$-th maximal
 bi-subalgebra of the bi-subalgebra $su(2^p)$,
 the $r$-th maximal bi-subalgebra $\mathfrak{B}^{[r]}$ of $\mathfrak{C}$
 is a $(p+r)$-th maximal bi-subalgebra of $su(2^p)$ and is
 denoted as $\mathcal{B}^{[p+r]}_{su}$ here~\cite{SuTsai1}.
 According to Lemma~\ref{lemnum1}, the subalgebra
 $\mathcal{B}^{[p+r]}_{su}=\mathfrak{B}^{[r]}$ consists of $2^{p-r}$
 spinor generators and is spanned by $p-r$ independent ones
 under bi-addition.
 This implies that there have a number $2^{p+r}$ of spinors
 commuting with $\mathcal{B}^{[p+r]}_{su}$ in $su(2^p)$.
 Furthermore, since the bi-additive ${\cal S}^{\zeta+\eta}_{\alpha+\beta}$
 commutes with $\mathcal{B}^{[p+r]}_{su}$ if
 $[{\cal S}^{\zeta}_{\alpha},\mathcal{B}^{[p+r]}_{su}]=
 [{\cal S}^{\eta}_{\beta},\mathcal{B}^{[p+r]}_{su}]=0$,
 these $2^{p+r}$ spinors form a $(p-r)$-th maximal bi-subalgebra $\mathcal{B}^{[p-r]}_{su}$ of $su(2^p)$
 with $\mathcal{B}^{[p-r]}_{su}\supset\mathcal{B}^{[p+r]}_{su}$.
 Being a bi-subalgebra of $\mathcal{B}^{[p-r]}_{su}$, the subalgebra
 $\mathcal{B}^{[p+r]}_{su}$ can generate a partition over
 $\mathcal{B}^{[p-r]}_{su}=\bigcup_{i\in{Z^{2r}_2}}\mathcal{B}^{[p+r,i]}_{su}$
 composed of $2^{2r}$ cosets $\mathcal{B}^{[p+r,i]}_{su}$
 for $\mathcal{B}^{[p+r]}_{su}=\mathcal{B}^{[p+r,\mathbf{0}]}_{su}$.
 According to Lemma~\ref{lemcosetparinC},
 a Cartan subalgebra
 $\mathfrak{C}'=\bigcup_{m\in{\tau}}\mathcal{B}^{[p+r,m]}_{su}$,
 which is a superset of $\mathcal{B}^{[p+r]}_{su}$,
 is a union of $2^r$ commuting cosets $\mathcal{B}^{[p+r,m]}_{su}$ of
 $\mathcal{B}^{[p+r]}_{su}$,
 here $\tau$ being an $r$-th maximal subgroup of $Z^{2r}_2$
 and
 $[\mathcal{B}^{[p+r,m]}_{su},\mathcal{B}^{[p+r,n]}_{su}]=0$ for all $m,n\in\tau$.
 Note that these $2^r$ commuting cosets obey the rule that
 $\mathcal{B}^{[p+r,m+n]}_{su}\subset\mathfrak{C}'$ for all
 $\mathcal{B}^{[p+r,m]}_{su}$ and
 $\mathcal{B}^{[p+r,n]}_{su}\subset\mathfrak{C}'$.
 Then, calculating the number of Cartan subalgebras
 containing $\mathcal{B}^{[p+r]}_{su}=\mathfrak{B}^{[r]}$ is equivalent to counting the
 number of the distinct unions, each of which is constituted by a group of $2^r$
 commuting cosets $\{\mathcal{B}^{[p+r,m]}_{su}:m\in\tau\}$
 from the $2^{2r}$ cosets of $\mathcal{B}^{[p+r]}_{su}$ in $\mathcal{B}^{[p-r]}_{su}$.

 Initially,
 an arbitrary coset $\mathcal{B}^{[p+r,i_1]}_{su}\subset\mathcal{B}^{[p-r]}_{su}-\mathcal{B}^{[p+r]}_{su}$
 is chosen to form an $(p+r-1)$-th maximal bi-subalgebra
 $\mathcal{B}^{[p+r-1]}_{su}=\mathcal{B}^{[p+r]}_{su}\cup\mathcal{B}^{[p+r,i_1]}_{su}$.
 The number of options to pick a such coset is $2^{2r}-1$.
 Next, another coset
 $\mathcal{B}^{[p+r,i_2]}_{su}\subset\mathcal{B}^{[p-r]}_{su}-\mathcal{B}^{[p+r-1]}_{su}$
 commuting with $\mathcal{B}^{[p+r-1]}_{su}$ is chosen.
 Since the subalgebra $\mathcal{B}^{[p+r-1]}_{su}$ commutes with $2^{p+r-1}$ spinor generators
 and these spinors form a $(p-r+1)$-th maximal bi-subalgebra $\mathcal{B}^{[p-r+1]}_{su}$,
 there have $2^{2r-1}$ cosets of $\mathcal{B}^{[p+r-1]}_{su}$ in
 $\mathcal{B}^{[p-r+1]}_{su}$ and the number of choices for the 2nd coset is $2^{2r-1}-1$.
 With a third one $\mathcal{B}^{[p+r,i_3]}_{su}$ for $i_3=i_1+i_2$ being determined,
 the union
 $\mathcal{B}^{[p+r-2]}_{su}=\mathcal{B}^{[p+r-1]}_{su}\cup\mathcal{B}^{[p+r,i_2]}_{su}\cup\mathcal{B}^{[p+r,i_3]}_{su}$
 is a $(p+r-2)$-th maximal bi-subalgebra of $su(2^p)$.
 Moreover, the number of options to pick
 $\mathcal{B}^{[p+r,i_1]}_{su}$ and $\mathcal{B}^{[p+r,i_2]}_{su}$ with no redundancy is
 $\prod^2_{u=0}\frac{(2^{2r-u}-1)}{(2^{u+1}-1)}=\frac{(2^{2r}-1)}{1}\cdot\frac{(2^{2r-1}-1)}{(2^2-1)}$.
 The procedure is exercised up to the $(r-1)$-th steps
 when the $p$-th maximal bi-subalgebra
 $\mathcal{B}^{[p]}_{su}=\bigcup^{r-1}_{u=0}\mathcal{B}^{[p+r,i_u]}_{su}$ of $su(2^p)$
 is constructed, which is a Cartan subalgebra including
 $\mathcal{B}^{[p+r,i_0]}_{su}=\mathcal{B}^{[p+r]}_{su}=\mathfrak{B}^{[r]}$,
 and the number $\prod^{r-1}_{u=0}\frac{(2^{2r-u}-1)}{(2^{u+1}-1)}=\prod^{r}_{s=1}(2^s+1)$
 of $\mathcal{B}^{[p]}_{su}$ is obtained.
 Therefore, this lemma is affirmed.
\end{proof}
\vspace{6pt}
 Notice that the special case $r=0$ in the above lemma
 is the number of all Cartan subalgebras of $su(2^p)$, {\em cf.}
 Corollary~6 in~\cite{SuTsai1}.

 Thus, in a Cartan subalgebra $\mathfrak{C}$, 
 there are in total $2^r-1$ $1$st maximal bi-subalgebras
 being a superset of a given $r$-th maximal bi-subalgebra $\mathfrak{B}^{[r]}$.
 Let $\mathcal{B}(\mathfrak{B}^{[r]})$ denote the set of all such bi-subalgebras
 in addition to $\mathfrak{C}$.
 As a corollary of Theorem~$2$ in~\cite{SuTsai1} asserting the group structure
 of the set $\mathcal{G}(\mathfrak{C})=\mathcal{B}(\mathfrak{B}^{[p]})$
 of all maximal bi-subalgebras of $\mathfrak{C}$,
 the reduced set $\mathcal{B}(\mathfrak{B}^{[r]})$
 inherits the same structure.
\vspace{6pt}
\begin{thm}\label{thmGroup}
 Given an $r$-th maximal bi-subalgebra $\mathfrak{B}^{[r]}$ of a
 Cartan subalgebra $\mathfrak{C}$, the set
 $\mathcal{B}(\mathfrak{B}^{[r]})=\{\mathfrak{B}_i:
 \text{$\mathfrak{B}_i$ is a maximal bi-subalgebra of $\mathfrak{C}$ and }
 \mathfrak{B}^{[r]}\subset\mathfrak{B}_i, 0\leq i<2^r\}$
 forms an abelian group under the $\sqcap$-operation:
 $\forall\hspace{2pt} \mathfrak{B}_i,\mathfrak{B}_j\in\mathcal{B}(\mathfrak{B}^{[r]})$,
 $\mathfrak{B}_i\sqcap\mathfrak{B}_j\equiv(\mathfrak{B}_i\cap\mathfrak{B}_j)\cup(\mathfrak{B}^c_i\cap\mathfrak{B}^c_j)\in\mathcal{B}(\mathfrak{B}^{[r]})$,
 where $\mathfrak{B}^c_i=\mathfrak{C}-\mathfrak{B}_i$, $\mathfrak{B}^c_j=\mathfrak{C}-\mathfrak{B}_j$ and
 $\mathfrak{B}_0=\mathfrak{C}$ is the group identity, 
 $0\leq i,j< 2^r$.
\end{thm}
\vspace{3pt}
\begin{proof}
 The group closure is endorsed by the self-evident inclusion
 $\mathfrak{B}^{[r]}\subset\mathfrak{B}_i\sqcap\mathfrak{B}_j$ 
 for every pair $\mathfrak{B}_i$ and $\mathfrak{B}_j\in\mathcal{B}(\mathfrak{B}^{[r]})$,
 and leads to the fact that
 $\mathcal{B}(\mathfrak{B}^{[r]})$ is a subgroup of
 $\mathcal{G}(\mathfrak{C})$~\cite{SuTsai1}.
\end{proof}
\vspace{6pt}
 Since, according to the theorem above,
 a third one is always obtainable with the $\sqcap$-operation from two arbitrary members in
 $\mathcal{B}(\mathfrak{B}^{[r]})$, the group $\mathcal{B}(\mathfrak{B}^{[r]})$
 is generated by $r$ independent maximal bi-subalgebras
 being a superset of $\mathfrak{B}^{[r]}$.
 Notice that
 $\mathfrak{B}_1\cap\mathfrak{B}_2=
 \mathfrak{B}_1\cap\mathfrak{B}_2\cap(\mathfrak{B}_1\sqcap\mathfrak{B}_2)$
 and thus the bi-subalgebra produced from $\mathfrak{B}_1$ and $\mathfrak{B}_2$
 with the $\sqcap$-operation
 makes no contribution to the intersection.
 The following corollary assures the truth that the bi-subalgebra $\mathfrak{B}^{[r]}$
 is identical to the intersection of $r$ independent 
 members of the group.
 \vspace{6pt}
 \begin{cor}\label{coroV1}
 An $r$-th maximal bi-subalgebra $\mathfrak{B}^{[r]}$ of a Cartan
 subalgebra $\mathfrak{C}$ is the intersection of
 all maximal bi-subalgebras being a superset of $\mathfrak{B}^{[r]}$,
 i.e., 
 $\mathfrak{B}^{[r]}=\bigcap_{\mathfrak{B}_s\in\mathcal{B}(\mathfrak{B}^{[r]})}\mathfrak{B}_s$. 
 \end{cor}
 \vspace{3pt} \noindent
 \begin{proof}
 This corollary is trivally true for $r=1$ and will be proved by induction.
 As $r=2$, suppose that the group $\mathcal{B}(\mathfrak{B}^{[2]}_{intr})$ is generated
 by two maximal bi-subalgebras $\mathfrak{B}^{[1]}=\mathfrak{B}_1$ and $\mathfrak{B}_2$
 of the Cartan subalgebra $\mathfrak{C}$,
 both being a superset of $\mathfrak{B}^{[2]}_{intr}$.
 Consider the bisection of the bi-subalgebra $\mathfrak{B}_1$ rendered by $\mathfrak{B}_2$,
 $\mathfrak{B}_1=(\mathfrak{B}_1\cap\mathfrak{B}_2)\cup(\mathfrak{B}_1\cap\mathfrak{B}^c_2)
 =\mathfrak{B}^{[2]}_{intr}\cup(\mathfrak{B}_1-\mathfrak{B}^{[2]}_{intr})$.
 Whereas both $\mathfrak{B}_1\cap\mathfrak{B}_2$ and $\mathfrak{B}^{[2]}_{intr}$ are a
 maximal bi-subalgebra of $\mathfrak{B}_1$ and
 $\mathfrak{B}^{[2]}_{intr}\subset\mathfrak{B}_1\cap\mathfrak{B}_2$,
 it derives that $\mathfrak{B}^{[2]}_{intr}=\mathfrak{B}_1\cap\mathfrak{B}_2$.
 Now assume that an $r$-th maximal bi-subalgebra $\mathfrak{B}^{[r]}$
 is a maximal bi-subalgebra of an $(r-1)$-th maximal bi-subalgebra $\mathfrak{B}^{[r-1]}$
 and the latter
 equals to the intersection of $r-1$ independent maximal bi-subalgebras
 which generate $\mathcal{B}(\mathfrak{B}^{[r-1]})$, {\em i.e.},
 $\mathfrak{B}^{[r-1]}=\bigcap^{r-1}_{t=1}\mathfrak{B}_t$ and
 $\mathcal{B}(\mathfrak{B}^{[r-1]})=span\{\mathfrak{B}_t;t=1,2,\cdots,r-1\}$.
 Let a maximal bi-subalgebra $\mathfrak{B}_r$ be a superset of $\mathfrak{B}^{[r]}$
 but not included in $\mathcal{B}(\mathfrak{B}^{[r-1]})$. Then the group
 $\mathcal{B}(\mathfrak{B}^{[r]})$ can be generated by the set of the $r$ independent
 bi-subalgebras $\{\mathfrak{B}_t;t=1,2,\cdots,r-1\}\cup\{\mathfrak{B}_r\}$.
 Similarly, $\mathfrak{B}^{[r-1]}$ has the bisection
 $\mathfrak{B}^{[r-1]}=(\mathfrak{B}^{[r-1]}\cap\mathfrak{B}_r)\cup(\mathfrak{B}^{[r-1]}\cap\mathfrak{B}^c_r)
 =\mathfrak{B}^{[r]}\cup(\mathfrak{B}^{[r-1]}-\mathfrak{B}^{[r]})$.
 Since both $\mathfrak{B}^{[r]}$ and $\mathfrak{B}^{[r-1]}\cap\mathfrak{B}_r$ are a
 maximal bi-subalgebra of $\mathfrak{B}^{[r-1]}$ and
 $\mathfrak{B}^{[r]}\subset\mathfrak{B}^{[r-1]}\cap\mathfrak{B}_r$, the relation is reached that
 $\mathfrak{B}^{[r]}=\mathfrak{B}^{[r-1]}\cap\mathfrak{B}_r=
 \bigcap^{r-1}_{t=1}\mathfrak{B}_t\cap\mathfrak{B}_r=
 \bigcap_{\mathfrak{B}_s\in\mathcal{B}(\mathfrak{B}^{[r]})}\mathfrak{B}_s$.
 The induction completes.
 Remark that this lemma can as well be proved by contradiction yet which
 requires a comparable complexity of arguments.
 \end{proof}
 \vspace{6pt}\noindent

\section{Quotient Algebra Partition}\label{secParQA}
\renewcommand{\theequation}{\arabic{section}.\arabic{equation}}
\setcounter{equation}{0} \noindent
 Within a quotient algebra $\{{\cal Q}(\mathfrak{C};2^p-1)\}$ of a
 Cartan subalgebra $\mathfrak{C}\subset su(N)$, $2^{p-1}<N\leq 2^p$,
 a conjugate-pair subspace ${\cal W}_{}\subset su(N)$ is uniquely
 determined by a maximal bi-subalgebra $\mathfrak{B}\in\mathcal{G}(\mathfrak{C})$
 following the commutator rule $[{\cal W}_{},\mathfrak{B}]=0$.
 Let the Cartan subalgebra $\mathfrak{C}$ be
 partitioned into $2^r$ cosets by an $r$-th maximal bi-subalgebra
 $\mathfrak{B}^{[r]}\subset\mathfrak{C}$ as in Lemma~\ref{lemcosetparinC}.
 Corresponding to this partition, each conjugate pair ${\cal W}$
 divides into $2^r$ pairs to meet a specific condition of cosets.
 The following lemma is considered {\em the refined version} of Lemma~7 in~\cite{SuTsai1}.
\vspace{6pt}
\begin{lemma}\label{lemconjupair}
 The conjugate-pair subspace ${\cal W}\subset su(2^p)$ determined by a maximal
 bi-subalgebra $\mathfrak{B}\in\mathcal{G}(\mathfrak{C})$
 of a Cartan subalgebra $\mathfrak{C}\subset su(2^p)$ following
 the commutator rule $[{\cal W},\mathfrak{B}]=0$ can be partitioned into $2^r$
 disjoint subspaces ${\cal W}(\mathfrak{B},\mathfrak{B}^{[r]};i)$, namely 
 ${\cal W}=\bigcup_{i\in{Z^r_2}}{\cal W}(\mathfrak{B},\mathfrak{B}^{[r]};i)$
 obeying the coset rule:
 $\forall\hspace{2pt} {\cal S}^{\zeta}_{\alpha}\in {\cal W}(\mathfrak{B},\mathfrak{B}^{[r]};i)$,
 ${\cal S}^{\eta}_{\beta}\in {\cal W}(\mathfrak{B}\cap\mathfrak{B}^{[r]};j)$,
 $\exists !\ \mathfrak{B}^{[r,l]}$, such that
 ${\cal S}^{\zeta+\eta}_{\alpha+\beta}\in\mathfrak{B}^{[r,l]}$ and
 $i+j=l$, $\forall\hspace{2pt}i,j,l\in{Z^r_2}$,
 where $\mathfrak{B}^{[r,l]}$ is a coset in the partition of
 $\mathfrak{C}=\bigcup_{\ i\in{Z^r_2}}\mathfrak{B}^{[r,i]}$ generated
 by a given $r$-th maximal bi-subalgebra $\mathfrak{B}^{[r]}=\mathfrak{B}^{[r,\mathbf{0}]}$.
\end{lemma}
\vspace{3pt}
\begin{proof}
 This partition of a conjugate pair ${\cal W}$ is an implication of
 Lemma~7 in~\cite{SuTsai1} incorporating the partition prescribed in Lemma~\ref{lemcosetparinC}.
 The rule governing the division of spinor generators in the subspace ${\cal W}$ is an equivalence relation
 and thus leads to a partition of ${\cal W}$.
\end{proof}
\vspace{6pt}

 It will become clear shortly that each subspace
 ${\cal W}(\mathfrak{B},\mathfrak{B}^{[r]};i)$ is a conjugate pair {\em per se}.
 To differ from  ${\cal W}$, a subspace ${\cal W}(\mathfrak{B},\mathfrak{B}^{[r]};i)$ is
 called  {\em a conjugate pair of rank} $r$
 {\em determined by the doublet}
 $(\mathfrak{B},\mathfrak{B}^{[r]})$ or
 {\em a partitioned conjugate pair determined by} 
 the same doublet. 
 The Cartan subalgebra $\mathfrak{C}$, being the {\em degrade} conjugate-pair subspace of itself~\cite{SuTsai1},
 splits into $2^r$ {\em degrade partitioned conjugate pairs} ${\cal W}(\mathfrak{C},\mathfrak{B}^{[r]};i)$
 by replacing the conditioning doublet $(\mathfrak{B},\mathfrak{B}^{[r]})$ with $(\mathfrak{C},\mathfrak{B}^{[r]})$
 in Lemma~\ref{lemconjupair}.
 These pairs are just the cosets of $\mathfrak{B}^{[r]}$ in
 $\mathfrak{C}$, {\em i.e.}, ${\cal W}(\mathfrak{C},\mathfrak{B}^{[r]};i)=\mathfrak{B}^{[r,i]}$.
 As illustrated in~\cite{SuTsai3}, 
 the set of conjugate pairs
 $\{{\cal W}(\mathfrak{B},\mathfrak{B}^{[r]};i):\forall\hspace{2pt}\mathfrak{B}\in\mathcal{G}(\mathfrak{C}),\hspace{2pt}i\in{Z^r_2}\}$
 forms a {\em bi-subalgebra partition of order $p+r$}
 in $su(2^p)$ generated by a $(p+r)$-th maximal bi-subalgebra $\mathfrak{B}^{[r]}=\mathcal{B}^{[p+r]}_{su}$ of $su(2^p)$.
 There is a refined version of Lemma~7 in~\cite{SuTsai1} for the commutation relations
 of the partitioned conjugate pairs.
 \vspace{6pt}
\begin{lemma}\label{lemcommparconsub}
 For two partitioned conjugate pairs
 ${\cal W}(\mathfrak{B}_1,\mathfrak{B}^{[r]};i_1)$ and ${\cal W}(\mathfrak{B}_2,\mathfrak{B}^{[r]};i_2)$
 determined by the doublets $(\mathfrak{B}_1,\mathfrak{B}^{[r]})$
 and $(\mathfrak{B}_2,\mathfrak{B}^{[r]})$ respectively,
 $\mathfrak{B}_1,\mathfrak{B}_2\in\mathcal{G}(\mathfrak{C})$ and $i_1,i_2\in{Z^r_2}$,
 either these two pairs commute or the pair
 ${\cal W}(\mathfrak{B}_1,\mathfrak{B}^{[r]};i_1)=W_{12}\cup\hat{W}_{12}$
 can divide into two subspaces, such that an arbitrary generator of ${\cal W}(\mathfrak{B}_2,\mathfrak{B}^{[r]};i_2)$
 commutes with $W_{12}$ but anti-commutes with
 $\hat{W}_{12}$,
 here
 ${W}_{12}$ and $\hat{W}_{12}$ obeying the condition
 ${\cal S}^{\zeta+\eta}_{\alpha+\beta},{\cal S}^{\zeta'+\eta'}_{\alpha'+\beta'}\in\mathfrak{B}_2\sqcap\mathfrak{B}^{[r]}$
 and
 ${\cal S}^{\zeta'+\eta}_{\alpha'+\beta},{\cal S}^{\zeta+\eta'}_{\alpha+\beta'}
 \in\mathfrak{B}^c_2\sqcap\mathfrak{B}^{[r]}$
 with $\mathfrak{B}^c_2=\mathfrak{C}-\mathfrak{B}_2$
 for all
 ${\cal S}^{\zeta}_{\alpha},{\cal S}^{\eta}_{\beta}\in{W}_{12}$
 and
 ${\cal S}^{\zeta'}_{\alpha'},{\cal S}^{\eta'}_{\beta'}\in\hat{W}_{12}$.
\end{lemma}
\vspace{3pt}
\begin{proof}
 The proof follows the same procedure of Lemma~7 in~\cite{SuTsai1}
 except replacing $\mathfrak{B}_2$ with $\mathfrak{B}_2\sqcap\mathfrak{B}^{[r]}$
 and $\mathfrak{B}^c_2$ with
 $\mathfrak{B}^c_2\sqcap\mathfrak{B}^{[r]}$.
 The occasion that the two pairs are commuting occurs when $\mathfrak{B}_1\cap\mathfrak{B}_2$
 is a superset of $\mathfrak{B}^{[r]}$.
\end{proof}
\vspace{6pt}

 Since $\mathfrak{B}\cap\mathfrak{B}^{[r]}$ is a maximal bi-subalgebra of
 $\mathfrak{B}^{[r]}$, it admits an additional partition, actually a bisection,
 of the latter by the former bi-subalgebra, namely,
 $\mathfrak{B}^{[r]}=(\mathfrak{B}\cap\mathfrak{B}^{[r]})\cup(\mathfrak{B}^c\cap\mathfrak{B}^{[r]})$.
 Reflecting the bisection of $\mathfrak{B}^{[r]}$,
 each partitioned conjugate pair 
 is further bisected into two {\em conditioned subspaces}
 with the imposition of an appropriate coset rule.
\vspace{6pt}
\begin{defn}\label{defrankrcondsub}
 A subset $W\subset su(2^p)$ is a conditioned subspace of
 the doublet $(\mathfrak{B},\mathfrak{B}^{[r]})$,
 $\mathfrak{B}^{[r]}$ being an $r$-th maximal bi-subalgebra of a
 Cartan subalgebra $\mathfrak{C}\subset su(2^p)$ and
 $\mathfrak{B}\in\mathcal{G}(\mathfrak{C})$,
 if satisfying 
 the commutator rule $[W,\mathfrak{B}]=0$ and the coset rule
 ${\cal S}^{\zeta+\eta}_{\alpha+\beta}\in\mathfrak{B}\cap\mathfrak{B}^{[r]}$,
 $\forall\hspace{2pt} {\cal S}^{\zeta}_{\alpha},{\cal S}^{\eta}_{\beta}\in W$.
\end{defn}
\vspace{6pt}
  A conditioned subspace
  so defined is in practice a refined version of that in Definition~3 in~\cite{SuTsai1} and is
  termed {\em a conditioned subspace of rank} $r$.
  Yet, this rank specification is often omitted unless confusion may occur.
\vspace{6pt}
\begin{lemma}\label{lemcondsub}
 For a partitioned conjugate pair
 ${\cal W}(\mathfrak{B},\mathfrak{B}^{[r]};i)\subset su(2^p)$
 determined by the doublet $(\mathfrak{B},\mathfrak{B}^{[r]})$,
 $i\in Z^r_2$ and $\mathfrak{B}\in\mathcal{G}(\mathfrak{C})$,
 either ${\cal W}(\mathfrak{B},\mathfrak{B}^{[r]};i)$ itself is a
 conditioned subspace of $(\mathfrak{B},\mathfrak{B}^{[r]})$
 when $\mathfrak{B}\supset\mathfrak{B}^{[r]}$,
 or on the other occasion when $\mathfrak{B}\nsupseteq\mathfrak{B}^{[r]}$,
 there exist two non-null conditioned subspaces ${W}(\mathfrak{B},\mathfrak{B}^{[r]};i)$
 and $\hat{W}(\mathfrak{B},\mathfrak{B}^{[r]};i)
 ={\cal W}(\mathfrak{B},\mathfrak{B}^{[r]};i)-{W}(\mathfrak{B},\mathfrak{B}^{[r]};i)$,
 such that $\forall\hspace{2pt}{\cal S}^{\zeta}_{\alpha}\in {W}(\mathfrak{B},\mathfrak{B}^{[r]};i),
 {\cal S}^{\eta}_{\beta}\in \hat{W}(\mathfrak{B},\mathfrak{B}^{[r]};i)$,
 ${\cal S}^{\zeta+\eta}_{\alpha+\beta}\in\mathfrak{B}^c\cap\mathfrak{B}^{[r]}$.
\end{lemma}
\vspace{3pt}
\begin{proof}
 By Lemma~\ref{lemconjupair}, any {\em bi-additive} generator
 ${\cal S}^{\zeta+\zeta'}_{\alpha+\alpha'}$
 of ${\cal S}^{\zeta}_{\alpha}$ and
 ${\cal S}^{\zeta'}_{\alpha'}\in{\cal W}(\mathfrak{B},\mathfrak{B}^{[r]};i)$
 must be in either $\mathfrak{B}\cap\mathfrak{B}^{[r]}$ or
 $\mathfrak{B}^c\cap\mathfrak{B}^{[r]}$,
 $\mathfrak{B}^c=\mathfrak{C}-\mathfrak{B}$.
 When $\mathfrak{B}\supset\mathfrak{B}^{[r]}$,
 the intersection $\mathfrak{B}^c\cap\mathfrak{B}^{[r]}$ is null
 and ${\cal S}^{\zeta+\zeta'}_{\alpha+\alpha'}\in\mathfrak{B}\cap\mathfrak{B}^{[r]}$
 for all ${\cal S}^{\zeta}_{\alpha},{\cal S}^{\zeta}_{\alpha'}\in{\cal W}(\mathfrak{B},\mathfrak{B}^{[r]};i)$.
 This implies that ${\cal W}(\mathfrak{B},\mathfrak{B}^{[r]};i)$ itself is a
 conditioned subspace of $(\mathfrak{B},\mathfrak{B}^{[r]})$.
 For the other occasion when
 $\mathfrak{B}\nsupseteq\mathfrak{B}^{[r]}$,
 the proof goes exactly the same as that for Lemma~8 in~\cite{SuTsai1} except
 having $\mathfrak{B}\cap\mathfrak{B}^{[r]}$ substitute for
 the conditioning bi-subalgebra $\mathfrak{B}$ therein.
\end{proof}
\vspace{6pt}
 This lemma approves the bisection expression
 ${\cal W}(\mathfrak{B},\mathfrak{B}^{[r]};i)=
 \{{W}(\mathfrak{B},\mathfrak{B}^{[r]};i),
 \hat{W}(\mathfrak{B},\mathfrak{B}^{[r]};i)\}$
 for a partitioned conjugate pair,
 albeit some conditioned subspace may be null. 
 Not only are {\em regular} ones
 as $\mathfrak{B}\nsupseteq\mathfrak{B}^{[r]}$ are formulated in the assertion,
  but also {\em degrade} conditioned subspaces as $\mathfrak{B}\supset\mathfrak{B}^{[r]}$,
  specifically the cosets $\mathfrak{B}^{[r,l]}$ and the null space.
 In a latter pair, one of the degrade conditioned subspaces is null
    and the other one is twice the size of a regular conditioned subspace.

 It can be alternatively understood that,
 given a conjugate-pair subspace of rank zero
 ${\cal W}$ determined by a maximal bi-subalgebra $\mathfrak{B}$,
 these conditioned subspaces
 are obtained by first bisecting ${\cal W}$ into
 the pair $\{W(\mathfrak{B}),\hat{W}(\mathfrak{B})\}$
 with the imposition of
 {\em the coset rule of bisection} 
 ${\cal S}^{\zeta+\eta}_{\alpha+\beta}\in\mathfrak{B}$ for all
 ${\cal S}^{\zeta}_{\alpha}$ and ${\cal S}^{\eta}_{\beta}\in W(\mathfrak{B})$
 (or $\in\hat{W}(\mathfrak{B})$), and then 
 partitioning each of ${W}(\mathfrak{B})$ and $\hat{W}(\mathfrak{B})$ into
 $2^r$ conditioned subspaces of rank $r$ by {\em the coset rule of partition}
 prescribed in Lemma~\ref{lemconjupair}.
 Since these two coset rules are independent, an identical set of
 conditioned subspaces is generated regardless of the application order.
 However, the exposition here will stick to the order of first employing the coset rule
 of partition and then that of bisection.
\vspace{6pt}
\begin{lemma}\label{lemWpairconnetion}
 A conjugate-pair subspace ${\cal W}_{}$ 
 determined by a maximal bi-subalgebra $\mathfrak{B}$ 
 of a Cartan subalgebra $\mathfrak{C}$ 
 can divide  into $2^{r}$ partitioned
 conjugate pairs $\{{W}(\mathfrak{B},\mathfrak{B}^{[r]};i), 
 \hat{W}(\mathfrak{B},\mathfrak{B}^{[r]};i)\}$ 
 possessing the property:
 $\forall\hspace{2pt}{\cal S}^{\zeta}_{\alpha}\in{W}(\mathfrak{B},\mathfrak{B}^{[r]};i)$,
 ${\cal S}^{\hat{\zeta}}_{\hat{\alpha}}\in\hat{W}(\mathfrak{B},\mathfrak{B}^{[r]};i)$,
 ${\cal S}^{\eta}_{\beta}\in{W}(\mathfrak{B},\mathfrak{B}^{[r]};j)$ and
 ${\cal S}^{\hat{\eta}}_{\hat{\beta}}\in\hat{W}(\mathfrak{B},\mathfrak{B}^{[r]};j)$,
 $\exists !\hspace{2pt} \mathfrak{B}^{[r,l]}$, such that
 if one of the following four inclusions is true,
 ${\cal S}^{\zeta+\eta}_{\alpha+\beta},
 {\cal S}^{\hat{\zeta}+\hat{\eta}}_{\hat{\alpha}+\hat{\beta}}\in\mathfrak{B}\cap\mathfrak{B}^{[r,l]}$
 and
 ${\cal S}^{\hat{\zeta}+\eta}_{\hat{\alpha}+\beta},
 {\cal S}^{\zeta+\hat{\eta}}_{\alpha+\hat{\beta}}\in\mathfrak{B}^c\cap\mathfrak{B}^{[r,l]}$,
 so are the rest three,
 where $i+j=l$, $\forall\hspace{2pt}i,j,l\in{Z^r_2}$, $\mathfrak{B}\neq\mathfrak{C}$ and
 $\mathfrak{B}^{[r,l]}$ is a coset in the partition of
 $\mathfrak{C}$ 
 generated by a given $r$-th maximal bi-subalgebra
 $\mathfrak{B}^{[r]}=\mathfrak{B}^{[r,\mathbf{0}]}$.
\end{lemma}
\vspace{3pt}
\begin{proof}
 Since the rest three instances based on the other
 assumed inclusions can be similarly asserted,
 this proof is concerned with only the assumption that
 ${\cal S}^{\zeta+\eta}_{\alpha+\beta}\in\mathfrak{B}\cap\mathfrak{B}^{[r,l]}$.
 Owing to Lemma~\ref{lemcosetparinC}, it easily derives that
 the two generators
 ${\cal S}^{\hat{\zeta}+\eta}_{\hat{\alpha}+\beta}
 ={\cal S}^{\zeta+\eta+\zeta+\hat{\zeta}}_{\alpha+\beta+\alpha+\hat{\alpha}}$
 and
 ${\cal S}^{\zeta+\hat{\eta}}_{\alpha+\hat{\beta}}
 ={\cal S}^{\zeta+\eta+\eta+\hat{\eta}}_{\alpha+\beta+\beta+\hat{\beta}}$
 belong to the subspace $\mathfrak{B}^c\cap\mathfrak{B}^{[r,l]}$, because
 both ${\cal S}^{\zeta+\hat{\zeta}}_{\alpha+\hat{\alpha}}$ and
 ${\cal S}^{\eta+\hat{\eta}}_{\beta+\hat{\beta}}\in\mathfrak{B}^c\cap\mathfrak{B}^{[r]}$
 by Lemma~\ref{lemcondsub} and
 ${\cal S}^{\zeta+\eta}_{\alpha+\beta}\in\mathfrak{B}\cap\mathfrak{B}^{[r,l]}$ by assumption.
 While since
 ${\cal S}^{\zeta+\hat{\zeta}+\eta+\hat{\eta}}_{\alpha+\hat{\alpha}+\beta+\hat{\beta}}\in\mathfrak{B}\cap\mathfrak{B}^{[r]}$
 once more by Lemma~\ref{lemcosetparinC},
 the generator ${\cal S}^{\hat{\zeta}+\hat{\eta}}_{\hat{\alpha}+\hat{\beta}}=
 {\cal S}^{\zeta+\eta+\zeta+\hat{\zeta}+\eta+\hat{\eta}}_{\alpha+\beta+\alpha+\hat{\alpha}+\beta+\hat{\beta}}$
 belongs to $\mathfrak{B}\cap\mathfrak{B}^{[r,l]}$.
 Notice that the above argument is equally applicable regardless of whether
 $\mathfrak{B}^{[r]}\subset\mathfrak{B}$.
\end{proof}
\vspace{6pt}
 According to Lemma~\ref{lemconjupair}, the generator ${\cal S}^{\zeta+\eta}_{\alpha+\beta}$
 in this proof is included in either $\mathfrak{B}\cap\mathfrak{B}^{[r,l]}$
 or $\mathfrak{B}^c\cap\mathfrak{B}^{[r,l]}$.
 Hence in addition to the set above, there exists one more legitimate option of the inclusion set:
 ${\cal S}^{\zeta+\eta}_{\alpha+\beta},{\cal S}^{\hat{\zeta}+\hat{\eta}}_{\hat{\alpha}+\hat{\beta}}\in\mathfrak{B}^c\cap\mathfrak{B}^{[r,l]}$
 and
 ${\cal S}^{\hat{\zeta}+\eta}_{\hat{\alpha}+\beta},{\cal S}^{\zeta+\hat{\eta}}_{\alpha+\hat{\beta}}\in\mathfrak{B}\cap\mathfrak{B}^{[r,l]}$
 for generators ${\cal S}^{\zeta}_{\alpha},{\cal S}^{\hat{\zeta}}_{\hat{\alpha}},
 {\cal S}^{\eta}_{\beta}\text{ and }{\cal S}^{\hat{\eta}}_{\hat{\beta}}$
 respectively belonging to the same four conditioned subspaces.
 Only will the former set in Lemma~\ref{lemWpairconnetion}
 leads to the expected expressions of the conjugate partition and the condition of closure
 originated in~\cite{Su} and thus be adopted here.
 Also to maintain the coset relation in Lemma~\ref{lemcosetparinC},
 the set of $2^r$ degrade conjugate pairs ${\cal W}(\mathfrak{C},\mathfrak{B}^{[r]};i)$
 takes the format either
 $\{{W}(\mathfrak{C},\mathfrak{B}^{[r]};i)=\mathfrak{B}^{[r,i]},\hat{W}(\mathfrak{C},\mathfrak{B}^{[r]};i)=\{0\}\}$
 for all $i\in Z^r_2$ or all ${W}(\mathfrak{C},\mathfrak{B}^{[r]};i)=\{0\}$.

 Despite its simple reason, the consistence of this choice
 is endorsed by a wordy lemma.
\vspace{6pt}
\begin{lemma}\label{lemthreeW}
 Given three arbitrary pairs of generators
 ${\cal S}^{\zeta}_{\alpha}\in{W}(\mathfrak{B},\mathfrak{B}^{[r]};i_1)$,
 ${\cal S}^{\hat{\zeta}}_{\hat{\alpha}}\in\hat{W}(\mathfrak{B},\mathfrak{B}^{[r]};i_1)$,
 ${\cal S}^{\eta}_{\beta}\in{W}(\mathfrak{B},\mathfrak{B}^{[r]};i_2)$,
 ${\cal S}^{\hat{\eta}}_{\hat{\beta}}\in\hat{W}(\mathfrak{B},\mathfrak{B}^{[r]};i_2)$,
 ${\cal S}^{\xi}_{\gamma}\in{W}(\mathfrak{B},\mathfrak{B}^{[r]};i_3)$
 and
 ${\cal S}^{\hat{\xi}}_{\hat{\gamma}}\in\hat{W}(\mathfrak{B},\mathfrak{B}^{[r]};i_3)$
 respectively belonging to
 three partitioned conjugate pairs
 determined by the doublet $(\mathfrak{B},\mathfrak{B}^{[r]})$,
 if the following eight inclusions hold,
 ${\cal S}^{\zeta+\eta}_{\alpha+\beta},{\cal S}^{\hat{\zeta}+\hat{\eta}}_{\hat{\alpha}+\hat{\beta}}\in\mathfrak{B}\cap\mathfrak{B}^{[r,l]}$,
 ${\cal S}^{\hat{\zeta}+\eta}_{\hat{\alpha}+\beta},{\cal S}^{\zeta+\hat{\eta}}_{\alpha+\hat{\beta}}\in\mathfrak{B}^c\cap\mathfrak{B}^{[r,l]}$,
 ${\cal S}^{\eta+\xi}_{\beta+\gamma},{\cal S}^{\hat{\eta}+\hat{\xi}}_{\hat{\beta}+\hat{\gamma}}\in\mathfrak{B}\cap\mathfrak{B}^{[r,m]}$
 and
 ${\cal S}^{\hat{\eta}+\xi}_{\hat{\beta}+\gamma},{\cal S}^{\eta+\hat{\xi}}_{\beta+\hat{\gamma}}\in\mathfrak{B}^c\cap\mathfrak{B}^{[r,m]}$,
 so do the two pairs
 ${\cal S}^{\zeta+\xi}_{\alpha+\gamma},{\cal S}^{\hat{\zeta}+\hat{\xi}}_{\hat{\alpha}+\hat{\gamma}}\in\mathfrak{B}\cap\mathfrak{B}^{[r,n]}$
 and
 ${\cal S}^{\hat{\zeta}+\xi}_{\hat{\alpha}+\gamma},{\cal S}^{\zeta+\hat{\xi}}_{\alpha+\hat{\gamma}}\in\mathfrak{B}^c\cap\mathfrak{B}^{[r,n]}$,
 here $l=i_1+i_2$, $m=i_2+i_3$ and $n=i_1+i_3$, $\forall\hspace{2pt}i_1,i_2,i_3,l,m,n\in Z^r_2$,
 $\mathfrak{B}\neq\mathfrak{C}$,
 and $\mathfrak{B}^{[r,l]}$, $\mathfrak{B}^{[r,m]}$ and $\mathfrak{B}^{[r,n]}$
 are three cosets in the partition of $\mathfrak{C}$ generated by
 $\mathfrak{B}^{[r]}=\mathfrak{B}^{[r,\mathbf{0}]}$.
\end{lemma}
\vspace{3pt}
\begin{proof}
 Simply by Lemma~\ref{lemcosetparinC} again, the generators
 ${\cal S}^{\zeta+\xi}_{\alpha+\gamma}={\cal S}^{\zeta+\eta+\eta+\xi}_{\alpha+\beta+\beta+\gamma}$
 and
 ${\cal S}^{\hat{\zeta}+\hat{\xi}}_{\hat{\alpha}+\hat{\gamma}}
 ={\cal S}^{\hat{\zeta}+\hat{\eta}+\hat{\eta}+\hat{\xi}}_{\hat{\alpha}+\hat{\beta}+\hat{\beta}+\hat{\gamma}}$
 must belong to 
 $\mathfrak{B}\cap\mathfrak{B}^{[r,l+m]}=\mathfrak{B}\cap\mathfrak{B}^{[r,n]}$,
 while
 ${\cal S}^{\hat{\zeta}+\xi}_{\hat{\alpha}+\gamma}={\cal S}^{\hat{\zeta}+\hat{\eta}+\hat{\eta}+\xi}_{\hat{\alpha}+\hat{\beta}+\hat{\beta}+\gamma}$
 and
 ${\cal S}^{\zeta+\hat{\xi}}_{\alpha+\hat{\gamma}}={\cal S}^{\zeta+\eta+\eta+\hat{\xi}}_{\alpha+\beta+\beta+\hat{\gamma}}$
 are included in
 $\mathfrak{B}^c\cap\mathfrak{B}^{[r,l+m]}=\mathfrak{B}^c\cap\mathfrak{B}^{[r,n]}$,
 owing to the inclusions ${\cal S}^{\zeta+\eta}_{\alpha+\beta},{\cal S}^{\hat{\zeta}+\hat{\eta}}_{\hat{\alpha}+\hat{\beta}}\in\mathfrak{B}\cap\mathfrak{B}^{[r,l]}$,
 ${\cal S}^{\eta+\xi}_{\beta+\gamma},{\cal S}^{\hat{\eta}+\hat{\xi}}_{\hat{\beta}+\hat{\gamma}}\in\mathfrak{B}\cap\mathfrak{B}^{[r,m]}$
 and
 ${\cal S}^{\hat{\eta}+\xi}_{\hat{\beta}+\gamma},{\cal S}^{\eta+\hat{\xi}}_{\beta+\hat{\gamma}}\in\mathfrak{B}^c\cap\mathfrak{B}^{[r,m]}$,
 here $l+m=i_1+i_3=n$.
\end{proof}
\vspace{6pt}
 This consistence faithfully demonstrates the orderless
 imposition of the two coset rules aforesaid.
 Conversely speaking, with the choice of the inclusion set in Lemma~\ref{lemWpairconnetion},
 the conditioned subspace $W(\mathfrak{B})$ is recovered when collecting altogether the
 {\em partitioned} conditioned subspaces ${W}(\mathfrak{B},\mathfrak{B}^{[r]};i)$,
 {\em i.e.}, $W(\mathfrak{B})=\bigcup_{i\in Z^r_2}{W}(\mathfrak{B},\mathfrak{B}^{[r]};i)$,
 and
 $\hat{W}(\mathfrak{B})=\bigcup_{i\in Z^r_2}\hat{W}(\mathfrak{B},\mathfrak{B}^{[r]};i)$ likewise.
 Alternatively, the subspaces $W(\mathfrak{B})$ and $\hat{W}(\mathfrak{B})$ respectively
 comprise a half union of ${W}(\mathfrak{B},\mathfrak{B}^{[r]};i)$ and the other
 half of $\hat{W}(\mathfrak{B},\mathfrak{B}^{[r]};i)$
 if the inclusion set of the $2$nd option is assigned.

 Thanks to the former choice, the non-commuting feature of
 conditioned subspaces of a doublet
 is retained in the same manner as that of Lemma~9 in~\cite{SuTsai1}.
\vspace{6pt}
\begin{lemma}\label{lemcondcomm}
 Two non-null conditioned subspaces ${W}(\mathfrak{B},\mathfrak{B}^{[r]};i)$
 and $\hat{W}(\mathfrak{B},\mathfrak{B}^{[r]};j)$ of the doublet
 $(\mathfrak{B},\mathfrak{B}^{[r]})$ anti-commute as $\mathfrak{B}\neq\mathfrak{C}$,
 namely $\{{\cal S}^{\zeta}_{\alpha},{\cal S}^{\eta}_{\beta}\}=0$
 for every pair ${\cal S}^{\zeta}_{\alpha}\in {W}(\mathfrak{B},\mathfrak{B}^{[r]};i)$
 and ${\cal S}^{\eta}_{\beta}\in\hat{W}(\mathfrak{B},\mathfrak{B}^{[r]};j)$,
 $\forall\hspace{2pt}i,j\in Z^r_2$.
\end{lemma}
\vspace{3pt}
\begin{proof}
 The proof is slightly extended from that of Lemma~9 in~\cite{SuTsai1}.
 Assume that there exists a vanishing commutator $[{\cal S}^{\zeta}_{\alpha},{\cal S}^{\eta}_{\beta}]=0$
 for ${\cal S}^{\zeta}_{\alpha}\in {W}(\mathfrak{B},\mathfrak{B}^{[r]};i)$
 and ${\cal S}^{\eta}_{\beta}\in\hat{W}(\mathfrak{B},\mathfrak{B}^{[r]};j)$.
 Their bi-additive generator ${\cal S}^{\zeta+\eta}_{\alpha+\beta}$ commutes
 with both ${\cal S}^{\zeta}_{\alpha}$ and ${\cal S}^{\eta}_{\beta}$
 and thus belongs to $\mathfrak{B}$ by Lemma~3 in~\cite{SuTsai1}.
 This result contradicts the inclusion
 ${\cal S}^{\zeta+\eta}_{\alpha+\beta}\in\mathfrak{B}^c=\mathfrak{C}-\mathfrak{B}$
 adopted in Lemma~\ref{lemWpairconnetion}. 
\end{proof}
\vspace{6pt}


  The bit-wise addition linking labelling indices is enjoyed in
  the refined version of the conjugate partition.
\vspace{6pt}
\begin{lemma}\label{lemconjupar}
 The property of the conjugate partition 
 $[{W}(\mathfrak{B},\mathfrak{B}^{[r]};i),\mathfrak{B}^{[r,l]}]\subset\hat{W}(\mathfrak{B},\mathfrak{B}^{[r]};j)$,
 $[\hat{W}(\mathfrak{B},\mathfrak{B}^{[r]};j),\mathfrak{B}^{[r,l]}]\subset {W}(\mathfrak{B},\mathfrak{B}^{[r]};i)$
 and
 $[{W}(\mathfrak{B},\mathfrak{B}^{[r]};i),\hat{W}(\mathfrak{B},\mathfrak{B}^{[r]};j)]\subset\mathfrak{B}^{[r,l]}$
 is fulfilled for a pair of conditioned subspaces ${W}(\mathfrak{B},\mathfrak{B}^{[r]};i)$
 and $\hat{W}(\mathfrak{B},\mathfrak{B}^{[r]};j)$ of the doublet $(\mathfrak{B},\mathfrak{B}^{[r]})$,
 where $i+j=l$, $\forall\hspace{2pt}i,j,l\in Z^r_2$,
 $\mathfrak{B}\in\mathcal{G}(\mathfrak{C})$
 and $\mathfrak{B}^{[r,l]}$ is a coset in the partition of $\mathfrak{C}$ generated by
 $\mathfrak{B}^{[r]}=\mathfrak{B}^{[r,\mathbf{0}]}$.
\end{lemma}
\vspace{3pt}
\begin{proof}
 The inclusions trivially hold as $\mathfrak{B}=\mathfrak{C}$.
 While $\mathfrak{B}\neq\mathfrak{C}$,
 for any
 ${\cal S}^{\zeta}_{\alpha}\in {W}(\mathfrak{B},\mathfrak{B}^{[r]};i)\subset{\cal W}(\mathfrak{B},\mathfrak{B}^{[r]};i)$
 and ${\cal S}^{\xi}_{\gamma}\in \mathfrak{B}^c\cap\mathfrak{B}^{[r,l]}$,
 the vanishing anti-commutator $\{{\cal S}^{\zeta}_{\alpha},{\cal S}^{\xi}_{\gamma}\}=0$
 produces the generator ${\cal S}^{\zeta+\xi}_{\alpha+\gamma}$ belonging to neither
 ${W}(\mathfrak{B},\mathfrak{B}^{[r]};i)$ nor $\mathfrak{C}$.
 Due to the partition prescribed in Lemma~\ref{lemconjupair},
 for the given
 ${\cal S}^{\zeta}_{\alpha}\in{\cal W}(\mathfrak{B},\mathfrak{B}^{[r]};i)$,
 there exists a unique generator, {\em i.e.}, ${\cal S}^{\zeta+\xi}_{\alpha+\gamma}$ here,
 located in the partitioned conjugate pair
 ${\cal W}(\mathfrak{B},\mathfrak{B}^{[r]};j=i+l)$ such that
 ${\cal S}^{\xi}_{\gamma}={\cal S}^{\zeta+\zeta+\xi}_{\alpha+\alpha+\gamma}
 \in\mathfrak{B}^{[r,l]}$. 
 Accordingly, it is determined that
 ${\cal S}^{\zeta+\xi}_{\alpha+\gamma}\in{\cal W}(\mathfrak{B},\mathfrak{B}^{[r]};j)$
 and $i+j=l$.
 Moreover, since
 ${\cal S}^{\xi}_{\gamma}\in \mathfrak{B}^c\cap\mathfrak{B}^{[r,l]}$
 and ${\cal S}^{\zeta}_{\alpha}\in {W}(\mathfrak{B},\mathfrak{B}^{[r]};i)$,
 based on the inclusion rule adopted in Lemma~\ref{lemWpairconnetion}, it further deduces that
 ${\cal S}^{\zeta+\xi}_{\alpha+\gamma}\in
 {\cal W}(\mathfrak{B},\mathfrak{B}^{[r]};j)-{W}(\mathfrak{B},\mathfrak{B}^{[r]};j)
 =\hat{W}(\mathfrak{B},\mathfrak{B}^{[r]};j)$, in other words,
 $[{W}(\mathfrak{B},\mathfrak{B}^{[r]};i),\mathfrak{B}^{[r,l]}]\subset\hat{W}(\mathfrak{B},\mathfrak{B}^{[r]};j)$
 with the addition relation $i+j=l$;
 similarly,
 $[\hat{W}(\mathfrak{B},\mathfrak{B}^{[r]};j),\mathfrak{B}^{[r,l]}]\subset {W}(\mathfrak{B},\mathfrak{B}^{[r]};i)$.
 The inclusion of the commutator
 $[{W}(\mathfrak{B},\mathfrak{B}^{[r]};i),\hat{W}(\mathfrak{B},\\
 \mathfrak{B}^{[r]};j)]\subset\mathfrak{B}^{[r,l]}$
 is an immediate result of Lemma~\ref{lemWpairconnetion}.
\end{proof}
\vspace{6pt}

  All nice characteristics of conditioned subspaces described in \cite{SuTsai1} are
  as well preserved. 
\vspace{6pt}
\begin{lemma}\label{lemabel}
 Every conditioned subspace of rank $r$ is abelian.
\end{lemma}
\vspace{3pt}
\begin{proof}
 As in Lemma~12 of~\cite{SuTsai1}, the abelianness of a conditioned subspace
 is a straightforward consequence of the coset and the commutator rules.
\end{proof}

\vspace{6pt}
\begin{lemma}\label{lemOclose1}
 The commutator vanishes
 \hspace{2pt} $[[W_1,W_2],\hspace{1pt}\mathfrak{B}_1\sqcap\mathfrak{B}_2]=0$
 for conditioned subspaces $W_1$ and $W_2$ of
 the doublets $(\mathfrak{B}_1,\mathfrak{B}^{[r]})$
 and $(\mathfrak{B}_2,\mathfrak{B}^{[r]})$ respectively,
 $\mathfrak{B}_1$ and $\mathfrak{B}_2\in\mathcal{G}(\mathfrak{C})$.
\end{lemma}
\vspace{3pt}
\begin{proof}
 Since the commutator rule stays unchanged regardless of the choice of the center subalgebra,
 the required proof is the same as that for Lemma~13 of~\cite{SuTsai1}.
\end{proof}

\vspace{6pt}
\begin{lemma}\label{lemOclose2}
 Given ${\cal S}^{\zeta_1}_{\alpha_1},{\cal S}^{\zeta_2}_{\alpha_2}\in W_1$ and
 ${\cal S}^{\eta_1}_{\beta_1},{\cal S}^{\eta_2}_{\beta_2}\in W_2$
 where $W_1$ and $W_2$ are a non-null conditioned subspace of
 the doublets $(\mathfrak{B}_1,\mathfrak{B}^{[r]})$
 and $(\mathfrak{B}_2,\mathfrak{B}^{[r]})$ respectively,
 $\mathfrak{B}_1$ and $\mathfrak{B}_2\in\mathcal{G}(\mathfrak{C})$,
 there keeps the inclusion
 ${\cal S}^{\zeta_1+\zeta_2+\eta_1+\eta_2}_{\alpha_1+\alpha_2+\beta_1+\beta_2}
 \in(\mathfrak{B}_1\sqcap\mathfrak{B}_2)\cap\mathfrak{B}^{[r]}$
 if $[{\cal S}^{\zeta_t}_{\alpha_t},{\cal S}^{\eta_t}_{\beta_t}]\neq 0$, $t=1,2$.
\end{lemma}
\vspace{3pt}
\begin{proof}
  It is obvious that
  ${\cal S}^{\zeta_1+\zeta_2+\eta_1+\eta_2}_{\alpha_1+\alpha_2+\beta_1+\beta_2}\in\mathfrak{B}^{[r]}$
  owing to the inclusions both ${\cal S}^{\zeta_1+\zeta_2}_{\alpha_1+\alpha_2}$
  and ${\cal S}^{\eta_1+\eta_2}_{\beta_1+\beta_2}\in\mathfrak{B}^{[r]}$.
  The rest is to deduce the relation
  ${\cal S}^{\zeta_1+\zeta_2+\eta_1+\eta_2}_{\alpha_1+\alpha_2+\beta_1+\beta_2}\in\mathfrak{B}$,
  which however is already asserted in Lemma~14 of \cite{SuTsai1}.
\end{proof}
\vspace{6pt}

  Lemmas~\ref{lemOclose1} and~\ref{lemOclose2} have validated
  the condition of closure for refined conditioned subspaces.
  It takes further steps to verify the index relation among subspaces. 
\vspace{6pt}
\begin{lemma}\label{lempairclose}
 For two pairs of generators
 ${\cal S}^{\zeta_1}_{\alpha_1},{\cal S}^{\zeta_2}_{\alpha_2}\in{\cal W}(\mathfrak{B}_1,\mathfrak{B}^{[r]};i)$ and
 ${\cal S}^{\eta_1}_{\beta_1},{\cal S}^{\eta_2}_{\beta_2}\in{\cal W}(\mathfrak{B}_2,\mathfrak{B}^{[r]};j)$
 where the partitioned conjugate pairs
 ${\cal W}(\mathfrak{B}_1,\mathfrak{B}^{[r]};i)$ and ${\cal W}(\mathfrak{B}_2,\mathfrak{B}^{[r]};j)$
 are
 respectively determined by the doublets $(\mathfrak{B}_1,\mathfrak{B}^{[r]})$
 and $(\mathfrak{B}_2,\mathfrak{B}^{[r]})$, $\mathfrak{B}_1$ and $\mathfrak{B}_2\in\mathcal{G}(\mathfrak{C})$,
 the relations hold that
 $[{\cal S}^{\zeta_t+\eta_t}_{\alpha_t+\beta_t},\mathfrak{B}_1\sqcap\mathfrak{B}_2]=0$, $t=1,2$,
 and ${\cal S}^{\zeta_1+\zeta_2+\eta_1+\eta_2}_{\alpha_1+\alpha_2+\beta_1+\beta_2}\in\mathfrak{B}^{[r]}$.
\end{lemma}
\vspace{3pt}
\begin{proof}
 In this proof, two occasions have to be concerned that
 one of $\mathfrak{B}_1$ and $\mathfrak{B}_2$
 is a Cartan subalgebra $\mathfrak{C}$ and the other that neither of them is
 a $\mathfrak{C}$.
 In the former case, say $\mathfrak{B}_1=\mathfrak{C}$,
 there has the commutation relation
 $[{\cal S}^{\zeta_t+\eta_t}_{\alpha_t+\beta_t},\mathfrak{B}_1\sqcap\mathfrak{B}_2]=0$, $t=1,2$,
 because of the parity identities from the vanishing
 commutators $[{\cal S}^{\zeta_t}_{\alpha_t},\mathfrak{C}]=[{\cal S}^{\eta_t}_{\beta_t},\mathfrak{B}_2]=0$
 and the fact
 $\mathfrak{B}_1\sqcap\mathfrak{B}_2=\mathfrak{C}\sqcap\mathfrak{B}_2=\mathfrak{B}_2$.
 While $\mathfrak{B}_1\neq\mathfrak{C}$ and
 $\mathfrak{B}_2\neq\mathfrak{C}$,
 owing to the parity identities obtained from the vanishing
 commutators and anti-commutators
 $[{\cal S}^{\zeta_t}_{\alpha_t},\mathfrak{B}_1]=0$, $[{\cal S}^{\eta_t}_{\beta_t},\mathfrak{B}_2]=0$,
 $\{{\cal S}^{\zeta_t}_{\alpha_t},\mathfrak{B}^c_1\}=0$ and $\{{\cal S}^{\eta_t}_{\beta_t},\mathfrak{B}^c_2\}=0$,
 it is easy to confirm that
 $[{\cal S}^{\zeta_t+\eta_t}_{\alpha_t+\beta_t},\mathfrak{B}_1\sqcap\mathfrak{B}_2]=0$, $t=1,2$.
 Moreover, in either case, both the generators ${\cal S}^{\zeta_1+\zeta_2}_{\alpha_1+\alpha_2}$
 and ${\cal S}^{\eta_1+\eta_2}_{\beta_1+\beta_2}$ belong to $\mathfrak{B}^{[r]}$ by Lemma~\ref{lemconjupair},
 which yields the inclusion
 ${\cal S}^{\zeta_1+\zeta_2+\eta_1+\eta_2}_{\alpha_1+\alpha_2+\beta_1+\beta_2}
  \in\mathfrak{B}^{[r]}$.
\end{proof}
\vspace{6pt}
 As an implication of this lemma,
 for any ${\cal S}^{\zeta}_{\alpha}\in{\cal W}(\mathfrak{B}_1,\mathfrak{B}^{[r]};i)$ and
 ${\cal S}^{\eta}_{\beta}\in{\cal W}(\mathfrak{B}_2,\mathfrak{B}^{[r]};j)$,
 their bi-additive generator ${\cal S}^{\zeta+\eta}_{\alpha+\beta}$ must be included in
 a unique partitioned conjugate pair
 ${\cal W}(\mathfrak{B}_1\sqcap\mathfrak{B}_2,\mathfrak{B}^{[r]};l)$
 determined by the doublet $(\mathfrak{B}_1\sqcap\mathfrak{B}_2,\mathfrak{B}^{[r]})$.
 Notice that it returns to Lemma~\ref{lemconjupair}
  as $\mathfrak{B}_1=\mathfrak{B}_2$.
 The following lemma  shows that still the addition relation of great convenience
 $i+j=l$, $\forall\hspace{2pt}i,j,l\in Z^r_2$,
 stays valid among the labelling indices as long as the {\em alignment} for those
 partitioned conjugate pairs indexed by the all-zero string $\mathbf{0}$ is set.
\vspace{6pt}
\begin{lemma}\label{lemclose2}
 Given
 ${\cal S}^{\zeta}_{\alpha}\in{\cal W}(\mathfrak{B}_1,\mathfrak{B}^{[r]};i)$ and
 ${\cal S}^{\eta}_{\beta}\in{\cal W}(\mathfrak{B}_2,\mathfrak{B}^{[r]};j)$ where
 ${\cal W}(\mathfrak{B}_1,\mathfrak{B}^{[r]};i)$ and
 ${\cal W}(\mathfrak{B}_2,\mathfrak{B}^{[r]};j)$ are a partitioned conjugate pair
 respectively determined by the doublets $(\mathfrak{B}_1,\mathfrak{B}^{[r]})$
 and $(\mathfrak{B}_2,\mathfrak{B}^{[r]})$
 with $\mathfrak{B}_1$ and $\mathfrak{B}_2\in\mathcal{G}(\mathfrak{C})$,
 the generator 
 ${\cal S}^{\zeta+\eta}_{\alpha+\beta}$ belongs uniquely to
 ${\cal W}(\mathfrak{B}_1\sqcap\mathfrak{B}_2,\mathfrak{B}^{[r]};l=i+j)$,
 provided the inclusion
 ${\cal S}^{\zeta_0+\eta_0}_{\alpha_0+\beta_0}\in
  {\cal W}(\mathfrak{B}_1\sqcap\mathfrak{B}_2,\mathfrak{B}^{[r]};\mathbf{0})$
 is initiated with any
 ${\cal S}^{\zeta_0}_{\alpha_0}\in{\cal W}(\mathfrak{B}_1,\mathfrak{B}^{[r]};\mathbf{0})$
 and
 ${\cal S}^{\eta_0}_{\beta_0}\in{\cal W}(\mathfrak{B}_2,\mathfrak{B}^{[r]};\mathbf{0})$,
 $\forall\hspace{2pt}i,j,l\in Z^r_2$.
\end{lemma}
\vspace{3pt}
\begin{proof}
 Two inclusions
 ${\cal S}^{\zeta_0+\zeta}_{\alpha_0+\alpha}\in\mathfrak{B}^{[r,i]}$ and
 ${\cal S}^{\eta_0+\eta}_{\beta_0+\beta}\in\mathfrak{B}^{[r,j]}$
 are acquired due to Lemma~\ref{lemconjupair},
 which by Lemma~\ref{lemcosetparinC} lead to the result
 ${\cal S}^{\zeta_0+\zeta+\eta_0+\eta}_{\alpha_0+\alpha+\beta_0+\beta}\in\mathfrak{B}^{[r,i+j]}$.
 On the other hand, there exists a unique $l\in Z^r_2$ by Lemma~\ref{lempairclose}
 to endorse the inclusion
 ${\cal S}^{\zeta+\eta}_{\alpha+\beta}\in
 {\cal W}(\mathfrak{B}_1\sqcap\mathfrak{B}_2,\mathfrak{B}^{[r]};l)$.
 Once more by Lemma~\ref{lemconjupair}, it deduces that
 ${\cal S}^{\zeta_0+\eta_0+\zeta+\eta}_{\alpha_0+\beta_0+\alpha+\beta}\in\mathfrak{B}^{[r,l]}$,
 for ${\cal S}^{\zeta_0+\eta_0}_{\alpha_0+\beta_0}\in
 {\cal W}(\mathfrak{B}_1\sqcap\mathfrak{B}_2,\mathfrak{B}^{[r]};\mathbf{0})$
 by assumption.
 Since
 ${\cal S}^{\zeta_0+\zeta+\eta_0+\eta}_{\alpha_0+\alpha+\beta_0+\beta}=
  {\cal S}^{\zeta_0+\eta_0+\zeta+\eta}_{\alpha_0+\beta_0+\alpha+\beta}$,
 the relation $l=i+j$ is asserted.
 \end{proof}
  \vspace{6pt}\noindent
 \noindent
 Concluding lemmas above, the condition of closure relating conditioned subspaces
 by the bitwise addition of labelling indices
 within partitioned conjugate pairs is established.
\vspace{6pt}
\begin{lemma}\label{lemW2}
 For two arbitrary partitioned conjugate pairs
 $\{{W}(\mathfrak{B}_1,\mathfrak{B}^{[r]};i),\hat{W}(\mathfrak{B}_1,\mathfrak{B}^{[r]};i)\}$
 and
 $\{{W}(\mathfrak{B}_2,\mathfrak{B}^{[r]};j),\hat{W}(\mathfrak{B}_2,\mathfrak{B}^{[r]};j)\}$
 determined by
 the doublets $(\mathfrak{B}_1,\mathfrak{B}^{[r]})$
 and $(\mathfrak{B}_2,\mathfrak{B}^{[r]})$ respectively,
 if one of the following commutation relations is true,
 $[{W}(\mathfrak{B}_1,\mathfrak{B}^{[r]};i),{W}(\mathfrak{B}_2,\mathfrak{B}^{[r]};j)] \\
 \subset\hat{W}(\mathfrak{B}_1\sqcap\mathfrak{B}_2,\mathfrak{B}^{[r]};l)$,
 $[{W}(\mathfrak{B}_1,\mathfrak{B}^{[r]};i),\hat{W}(\mathfrak{B}_2,\mathfrak{B}^{[r]};j)]
 \subset{W}(\mathfrak{B}_1\sqcap\mathfrak{B}_2,\mathfrak{B}^{[r]};l)$,
 $[\hat{W}(\mathfrak{B}_1,\mathfrak{B}^{[r]};i),\\ {W}(\mathfrak{B}_2,\mathfrak{B}^{[r]};j)]
 \subset{W}(\mathfrak{B}_1\sqcap\mathfrak{B}_2,\mathfrak{B}^{[r]};l)$
 \hspace{0pt}and\hspace{0pt}
 $[\hat{W}(\mathfrak{B}_1,\mathfrak{B}^{[r]};i),\hat{W}(\mathfrak{B}_2,\mathfrak{B}^{[r]};j)]
 \subset\hat{W}(\mathfrak{B}_1\sqcap\mathfrak{B}_2,\mathfrak{B}^{[r]};l)$,
 so are the rest three, where $i+j=l$,
 $\forall\hspace{2pt}i,j,l\in{Z^r_2}$, $\mathfrak{B}_1\neq\mathfrak{C}$ and $\mathfrak{B}_2\neq\mathfrak{C}$
 are a maximal bi-subalgebra of a Cartan subalgebra
 $\mathfrak{C}$,
 and
 $\{{W}(\mathfrak{B}_1\sqcap\mathfrak{B}_2,\mathfrak{B}^{[r]};l), \hat{W}(\mathfrak{B}_1\sqcap\mathfrak{B}_2,\mathfrak{B}^{[r]};l)\}$
 is a partitioned conjugate pair determined by
 $(\mathfrak{B}_1\sqcap\mathfrak{B}_2,\mathfrak{B}^{[r]})$.
\end{lemma}
\vspace{3pt}
\begin{proof}
 The proof proceeds exactly following the same procedure as given in Lemma~15 of \cite{SuTsai1},
 except replacing the maximal bi-subalgebras $\mathfrak{B}_1$, $\mathfrak{B}_2$ and
 $\mathfrak{B}_1\sqcap\mathfrak{B}_2$ therein respectively by the doublets
 $(\mathfrak{B}_1,\mathfrak{B}^{[r]})$, $(\mathfrak{B}_2,\mathfrak{B}^{[r]})$ and
 $(\mathfrak{B}_1\sqcap\mathfrak{B}_2,\mathfrak{B}^{[r]})$.
 Note that these relations are valid regardless of whether conditioned subspaces in consideration are
 non-null.
\end{proof}
\vspace{6pt}

 The condition of closure is phrased in the other set of commutation relations
 $[{W}(\mathfrak{B}_1,\mathfrak{B}^{[r]};i),\\ {W}(\mathfrak{B}_2,\mathfrak{B}^{[r]};j)]
 \subset{W}(\mathfrak{B}_1\sqcap\mathfrak{B}_2,\mathfrak{B}^{[r]};l)$,
 $[{W}(\mathfrak{B}_1,\mathfrak{B}^{[r]};i),\hat{W}(\mathfrak{B}_2,\mathfrak{B}^{[r]};j)]
 \subset\hat{W}(\mathfrak{B}_1\sqcap\mathfrak{B}_2,\mathfrak{B}^{[r]};l)$,
 $[\hat{W}(\mathfrak{B}_1,\mathfrak{B}^{[r]};i),{W}(\mathfrak{B}_2,\mathfrak{B}^{[r]};j)]
 \subset\hat{W}(\mathfrak{B}_1\sqcap\mathfrak{B}_2,\mathfrak{B}^{[r]};l)$
 and
 $[\hat{W}(\mathfrak{B}_1,\mathfrak{B}^{[r]};i),\hat{W}(\mathfrak{B}_2,\mathfrak{B}^{[r]};j)]
 \subset{W}(\mathfrak{B}_1\sqcap\mathfrak{B}_2,\mathfrak{B}^{[r]};l)$,
 if the $2$nd option of the inclusion set of Lemma~\ref{lemWpairconnetion} is adopted.
 However, this expression is not favored and only the original set
 is assumed throughout the exposition.
 Thus, each of the $2^r$ degrade conjugate pairs ${\cal W}(\mathfrak{C},\mathfrak{B}^{[r]};i)$
 chooses the format
 $\{{W}(\mathfrak{C},\mathfrak{B}^{[r]};i)=\mathfrak{B}^{[r,i]}, \hat{W}(\mathfrak{C},\mathfrak{B}^{[r]};i)=\{0\}\}$
 such that
 $[{W}(\mathfrak{B},\mathfrak{B}^{[r]};i),\hat{W}(\mathfrak{B},\mathfrak{B}^{[r]};j)]
 \subset{W}(\mathfrak{C},\mathfrak{B}^{[r]};l)=\mathfrak{B}^{[r,l]}$ and
 $i+j=l$ for all $i,j,l\in{Z^r_2}$.
 Moreover, by introducing an appropriate parity to
 each subspace, the conditioned subspaces of the doublet
 $(\mathfrak{B},\mathfrak{B}^{[r]})$ are redenoted in the forms
 $\hat{W}(\mathfrak{B},\mathfrak{B}^{[r]};i)={W}^0(\mathfrak{B},\mathfrak{B}^{[r]};i)$
 and
 ${W}(\mathfrak{B},\mathfrak{B}^{[r]};i)={W}^1(\mathfrak{B},\mathfrak{B}^{[r]};i)$.
 Accordingly, these closures   
 can be encapsulated 
 in as compact as the following formulation.
\vspace{6pt}
\begin{thm}\label{thmgenWcommrankr}
  With an $r$-th maximal bi-subalgebra $\mathfrak{B}^{[r]}$ of a Cartan subalgebra
  $\mathfrak{C}\subset{su(N)}$ and the abelian group $\mathcal{G}(\mathfrak{C})$
  comprising all maximal bi-subalgebras of $\mathfrak{C}$,
  $2^{p-1}<N\leq 2^p$,
  the Lie algebra $su(N)$ can be partitioned into $2^{p+r+1}$
  conditioned subspaces 
  realising the commutation relation,
  $\forall\hspace{2pt}\epsilon,\sigma\in Z_2$, $i,j\in Z^r_2$ and
  $\mathfrak{B},\mathfrak{B}'\in\mathcal{G}(\mathfrak{C})$,
  \begin{align}\label{eqgenWcommrankr}
  [{W}^{\epsilon}(\mathfrak{B},\mathfrak{B}^{[r]};i), {W}^{\sigma}(\mathfrak{B}',\mathfrak{B}^{[r]};j)]\subset
  {W}^{\epsilon+\sigma}(\mathfrak{B}\sqcap\mathfrak{B}',\mathfrak{B}^{[r]};i+j),
  \end{align}
  where
  ${W}^0(\mathfrak{B},\mathfrak{B}^{[r]};i)=\hat{W}(\mathfrak{B},\mathfrak{B}^{[r]};i)$
  and
  ${W}^1(\mathfrak{B},\mathfrak{B}^{[r]};i)={W}(\mathfrak{B},\mathfrak{B}^{[r]};i)$
  are conditioned subspaces of the doublet $(\mathfrak{B},\mathfrak{B}^{[r]})$ as $\mathfrak{B}\neq\mathfrak{C}$,
  and
  ${W}^0(\mathfrak{B},\mathfrak{B}^{[r]};i)=\{0\}$ and
  ${W}^1(\mathfrak{B},\mathfrak{B}^{[r]};i)=\mathfrak{B}^{[r,i]}$
  are cosets of $\mathfrak{B}^{[r]}$ in $\mathfrak{C}$ as $\mathfrak{B}=\mathfrak{C}$.
 \end{thm}
 \vspace{6pt}

  The commutation relation of Eq.~\ref{eqgenWcommrankr} known
  as {\em the quaternion condition of closure of rank r} is the
  refined version of the condition in Theorem~4 in~\cite{SuTsai1}.
  Denoted as $\{\mathcal{P}_{\mathcal{Q}}(\mathfrak{B}^{[r]})\}$,
  the set of the conditioned subspaces with this relation corresponds to
  {\em the quotient algebra partition of rank $r$ determined by} $\mathfrak{B}^{[r]}$
  over the Lie algebra $su(2^p)$.
  As a rephrasing of the above theorem, of great significance is that
  a such partition is endowed with an abelian-group structure through an addition operation defined as follows.
  This reading of the quotient algebra partition is crucial to deciding algebra decompositions
  discussed in Section~\ref{sectypesC}.
 \vspace{6pt}
 \begin{cor}\label{coroQAPGrankr}
  The quotient algebra partition of rank $r$
  $\{\mathcal{P}_{\mathcal{Q}}(\mathfrak{B}^{[r]})\}=
  \{{W}^{\epsilon}(\mathfrak{B},\mathfrak{B}^{[r]};i):\forall\hspace{2pt}\mathfrak{B}\in\mathcal{G}(\mathfrak{C}),
  \hspace{2pt}\epsilon\in{Z_2}\text{ and }i\in{Z^r_2}\}$ generated by an $r$-th maximal bi-subalgebra
  $\mathfrak{B}^{[r]}$ of a Cartan subalgebra $\mathfrak{C}$ forms an abelian group isomorphic to $Z^{p+r+1}_2$ under the
  tri-addition $\circledcirc$:
  $\forall\hspace{2pt}{W}^{\epsilon}(\mathfrak{B}_1,\mathfrak{B}^{[r]};i),\\
  {W}^{\sigma}(\mathfrak{B}_2,\mathfrak{B}^{[r]};j)
  \in\{\mathcal{P}_{\mathcal{Q}}(\mathfrak{B}^{[r]})\}$,
  ${W}^{\epsilon}(\mathfrak{B}_1,\mathfrak{B}^{[r]};i)\circledcirc{W}^{\sigma}(\mathfrak{B}_2,\mathfrak{B}^{[r]};j)
  ={W}^{\epsilon+\sigma}(\mathfrak{B}_1\sqcap\mathfrak{B}_2,\mathfrak{B}^{[r]};i+j)\in\{\mathcal{P}_{\mathcal{Q}}(\mathfrak{B}^{[r]})\}$,
  where $\mathfrak{B}_1,\mathfrak{B}_2\in\mathcal{G}(\mathfrak{C})$,
  $\epsilon,\sigma\in{Z_2}$, $i,j\in{Z^r_2}$ and
  ${W}^{0}(\mathfrak{C},\mathfrak{B}^{[r]};\mathbf{0})=\{0\}$ is the
  group identity.
 \end{cor}
 \vspace{6pt}\noindent
  Remid Theorem~2 in~\cite{SuTsai1} that, given two maximal bi-subalgebras
  $\mathfrak{B}_{\alpha}$ and $\mathfrak{B}_{\beta}$ of a Cartan subalgebra
  with $\alpha$ and $\beta\in Z^p_2$, the derivation
  of the 3rd member $\mathfrak{B}_{\alpha+\beta}=\mathfrak{B}_{\alpha}\sqcap\mathfrak{B}_{\beta}$
  is an addition too. The above operation are thus regarded as an action
  of {\em tri-addition}.
  Yielded from this operation, the conditioned subspace
  ${W}^{\epsilon+\sigma}(\mathfrak{B}_1\sqcap\mathfrak{B}_2,\mathfrak{B}^{[r]};i+j)$
  may be called the {\em tri-additive} of the two subspaces
  ${W}^{\epsilon}(\mathfrak{B}_1,\mathfrak{B}^{[r]};i)$
  and ${W}^{\sigma}(\mathfrak{B}_2,\mathfrak{B}^{[r]};j)$.

  Complementary to Theorem~\ref{thmgenWcommrankr},
  also the following assertion concerned with commuting spinors
  will help during decomposing algebras. 
 \vspace{6pt}
 \begin{cor}\label{coroanticommclose}
  The bi-additive generator ${\cal S}^{\zeta+\eta}_{\alpha+\beta}$
  of two spinors ${\cal S}^{\zeta}_{\alpha}\in{W}^{\epsilon}(\mathfrak{B},\mathfrak{B}^{[r]};i)$
  and ${\cal S}^{\eta}_{\beta}\in{W}^{\sigma}(\mathfrak{B}',\mathfrak{B}^{[r]};j)$
  belongs to ${W}^{1+\epsilon+\sigma}(\mathfrak{B}\sqcap\mathfrak{B}',\mathfrak{B}^{[r]};i+j)$
  if $[{\cal S}^{\zeta}_{\alpha},{\cal S}^{\eta}_{\beta}]=0$,
  here ${W}^{\epsilon}(\mathfrak{B},\mathfrak{B}^{[r]};i)$
  and ${W}^{\sigma}(\mathfrak{B}',\mathfrak{B}^{[r]};j)$
  being a non-null conditioned subspace of the
  doublets $(\mathfrak{B},\mathfrak{B}^{[r]})$ and $(\mathfrak{B}',\mathfrak{B}^{[r]})$
  respectively, $\epsilon,\sigma\in{Z_2}$, $i,j\in{Z^r_2}$ and
  $\mathfrak{B},\mathfrak{B}'\in\mathcal{G}(\mathfrak{C})$;
  an anti-commutator serves to relate commuting spinors from two
  conditioned subspaces
  \begin{align}\label{eqgenWanticommrankr}
  \{{W}^{\epsilon}(\mathfrak{B},\mathfrak{B}^{[r]};i), {W}^{\sigma}(\mathfrak{B}',\mathfrak{B}^{[r]};j)\}\subset
  {W}^{1+\epsilon+\sigma}(\mathfrak{B}\sqcap\mathfrak{B}',\mathfrak{B}^{[r]};i+j).
  \end{align}
 \end{cor}
 \vspace{2pt}
 \begin{proof}
  By Lemma~\ref{lemclose2}, the generator ${\cal S}^{\zeta+\eta}_{\alpha+\beta}$
  must be included in either ${W}^{\epsilon+\sigma}(\mathfrak{B}\sqcap\mathfrak{B}',\mathfrak{B}^{[r]};i+j)$
  or
  ${W}^{1+\epsilon+\sigma}(\mathfrak{B}\sqcap\mathfrak{B}',\mathfrak{B}^{[r]};i+j)$.
  Suppose that
  ${\cal S}^{\zeta+\eta}_{\alpha+\beta}\in{W}^{\epsilon+\sigma}(\mathfrak{B}\sqcap\mathfrak{B}',\mathfrak{B}^{[r]};i+j)$
  and let $[{\cal S}^{\zeta}_{\alpha},{\cal S}^{\eta'}_{\beta'}]\neq 0$ for some ${\cal S}^{\eta'}_{\beta'}\in{W}^{\sigma}(\mathfrak{B}',\mathfrak{B}^{[r]};j)$.
  Then, the abelianness of a conditioned subspace 
  is violated that
  $[{\cal S}^{\zeta+\eta}_{\alpha+\beta},{\cal S}^{\zeta+\eta'}_{\alpha+\beta'}]\neq 0$
  owing to the fact ${\cal S}^{\zeta+\eta'}_{\alpha+\beta'}\in{W}^{\epsilon+\sigma}(\mathfrak{B}\sqcap\mathfrak{B}',\mathfrak{B}^{[r]};i+j)$
  by Theorem~\ref{thmgenWcommrankr}.
 \end{proof}
 \vspace{6pt}
 Similar to Corollary~\ref{coroQAPGrankr},
  the group structure of a quotient-algebra partition can be viewed
  through by defining the group operation for the anti-commutation
  relation.

\vspace{6pt} \noindent{\bf Corollary 3'}
 {\em  The
 quotient-algebra partition of rank $r$
  $\{\mathcal{P}_{\mathcal{Q}}(\mathfrak{B}^{[r]})\}=
  \{{W}^{\epsilon}(\mathfrak{B},\mathfrak{B}^{[r]};i):\forall\hspace{2pt}\mathfrak{B}\in\mathcal{G}(\mathfrak{C}),
  \hspace{2pt}\epsilon\in{Z_2}\text{ and }i\in{Z^r_2}\}$ generated by an $r$-th maximal bi-subalgebra
  $\mathfrak{B}^{[r]}$ of a Cartan subalgebra $\mathfrak{C}$ forms an abelian group isomorphic to $Z^{p+r+1}_2$ under the
  anti-triaddition $\circledast$:
  $\forall\hspace{2pt}{W}^{\epsilon}(\mathfrak{B}_1,\mathfrak{B}^{[r]};i),
  {W}^{\sigma}(\mathfrak{B}_2,\mathfrak{B}^{[r]};j)
  \in\{\mathcal{P}_{\mathcal{Q}}(\mathfrak{B}^{[r]})\}$,
  ${W}^{\epsilon}(\mathfrak{B}_1,\mathfrak{B}^{[r]};i)\circledast{W}^{\sigma}(\mathfrak{B}_2,\mathfrak{B}^{[r]};j)
  ={W}^{1+\epsilon+\sigma}(\mathfrak{B}_1\sqcap\mathfrak{B}_2,\mathfrak{B}^{[r]};i+j)\in\{\mathcal{P}_{\mathcal{Q}}(\mathfrak{B}^{[r]})\}$,
  where $\mathfrak{B}_1,\mathfrak{B}_2\in\mathcal{G}(\mathfrak{C})$,
  $\epsilon,\sigma\in{Z_2}$, $i,j\in{Z^r_2}$ and
  ${W}^{0}(\mathfrak{C},\mathfrak{B}^{[r]};\mathbf{0})=\{0\}$ is the
  group identity.}
\vspace{6pt}

\noindent That is, the quotient-algebra partition (QAP) endorses
 both the commutation and anti-commutation relations that can be
 represented by the tri-addition and anti-triaddition respectively.

\section{Quotient and Co-Quotient Algebras of Rank $r$}\label{secQAcoQArankr}
\renewcommand{\theequation}{\arabic{section}.\arabic{equation}}
\setcounter{equation}{0} \noindent
 Similar to the case of rank zero~\cite{SuTsai1},
 there breed two types of structures as an immediate result of Eq.~\ref{eqgenWcommrankr} in a quotient-algebra partition of rank $r$:
 {\em a quotient algebra of rank $r$} as an $r$-th
 maximal bi-subalgebra $\mathfrak{B}^{[r]}$ of a Cartan subalgebra
 $\mathfrak{C}$ serving as the center subalgebra, and
 {\em a co-quotient algebra of rank $r$} as a non-null conditioned
 subspace other than $\mathfrak{B}^{[r]}$ playing the role
 instead.
\vspace{6pt}
\begin{thm}\label{thmQAcoQA}
 In the quotient-algebra partition of rank $r$ $\{\mathcal{P}_{\mathcal{Q}}(\mathfrak{B}^{[r]})\}$ generated by
 an $r$-th maximal bi-subalgebra $\mathfrak{B}^{[r]}$ of a Cartan subalgebra
 $\mathfrak{C}\subset{su(N)}$, $2^{p-1}<N\leq 2^p$,
 there determines a quotient algebra of rank $r$ given by $\mathfrak{B}^{[r]}$,
 denoted as $\{{\cal Q}(\mathfrak{B}^{[r]};2^{p+r}-1)\}$ and as illustrated in Fig.~\ref{figQAandCo-QArankr}(a),
 when $\mathfrak{B}^{[r]}$ is taken as the center subalgebra,
 or a co-quotient algebra of rank $r$ given by a non-null conditioned subspace
 ${W}^{\tau}(\mathfrak{B}_1,\mathfrak{B}^{[r]};l)\in\{\mathcal{P}_{\mathcal{Q}}(\mathfrak{B}^{[r]})\}$
 else from $\mathfrak{B}^{[r]}$,
 denoted as $\{{\cal Q}({W}^{\tau}(\mathfrak{B}_1,\mathfrak{B}^{[r]};l);q)\}$
 and as illustrated in Fig.~\ref{figQAandCo-QArankr}(b),
 when ${W}^{\tau}(\mathfrak{B}_1,\mathfrak{B}^{[r]};l)$
 as the center subalgebra,
 here $\mathfrak{B}_1\in\mathcal{G}(\mathfrak{C})$, $\tau\in{Z_2}$,
 $l\in{Z^r_2}$, and
 $q=2^{p+r}-2^{2r-2}$ as $\mathfrak{B}_1\supset\mathfrak{B}^{[r]}$ or
 $q=2^{p+r}-1$ as $\mathfrak{B}_1\nsupseteq\mathfrak{B}^{[r]}$.
\end{thm}
\vspace{6pt}
 \begin{figure}[!ht]
 \begin{center}
\[\begin{array}{cc}
      \hspace{-15pt}
      \begin{array}{c}
             \hspace{-180pt}(a)\\
             \\
             \begin{array}{c}
             \mathfrak{B}^{[r]}=W^1(\mathfrak{C},\mathfrak{B}^{[r]};\mathbf{0})
             \end{array}\\
                        \\
             \begin{array}{cc}
             W^{\tau}(\hspace{0pt}\mathfrak{B}_1,\hspace{0pt}\mathfrak{B}^{[r]};\hspace{0pt}l)
             &
             W^{\hat{\tau}}(\hspace{0pt}\mathfrak{B}_1,\hspace{0pt}\mathfrak{B}^{[r]};\hspace{0pt}l)\\
             &\\
             W^\epsilon\hspace{0pt}(\mathfrak{B}_m,\mathfrak{B}^{[r]};\hspace{0pt}i)
             &
             W^{\hat{\epsilon}}\hspace{0pt}(\mathfrak{B}_m,\mathfrak{B}^{[r]};\hspace{0pt}i)\\
             &\\
             W^\sigma(\mathfrak{B}_n,\mathfrak{B}^{[r]};\hspace{0pt}j)
             &
             W^{\hat{\sigma}}(\mathfrak{B}_n,\mathfrak{B}^{[r]};\hspace{0pt}j)\\
             \end{array}
      \end{array}
&\hspace{5pt}
     \hspace{-15pt}
      \begin{array}{c}
             \hspace{-180pt}(b)\\
             \\
             \begin{array}{c}
              W^{\tau}(\hspace{0pt}\mathfrak{B}_1,\hspace{0pt}\mathfrak{B}^{[r]};\hspace{0pt}l)
             \end{array}\\
                        \\
             \begin{array}{cc}
             \mathfrak{B}^{[r]}=W^1(\mathfrak{C},\mathfrak{B}^{[r]};\mathbf{0})
             &
             W^{\hat{\tau}}(\hspace{0pt}\mathfrak{B}_1,\hspace{0pt}\mathfrak{B}^{[r]};\hspace{0pt}l)=\{0\}\\
             &\\
             W^\epsilon\hspace{0pt}(\mathfrak{B}_m,\mathfrak{B}^{[r]};\hspace{0pt}i)
             &
             W^{\hat{\sigma}}\hspace{0pt}(\mathfrak{B}_n,\mathfrak{B}^{[r]};\hspace{0pt}j)\\
             &\\
             W^\sigma(\mathfrak{B}_n,\mathfrak{B}^{[r]};\hspace{0pt}j)
             &
             W^{\hat{\epsilon}}\hspace{0pt}(\mathfrak{B}_m,\mathfrak{B}^{[r]};\hspace{0pt}i)\\
             \end{array}
      \end{array}
\end{array}\]
 \end{center}
 \fcaption{In (a) a quotient algebra of rank $r$
  $\{{\cal Q}(\mathfrak{B}^{[r]})\}$ given by an $r$-th
  maximal bi-subalgebra $\mathfrak{B}^{[r]}$ of a Cartan
  subalgebra $\mathfrak{C}\subset{su(N)}$, and in (b) a co-quotient
  algebra of rank $r$ $\{{\cal Q}(W^{\tau}(\mathfrak{B}_1,\mathfrak{B}^{[r]};l))\}$
  given by a non-null conditioned subspace
  $W^{\tau}(\mathfrak{B}_1,\mathfrak{B}^{[r]};l)\in\{{\cal Q}(\mathfrak{B}^{[r]})\}$;
  here
  $\mathfrak{B}_1,\mathfrak{B}_m,\mathfrak{B}_n\in\mathcal{G}(\mathfrak{C})$,
  $i,j,l\in{Z^r_2}$, $\tau,\epsilon,\sigma,\hat{\tau},\hat{\epsilon},\hat{\sigma}\in{Z_2}$,
  $\mathfrak{B}_1=\mathfrak{B}_m\sqcap\mathfrak{B}_n$,
  $l=i+j$, $\tau=\epsilon+\hat{\sigma}$, $\hat{\tau}=1+\tau$, $\hat{\epsilon}=1+\epsilon$,
  $\hat{\sigma}=1+\sigma$, and $1< m,n<2^p$.
 \label{figQAandCo-QArankr}}
 \end{figure}

  A such quotient (or co-quotient) algebra can be generated
  employing the algorithm introduced in~\cite{Su} with the assignment
  of the center subalgebra $\mathfrak{B}^{[r]}$
  (or $W^{\tau}(\mathfrak{B}_1,\mathfrak{B}^{[r]};l)$).
  While taking advantage of the commutation relation of
  Eq.~\ref{eqgenWcommrankr},
  an alternative procedure becomes easier.
  Portraying the construction,
  the general forms of the two structures
  are presented in Figs.~\ref{figQAandCo-QArankr}(a) and~\ref{figQAandCo-QArankr}(b),
  where the maximal bi-subalgebras keep the closure
  $\mathfrak{B}_1=\mathfrak{B}_m\sqcap\mathfrak{B}_n$,
  $1<m,n<2^p$,
  the relation holds for the subspace indices
  $l=i+j$, $i,j,l\in{Z^r_2}$, and the parities of conditioned subspaces meet the
  identities $\tau=1+\hat{\tau}=1+\epsilon+\sigma=\epsilon+\hat{\sigma}=\hat{\epsilon}+\sigma$
  for $\epsilon,\sigma,\tau,\hat{\epsilon},\hat{\sigma},\hat{\tau}\in{Z_2}$.
  Once the center subalgebra is decided, the construction is as straightforward
  as letting
  every conditioned subspace in the 
  partition $\{\mathcal{P}_{\mathcal{Q}}(\mathfrak{B}^{[r]})\}$
  find its {\em conjugate partner} to
  form a {\em conjugate pair} simply following Eq.~\ref{eqgenWcommrankr}.
  The quotient algebra of rank $r$ $\{{\cal Q}(\mathfrak{B}^{[r]})\}$
  can alternatively be obtained from that of rank zero $\{{\cal Q}(\mathfrak{C})\}$
  when employing the order of applying the coset rule of partition to a conjugate-pair subspace
  within $\{{\cal Q}(\mathfrak{C})\}$ and then the coset rule of bisection to each refined conjugate-pair subspace,
  referring to the figures in Appendix~\ref{appQAPDisplay}.
  Notice that two kinds of co-quotient algebras are permitted in $\{\mathcal{P}_{\mathcal{Q}}(\mathfrak{B}^{[r]})\}$
  due to the choice of the center subalgebra ${W}^{\tau}(\mathfrak{B}_1,\mathfrak{B}^{[r]};l)$
  being a {\em degrade} conditioned subspace as $\mathfrak{B}_1\supset\mathfrak{B}^{[r]}$
  or a {\em regular} one as
  $\mathfrak{B}_1\nsupseteq\mathfrak{B}^{[r]}$.
  In order to further discern structures of quotient and co-quotient algebras,
  it need examine the {\em types} of their conjugate pairs.

  According to Lemma~\ref{lemcondsub}, there exist three kinds of conditioned subspaces
  in a quotient-algebra partition of rank $r$:
  null degrade, non-null degrade and regular conditioned subspaces;
  remind that a non-null degrade conditioned subspace is twice the size of a regular one.
  This leads to the production of six types of
  conjugate pairs as listed in Eq.~\ref{eqtypesofCPs}.
  The pair ${\cal W}_{deg\text{-}\Omega}$
  is called a {\em degrade} conjugate pair carrying at least one null conditioned subspace
  and
  ${\cal W}_{reg\text{-}\Omega}$
  is a {\em regular} pair consisting of two non-null subspaces,
  here $\Omega=$I, II, III.
   \vspace{6pt}
   \begin{align}\label{eqtypesofCPs}
             &
             {\cal W}_{deg\text{-}I}
               =\{W^{\mu}\hspace{0pt}(\mathfrak{B},\mathfrak{B}^{[r]};\hspace{0pt}s),
                  W^{\nu}\hspace{0pt}(\mathfrak{B}^{\S},\mathfrak{B}^{[r]};\hspace{0pt}t)=\{0\}\}
               \text{ as }\mathfrak{B}^{[r]}\subset\mathfrak{B}\text{ and }\mathfrak{B}^{\S}\hspace{1pt};
             \notag\\
             &
             {\cal W}_{deg\text{-}II}
             =\{W^{\mu}\hspace{0pt}(\mathfrak{B}^{\dag},\mathfrak{B}^{[r]};\hspace{0pt}s)=\{0\},
                W^{\nu}\hspace{0pt}(\mathfrak{B}^{\ddag},\mathfrak{B}^{[r]};\hspace{0pt}t)=\{0\}\}
                \text{ as }\mathfrak{B}^{[r]}\cup\mathfrak{B}^{[r,s+t]}\subset\mathfrak{B}^{\dag}\text{ and }\mathfrak{B}^{\ddag}\hspace{1pt};
             \notag\\
             &
             {\cal W}_{deg\text{-}III}=\{W^{\mu}\hspace{0pt}(\mathfrak{B}^{\natural},\mathfrak{B}^{[r]};\hspace{0pt}s)=\{0\},
             W^{\nu}\hspace{0pt}(\breve{\mathfrak{B}},\mathfrak{B}^{[r]};\hspace{0pt}t)\}
             \text{ as }\mathfrak{B}^{[r]}\subset\mathfrak{B}^{\natural}\text{ and }\mathfrak{B}^{[r]}\nsubseteq\breve{\mathfrak{B}}\hspace{1pt};
             \notag\\
             &
             {\cal W}_{reg\text{-}I}
             =\{W^{\mu}\hspace{0pt}(\hat{\mathfrak{B}},\mathfrak{B}^{[r]};\hspace{0pt}s),
                W^{\nu}\hspace{0pt}(\breve{\mathfrak{B}},\mathfrak{B}^{[r]};\hspace{0pt}t)\}
             \text{ as }\mathfrak{B}^{[r]}\nsubseteq\hat{\mathfrak{B}}\text{ or }\breve{\mathfrak{B}};
             \notag\\
             &
             {\cal W}_{reg\text{-}II}=\{W^{\mu}\hspace{0pt}(\mathfrak{B}^{\top},\mathfrak{B}^{[r]};\hspace{0pt}s),
             W^{\nu}\hspace{0pt}(\mathfrak{B}^{\bot},\mathfrak{B}^{[r]};\hspace{0pt}t)\}
             \text{ as }\mathfrak{B}^{[r]}\cup\mathfrak{B}^{[r,s+t]}\subset\mathfrak{B}^{\top}\text{ and }\mathfrak{B}^{\bot}\hspace{1pt};
             \notag\\
             &
             {\cal W}_{reg\text{-}III}
             =\{W^{\mu}\hspace{0pt}(\mathfrak{B}^{\sharp},\mathfrak{B}^{[r]};\hspace{0pt}s),
             W^{\nu}\hspace{0pt}(\breve{\mathfrak{B}},\mathfrak{B}^{[r]};\hspace{0pt}t)\}
             \text{ as }\mathfrak{B}^{[r]}\subset\mathfrak{B}^{\sharp}\text{ and }\mathfrak{B}^{[r]}\nsubseteq\breve{\mathfrak{B}}.
  \end{align}
  The conjugate pair ${\cal W}_{deg\text{-}I}$ of Eq.~\ref{eqtypesofCPs}
  is type {\em degrade-I} formed of a non-null degrade conditioned subspace
  $W^{\mu}(\mathfrak{B},\mathfrak{B}^{[r]};s)$ and a null one
  $W^{\nu}(\mathfrak{B}^{\S},\mathfrak{B}^{[r]};t)$,
  for the maximal bi-subalgebras
  $\mathfrak{B}\supset\mathfrak{B}^{[r]}$ and $\mathfrak{B}^{\S}\supset\mathfrak{B}^{[r]}$ in
  $\mathcal{G}(\mathfrak{C})$, $\mu,\nu\in{Z_2}$
  and $s,t\in{Z^r_2}$.
  Comprising two null degrade conditioned subspaces
  $W^{\mu}(\mathfrak{B}^{\dag},\mathfrak{B}^{[r]};s)=W^{\nu}(\mathfrak{B}^{\ddag},\mathfrak{B}^{[r]};t)=\{0\}$,
  the pair ${\cal W}_{deg\text{-}II}$ is a {\em degrade-II},
  here the maximal bi-subalgebras
  $\mathfrak{B}^{\dag}$ and $\mathfrak{B}^{\ddag}\in\mathcal{G}(\mathfrak{C})$
  being a superset of both the bi-subalgebra $\mathfrak{B}^{[r]}$ and the coset
  $\mathfrak{B}^{[r,s+t]}$ of $\mathfrak{B}^{[r]}$ in $\mathfrak{C}$.
  A {\em degrade-III} pair ${\cal W}_{deg\text{-}III}$
  is made of a null degrade conditioned subspace
  $W^{\mu}(\mathfrak{B}^{\natural},\mathfrak{B}^{[r]};s)=\{0\}$
  and a regular
  $W^{\nu}(\breve{\mathfrak{B}},\mathfrak{B}^{[r]};t)$,
  because $\mathfrak{B}^{\natural}\supset\mathfrak{B}^{[r]}$ and $\breve{\mathfrak{B}}\nsupseteq\mathfrak{B}^{[r]}$.
  Two regular conditioned subspaces
  $W^{\mu}\hspace{0pt}(\hat{\mathfrak{B}},\mathfrak{B}^{[r]};\hspace{0pt}s)$
  and
  $W^{\nu}\hspace{0pt}(\breve{\mathfrak{B}},\mathfrak{B}^{[r]};\hspace{0pt}t)$
  form a {\em regular-I} pair ${\cal W}_{reg\text{-}I}$, when
  neither $\hat{\mathfrak{B}}$ nor $\breve{\mathfrak{B}}$
  is a superset of $\mathfrak{B}^{[r]}$.
  Constituted of two non-null degrade conditioned subspaces,
  the conjugate pair
  ${\cal W}_{reg\text{-}II}
  =\{W^{\mu}(\mathfrak{B}^\top,\mathfrak{B}^{[r]};s),W^{\nu}(\mathfrak{B}^\bot,\mathfrak{B}^{[r]};t)\}$
  is a {\em regular-II} type.
  Lastly ${\cal W}_{reg\text{-}III}$, type {\em regular-III}, 
  is composed of a non-null degrade conditioned subspace
  $W^{\mu}(\mathfrak{B}^{\sharp},\mathfrak{B}^{[r]};s)$
  and a regular $W^{\nu}(\breve{\mathfrak{B}},\mathfrak{B}^{[r]};t)$.
  Quotient and co-quotient algebras exhibit different ensembles
  in types of conjugate pairs.

  Within the quotient algebra of rank $r$
  $\{{\cal Q}(\mathfrak{B}^{[r]})\}$ in
  Fig.~\ref{figQAandCo-QArankr}(a),
  a conjugate pair
  is a union of two conditioned subspaces of
  the doublet $(\mathfrak{B},\mathfrak{B}^{[r]})$ for
  $\mathfrak{B}\in\mathcal{G}(\mathfrak{C})$.
  The pair is a degrade-I type
  as $\mathfrak{B}\supset\mathfrak{B}^{[r]}$
  or a regular-I
  as $\mathfrak{B}\nsupseteq\mathfrak{B}^{[r]}$.
  Recall in~\cite{SuTsai1} that a rank-zero quotient algebra
  $\{\mathcal{Q}(\mathfrak{C})\}$ given by a Cartan subalgebra $\mathfrak{C}$
  contains exclusively regular-I pairs because none of maximal bi-subalgebras
  of $\mathcal{G}(\mathfrak{C})-\mathfrak{C}$ is a superset of the zeroth maximal bi-subalgebra
  $\mathfrak{B}^{[0]}=\mathfrak{C}$.
  Yet the both types appear in a quotient algebra of rank $r\geq 1$.
  Since there have $2^r$ conjugate pairs for every maximal
  bi-subalgebra of $\mathcal{G}(\mathfrak{C})$ and
  the same number of maximal bi-subalgebras being a superset of
  $\mathfrak{B}^{[r]}$,
  the algebra $\{{\cal Q}(\mathfrak{B}^{[r]};2^{p+r}-1)\}$
  consists of a number $2^{2r}-1$ of degrade-I and $2^{p+r}-2^{2r}$ regular-I pairs.

  Two kinds of co-quotient algebras are acquirable in
  $\{\mathcal{P}_{\mathcal{Q}}(\mathfrak{B}^{[r]})\}$
  owing to the fact that the center subalgebra
  $W^{\tau}(\mathfrak{B}_1,\mathfrak{B}^{[r]};l)$ in
  Fig.~\ref{figQAandCo-QArankr}(b)
  is either a degrade conditioned subspace as $\mathfrak{B}_1\supset\mathfrak{B}^{[r]}$
  or a regular as
  $\mathfrak{B}_1\nsupseteq\mathfrak{B}^{[r]}$.
  On the occasion $\mathfrak{B}_1\supset\mathfrak{B}^{[r]}$,
  conjugate pairs of type degrade-I, degrade-II, regular-I and regular-II
  form the co-quotient algebra.
  As shown in the figure, 
  the bi-subalgebra $\mathfrak{B}^{[r]}$ and its null partner
  $W^{\hat{\tau}}(\mathfrak{B}_1,\mathfrak{B}^{[r]};l)=\{0\}$ make a degrade-II pair.
  Since 
  $\mathfrak{B}_1=\mathfrak{B}_m\sqcap\mathfrak{B}_n$ and $\mathfrak{B}_1\supset\mathfrak{B}^{[r]}$,
  both the maximal bi-subalgebras $\mathfrak{B}_m$ and $\mathfrak{B}_n\in\mathcal{G}(\mathfrak{C})$
  are a superset of $\mathfrak{B}^{[r]}$
  or neither of them is so, $1< m,n<2^p$.
  Each of the two conjugate pairs
  $\{W^\epsilon(\mathfrak{B}_m,\mathfrak{B}^{[r]};i), W^{\hat{\sigma}}(\mathfrak{B}_n,\mathfrak{B}^{[r]};,j)\}$
  and
  $\{W^\sigma(\mathfrak{B}_n,\mathfrak{B}^{[r]};j),W^{\hat{\epsilon}}(\mathfrak{B}_m,\mathfrak{B}^{[r]};i)\}$
  is a regular-I type when $\mathfrak{B}_m\nsupseteq\mathfrak{B}^{[r]}$
  and $\mathfrak{B}_n\nsupseteq\mathfrak{B}^{[r]}$.
  However when
  $\mathfrak{B}_m\supseteq\mathfrak{B}^{[r]}$
  and $\mathfrak{B}_n\supseteq\mathfrak{B}^{[r]}$,
  it is necessary to consider the two separate cases that
  both $\mathfrak{B}_m$ and $\mathfrak{B}_n$
  are a superset of $\mathfrak{B}^{[r,l]}$
  and neither of them is so
  due to the inclusion $\mathfrak{B}_1\supset\mathfrak{B}^{[r,l]}$,
  here $\mathfrak{B}^{[r,l]}=W^1(\mathfrak{C},\mathfrak{B}^{[r]};l)$ being a coset of $\mathfrak{B}^{[r]}$
  commuting with center subalgebra
  $W^{\tau}(\mathfrak{B}_1,\mathfrak{B}^{[r]};l)$.
  In the former case these two pairs are both a degrade-I type,
  while one of them is a degrade-II and the other is a regular-II in the latter case.
  On account of a half of the $2^r$ maximal bi-subalgebras
  $\{\mathfrak{B}:\mathfrak{B}\in\mathcal{G}(\mathfrak{C}),\mathfrak{B}\supset\mathfrak{B}^{[r]}\}$
  being a superset of $\mathfrak{B}^{[r,l]}$,
  there create $2^{2r-1}$ degrade-I, $2^{2r-2}$ degrade-II
  and $2^{2r-2}$ regular-II pairs.
  Removing the null redundancy of the type degrade-II, the co-quotient algebra
  $\{{\cal Q}({W}^{\tau}(\mathfrak{B}_1,\mathfrak{B}^{[r]};s);2^{p+r}-2^{2r-2})\}$
  with $\mathfrak{B}_1\supset\mathfrak{B}^{[r]}$
  has a reduced number $2^{p+r}-2^{2r-2}$ of conjugate pairs
  including one null, $2^{2r-1}$ degrade-I, $2^{2r-2}$ regular-II
  and $2^{p+r}-3\cdot 2^{2r-2}$ regular-I pairs.
  Note that the quotient-algebra partition of rank zero
  $\{\mathcal{P}_{\mathcal{Q}}(\mathfrak{C})\}$
  admits no co-quotient algebra given by a non-null degrade conditioned subspace,
  for the Cartan subalgebra $\mathfrak{C}$ being the unique non-null degrade subspace
  in this partition.

  On the occasion $\mathfrak{B}_1\nsupseteq\mathfrak{B}^{[r]}$,
  the co-quotient algebra possesses three types of conjugate pairs:
  the degrade-III, regular-I and regular-III.
  The pair $\{\mathfrak{B}^{[r]},W^{\hat{\tau}}(\mathfrak{B}_1,\mathfrak{B}^{[r]};l)\}$,
  a regular-III,
  is formed of the degrade conditioned subspace $\mathfrak{B}^{[r]}$
  and a regular $W^{\hat{\tau}}(\mathfrak{B}_1, \mathfrak{B}^{[r]};l)$
  with a half size of the former.
  Similarly since
  $\mathfrak{B}_1=\mathfrak{B}_m\sqcap\mathfrak{B}_n$
  and $\mathfrak{B}_1\nsupseteq\mathfrak{B}^{[r]}$, the two cases are considered that
  neither of the maximal bi-subalgebras
  $\mathfrak{B}_m$ and $\mathfrak{B}_n\in\mathcal{G}(\mathfrak{C})$
  is a superset of $\mathfrak{B}^{[r]}$,
  and only one of them is so.
  When $\mathfrak{B}_m\nsupseteq\mathfrak{B}^{[r]}$
  and $\mathfrak{B}_n\nsupseteq\mathfrak{B}^{[r]}$,
  both the pairs
  $\{W^\epsilon(\mathfrak{B}_m,\mathfrak{B}^{[r]};i), W^{\hat{\sigma}}(\mathfrak{B}_n,\mathfrak{B}^{[r]};,j)\}$
  and
  $\{W^\sigma(\mathfrak{B}_n,\mathfrak{B}^{[r]};j),W^{\hat{\epsilon}}(\mathfrak{B}_m,\mathfrak{B}^{[r]};i)\}$
  are a regular-I type.
  While when $\mathfrak{B}_m\supset\mathfrak{B}^{[r]}$
  and $\mathfrak{B}_n\nsupseteq\mathfrak{B}^{[r]}$
  (or $\mathfrak{B}_m\nsupseteq\mathfrak{B}^{[r]}$
  and $\mathfrak{B}_n\supset\mathfrak{B}^{[r]}$),
  one of the two pairs is type degrade-III and the other is regular-III.
  Thus, 
  $2^{2r}$ degrade-III, $2^{2r}$ regular-I and $2^{p+r}-2^{2r+1}$ regular-III pairs
  comprise the co-quotient algebra
  $\{{\cal Q}(W^\tau(\mathfrak{B}_1,\mathfrak{B}^{[r]};l);2^{p+r}-1)\}$.

  Importantly, the two kinds of co-quotient algebras differ in an essential feature.
  That is, the structure
  $\{{\cal Q}({W}^{\tau}(\mathfrak{B}_1,\mathfrak{B}^{[r]};l))\}$
  allows a step of {\em merging} as $\mathfrak{B}_1\supset\mathfrak{B}^{[r]}$
  or a step of {\em detaching} as
  $\mathfrak{B}_1\nsupseteq\mathfrak{B}^{[r]}$.
  By merging the abelian subspaces except ${W}^{\tau}(\mathfrak{B}_1,\mathfrak{B}^{[r]};l)$
  with $\mathfrak{B}_1\supset\mathfrak{B}^{[r]}$,
  the co-quotient algebra of rank $r$
  $\{{\cal Q}({W}^{\tau}(\mathfrak{B}_1,\mathfrak{B}^{[r]};l);2^{p+r}-2^{2r-2})\}$
  returns to a rank $r-1$ $\{{\cal Q}({W}^{\tau}(\mathfrak{B}_1,\mathfrak{B}^{[r]};l);2^{p+r-1}-1)\}$
  given by the same center subalgebra. 
  On the other hand,
  since every conditioned subspace
  in $\{{\cal Q}({W}^{\tau}(\mathfrak{B}_1,\mathfrak{B}^{[r]};l))\}$
  with $\mathfrak{B}_1\nsupseteq\mathfrak{B}^{[r]}$
  can be further divided into two conditioned subspaces by the coset rule of partition,
  the co-quotient algebra of rank $r$
  $\{{\cal Q}({W}^\tau(\mathfrak{B}_1,\mathfrak{B}^{[r]};l);2^{p+r}-1)\}$
  turns into a rank $r+1$
  $\{{\cal Q}({W}^\tau(\mathfrak{B}_1,\mathfrak{B}^{[r]};l);2^{p+r+1}-2^{2r+2})\}$
  undergoing a such detaching. 
  Refer to~\cite{SuTsai3}
  for more details of the two procedures.
  In other words, 
  a quotient-algebra partition of rank $r$
  admits not only quotient and co-quotient algebras
  of the same rank as in Theorem~\ref{thmQAcoQA},
  but co-quotient algebras of ranks $r-1$ and $r+1$.

\section{Proof of Main Theorem}\label{secmainthm}
\renewcommand{\theequation}{\arabic{section}.\arabic{equation}}
\setcounter{equation}{0} \noindent
 In the $s$-representation, a subset of a Cartan subalgebra $\mathfrak{C}\subset su(2^p)$
 chosen to be the center subalgebra and
 producing a quotient or a co-quotient algebra
 must be either an $r$-th maximal bi-subalgebra
 $\mathfrak{B}^{[r]}$ of $\mathfrak{C}$, $0\leq r<p$,
 or a coset in the partition of $\mathfrak{C}$ generated by $\mathfrak{B}^{[r]}$.
 Before proceeding to Main Theorem, this important statement will be validated by first
 showing that the conjugate partition leads to the abelianness of the center subalgebra
 as in the following lemma,
 and then that the condition of closure requires the center subalgebra
 to be either a bi-subalgebra or a coset of a bi-subalgebra.
 \vspace{6pt}
 \begin{lemma}\label{lemconjuparabel}
 Within a partition of the Lie algebra
 $su(2^p)={\cal A}\oplus{W}_1\oplus\hat{W}_1\oplus\cdots\oplus{W}_m\oplus\hat{W}_m\oplus\cdots\oplus{W}_q\oplus\hat{W}_q$,
 $p>1$ and $1\leq m\leq q$,
 consisting of a subspace ${\cal A}$ and a number q of conjugate pairs,
 ${\cal A}$ is abelian
 if the conjugate partition is fulfilled, namely
 $[{W}_m,{\cal A}]\subset\hat{W}_m$, $[\hat{W}_m,{\cal A}]\subset{W}_m$ and $[{W}_m,\hat{W}_m]\subset{\cal A}$.
 \end{lemma}
 \vspace{3pt}
 \begin{proof}
 Let the abelianness of ${\cal A}$ be proved by contradiction.
 Assume that ${\cal A}$ is not abelian and is a proper subset of $su(2^p)$.
 Then there exist a pair of spinors
 ${\cal S}^{\zeta}_{\alpha}$ and ${\cal S}^{\eta}_{\beta}\in {\cal A}$ and
 $[{\cal S}^{\zeta}_{\alpha},{\cal S}^{\eta}_{\beta}]\neq 0$.
 The noncommuting breeds the generator ${\cal S}^{\zeta+\eta}_{\alpha+\beta}$
 which falls either in ${\cal A}$ or in $su(2^p)-{\cal A}$.
 It is noticed that for every spinor  ${\cal S}^{\zeta}_{\alpha}\in {\cal A}$,
 there must have at least a noncommuting spinor  ${\cal S}^{\xi}_{\gamma}\in su(2^p)-{\cal A}$,
 {\em i.e.,} $\{{\cal S}^{\zeta}_{\alpha},{\cal S}^{\xi}_{\gamma}\}=0$.
 The reason is as simple as follows. If no any such  spinor exists.
 then all spinors anti-commuting with ${\cal S}^{\zeta}_{\alpha}$ are
 contained in the subspace ${\cal A}$.
 Consider a spinor ${\cal S}^{\zeta'}_{\alpha'}$ respectively anti-commuting
 with ${\cal S}^{\zeta}_{\alpha}$ and also with a spinor
 ${\cal S}^{\xi'}_{\gamma'}\in{W}_n\subset{su(2^p)-{\cal A}}$,
 by the permission of the equation
 $\zeta'\cdot\alpha+\zeta\cdot\alpha'=\zeta'\cdot\gamma'+\xi'\cdot\alpha'=1$
 as $p>1$.
 Thus the spinor ${\cal S}^{\zeta'}_{\alpha'}$
 belongs to ${\cal A}$.
 From the commutator
 $[{\cal S}^{\zeta'}_{\alpha'},{\cal S}^{\xi'}_{\gamma'}]\neq 0$,
 the bi-additive ${\cal S}^{\zeta'+\xi'}_{\alpha'+\gamma'}$
 is in $\hat{W}_n\subset su(2^p)-{\cal A}$ due to the conjugate partition.
 While since 
 anti-commuting with ${\cal S}^{\zeta}_{\alpha}$,
 the spinor ${\cal S}^{\zeta'+\xi'}_{\alpha'+\gamma'}$ is obliged to be
 in ${\cal A}$ too, which is forbidden in the partition.

 Now assume ${\cal S}^{\zeta+\eta}_{\alpha+\beta}\in{\cal A}$ and
 let some spinor ${\cal S}^{\xi}_{\gamma}\in W_m\subset{su(2^p)-{\cal A}}$
 anti-commuting with ${\cal S}^{\zeta}_{\alpha}$ be assigned.
 Accordingly, via the conjugate partition yields another bi-additive
 ${\cal S}^{\zeta+\xi}_{\alpha+\gamma}\in\hat{W}_m$.
 With the additional assumption $[{\cal S}^{\eta}_{\beta},{\cal S}^{\xi}_{\gamma}]=0$
 (or $[{\cal S}^{\eta}_{\beta},{\cal S}^{\xi}_{\gamma}]\neq 0$),
 the commutators
 $[{\cal S}^{\xi}_{\gamma},{\cal S}^{\zeta+\eta}_{\alpha+\beta}]\neq 0$
 and $[{\cal S}^{\eta}_{\beta},{\cal S}^{\zeta+\xi}_{\alpha+\gamma}]\neq 0$
 (or $[{\cal S}^{\eta}_{\beta},{\cal S}^{\xi}_{\gamma}]\neq 0$ and
 $[{\cal S}^{\zeta+\eta}_{\alpha+\beta},{\cal S}^{\zeta+\xi}_{\alpha+\gamma}]\neq 0$)
 bring in 
 ${\cal S}^{\zeta+\eta+\xi}_{\alpha+\beta+\gamma}$ (or ${\cal S}^{\eta+\xi}_{\beta+\gamma}$)
 belonging to both $W_m$ and $\hat{W}_m$ by the conjugate partition.
 This discords with the partition property.

 On the other hand, suppose ${\cal S}^{\zeta+\eta}_{\alpha+\beta}\notin{\cal A}$
 and let it be in a subspace ${W}_k$ (or $\hat{W}_k)$.
 Resulting from the nonvanishing commutators
 $[{\cal S}^{\zeta}_{\alpha},{\cal S}^{\zeta+\eta}_{\alpha+\beta}]\neq 0$
 and $[{\cal S}^{\eta}_{\beta},{\cal S}^{\zeta+\eta}_{\alpha+\beta}]\neq 0$,
 the spinors ${\cal S}^{\eta}_{\beta}={\cal S}^{\zeta+\zeta+\eta}_{\alpha+\alpha+\beta}$ and
 ${\cal S}^{\zeta}_{\alpha}={\cal S}^{\eta+\zeta+\eta}_{\beta+\alpha+\beta}$
 are coerced into $\hat{W}_k$ (or ${W}_k$) again via the conjugate partition,
 which contradicts the early assumption that ${\cal S}^{\zeta}_{\alpha}$
 and ${\cal S}^{\eta}_{\beta}\in{\cal A}$.
 As a consequence of requiring the conjugate partition, whether ${\cal S}^{\zeta+\eta}_{\alpha+\beta}\in{\cal A}$ or $\notin{\cal A}$,
 a contradiction is attained by assuming the nonabelianness of ${\cal A}$.
 Hence the assumption is denied.
 \end{proof}

 \vspace{6pt}
 Based on the conjugate partition, there follows the $2$nd assertion.

 \vspace{6pt}
 \begin{lemma}\label{lemWabelcondclose}
  Given a conjugate partition of the Lie algebra $su(2^p)$, $p>1$,
  consisting of the center subalgebra ${\cal A}$ and a number $q$ of conjugate pairs
  $\{{W}_m,\hat{W}_m\}$, 
  $1\leq m\leq q$,
  each subspace of the pair ${W}_m$ and $\hat{W}_m$ is abelian if the condition of closure holds, that is,
  there exisiting a unique conjugate pair $\{{W}_s,\hat{W}_s\}$
  for two arbitrary conjugate pairs $\{{W}_m,\hat{W}_m\}$ and $\{{W}_n,\hat{W}_n\}$
  such that
  $[{W}_m,{W}_n]\subset\hat{W}_s$, $[{W}_m,\hat{W}_n]\subset{W}_s$,
  $[\hat{W}_m,{W}_n]\subset{W}_s$ and $[\hat{W}_m,\hat{W}_n]\subset\hat{W}_s$,
  $1\leq m,n,s\leq q$.
 \end{lemma}
 \vspace{3pt}
 \begin{proof}
 A contradiction proof will affirm the abelianness of ${W}_m$;
 although omitted, a similar argument is applicable to $\hat{W}_m$.
 There have at least a pair of noncommuting spinors ${\cal S}^{\zeta}_{\alpha}$
 and ${\cal S}^{\eta}_{\beta}\in{W}_m$ by assuming ${W}_m$ to be nonabelian.
 The bi-additive ${\cal S}^{\zeta+\eta}_{\alpha+\beta}$
 is either in ${\cal A}$ or in ${su(2^p)}-{\cal A}$.
 As ${\cal S}^{\zeta+\eta}_{\alpha+\beta}\in{\cal A}$,
 granted by the commutators
 $[{\cal S}^{\zeta}_{\alpha},{\cal S}^{\zeta+\eta}_{\alpha+\beta}]\neq 0$
 and
 $[{\cal S}^{\eta}_{\beta},{\cal S}^{\zeta+\eta}_{\alpha+\beta}]\neq 0$,
 the subspace $\hat{W}_m$ owns the two spinors
 ${\cal S}^{\zeta}_{\alpha}={\cal S}^{\eta+\zeta+\eta}_{\beta+\alpha+\beta}$
 and
 ${\cal S}^{\eta}_{\beta}={\cal S}^{\zeta+\zeta+\eta}_{\alpha+\alpha+\beta}$
 according to the conjugate partition.
 This spoils the assumed partition over $su(2^p)$.

 While as ${\cal S}^{\zeta+\eta}_{\alpha+\beta}\in su(2^p)-{\cal A}$,
 three cases have to be considered that
 ${\cal S}^{\zeta+\eta}_{\alpha+\beta}\in{W}_m$, ${\cal S}^{\zeta+\eta}_{\alpha+\beta}\in\hat{W}_m$
 and ${\cal S}^{\zeta+\eta}_{\alpha+\beta}\in{W}_n$ (or $\hat{W}_n$),
 here $m\neq n$.
 Note that ${\cal S}^{\zeta+\eta}_{\alpha+\beta}$ commutes with
 neither ${\cal S}^{\zeta}_{\alpha}$ nor ${\cal S}^{\eta}_{\beta}$
 because $[{\cal S}^{\zeta}_{\alpha},{\cal S}^{\eta}_{\beta}]\neq 0$.
 The examination on the case ${\cal S}^{\zeta+\eta}_{\alpha+\beta}\in{W}_m$
 can follow the same procedure in the proof of Lemma~\ref{lemconjuparabel}
 except replacing ${\cal A}$ with ${W}_m$.
 For the 2nd case ${\cal S}^{\zeta+\eta}_{\alpha+\beta}\in\hat{W}_m$,
 the subspace ${\cal A}$ absorbs both the spinors ${\cal S}^{\zeta}_{\alpha}$
 and ${\cal S}^{\eta}_{\beta}$ via the conjugate partition, which is obviously
 in contradiction to the partition.
 Finally, the spinors
 ${\cal S}^{\zeta}_{\alpha}$
 and
 ${\cal S}^{\eta}_{\beta}$
 are turned to the subspace $\hat{W}_s$ (or $W_s$) in virtue of the condition of closure
 in the case ${\cal S}^{\zeta+\eta}_{\alpha+\beta}\in{W}_n$ (or $\hat{W}_n$),
 which again violates the partition assumption.
 The proof completes.
 \end{proof}
 \vspace{6pt}

 \vspace{6pt}
 \begin{lemma}\label{lemcondclose}
 Given a conjugate partition of the Lie algebra $su(2^p)$, $p>1$,
 consisting of the center subalgebra ${\cal A}$ and a number $q$ of conjugate pairs
 $\{{W}_m,\hat{W}_m\}$, 
 $1\leq m\leq q$,
 the abelian subspace ${\cal A}$ is either a bi-subalgebra $\mathcal{B}\subset su(2^p)$
 or a coset in the partition of $su(2^p)$ generated by $\mathcal{B}$ if
 the condition of closure holds, that is,
 there exisiting a unique conjugate pair $\{{W}_s,\hat{W}_s\}$
 for two arbitrary conjugate pairs $\{{W}_m,\hat{W}_m\}$ and $\{{W}_n,\hat{W}_n\}$
 such that
 $[{W}_m,{W}_n]\subset\hat{W}_s$, $[{W}_m,\hat{W}_n]\subset{W}_s$,
 $[\hat{W}_m,{W}_n]\subset{W}_s$ and $[\hat{W}_m,\hat{W}_n]\subset\hat{W}_s$,
 $1\leq m,n,s\leq q$.
 \end{lemma}
 \vspace{3pt}
 \begin{proof}
 Since a subspace containing single one or only two spinor generators
 is a bi-subalgebra or a coset of a bi-subalgebra {\em per se}, there suppose
 at least three spinors in ${\cal A}$.
 This lemma is to be proved by contradiction.
 Thus, assume that
 the abelian subspace ${\cal A}$ is neither a bi-subalgebra nor a coset
 of a bi-subalgebra under the bi-addition.
 The former part of the negation leads to the condition of ``${\cal A}$-nonclosure"
 that ${\cal S}^{\zeta_1+\zeta_2}_{\alpha_1+\alpha_2}\notin{\cal A}$ for at least
 a pair of spinors ${\cal S}^{\zeta_1}_{\alpha_1}$ and ${\cal S}^{\zeta_2}_{\alpha_2}\in{\cal A}$.
 Besides, two non-closure conditions follow the latter part of the negation
 in regard to an arbitrary bi-subalgebra, denoted  $\mathfrak{B}^{[r]}$, of the Cartan
 subalgebra $\mathfrak{C}\subset{su(2^p)}$.
 One condition is the ``$\mathfrak{B}^{[r]}$-nonclosure" that,
 for every bi-subalgebra $\mathfrak{B}^{[r]}$ of $\mathfrak{C}$,
 ${\cal S}^{\zeta_3+\zeta_4}_{\alpha_3+\alpha_4}\notin\mathfrak{B}^{[r]}$
 for at least a pair of spinors
 ${\cal S}^{\zeta_3}_{\alpha_3}$ and ${\cal S}^{\zeta_4}_{\alpha_4}\in{\cal A}$.
 The other is the ``${\cal A}$-$\mathfrak{B}^{[r]}$-nonclosure" that
 there exist two spinors ${\cal S}^{\zeta_5}_{\alpha_5}\in{\cal A}$
 and ${\cal S}^{\eta}_{\beta}\in\mathfrak{B}^{[r]}$
 obeying ${\cal S}^{\zeta_5+\eta}_{\alpha_5+\beta}\notin{\cal A}$
 for every bi-subalgebra $\mathfrak{B}^{[r]}$.
 As illustrated immediately below, a contradiction will be caused by the ${\cal A}$-nonclosure together
 with either one of the other two nonclosures.

 With the assumption of the ${\cal A}$- and
 $\mathfrak{B}^{[r]}$-nonclosures,
 here examine only the case that
 ${\cal S}^{\zeta+\zeta_1}_{\alpha+\alpha_1}\in{W}_m\subset{su(2^p)-{\cal A}}$,
 ${\cal S}^{\zeta+\zeta_3}_{\alpha+\alpha_3}\in{W}_n\subset{su(2^p)-\mathfrak{B}^{[r]}}$
 and ${\cal S}^{\zeta_1+\zeta_3}_{\alpha_1+\alpha_3}\in{\cal A}$
 for three spinors ${\cal S}^{\zeta}_{\alpha},{\cal S}^{\zeta_1}_{\alpha_1}$
 and ${\cal S}^{\zeta_3}_{\alpha_3}\in{\cal A}$, because
 similar contradictions will occur in the other cases.
 Given a generator ${\cal S}^{\xi}_{\gamma}\in{W}_s\subset su(2^p)-{\cal A}$,
 further assume the commutation relations
 $[{\cal S}^{\zeta}_{\alpha},{\cal S}^{\xi}_{\gamma}]=0$,
 $[{\cal S}^{\zeta_1}_{\alpha_1},{\cal S}^{\xi}_{\gamma}]\neq 0$ and
 $[{\cal S}^{\zeta_3}_{\alpha_3},{\cal S}^{\xi}_{\gamma}]\neq 0$;
 a same conclusion will be reached with the other assumptions of commutation relations.
 Accordingly, the nonvanishing commutators
 $[{\cal S}^{\zeta_i}_{\alpha_i},{\cal S}^{\xi}_{\gamma}]\neq 0$
 introduces the inclusions
 ${\cal S}^{\zeta_i+\xi}_{\alpha_i+\gamma}\in\hat{W}_s$
 via the conjugate partition, here $i=1,3$.
 Based on the assumption
 ${\cal S}^{\zeta+\zeta_1}_{\alpha+\alpha_1}\in{W}_m$
 and
 ${\cal S}^{\zeta+\zeta_3}_{\alpha+\alpha_3}\in{W}_n$
 as well as again the nonvanishing commutators
 $[{\cal S}^{\zeta_i}_{\alpha_i},{\cal S}^{\xi}_{\gamma}]\neq 0$,
 the fallacy is derived that
 the spinor ${\cal S}^{\zeta+\xi}_{\alpha+\gamma}={\cal S}^{\zeta+\zeta_i+\zeta_i+\xi}_{\alpha+\alpha_i+\alpha_i+\gamma}$
 is embraced in both the conditioned subspaces ${W}_m$ and ${W}_n$
 due to the condition of closure, which violates the assumed partition.

 While the conditions of ${\cal A}$- and
 ${\cal A}$-$\mathfrak{B}^{[r]}$-nonclosures are considered,
 likewise,
 only one case is discussed that
 ${\cal S}^{\zeta+\zeta_5}_{\alpha+\alpha_5}\in{W}_m\subset{su(2^p)-{\cal A}}$,
 ${\cal S}^{\zeta_5+\eta}_{\alpha_5+\beta}\in{W}_n\subset{su(2^p)-{\cal A}}$
 and ${\cal S}^{\zeta+\eta}_{\alpha+\beta}\in{\cal A}$
 for three spinors ${\cal S}^{\zeta}_{\alpha},{\cal S}^{\zeta_5}_{\alpha_5}\in{\cal A}$
 and ${\cal S}^{\eta}_{\beta}\in\mathfrak{B}^{[r]}$.
 Since ${\cal A}$ is abelian by Lemma~\ref{lemconjuparabel},
 there have the vanishing commutators
 $[{\cal S}^{\zeta}_{\alpha},{\cal S}^{\zeta_5}_{\alpha_5}]
 =[{\cal S}^{\zeta}_{\alpha},{\cal S}^{\zeta+\eta}_{\alpha+\beta}]
 =[{\cal S}^{\zeta_5}_{\alpha_5},{\cal S}^{\zeta+\eta}_{\alpha+\beta}]=0$,
 which lead to the other two
 $[{\cal S}^{\zeta}_{\alpha},{\cal S}^{\eta}_{\beta}]
 =[{\cal S}^{\zeta_5}_{\alpha_5},{\cal S}^{\eta}_{\beta}]=0$.
 Let ${\cal S}^{\xi}_{\gamma}$ be a spinor in $\hat{W}_s$
 and further assume the instance
 $[{\cal S}^{\zeta}_{\alpha},{\cal S}^{\xi}_{\gamma}]=[{\cal S}^{\eta}_{\beta},{\cal S}^{\xi}_{\gamma}]=0$
 and $[{\cal S}^{\zeta_5}_{\alpha_5},{\cal S}^{\xi}_{\gamma}]\neq 0$;
 similar contradictions will be obtained in the other instances.
 Bred from the noncommuting of the two spinors
 ${\cal S}^{\zeta_5+\eta}_{\alpha_5+\beta}\in{W}_n$
 and ${\cal S}^{\xi}_{\gamma}\in\hat{W}_s$
 owing to the commutators
 $[{\cal S}^{\eta}_{\beta},{\cal S}^{\xi}_{\gamma}]=0$
 and $[{\cal S}^{\zeta_5}_{\alpha_5},{\cal S}^{\xi}_{\gamma}]\neq 0$,
 the bi-additive ${\cal S}^{\zeta_5+\eta+\xi}_{\alpha_5+\beta+\gamma}$
 belongs to ${W}_m$ via the condition of closure.
 Moreover, the two spinors ${\cal S}^{\zeta_5+\eta+\xi}_{\alpha_5+\beta+\gamma}$
 and ${\cal S}^{\zeta+\zeta_5}_{\alpha+\alpha_5}\in{W}_m$
 do not commute because
 $[{\cal S}^{\zeta}_{\alpha},{\cal S}^{\eta}_{\beta}]
 =[{\cal S}^{\zeta}_{\alpha},{\cal S}^{\xi}_{\gamma}]
 =[{\cal S}^{\zeta_5}_{\alpha_5},{\cal S}^{\eta}_{\beta}]=0$
 and
 $[{\cal S}^{\zeta_5}_{\alpha_5},{\cal S}^{\xi}_{\gamma}]\neq 0$.
 It contradictorily implies the nonabelianness of the subspace ${W}_m$,
 {\em cf.} Lemma~\ref{lemWabelcondclose}.
 This ends the proof.
 \end{proof}
 \vspace{6pt}
 The eligibility of a center subalgebra is thereby assured.
 \vspace{6pt}
 \begin{cor}\label{corocentersub}
 A subset of the Lie algebra $su(2^p)$ in the s-representation which can be chosen as
 the center subalgebra and generate a quotient
 or a co-quotient algebra  
 is either a bi-subalgebra of a Cartan subalgebra $\mathfrak{C}\subset su(2^p)$
 or a coset of a bi-subalgebra in $\mathfrak{C}$.
 \end{cor}
 \vspace{6pt}
 This theorem echos {\em the quaternion democracy}~\cite{SuTsai1} exhibited
 in Eq.~\ref{eqgenWcommrankr}.
 There do exist quotient algebras with center algebras spanned by
 non-commuting spinors. Nevertheless, up to subscript permutations in
 the $\lambda$-representation as shown in~\cite{Su},
 they are equivalent to those generated by bi-subalgebras
 or cosets in Cartan subalgebras.

 Being a major clause of Main Theorem, the following lemma endorses the
 isomorphism of quotient-algebra partitions of different dimensions
 with 
 the help of the removing process~\cite{Su}.
 \vspace{6pt}
 \begin{lemma}\label{lemQArankforsuN}
 The quotient-algebra partitions of rank $r$ for the Lie algebra
 $su(N)$ are respectively isomorphic to those of $su(2^p)$,
 where $2^{p-1}<N\leq 2^p$, $0\leq r\leq r_0<p$, and
 the dimension has the factorization $N=2^{r_0}N'$ with $N'$ being odd.
 \end{lemma}
 \vspace{3pt}
 \begin{proof}
 Similar to specifying the intrinsic quotient algebra during asserting Theorem~1 of~\cite{Su},
 the proof here is conducted in a designated form, {\em i.e.}, the {\em canonical quotient-algebra partition}
 $\{\mathcal{P}_{\mathcal{Q}}(\mathfrak{B}^{[r]}_{can})\}$
 given by the {\em canonical bi-subalgebra} $\mathfrak{B}^{[r]}_{can}$.
 Every another quotient-algebra partition of the same rank is related to this selection by
 a conjugate transformation, 
 referring to Appendix~\ref{appTransOAPs} for the explicit form.
 The canonical bi-subalgebra
 $\mathfrak{B}^{[r]}_{can}=\{{\cal S}^{\nu_{\mathbf{0}}}_{\mathbf{0}}:
 \forall\hspace{2pt} \nu_{\mathbf{0}}\in{Z^p_2},\hspace{2pt}\rho_i\in{Z_2},\hspace{2pt}1<i\leq p,\hspace{2pt}
 \nu_l=\rho_{1}\rho_{2}\cdots\rho_p\text{ and }\rho_{p-r+1}=\rho_{p-r+2}=\cdots=\rho_p=0\}$
 so chosen is an $r$-th maximal bi-subalgebra of the intrinsic Cartan subalgebra
 $\mathfrak{C}_{[\mathbf{0}]}=\{{\cal S}^{\nu_0}_{\mathbf{0}}:\nu_0\in Z^p_2\}\subset su(2^p)$;
 note that the bi-subalgebra $\mathfrak{B}^{[0]}_{can}=\mathfrak{C}_{[\mathbf{0}]}$ 
 generates the canonical quotient-algebra partition of rank zero.
 With the notation in Theorem~\ref{thmgenWcommrankr},
 the conditioned subspaces within the canonical quotient-algebra partition of rank $r$
 $\{\mathcal{P}_{\mathcal{Q}}(\mathfrak{B}^{[r]}_{can})\}$ take the form,
 $\forall\hspace{2pt}\epsilon\in{Z_2}$, $l\in{Z^r_2}$ and $\alpha\in{Z^p_2}$,
 \begin{align}\label{eqgenWcanQAr}
 &{W}^\epsilon(\mathfrak{B}_\alpha,\mathfrak{B}^{[r]}_{can};l)\notag\\
 &\hspace{-3pt}=\{{\cal S}^{\zeta}_{\alpha}:
 \forall\hspace{2pt} \zeta\in{Z^p_2},\hspace{2pt}\rho_i\in{Z_2},\hspace{2pt}p-r< i\leq p,\hspace{2pt}
 \zeta=\rho_{1}\rho_{2}\cdots\rho_{p-r}\circ l\text{ and }\zeta\cdot\alpha=1+\epsilon\},
 \end{align}
 and are listed more explicitly as follows,
 \begin{align}\label{eqcanQAexpW}
 &{W}^0(\mathfrak{B}_{\alpha=\mathbf{0}},\mathfrak{B}^{[r]}_{can};l)=\{0\}\notag\\
 &{W}^1(\mathfrak{B}_{\alpha=\mathbf{0}},\mathfrak{B}^{[r]}_{can};l)=\mathfrak{B}^{[r,l]}_{can}\notag\\
 &\hspace{-3pt}=\{{\cal S}^{\nu_l}_{\mathbf{0}}:
 \forall\hspace{2pt} \nu_l\in{Z^p_2},\hspace{2pt}\rho_i\in{Z_2},\hspace{2pt}p-r<i\leq p,\hspace{2pt}
 \nu_l=\rho_{1}\rho_{2}\cdots\rho_{p-r}\circ l\},\notag\\
 &{W}^0(\mathfrak{B}_{\alpha\neq\mathbf{0}},\mathfrak{B}^{[r]}_{can};l)=\hat{W}(\mathfrak{B}_{\alpha\neq\mathbf{0}},\mathfrak{B}^{[r]}_{can};l)\notag\\
 &\hspace{-3pt}=\{{\cal S}^{\hat{\zeta}}_{\alpha}:
 \forall\hspace{2pt} \hat{\zeta}\in{Z^p_2},\hspace{2pt}\hat{\tau}_i\in{Z_2},\hspace{2pt}p-r<i\leq p,\hspace{2pt}
 \hat{\zeta}=\hat{\tau}_{1}\hat{\tau}_{2}\cdots\hat{\tau}_{p-r}\circ l\text{ and }\hat{\zeta}\cdot\alpha=1\},\text{ and}\notag \\
 &{W}^1(\mathfrak{B}_{\alpha\neq\mathbf{0}},\mathfrak{B}^{[r]}_{can};l)={W}(\mathfrak{B}_{\alpha},\mathfrak{B}^{[r]}_{can};l)\notag \\
 &\hspace{-3pt}=\{{\cal S}^{\zeta}_{\alpha}:
 \forall\hspace{2pt} \zeta\in{Z^p_2},\hspace{2pt}\tau_i\in{Z_2},\hspace{2pt}p-r< i\leq p,\hspace{2pt}
 \zeta=\tau_{1}\tau_{2}\cdots\tau_{p-r}\circ l\text{ and }\zeta\cdot\alpha=0\},
 \end{align}
 where
 $\mathfrak{B}_{\alpha}=\{{\cal S}^{\nu}_{\mathbf{0}}:\forall\hspace{2pt} \nu\in{Z^p_2},\hspace{2pt}\nu\cdot\alpha=0\}$
 is a maximal bi-subalgebra of
 $\mathfrak{B}_{\mathbf{0}}=\mathfrak{C}_{[\mathbf{0}]}$.
 The symbol ``$\circ$" denotes the concatenation of two strings, namely
 the last $r$ digits of the strings $\nu_l$ and $\zeta$ (or $\hat{\zeta}$) coinciding with
 the index string $l$.
 This index labelling is assigned merely for convenience among
 a good number of alternatives, {\em cf.} Lemma~\ref{lemcosetparinC}.
 In addition,  the generators of
 ${W}^\epsilon(\mathfrak{B}_\alpha,\mathfrak{B}^{[r]}_{can};l)$ share the same
 binary-partitioning $\alpha$ and self parity $1+\epsilon$.
 Some examples are illustrated in Appendix~\ref{appQAFigs}.

 To obtain quotient-algebra partition for $su(N)$ of dimension $2^{p-1}<N<2^p$,
 the removing process introduced in~\cite{Su} is employed.
 The process is to 
 rewrite the spinors of $\{\mathcal{P}_{\mathcal{Q}}(\mathfrak{B}^{[r]}_{can})\}$
 in the $\lambda$-representation, followed by deleting
 the $\lambda$-generators $\lambda_{ij},\hat{\lambda}_{ij}\text{ and }d_{ij}\in su(2^p)$
 iff $i\text{ or }j>N$.
 As $r>r_0$, it is easy to verify that
 every pair of the degrade conditioned subspaces
 $\mathfrak{B}^{[r_0+1,t]}_{can}=\bigcup_{l\in{K}}\mathfrak{B}^{[r,l]}_{can}$
 and $\mathfrak{B}^{[r_0+1,\hat{t}]}_{can}=\bigcup_{\hat{l}\in{\hat{K}}}\mathfrak{B}^{[r,\hat{l}]}_{can}$
 at least share the generator $\sum^{N-1}_{i=2^{p-1}+1}\upsilon_i\ket{i}\bra{i}$ in common,
 where the coefficient $\upsilon_i$ is the parity $1$ or $-1$ for $2^{p-1}<i\leq N$,
 the addition $t+\hat{t}=10\cdots 0$ of the two strings $t\text{ and }\hat{t}\in Z^{r_0+1}_2$ 
 has only one single bit $1$ in the leftmost digit,
 and the last $r_0+1$ digits of the sets
 $K=\{\hspace{2pt}l:\forall\hspace{2pt}l\in Z^r_2,\rho_i\in Z_2,1\leq i\leq r,
 \hspace{2pt}l=\rho_1\rho_2\cdots\rho_{r-r_0-1}\circ{t}\hspace{2pt}\}$
 and
 $\hat{K}=\{\hspace{2pt}\hat{l}:\forall\hspace{2pt}l\in Z^r_2,\hat{\rho}_i\in Z_2,1\leq i\leq r,
 \hspace{2pt}l=\hat{\rho}_1\hat{\rho}_2\cdots\hat{\rho}_{r-r_0-1}\circ\hat{t}\hspace{2pt}\}$
 are identical to the two subspace indices respectively.
 This implies that the degrade subspaces $\mathfrak{B}^{[r,l]}_{can}\subset\mathfrak{C}_{[\mathbf{0}]}$
 are no longer disjoint 
 and then there exists no corresponding quotient-algebra partition
 of rank $r$ for $su(N)$.
 For example, the $3$rd maximal bi-subalgebra
 $\mathfrak{B}^{[3,000]}_{can}$ and the degrade subspace $\mathfrak{B}^{[3,\hspace{1pt}100]}_{can}$
 in $su(12)$ with $r_0=2$ as in Fig.~\ref{figsu12QArank0intr} share a common generator
 $\sum^{12}_{i=9}\ket{i}\bra{i}
 =\frac{1}{3}I\otimes {\cal S}^{00}_{00}-\frac{2}{\sqrt{3}}\mu_8\otimes{\cal S}^{00}_{00}$.

 While $0\leq r\leq r_0$ on the other hand,
 a generator in the non-null conditioned subspace
 ${W}^{\epsilon}(\mathfrak{B}_{\alpha},\mathfrak{B}^{[r]}_{can};\zeta)$
 can take the form
 ${\cal L}^{\hspace{1pt}\epsilon}_{\hspace{1pt}\omega,\alpha}\otimes{\cal S}^{\zeta}_{\hat{\alpha}}$,
 where 
 $\hat{\alpha}=a_{p-r+1}a_{p-r+2}\cdots a_p\in Z^r_2$
 is a string of the last $r$ digits of $\alpha=a_1a_2\cdots a_p$,
 and
 ${\cal L}^{\hspace{1pt}\epsilon}_{\hspace{1pt}\omega,\alpha}=\lambda_{ij},\hat{\lambda}_{ij}$
 or $d_{ij}$
 is a $\lambda$-generator in $su(N/2^{r})$ with $\omega\in Z^p_2$,
 $i-1=\omega$ and $j-1=\omega+\alpha$ for $1\leq i,j\leq N/2^{r}$, 
 referring to Appendix~B of~\cite{Su} for details.
 This demonstrates that all these conditioned subspaces in the partition
 remain non-null and disjoint after the application of the removing process.
 Surely a generator allows other versions of {\em factorizations}, {\em e.g.},
 ${\cal L}^{\hspace{1pt}\epsilon}_{\hspace{1pt}{\omega}',{\alpha}'}\otimes{\cal S}^{{\zeta}'}_{\hat{\alpha}'}$
 with ${\omega}',{\alpha}'\in Z^{p-r_0}_2$ and ${\zeta}',\hat{\alpha}'\in Z^{r_0}_2$,
 yet only the current form bears the apparent association with the conditioned subspace.

 For two arbitrary generators
 ${\cal L}^{\hspace{1pt}\epsilon}_{\hspace{1pt}\omega,\alpha}\otimes{\cal S}^{\zeta}_{\hat{\alpha}}$
 and
 ${\cal L}^{\hspace{1pt}\sigma}_{\hspace{1pt}\tau,\beta}\otimes{\cal S}^{\eta}_{\hat{\beta}}\in su(N)$,
 then there holds the commutation relation
 $[{\cal L}^{\hspace{1pt}\epsilon}_{\hspace{1pt}\omega,\alpha}\otimes{\cal S}^{\zeta}_{\hat{\alpha}},
 {\cal L}^{\hspace{1pt}\sigma}_{\hspace{1pt}\tau,\beta}\otimes{\cal S}^{\eta}_{\hat{\beta}}]=
 (-1)^{\eta\cdot\hat{\alpha}}[{\cal L}^{\hspace{1pt}\epsilon}_{\hspace{1pt}\omega,\alpha},{\cal L}^{\hspace{1pt}\sigma}_{\hspace{1pt}\tau,\beta}]
 \otimes{\cal S}^{\zeta+\eta}_{\hat{\alpha}+\hat{\beta}}
 =(-1)^{\eta\cdot\hat{\alpha}}(\delta_{\omega+\alpha,\tau}+(-1)^{\sigma}\delta_{\omega+\alpha,\tau+\beta}){\cal L}^{\hspace{1pt}\epsilon+\sigma}_{\hspace{1pt}\omega,\alpha+\beta}\otimes{\cal S}^{\zeta+\eta}_{\hat{\alpha}+\hat{\beta}}
 +(-1)^{\epsilon+\eta\cdot\hat{\alpha}}(\delta_{\omega,\tau}+(-1)^{\sigma}\delta_{\omega,\tau+\beta}){\cal L}^{\hspace{1pt}\epsilon+\sigma}_{\hspace{1pt}\omega+\alpha,\alpha+\beta}\otimes{\cal S}^{\zeta+\eta}_{\hat{\alpha}+\hat{\beta}}$
 as $\eta\cdot\hat{\alpha}+\zeta\cdot\hat{\beta}=0$,
 or
 $[{\cal L}^{\hspace{1pt}\epsilon}_{\hspace{1pt}\omega,\alpha}\otimes{\cal S}^{\zeta}_{\hat{\alpha}},
 {\cal L}^{\hspace{1pt}\sigma}_{\hspace{1pt}\tau,\beta}\otimes{\cal S}^{\eta}_{\hat{\beta}}]
 =(-1)^{\eta\cdot\hat{\alpha}}\{{\cal L}^{\hspace{1pt}\epsilon}_{\hspace{1pt}\omega,\alpha},{\cal L}^{\hspace{1pt}\sigma}_{\hspace{1pt}\tau,\beta}\}
 \otimes{\cal S}^{\zeta+\eta}_{\hat{\alpha}+\hat{\beta}}=
 (-1)^{\eta\cdot\hat{\alpha}}(\delta_{\omega+\alpha,\tau}-(-1)^{\sigma}\delta_{\omega+\alpha,\tau+\beta}){\cal L}^{\hspace{1pt}1+\epsilon+\sigma}_{\hspace{1pt}\omega,\alpha+\beta}\otimes{\cal S}^{\zeta+\eta}_{\hat{\alpha}+\hat{\beta}}
 -(-1)^{\epsilon+\eta\cdot\hat{\alpha}}(\delta_{\omega,\tau}-(-1)^{\sigma}\delta_{\omega,\tau+\beta}){\cal L}^{\hspace{1pt}1+\epsilon+\sigma}_{\hspace{1pt}\omega+\alpha,\alpha+\beta}\otimes{\cal S}^{\zeta+\eta}_{\hat{\alpha}+\hat{\beta}}$
 as $\eta\cdot\hat{\alpha}+\zeta\cdot\hat{\beta}=1$,
 where
 $[{\cal L}^{\hspace{1pt}\epsilon}_{\hspace{1pt}\omega,\alpha},{\cal L}^{\hspace{1pt}\sigma}_{\hspace{1pt}\tau,\beta}]$
 and
 $\{{\cal L}^{\hspace{1pt}\epsilon}_{\hspace{1pt}\omega,\alpha},{\cal L}^{\hspace{1pt}\sigma}_{\hspace{1pt}\tau,\beta}\}$
 are the commutator and the anti-commutator of
 the $\lambda$-generators of $su(N/2^r)$, {\em cf.} Eqs.~B.6 and~B.7 in~\cite{Su}.
 The linking among the member subspaces
 within $\{\mathcal{P}_{\mathcal{Q}}(\mathfrak{B}^{[r]}_{can})\}$ is thus derived that
 the commutator of the two generators
 ${\cal L}^{\hspace{1pt}\epsilon}_{\hspace{1pt}\omega,\alpha}\otimes{\cal S}^{\zeta}_{\hat{\alpha}}
 \in{W}^{\epsilon}(\mathfrak{B}_{\alpha},\mathfrak{B}^{[r]}_{can};\zeta)$
 and
 ${\cal L}^{\hspace{1pt}\sigma}_{\hspace{1pt}\tau,\beta}\otimes{\cal S}^{\eta}_{\hat{\beta}}
 \in{W}^{\sigma}(\mathfrak{B}_{\beta},\mathfrak{B}^{[r]}_{can};\eta)$
 produces the third generator
 belonging to
 ${W}^{\epsilon+\sigma}(\mathfrak{B}_{\alpha+\beta},\mathfrak{B}^{[r]}_{can};\zeta+\eta)$,
 which revives the commutation relation 
 \begin{align}\label{eqncommcan}
 [{W}^{\epsilon}(\mathfrak{B}_{\alpha},\mathfrak{B}^{[r]}_{can};\zeta),
 {W}^{\sigma}(\mathfrak{B}_{\beta},\mathfrak{B}^{[r]}_{can};\eta)]\subset
 {W}^{\epsilon+\sigma}(\mathfrak{B}_{\alpha+\beta},\mathfrak{B}^{[r]}_{can};\zeta+\eta).
 \end{align}
 This relation fulfills the goal that the properties of the conjugate partition and
 the condition of closure withstand the removing process applied to such partition of $su(N)$.
 The isomorphism of the quotient-algebra partitions of the same rank for
 $su(2^p)$ and $su(N)$ is therefore affirmed.
 \end{proof}
 \vspace{6pt}
 Examples of partitions are provided in from Figs.~\ref{figsu5QArank0intr} to \ref{figsu7QArank0inC} for
 $su(N)$, $4<N<8$,
 and in from Figs.~\ref{figsu12canQArank1} to \ref{figsu12cancoQArank2} for $su(12)$.

 Compiling Theorem~\ref{thmQAcoQA} with Lemma~\ref{lemQArankforsuN},
 Main Theorem concludes the existence of quotient and co-quotient algebras
 determined within quotient-algebra partitions of $su(N)$.
\paragraph{Main Theorem}
 {\em Every Lie algebra $su(N)$ admits structures of quotient and co-quotient algebras up to rank $r_0$
 and its quotient and co-quotient algebras of rank $r$
 are respectively isomorphic to those of $su(2^p)$,
 where $2^{p-1}<N\leq 2^p$, $0\leq r\leq r_0\leq p$, and the dimension has the factorization $N=2^{r_0}N'$
 with $N'$ being an odd integer.}

 \section{QAP Preserving}\label{secQAPPerserve}
  This section focuses on QAP preserving transformations.
 \vspace{6pt}
 \begin{lemma}\label{lemS-rot}
  An $s$-rotation ${\cal R}^{\zeta}_{\alpha}(\theta)=e^{i\theta(-i)^{\zeta\cdot\alpha}{\cal S}^{\zeta}_{\alpha}}\in{SU(2^p)}$
  of a spinor $(-i)^{\zeta\cdot\alpha}{\cal S}^{\zeta}_{\alpha}$
  has the expression
 \begin{align}\label{s-rots}
 e^{i\theta(-i)^{\zeta\cdot\alpha}{\cal S}^{\zeta}_{\alpha}}
 =cos\theta\hspace{1pt}{\cal S}^{\mathbf{0}}_{\mathbf{0}}+isin\theta\hspace{1pt}(-i)^{\zeta\cdot\alpha}{\cal S}^{\zeta}_{\alpha}
 \end{align}
  with the identity ${\cal S}^{\mathbf{0}}_{\mathbf{0}}\in{su(2^n)}$, $0\leq\theta <2\pi$.
 \end{lemma}
 \vspace{2pt}
 \begin{proof}
  Refer to~\cite{Su}
  for the derivation.
 \end{proof}
 \vspace{6pt}
 The $s$-rotation
 $e^{i\theta(-i)^{\zeta\cdot\alpha}{\cal S}^{\zeta}_{\alpha}}$
 is an exponential transformation of $(-i)^{\zeta\cdot\alpha}{\cal S}^{\zeta}_{\alpha}$,
 a rotation acting on the space of $p$-qubit states,
 albeit not a rotation about the axis along this spinor.
 Remind that a spinor serves as an algebraic generator
 and also a group action.
 \vspace{6pt}
 \begin{lemma}\label{lemS-rotStoSMap}
 An $s$-rotation ${\cal R}^{\zeta}_{\alpha}(\theta)\in{SU(2^p)}$
 is a spinor-to-spinor mapping as
 $\theta=\pm\frac{\pi}{2},\pm\frac{\pi}{4}$.
 \end{lemma}
 \vspace{2pt}
 \begin{proof}
 By applying ${\cal R}^{\zeta}_{\alpha}(\theta)$ to a spinor $(-i)^{\eta\cdot\beta}{\cal S}^{\eta}_{\beta}$,
  it obtains
  \begin{align}\label{spintospinTrans}
  &{\cal R}^{\zeta\dag}_{\alpha}(\theta)\hspace{2pt}(-i)^{\eta\cdot\beta}{\cal S}^{\eta}_{\beta}\hspace{2pt}{\cal R}^{\zeta}_{\alpha}(\theta)\notag\\
  =&\hspace{10pt}(cos^2\theta+(-1)^{\eta\cdot\alpha+\zeta\cdot\beta}sin^2\theta)\hspace{2pt}(-i)^{\eta\cdot\beta}{\cal S}^{\eta}_{\beta}\notag\\
   &+\frac{i}{2}(-1)^{\zeta\cdot\beta}(-i)^{\zeta\cdot\alpha+\eta\cdot\beta}
    sin2\theta\hspace{1pt}(1-(-1)^{\eta\cdot\alpha+\zeta\cdot\beta}){\cal  S}^{\zeta+\eta}_{\alpha+\beta}.
 \end{align}
 For the detailed derivation, refer to~\cite{Su}.
 The operator remains invariant if $\eta\cdot\alpha+\zeta\cdot\beta=0,2$,
 that is,
 \begin{align}\label{spintospinTransComm}
  \hspace{-9pt}{\cal R}^{\zeta\dag}_{\alpha}(\theta)\hspace{2pt}(-i)^{\eta\cdot\beta}{\cal S}^{\eta}_{\beta}\hspace{2pt}{\cal R}^{\zeta}_{\alpha}(\theta)
 =(-i)^{\eta\cdot\beta}{\cal S}^{\eta}_{\beta}.
 \end{align}
 If $\eta\cdot\alpha+\zeta\cdot\beta=1,3$,
  this spinor is transformed into
 \begin{numcases}{({\cal R}^{\zeta}_{\alpha}(\theta)^{\dag}
 \hspace{2pt}(-i)^{\eta\cdot\beta}{\cal S}^{\eta}_{\beta}\hspace{2pt}
 {\cal R}^{\zeta}_{\alpha}(\theta)=}
  \hspace{46pt}-(-i)^{\eta\cdot\beta}\hspace{1pt}{\cal S}^{\eta}_{\beta}
  & as  \hspace{6pt}$\theta=\pm\frac{\pi}{2}$; \label{eqsRotpi/2}\\
  \pm \varrho\cdot(-i)^{(\zeta+\eta)\cdot(\alpha+\beta)}{\cal S}^{\zeta+\eta}_{\alpha+\beta}
  &  as $\hspace{6pt}\theta=\pm\frac{\pi}{4}$,\label{eqsRotpi/4}
\end{numcases}
 the coefficient
 $\varrho=i(-1)^{\zeta\cdot\beta}(-i)^{\zeta\cdot\alpha+\eta\cdot\beta}(i)^{(\zeta+\eta)\cdot(\alpha+\beta)}=\pm 1$.
 \end{proof}
 \vspace{6pt}
 The hermiticity
 is maintained by the arithmetic that
 the inner product $\zeta\cdot\alpha$ counts the number of
 the 1-qubit component ${\cal S}^1_1$ occurring in
 an $n$-qubit spinor ${\cal S}^{\zeta}_{\alpha}$,
 and the accumulated exponent $\zeta\cdot\alpha$
 of a phase $(\pm i)^{\zeta\cdot\alpha}$  is modulo $4$.

  Given every unitary action being recursively factorized into a composition of $s$-rotations
  according to the $KAK$ theorem~\cite{Su,SuTsai3},
  the computational universality is realized by selecting a set of  $s$-rotations.
 \vspace{6pt}
 \begin{lemma}\label{Eachs-rotfactorzable}
  An $n$-qubit $s$-rotation ${\cal R}^{\zeta}_{\alpha}(\theta)\in{SU(2^p)}$
  with an angle $\theta\in[0,2\pi]$, $n\leq p$, is factorizable into a
  composition of a $1$-qubit $s$-rotation with the same anlge
  and a serial of spinor-to-spinor $s$-rotations of qubits $< n$.
 \end{lemma}
 \vspace{2pt}
 \begin{proof}
  This lemma is proved
  by repeatedly applying Eq.~\ref{eqsRotpi/4} to
  ${\cal R}^{\zeta}_{\alpha}(\theta)=e^{i\theta(-i)^{\zeta\cdot\alpha}{\cal S}^{\zeta}_{\alpha}}$.
  Explicitly, there deduces
  \begin{align}\label{eqs-rotFatcorrepeat}
  {\cal R}^{\zeta}_{\alpha}(\theta)
  =({\cal R}_1{\cal R}_2\cdots{\cal R}_m)^{\dag}{\cal R}^{\eta}_{\beta}(\theta)({\cal R}_1{\cal R}_2\cdots{\cal R}_m),
  \end{align}
  here
  ${\cal R}_h={\cal R}^{\zeta_h}_{\alpha_h}(\pm\frac{\pi}{4})=e^{i\frac{\pi}{4}(-i)^{\zeta_h\cdot\alpha_h}{\cal S}^{\zeta_h}_{\alpha_h}}$
  being an $s$-rotation of qubit $<n$,
  $[{\cal S}^{\zeta_1}_{\alpha_1},{\cal S}^{\zeta}_{\alpha}]\neq 0$,
  $[{\cal S}^{\zeta_{h}}_{\alpha_{h}},{\cal S}^{\zeta+\zeta_1+\cdots+\zeta_{h-1}}_{\alpha+\alpha_1+\cdots+\alpha_{h-1}}]\neq 0$,
  $1< h\leq m< p$,
  and ${\cal R}^{\eta}_{\beta}(\theta)$ being of $1$ qubit
  with $\eta=\zeta+\zeta_1+\cdots+\zeta_m$
  and $\beta=\alpha+\alpha_1+\cdots+\alpha_m$.
  Specifically,
  it allows that each ${\cal R}_h$ is a $2$-qubit
  $s$-rotation.
 \end{proof}
 \vspace{6pt}
   \vspace{6pt}
 \begin{thm}\label{computeuniversalS-rot}
  Every unitary action in $SU(2^p)$ can be written in a
  composition of a serial of $1$-qubit 
  and spinor-to-spinor $s$-rotations.
 \end{thm}
 \vspace{2pt}
 \begin{proof}
  Since every unitary action is a composition
  of $s$-rotations by the $KAK$ theorem~\cite{Su,SuTsai3},
  this theorem is affirmed based on Lemma~\ref{Eachs-rotfactorzable}.
 \end{proof}
 \vspace{6pt}

  \vspace{6pt}
 \begin{lemma}\label{QpreserveBiadd}
  Every spinor-to-spinor mapping preserves the bi-addition
  respecting the relation
  \begin{align}\label{biaddsRelation}
  Q\hspace{1pt}(-i)^{(\xi+\mu)\cdot(\gamma+\nu)}{\cal S}^{\xi+\mu}_{\gamma+\nu}\hspace{1pt}Q^{\dag}
  =\sigma\cdot(-i)^{(\bar{\xi}+\bar{\mu})\cdot(\bar{\gamma}+\bar{\nu})}{\cal S}^{\bar{\xi}+\bar{\mu}}_{\bar{\gamma}+\bar{\nu}},
  \end{align}
  $\sigma=\pm 1$,
  for two spinors
  $(-i)^{\xi\cdot\gamma}{\cal S}^{\xi}_{\gamma}$ and $(-i)^{\mu\cdot\nu}{\cal S}^{\mu}_{\nu}\in{su(2^n)}$,
  and the pair
  $(-i)^{\bar{\xi}\cdot\bar{\gamma}}{\cal S}^{\bar{\xi}}_{\bar{\gamma}}=Q(-i)^{\xi\cdot\gamma}{\cal S}^{\xi}_{\gamma}Q^{\dag}$
  and
  $(-i)^{\bar{\mu}\cdot\bar{\nu}}{\cal S}^{\bar{\mu}}_{\bar{\nu}}=Q(-i)^{\mu\cdot\nu}{\cal S}^{\mu}_{\nu}Q^{\dag}$
  transformed by a spinor-to-spinor mapping $Q\in{SU(2^p)}$.
 \end{lemma}
 \vspace{2pt}
 \begin{proof}
  This lemma is verified by equaling the two identities
 \begin{align}\label{biaddsBeta}
   &\hspace{0pt}Q(-i)^{\xi\cdot\gamma}{\cal S}^{\xi}_{\gamma}Q^{\dag}\cdot Q(-i)^{\mu\cdot\nu}{\cal S}^{\mu}_{\nu}Q^{\dag}\notag\\
  =& Q(-i)^{\xi\cdot\gamma}{\cal S}^{\xi}_{\gamma}\cdot(-i)^{\mu\cdot\nu}{\cal S}^{\mu}_{\nu}Q^{\dag}\notag\\
  =&\hspace{0pt}(-1)^{\mu\cdot\gamma}(-i)^{\xi\cdot\gamma+\mu\cdot\nu}(i)^{(\xi+\mu)\cdot(\gamma+\nu)}
               Q(-i)^{(\xi+\mu)\cdot(\gamma+\nu)}{\cal S}^{\xi+\mu}_{\gamma+\nu}Q^{\dag}
  \end{align}
  and
  \begin{align}\label{biaddsAlpha}
   \hspace{0pt}&Q(-i)^{\xi\cdot\gamma}{\cal S}^{\xi}_{\gamma}Q^{\dag}\cdot Q(-i)^{\mu\cdot\nu}{\cal S}^{\mu}_{\nu}Q^{\dag}\notag\\
  =\hspace{0pt}&(-i)^{\bar{\xi}\cdot\bar{\gamma}}{\cal S}^{\bar{\xi}}_{\bar{\gamma}}
                \cdot(-i)^{\bar{\mu}\cdot\bar{\nu}}{\cal S}^{\bar{\mu}}_{\bar{\nu}}\notag\\
  =\hspace{0pt}&(-1)^{\bar{\mu}\cdot\bar{\gamma}}(-i)^{\bar{\xi}\cdot\bar{\gamma}+\bar{\mu}\cdot\bar{\nu}}(i)^{(\bar{\xi}+\bar{\mu})\cdot(\bar{\gamma}+\bar{\nu})}
    (-i)^{(\bar{\xi}+\bar{\mu})\cdot(\bar{\gamma}+\bar{\nu})}{\cal S}^{\bar{\xi}+\bar{\mu}}_{\bar{\gamma}+\bar{\nu}},
  \end{align}
  $\sigma=(-1)^{\mu\cdot\gamma+\bar{\mu}\cdot\bar{\gamma}}
  (i)^{\xi\cdot\gamma+\mu\cdot\nu}(-i)^{(\xi+\mu)\cdot(\gamma+\nu)}
  (-i)^{\bar{\xi}\cdot\bar{\gamma}+\bar{\mu}\cdot\bar{\nu}}
  (i)^{(\bar{\xi}+\bar{\mu})\cdot(\bar{\gamma}+\bar{\nu})}=\pm 1$.
 \end{proof}
 \vspace{6pt}
  \vspace{6pt}
 \begin{thm}\label{QpreserveQAP}
 Every spinor-to-spinor mapping is QAP preserving.
 \end{thm}
 \vspace{2pt}
 \begin{proof}
 Attributed to the bi-addition preserving asserted in Lemma~\ref{QpreserveBiadd},
 a spinor-to-spinor mapping is coset preserving and upholds the structure of bi-subalgebra partition.
 Commuting with the bi-subalgebra $\mathfrak{B}^{[r]}$,
 every conditioned subspace
 $W^{\epsilon}(\mathfrak{B},\mathfrak{B}^{[r]};i)$
 with $\mathfrak{B}\supset\mathfrak{B}^{[r]}$
 in the partition generated by $\mathfrak{B}^{[r]}$ is maintained by this transformation,
 {\em cf.} Theorem~\ref{thmQAcoQA}.
 Moreover, the unitarity of the transformation protects the same commutation relation of every pair of spinors.
 Hence, the QAP structure is preserved against a spinor-to-spinor mapping.
 \end{proof}
 \vspace{6pt}

  The succeeding assertion details the procedure of
  constructing spinor-to-spinor mappings for the QAP transformation.
\vspace{6pt}
 \begin{thm}\label{s-to-smap2nindepspins}
  Given two ordered sets of $2p$ independent spinors
  $\mathbf{S}_1=\{ (-i)^{\xi_u\cdot\gamma_u}{\cal S}^{\xi_u}_{\gamma_u} \}$ and
  $\mathbf{S}_2=\{ (-i)^{\mu_v\cdot\nu_v}{\cal S}^{\mu_v}_{\nu_v} \}\subset{su(2^p)}$
  sharing the identical commutation relations
  $\xi_u\cdot\gamma_v+\xi_v\cdot\gamma_u=\mu_u\cdot\nu_v+\mu_v\cdot\nu_u$
  for the $u$-th spinor
  ${\cal S}^{\xi_u}_{\gamma_u}$
  and  the $v$-th ${\cal S}^{\xi_v}_{\gamma_v}$
  in $\mathbf{S}_1$
  as well as
  ${\cal S}^{\mu_u}_{\nu_u}$
  and
  ${\cal S}^{\mu_v}_{\nu_v}$
  in $\mathbf{S}_2$,
  there exists a spinor-to-spinor transformation $Q\in{SU(2^p)}$
  mapping the $p$-th member in $\mathbf{S}_1$
  to her counterpart in $\mathbf{S}_2$,
  i.e.,
  $(-i)^{\mu_u\cdot\nu_u}{\cal S}^{\mu_u}_{\nu_u}
  =Q^{\dag}(-i)^{\xi_u\cdot\gamma_u}{\cal S}^{\xi_u}_{\gamma_u}Q$,
  $u,v=1,2,\cdots,2p$.
 \end{thm}
 \vspace{2pt}
 \begin{proof}
  The key is to alter spinors in the ordered set  $\mathbf{S}_1$
  one by one but keep the preceding members invariant.
  In brief,
  after applications of the first $u-1$ evolutions $Q_r$ of the sequential mapping $Q=Q_1Q_2\cdots Q_{2p}$,
  $1\leq r<u\leq 2p$,
  by which
  the $r$-th member ${\cal S}^{\xi_r}_{\gamma_r}\in\mathbf{S}_1$
  is converted into ${\cal S}^{\mu_r}_{\nu_r}\in\mathbf{S}_2$
  and
  the $u$-th
  ${\cal S}^{\xi_u}_{\gamma_u}\in\mathbf{S}_1$ into $\pm{\cal S}^{\hspace{2pt}\iota_u}_{\varpi_u}$,
  the $u$-th operation $Q_u$ maps $\pm{\cal S}^{\hspace{2pt}\iota_u}_{\varpi_u}$
  to ${\cal S}^{\mu_u}_{\nu_u}\in\mathbf{S}_2$ and safeguards the preceding ${\cal S}^{\mu_r}_{\nu_r}$.
  Being unitary and formed in $s$-rotations ${\cal R}^{\xi}_{\gamma}(\theta)$
  as of Eqs.~\ref{eqsRotpi/2} and~\ref{eqsRotpi/4},
  every operation $Q_u$
  preserves the commutation relations
  among the $2p$ spinors respectively in
  $\mathbf{S}_1$ and $\mathbf{S}_2$.
  For convenience without confusion, ${\cal S}^{\xi}_{\gamma}$
  denotes a hermitian spinor by ignoring the phase $(-i)^{\xi\cdot\gamma}$.

  To build $Q_u$,
  three occasions are considered.
  The pair of spinors
  ${\cal S}^{\hspace{2pt}\iota_u}_{\varpi_u}$ and ${\cal S}^{\mu_u}_{\nu_u}$
  are identical apart from a sign $\pm 1$ on the 1st occasion,
  anticommuting secondly,
  and on the 3rd commuting but unequal.
  As $u=1$,
  the linear equations $\varsigma_u\cdot\nu_r+\mu_r\cdot\tau_u=0$
  of the parity constraint are of no effect because of the absence of preceding ${\cal S}^{\mu_r}_{\nu_r}$.

  On the 1st occasion,
  there have a number $2^{2p-u}$ of solutions
  $Q_u={\cal R}^{\varsigma_u}_{\tau_u}(\frac{\pi}{2})$
  under the condition consisting of a number $p$ of linearly independent equations of parity constraint I
  $\varsigma_u\cdot\nu_u+\mu_u\cdot\tau_u=1$ and $\varsigma_u\cdot\nu_r+\mu_r\cdot\tau_u=0$,
  such that ${\cal S}^{\mu_u}_{\nu_u}=Q^{\dag}_u(-{\cal S}^{\mu_u}_{\nu_u})Q_u$
  and ${\cal S}^{\mu_r}_{\nu_r}=Q^{\dag}_p{\cal S}^{\mu_r}_{\nu_r}Q_p$ according to Eq.~\ref{eqsRotpi/2},
  $1\leq r<u\leq 2n$.
  On the 2nd occasion,
  the operation $Q_u={\cal R}^{\iota_u+\mu_u}_{\varpi_u+\nu_u}(\pm\frac{\pi}{4})$
  turns $\pm{\cal S}^{\hspace{2pt}\iota_u}_{\varpi_u}$ into ${\cal S}^{\mu_u}_{\nu_u}$
  by Eq.~\ref{eqsRotpi/4} with an appropriate angle $\pm\frac{\pi}{4}$
  and retains the preceding spinors ${\cal S}^{\mu_r}_{\nu_r}$
  given the conditions
  $\mu_r\cdot(\varpi_u+\nu_u)+(\iota_u+\mu_u)\cdot\nu_r=0$
  granted from the assumption of identical commutation relations
  $\mu_r\cdot\varpi_u+\iota_u\cdot\nu_r
  =\mu_r\cdot\nu_u+\mu_u\cdot\nu_r$.

  Now, come to $Q_{u}=Q_{u1}Q_{u2}$ on the 3rd occasion.
  There exist a number $2^{2p-u-1}$ of candidates
  $Q_{u1}={\cal R}^{\varsigma_u}_{\tau_u}(\frac{\pi}{4})$
  as in Eq.~\ref{eqsRotpi/4},
  each of which transforms $\pm{\cal S}^{\hspace{2pt}\iota_u}_{\varpi_u}$ to $\pm{\cal S}^{\iota_u+\varsigma_u}_{\varpi_u+\tau_u}$
  and leaves the preceding ${\cal S}^{\mu_r}_{\nu_r}$ unamended
  thanks to a number $u+1$ of linearly independent equations of parity constraint II
   \hspace{2pt}$\varsigma_p\cdot\varpi_p+\iota_p\cdot\tau_p=1$, $\varsigma_p\cdot\nu_p+\mu_p\cdot\tau_p=1$
  and $\varsigma_p\cdot\nu_r+\mu_r\cdot\tau_p=0$, $1\leq r<u\leq 2n$.
  Successively,
  the spinor
  $\pm{\cal S}^{\iota_u+\varsigma_u}_{\varpi_u+\tau_u}$
  is mapped to ${\cal S}^{\mu_u}_{\nu_u}$
  via
  $Q_{u2}={\cal R}^{\iota_u+\varsigma_u+\mu_u}_{\varpi_u+\tau_u+\nu_u}(\pm\frac{\pi}{4})$
  with an appropriate angle $\pm\frac{\pi}{4}$
  due to
  $\mu_u\cdot(\varpi_u+\tau_u+\nu_u)+(\iota_u+\varsigma_u+\mu_u)\cdot\nu_u=1$.
  Since
  $\mu_r\cdot(\varpi_u+\tau_u+\nu_u)+(\iota_u+\varsigma_u+\mu_u)\cdot\nu_r=0$,
  the preceding ${\cal S}^{\mu_r}_{\nu_r}$ remain intact.

  At the $(2p-1)$-th step, only the $3$rd occasion needs to be concerned, $u=2p-1\equiv  w$.
  Although there are in total $2p$ linear equations subject to the parity constraint II,
  one of them is derivable from the other independent $2p-1$.
  To assert this fact, it requires showing that
  ${\cal S}^{\iota_w}_{\varpi_w}$ belongs
  to the bi-subalgebra $\mathcal{B}$
  spanned by the $2p-1$ independent spinors
  ${\cal S}^{\mu_{r'}}_{\nu_{r'}}$, $1\leq r'< w$, and ${\cal S}^{\mu_w}_{\nu_w}$.
  Assume that ${\cal S}^{\iota_w}_{\varpi_w}\notin\mathcal{B}$.
  The identities
  $(\iota_w+\mu_w)\cdot\nu_{r'}+\mu_{r'}\cdot(\varpi_w+\nu_w)=0$
  are ascribed to the commutation relations
  $\iota_w\cdot\nu_{r'}+\mu_{r'}\cdot\varpi_w=\mu_w\cdot\nu_{r'}+\mu_{r'}\cdot\nu_w$.
  Next, in view of the commuting of
  ${\cal S}^{\iota_w}_{\varpi_w}$ and ${\cal S}^{\mu_w}_{\nu_w}$,
  there attain the two identities
  $(\iota_w+\mu_w)\cdot\varpi_w+\iota_w\cdot(\varpi_w+\nu_w)=0$
  and
  $(\iota_w+\mu_w)\cdot\nu_w+\mu_w\cdot(\varpi_w+\nu_w)=0$.
  Accordingly, these identities induce the implication that
  the bi-additive ${\cal S}^{\iota_w+\mu_w}_{\varpi_w+\nu_w}$
  commutes with the $2n$ independent spinors
  $\{  {\cal S}^{\mu_{r'}}_{\nu_{r'}},{\cal S}^{\mu_w}_{\nu_w},{\cal S}^{\iota_w}_{\varpi_w}:
  1\leq r'< w\}$,
  namely
  ${\cal S}^{\iota_w+\mu_w}_{\varpi_w+\nu_w}=\pm{\cal S}^{\mathbf{0}}_{\mathbf{0}}$,
  which contradicts the assumption ${\cal S}^{\iota_w}_{\varpi_w}\neq \pm{\cal S}^{\mu_w}_{\nu_w}$.
  This affirms ${\cal S}^{\iota_w}_{\varpi_w}\in\mathcal{B}$
  and the fact that the equation
  $\varsigma_w\cdot\varpi_w+\iota_w\cdot\tau_w=1$
  of constraint II can be derived from the other independent $2p-1$.
  As a result, the evolution $Q_{w1}$ of $Q_{w}=Q_{w1}Q_{w2}$ has two solutions.

  Finally, at step $2p$,
  the operation $Q_{2p}$ on the first two occasions respectively is produced
  through the same procedure as at the $u$-th step.
  On the 3rd occasion,
  the identities
  $(\iota_{2p}+\nu_{2p})\cdot\nu_{r''}+\mu_{r''}\cdot(\varpi_{2p}+\nu_{2p})=0$,
  $1\leq r''<2p$,
  are earned from the commutation relations
  $\iota_{2p}\cdot\nu_{r''}+\mu_{r''}\cdot\varpi_{2p}
  =\mu_{2p}\cdot\nu_{r''}+\mu_{r''}\cdot\nu_{2p}$.
  In addition, the assumption of the commuting
  ${\cal S}^{\iota_{2p}}_{\varpi_{2p}}$ and ${\cal  S}^{\mu_{2p}}_{\nu_{2p}}$
  yields the identity
  $(\iota_{2p}+\nu_{2p})\cdot\nu_{2p}+\mu_{2p}\cdot(\varpi_{2p}+\nu_{2p})=0$.
  Likewise, these $2p$ identities
  lead to the fact
  ${\cal S}^{\iota_{2p}+\mu_{2p}}_{\varpi_{2p}+\nu_{2p}}=\pm{\cal S}^{\mathbf{0}}_{\mathbf{0}}$.
  That is, the 3rd occasion reduces to the 1st.
  \end{proof}
 \vspace{6pt}
  \vspace{6pt}
 \begin{cor}\label{2nSindinnkC}
  Given an ordered set of $p-r$ independent spinors
  $\mathbf{S}_{\mathfrak{B}^{[r]}}=\{ {\cal S}^{\xi_g}_{\gamma_g}:g=1,2,\cdots,p-r \}$
  from a bi-subalgebra
  $\mathfrak{B}^{[r]}\subset{su(2^p)}$,
  there exist a number $k$ of spinors
  $\{ {\cal S}^{\xi_m}_{\gamma_m}:m=p-r+1,\cdots,p \}$
  in $su(2^p)-\mathfrak{B}^{[r]}$
  to span a Cartan subalgebra $\mathfrak{C}$ with $\mathbf{S}_{\mathfrak{B}^{[r]}}$,
  and a number $p$ of ${\cal S}^{\mu_t}_{\nu_t}\in{su(2^p)-\mathfrak{C}}$
  are independent under the bi-addition iff
  the associated strings
  $\omega_t=\epsilon_{t1}\epsilon_{t2}\cdots\epsilon_{tp}$
  are independent under the bitwise addition,
  $\epsilon_{ts}=\mu_t\cdot\gamma_s+\xi_s\cdot\nu_t\in{Z_2}$
  being the parities of ${\cal S}^{\mu_t}_{\nu_t}$
  and the former $p$ members ${\cal S}^{\xi_s}_{\gamma_s}$,
  $s,t=1,2,\cdots, p$.
 \end{cor}
 \vspace{2pt}
 \begin{proof}
   By solving a number $n-k$ of independent linear equations
  of parities
  $\xi_m\cdot\gamma_{g}+\xi_{g}\cdot\gamma_m=0$ for all
  ${\cal S}^{\xi_g}_{\gamma_g}\in\mathbf{S}_{\mathfrak{B}^{[r]}}$,
  $1\leq g\leq p-r$ and $p-r+1\leq m\leq p$,
  there exist multiple solutions of $r$ independent spinors
  ${\cal S}^{\xi_m}_{\gamma_m}$
  to span a Cartan subalgebra $\mathfrak{C}$
  with $\mathbf{S}_{\mathfrak{B}^{[r]}}$,
  referring to~\cite{Su,SuTsai1}.
  It is evident that every bi-additive ${\cal S}^{\mu_t+\mu_h}_{\nu_t+\nu_h}$
  of ${\cal S}^{\mu_t}_{\nu_t}$ and ${\cal S}^{\mu_h}_{\nu_h}$
  corresponds to the bitwise addition
  $\omega_t+\omega_h$ of two strings $\omega_t$ and $\omega_h$,
  $1\leq t,h\leq p$.
  Thus, the implication is validated that the set of latter $p$ spinors
  is independent iff so is the set of their associated strings.

  A preferred selection of an ordered set of $2n$ independent spinors contains all ${\cal S}^{\xi_s}_{\gamma_s}$
  and the 2nd half ${\cal S}^{\mu_t}_{\nu_t}$
  composed of $k$ members from the same number of independent cosets in
  $\Gamma_{\mathbf{0}}-{\cal C}$ and $p-r$ independent members
  respectively from distinct blocks $\Gamma_{\tau\neq\mathbf{0}}$.
  A such ordered set in the partition
  $\{\mathcal{P}_{{\cal Q}}(\mathfrak{B}^{[r]}_{intr})\}$
  generated by the intrinsic bi-subalgebra
  $\mathfrak{B}^{[r]}_{intr}$
  is suggested as
  \begin{align}\label{int2nIndpSpin}
  \hat{\mathbf{S}}
  =\{ & \hat{S}_{\hspace{1pt}l}={\cal S}^{\zeta_l}_{\mathbf{0}}\otimes{\cal S}^{\mathbf{0}}_{\mathbf{0}},
      \hspace{2pt}
      \hat{S}_{p-r+u}={\cal S}^{\mathbf{0}}_{\mathbf{0}}\otimes{\cal S}^{\varsigma_u}_{\mathbf{0}},
      \hspace{2pt}
      \hat{S}_{p+u}={\cal S}^{\mathbf{0}}_{\mathbf{0}}\otimes{\cal S}^{\mathbf{0}}_{\kappa_u}, \notag\\
      &\hat{S}_{p+r+l}={\cal S}^{\mathbf{0}}_{\tau_l}\otimes{\cal S}^{\mathbf{0}}_{\mathbf{0}}:
      1\leq l\leq p-r,1\leq u\leq r \},
  \end{align}
  here $\zeta_{\hspace{1pt}l}=\varrho_{l,1}\varrho_{l,2}\cdots\varrho_{l,p-r}\in{Z^{p-r}_2}$
  obeying $\varrho_{\hspace{1pt}l,l'}=\delta_{\hspace{1pt}ll'}$
  for $1\leq l'\leq p-r$,
  $\varsigma_u=\iota_{u,1}\iota_{u,2}\cdots\iota_{u,r}\in{Z^r_2}$ fulfilling $\iota_{u,u'}=\delta_{uu'}$ for $1\leq u'\leq r$,
  $\kappa_{u}=\varsigma_{u-r+1}$
  and $\tau_l=\zeta_{p-r-l+1}$.
  The first $n-k$ independent members $\hat{S}_{\hspace{1pt}l}$ are picked from $\hat{{\cal C}}$,
  the $2r$ $\hat{S}_{p-r+u}$ and $\hat{S}_{p+u}$ from independent cosets in
  $\hat{\Gamma}_{\mathbf{0}}-\hat{{\cal C}}$,
  and the rest $p-r$ $\hat{S}_{p+r+l}$ from independent blocks $\Gamma_{\tau_l\neq\mathbf{0}}$.
  In this set, the first $p$ spinors span the intrinsic Cartan subalgebra of $su(2^p)$.
  As to the last $p$ members,
  each $\hat{S}_{p+u}$ is associated with the string
  $\omega_u=\epsilon_1\epsilon_2\cdots\epsilon_p\in{Z^p_2}$
  carrying $\epsilon_{p-u+1}=1$ and $\epsilon_h=0$ if $h\neq p-u+1$,
  and every $\hat{S}_{p+r+l}$ of the string
  $\omega_{r+l}=\varrho_1\varrho_2\cdots\varrho_p$
  possessing $\varrho_{p-r-l+1}=1$ and $\varrho_h=0$ if $h\neq p-r-l+1$.
 \end{proof}
 \vspace{6pt}
  Maneuvering a QAP transformation begins with the assignment
 $\mathbf{S}_2=\hat{\mathbf{S}}$ of the suggested set
 of Eq.~\ref{int2nIndpSpin} in the intrinsic coordinate.
 Let the other ordered set $\mathbf{S}_1$ of $2p$ independent spinors
 that enjoys the identical commutation relations as those of $\mathbf{S}_2$
 and respects the independence condition of
 Corollary~\ref{2nSindinnkC}
 be drawn from the partition $\{\mathcal{P}_{{\cal Q}}(\mathfrak{B}^{[r]})\}$
 by solving linear equations of parities of
 Theorem~\ref{s-to-smap2nindepspins},
 referring to the proof of
 Corollary~\ref{QcmapsEncoding} for
 its existence.
 According to Theorem~\ref{s-to-smap2nindepspins} again,
 there provide a number $2p$ of sequential spinor-to-spinor
 operations
 $\{Q^{\dag}_{h}:h=1,2,\cdots, 2p\}$
 recursively changing each member of $\mathbf{S}_1$
 to her opposite in $\mathbf{S}_2$,
 and thus mapping $\{\mathcal{P}_{{\cal Q}}(\mathfrak{B}^{[r]})\}$
 to $\{\mathcal{P}_{{\cal Q}}(\mathfrak{B}^{[r]}_{intr})\}$.

 Nonetheless,
 it is always achievable to transform
 $\{\mathcal{P}_{{\cal Q}}(\mathfrak{B}^{[r]}_{intr})\}$
 into
 $\{\mathcal{P}_{{\cal Q}}(\mathfrak{B}^{[r]})\}$
 by simply converting $\mathfrak{B}^{[r]}_{intr}$ to $\mathfrak{B}^{[r]}$,
 namely applying only the last $p-r$ evolutions $Q_{p-r}$, $\cdots$, $Q_{1}$ of the sequential
 operations $\{Q_{\hat{h}}:\hat{h}=2p,2p-1,\cdots,1\}$.
 \vspace{6pt}
 \begin{cor}\label{QcmapsEncoding}
 Transforming the partition
 $\{ \mathcal{P}_{{\cal Q}}(\mathfrak{B}^{[r]}_{intr}) \}$
 into
 $\{ \mathcal{P}_{{\cal Q}}(\mathfrak{B}^{[r]}) \}$
 is achievable by a spinor-to-spinor mapping
 $Q_{{\cal B}}\in{SU(2^p)}$
 that converts the intrinsic bi-subalgebra
 $\mathfrak{B}^{[r]}_{intr}\subset{su(2^p)}$
 to $\mathfrak{B}^{[r]}=Q_{{\cal B}}\mathfrak{B}^{[r]}_{intr}Q^{\dag}_{{\cal B}}$.
 \end{cor}
 \vspace{2pt}
 \begin{proof}
 A major clue is attributed to the truth that
 a partition is fully determined by its generating bi-subalgebra.
 With the designation of
 the ordered set
 $\hat{\mathbf{S}}_{\mathfrak{B}^{[r]}_{intr}}
 =\{ {\cal S}^{\zeta_h}_{\mathbf{0}}\otimes{\cal S}^{\mathbf{0}}_{\mathbf{0}}:
 h=1,2,\cdots,p-r \}$
 from $\mathfrak{B}^{[r]}_{intr}$,
 the mapping $Q_{{\cal B}}$ transforms
 $\{ \mathcal{P}_{{\cal Q}}(\mathfrak{B}^{[r]}_{intr}) \}$
 into
 $\{ \mathcal{P}_{{\cal Q}}(\mathfrak{B}^{[r]}) \}$
 by converting
 $\mathfrak{B}^{[r]}_{intr}$
 to
 $\mathfrak{B}^{[r]}$,
 specifically by Theorem~\ref{s-to-smap2nindepspins} transmuting
 $\hat{\mathbf{S}}_{\mathfrak{B}^{[r]}_{intr}}$
 of $\mathfrak{B}^{[r]}_{intr}$ spinor by spinor
 into an independent ordered set $\mathbf{S}_{\mathfrak{B}^{[r]}}$ of
 $\mathfrak{B}^{[r]}$ as in Corollary~\ref{2nSindinnkC}.
 The immense freedom of selecting $\mathbf{S}_{\mathfrak{B}^{[r]}}$
 from $\mathfrak{B}^{[r]}$
 gives a combinatorially large number of all versions of such mappings
 ${\cal N}=\prod^{p-r-1}_{l=0}(2^{p-r}-2^{l})$.
 Meanwhile, there have at least the number ${\cal N}$ of candidates
 of an independent ordered set $\mathbf{S}_1$
 sharing the identical commutation relations
 with those of $\hat{\mathbf{S}}$,
 {\em cf.} Corollary~\ref{2nSindinnkC},
 for each of which is easily acquired by mapping $\hat{\mathbf{S}}$ to the partition
 $\{ \mathcal{P}_{{\cal Q}}(\mathfrak{B}^{[r]}) \}$
 via a transformation $Q_{{\cal B}}$.
 \end{proof}
 \vspace{6pt}

 \vspace{6pt}
 \begin{lemma}\label{CartanClsoedS-rot}
 A Cartan subalgebra of the $k$-th kind in $su(2^p)$,
 $k=1,2,\cdots,p-1$,
 can be transformed into another Cartan subalgebra of the same kind,
 the $(k-1)$-th kind or the $(k+1)$-th with a spinor-to-spinor $s$-rotation
 ${\cal R}^{\eta}_{\beta}(\pm\frac{\pi}{4})\in{SU(2^p)}$.
 \end{lemma}
 \vspace{2pt}
 \begin{proof}
 Assume a Cartan subalgebra of the $k$-th kind
 $\mathfrak{C}_k
 =\{{\cal S}^{\zeta_i}_{\alpha_i}:0\leq i<2^k,\alpha_i\in [\alpha]_k\text{ and }\zeta_i\in [\zeta]_q \}$
 and an $s$-rotation
 ${\cal R}^{\eta}_{\beta}(\pm\frac{\pi}{4})=e^{\pm\frac{\pi}{4}(-i)^{\eta\cdot\beta}{\cal S}^{\eta}_{\beta}}$
 obeying $[{\cal S}^{\eta}_{\beta},\mathfrak{C}_k]\neq 0$,
 $[\alpha]_k$ being a $k$-th maximal subgroup
 and $[\zeta]_q$ a coset of a $q$-th maximal subgroup of $Z^p_2$,
 $1\leq k,q\leq p$~\cite{SuTsai1}.
 Each spinor ${\cal S}^{\zeta}_{\alpha}\in\mathfrak{C}_k$
 not commuting with ${\cal S}^{\eta}_{\beta}$
 is mapped into ${\cal S}^{\zeta+\eta}_{\alpha+\beta}$ via ${\cal R}^{\eta}_{\beta}(\pm\frac{\pi}{4})$.

 One of three circumstances occurs for the bit string $\alpha+\beta$ of the transformed spinor
 ${\cal S}^{\zeta+\eta}_{\alpha+\beta}$ that
 $\alpha+\beta=\alpha$ if $\beta=\mathbf{0}$,
 $\alpha+\beta\in[\alpha]_{k-1}$ if $\beta\in[\alpha]_k$,
 and
 $\alpha+\beta\in[\alpha]_{k+1}$ if $\beta\notin[\alpha]_k$.
 That is,
 $\mathfrak{C}_k$
 is mapped into another Cartan subalgebra of the same kind
 in the 1st circumstance,
 into a $(k-1)$-th kind in the 2nd,
 and thirdly into a $(k+1)$-th kind.
 \end{proof}
 \vspace{6pt}
 \vspace{6pt}
 \begin{lemma}\label{k+1kindExistkkind}
 For every Cartan subalgebra of the $(k+1)$-th kind
 $\mathfrak{C}_{k+1}\subset{su(2^p)}$,
 $k=0,1,\cdots,p-1$,
 there exists a spinor-to-spinor $s$-rotation
 ${\cal R}^{\eta}_{\beta}(\theta)\in{SU(2^p)}$
 mapping $\mathfrak{C}_{k+1}$ into a $k$-th kind
 $\mathfrak{C}_k
 =({\cal R}^{\eta}_{\beta}(\theta))^{\dag}
 \mathfrak{C}_{k+1}
 {\cal R}^{\eta}_{\beta}(\theta)$,
 $\theta=\pm\frac{\pi}{4}$.
 \end{lemma}
 \vspace{2pt}
 \begin{proof}
 This lemma is affirmed by choosing an $s$-rotation
 ${\cal R}^{\eta}_{\beta}(\pm\frac{\pi}{4})=e^{\pm\frac{\pi}{4}(-i)^{\eta\cdot\beta}{\cal S}^{\eta}_{\beta}}$
 fulfilling $\beta\in[\alpha]_{k+1}$ and
 $[{\cal S}^{\eta}_{\beta},\mathfrak{C}_{k+1}]\neq 0$,
 {\em cf.} Lemma~\ref{CartanClsoedS-rot},
 $[\alpha]_{k+1}\subset{Z^p_2}$ being the group of bit strings in
 $\mathfrak{C}_{k+1}
 =\{{\cal S}^{\zeta_i}_{\alpha_i}:0\leq i<2^{k+1},\alpha_i\in [\alpha]_{k+1}\text{ and }\zeta_i\in [\zeta]_q \}$~\cite{SuTsai1}.
 That is, a Cartan subalgebra of the
 $k$-th kind is elicited from the intrinsic
 $\mathfrak{C}_{[\mathbf{0}]}$
 by a transformation composed of spinor-to-spinor $s$-rotations.
 \end{proof}
 \vspace{6pt}
  \vspace{6pt}
 \begin{lemma}\label{Bomb=SubalgExt}
  Every Cartan subalgebra derived in the
  subalgebra extension
  is obtainable through a spinor-to-spinor mapping over the intrinsic Cartan subalgebra
  $\mathfrak{C}_{[\mathbf{0}]}$.
 \end{lemma}
 \vspace{2pt}
 \begin{proof}
 This corollary is a direct consequence of
 Lemma~\ref{k+1kindExistkkind}.
 Also with Lemma~\ref{CartanClsoedS-rot},
 this has an immediate implication that
 the set of Cartan subalgebras generated in the subalgebra extension
 is identical to that obtained via exhaustive spinor-to-spinor mappings on $\mathfrak{C}_{[\mathbf{0}]}$.
 \end{proof}
 \vspace{6pt}
   \vspace{6pt}
 \begin{thm}\label{Alladmissble-Partitions}
 Every quotient algebra partition of rank $r$
 is derivable from the partition of the same rank generated by the intrinsic bi-subalgebra
 $\mathfrak{B}^{[r]}_{intr}$ with a spinor-to-spinor mapping.
 \end{thm}
 \vspace{2pt}
 \begin{proof}
 The key of the proof is to assert that every $r$-th maximal bi-subalgebra
 of a $k$-th-kind Cartan subalgebra $\mathfrak{C}\subset{su(2^p)}$
 is acquired from $\mathfrak{B}^{[r]}_{intr}$ through a spinor-to-spinor mapping.
 Since the intrinsic Cartan subalgebra
 $\mathfrak{C}_{[\mathbf{0}]}$
 is isomorphic to $Z^p_2$~\cite{SuTsai1},
 there find all $r$-th maximal bi-subalgebras of
 $\mathfrak{C}_{[\mathbf{0}]}$
 by exhaustively searching $r$-th maximal subgroups of $Z^p_2$.
 For each $r$-th maximal bi-subalgebra
 $\hat{\mathfrak{B}}^{[r]}$ in $\mathfrak{C}_{[\mathbf{0}]}$,
 there exists a spinor-to-spinor mapping
 $\hat{Q}$
 transforming
 $\mathfrak{B}^{[r]}_{intr}$
 into
 $\hat{\mathfrak{B}}^{[r]}=\hat{Q}\mathfrak{B}^{[r]}_{intr}\hat{Q}^{\dag}$,
 {\em cf.} the same procedure in
 Theorem~\ref{s-to-smap2nindepspins}.
 By applying a spinor-to-spinor transformation
 $Q$ as in Lemma~\ref{k+1kindExistkkind}
 from $\mathfrak{C}_{[\mathbf{0}]}$
 into $\mathfrak{C}=Q\mathfrak{C}_{[\mathbf{0}]}Q^{\dag}$,
 the subalgebra
 $\hat{\mathfrak{B}}^{[r]}$ is altered to an $r$-th maximal
 bi-subalgebra $\mathfrak{B}^{[r]}=Q\hat{\mathfrak{B}}^{[r]}Q^{\dag}$ of
 $\mathfrak{C}$.
 The operation $Q\hat{Q}$ is thus the required mapping from
 $\mathfrak{B}^{[r]}_{intr}$ to $\mathfrak{B}^{[r]}$.
 Therefore,
 every $r$-th maximal bi-subalgebra of $\mathfrak{C}$
 is attainable from $\mathfrak{B}^{[r]}_{intr}$ via a spinor-to-spinor mapping.
 \end{proof}
 \vspace{6pt}

 \section{Types of Cartan Decompositions}\label{sectypesC}
 \renewcommand{\theequation}{\arabic{section}.\arabic{equation}}
\setcounter{equation}{0} \noindent
  A Cartan decomposition of the Lie algebra $su(N)$ is referred to as a
  composition $su(N)=\mathfrak{t}\oplus\mathfrak{p}$, where the
  subalgebra $\mathfrak{t}$ and vector subspace $\mathfrak{p}$
  satisfy {\em the decomposition condition}
  $[\mathfrak{t},\mathfrak{t}]\subset\mathfrak{t}$,
  $[\mathfrak{t},\mathfrak{p}]\subset\mathfrak{p}$,
  $[\mathfrak{p},\mathfrak{p}]\subset\mathfrak{t}$ and
  ${\rm Tr}\{\mathfrak{t}\mathfrak{p}\}=0$.
  Based on the decomposition condition, two choices of
  {\em anti-commutation relations} for
  $\mathfrak{t}\oplus\mathfrak{p}$ can be discovered, which leads to two kinds of Cartan decompositions
  in $su(N)$.
 \vspace{6pt}
 \begin{lemma}\label{lemtpanticomm}
  According to  anti-commutation relations, 
  Cartan decompositions of the Lie algebra $su(2^p)$ 
  divide into two kinds, 
  the top Cartan decomposition $su(2^p)=\mathfrak{t}^{\top}\oplus\mathfrak{p}^{\top}$
  whose subalgebra $\mathfrak{t}^{\top}$ and subspace $\mathfrak{p}^{\top}$
  satisfy the anti-commutation relations
  $\{\mathfrak{t}^{\top},\mathfrak{t}^{\top}\}\subset\mathfrak{t}^{\top}$,
  $\{\mathfrak{t}^{\top},\mathfrak{p}^{\top}\}\subset\mathfrak{p}^{\top}$
  and
  $\{\mathfrak{p}^{\top},\mathfrak{p}^{\top}\}\subset\mathfrak{t}^{\top}$,
  and the bottom Cartan decomposition $su(N)=\mathfrak{t}^{\bot}\oplus\mathfrak{p}^{\bot}$
  whose $\mathfrak{t}^{\bot}$ and $\mathfrak{p}^{\bot}$
  hold the relations
  $\{\mathfrak{t}^{\bot},\mathfrak{t}^{\bot}\}\subset\mathfrak{p}^{\bot}$,
  $\{\mathfrak{t}^{\bot},\mathfrak{p}^{\bot}\}\subset\mathfrak{t}^{\bot}$
  and
  $\{\mathfrak{p}^{\bot},\mathfrak{p}^{\bot}\}\subset\mathfrak{p}^{\bot}$.
 \end{lemma}
 \vspace{3pt}
 \begin{proof}
  This lemma will be affirmed by verifying the inclusion relations in
  the following order.
  Firstly, a Cartan decomposition
  $\mathfrak{t}\oplus\mathfrak{p}$ admits
  the two choices of anti-commutator relations either $\{\mathfrak{p},\mathfrak{p}\}\subset\mathfrak{t}$
  or $\{\mathfrak{p},\mathfrak{p}\}\subset\mathfrak{p}$.
  The former inclusion corresponds to the top Cartan decomposition
  $\mathfrak{t}^{\top}\oplus\mathfrak{p}^{\top}$
  and the latter to the bottom decomposition
  $\mathfrak{t}^{\bot}\oplus\mathfrak{p}^{\bot}$.
  Given  $\{\mathfrak{p}^{\top},\mathfrak{p}^{\top}\}\subset\mathfrak{t}^{\top}$
  (or $\{\mathfrak{p}^{\bot},\mathfrak{p}^{\bot}\}\subset\mathfrak{p}^{\bot}$), there acquires
  the secondary relation $\{\mathfrak{t}^{\top},\mathfrak{p}^{\top}\}\subset\mathfrak{p}^{\top}$
  (or $\{\mathfrak{t}^{\bot},\mathfrak{p}^{\bot}\}\subset\mathfrak{t}^{\bot}$).
  Thirdly, these two assertions
  $\{\mathfrak{p}^{\top},\mathfrak{p}^{\top}\}\subset\mathfrak{t}^{\top}$
  and $\{\mathfrak{t}^{\top},\mathfrak{p}^{\top}\}\subset\mathfrak{p}^{\top}$
  (or $\{\mathfrak{p}^{\bot},\mathfrak{p}^{\bot}\}\subset\mathfrak{p}^{\bot}$
  and
  $\{\mathfrak{t}^{\bot},\mathfrak{p}^{\bot}\}\subset\mathfrak{t}^{\bot}$)
  leads to the last inclusion $\{\mathfrak{t}^{\top},\mathfrak{t}^{\top}\}\subset\mathfrak{t}^{\top}$
  (or $\{\mathfrak{t}^{\bot},\mathfrak{t}^{\bot}\}\subset\mathfrak{p}^{\bot}$).

  The 1st inclusion relation is proved by contradiction.
  Thus, assume that
  $\{\mathfrak{p},\mathfrak{p}\}\nsubseteq\mathfrak{t}$
  and
  $\{\mathfrak{p},\mathfrak{p}\}\nsubseteq\mathfrak{p}$.
  By this negation, suppose
  ${\cal S}^{\zeta_1+\zeta_2}_{\alpha_1+\alpha_2}\in\mathfrak{p}$
  and ${\cal S}^{\zeta_3+\zeta_4}_{\alpha_3+\alpha_4}\in\mathfrak{t}$
  for four spinors ${\cal S}^{\zeta_1}_{\alpha_1},{\cal S}^{\zeta_2}_{\alpha_2},{\cal S}^{\zeta_3}_{\alpha_3}$
  and ${\cal S}^{\zeta_4}_{\alpha_4}\in\mathfrak{p}$ with 
  $[{\cal S}^{\zeta_1}_{\alpha_1},{\cal S}^{\zeta_2}_{\alpha_2}]=
  [{\cal S}^{\zeta_3}_{\alpha_3},{\cal S}^{\zeta_4}_{\alpha_4}]=0$.
  Consider only the case
  $[{\cal S}^{\zeta_1}_{\alpha_1},{\cal S}^{\zeta_3}_{\alpha_3}]\neq 0$,
  $[{\cal S}^{\zeta_2}_{\alpha_2},{\cal S}^{\zeta_4}_{\alpha_4}]\neq 0$
  and
  $[{\cal S}^{\zeta_1}_{\alpha_1},{\cal S}^{\zeta_4}_{\alpha_4}]
  =[{\cal S}^{\zeta_2}_{\alpha_2},{\cal S}^{\zeta_3}_{\alpha_3}]=0$,
  because contradictions will be reached as well in the other cases
  of commutator assumptions.
  Due to the nonvanishing commutators
  $[{\cal S}^{\zeta_1}_{\alpha_1},{\cal S}^{\zeta_3}_{\alpha_3}]\neq 0$
  and
  $[{\cal S}^{\zeta_2}_{\alpha_2},{\cal S}^{\zeta_4}_{\alpha_4}]\neq 0$
  together with the rule of decomposition condition
  $[\mathfrak{p},\mathfrak{p}]\subset\mathfrak{t}$,
  both the bi-additives
  ${\cal S}^{\zeta_1+\zeta_3}_{\alpha_1+\alpha_3}$
  and
  ${\cal S}^{\zeta_2+\zeta_4}_{\alpha_2+\alpha_4}$
  should belong to $\mathfrak{t}$.
  The subspace $\mathfrak{p}$ contains the bi-additive
  ${\cal S}^{\zeta_1+\zeta_4}_{\alpha_1+\alpha_4}$
  owing to the nonvanishing commutator
  $[{\cal S}^{\zeta_1+\zeta_2}_{\alpha_1+\alpha_2},{\cal S}^{\zeta_2+\zeta_4}_{\alpha_2+\alpha_4}]\neq 0$
  and the rule $[\mathfrak{t},\mathfrak{p}]\subset\mathfrak{p}$.
  Then, by the commutator
  $[{\cal S}^{\zeta_1+\zeta_4}_{\alpha_1+\alpha_4},{\cal S}^{\zeta_2+\zeta_4}_{\alpha_2+\alpha_4}]\neq 0$
  and the rule $[\mathfrak{t},\mathfrak{p}]\subset\mathfrak{p}$,
  the bi-additive ${\cal S}^{\zeta_3+\zeta_4}_{\alpha_3+\alpha_4}$
  is turned into a spinor in $\mathfrak{p}$, which is fallacious. 
  Therefore, either choice of the inclusions
  $\{\mathfrak{p},\mathfrak{p}\}=\{\mathfrak{p}^{\top},\mathfrak{p}^{\top}\}\subset\mathfrak{t}^{\top}$
  and
  $\{\mathfrak{p},\mathfrak{p}\}=\{\mathfrak{p}^{\bot},\mathfrak{p}^{\bot}\}\subset\mathfrak{p}^{\bot}$
  is permitted. 
  In the rest of the proof, since similar arguments are applicable to the
  bottom $\mathfrak{t}^{\bot}\oplus\mathfrak{p}^{\bot}$,
  only instances for the top decomposition
  $\mathfrak{t}^{\top}\oplus\mathfrak{p}^{\top}$
  will be examined.

  Confirming the 2nd relation is equivalent to
  validating the assertion
  ${\cal S}^{\zeta'+\eta'}_{\alpha'+\beta'}\in\mathfrak{p}^{\top}$
  for all ${\cal S}^{\zeta'}_{\alpha'}\in\mathfrak{p}^{\top}$
  and ${\cal S}^{\eta'}_{\beta'}\in\mathfrak{t}^{\top}$.
  For two noncommuting spinors, {\em i.e.},
  $[{\cal S}^{\zeta'}_{\alpha'},{\cal S}^{\eta'}_{\beta'}]\neq 0$,
  this is obvious by the rule of the decomposition condition
  $[\mathfrak{t}^{\top},\mathfrak{p}^{\top}]\subset\mathfrak{p}^{\top}$.
  When the two spinors commute, 
  the identity ${\cal S}^{\bf 0}_{\hspace{.5pt}\bf 0}$ is
  in $\mathfrak{t}^{\top}$ through the proved anti-commutator relation
  $\{\mathfrak{p}^{\top},\mathfrak{p}^{\top}\}\subset\mathfrak{t}^{\top}$, and thus
  the discussion divides into for
  the two cases ${\cal S}^{\eta'}_{\beta'}={\cal S}^{\bf 0}_{\hspace{.5pt}\bf 0}$
  and
  ${\cal S}^{\eta'}_{\beta'}\neq{\cal S}^{\bf 0}_{\hspace{.5pt}\bf 0}$.
  The inclusion ${\cal S}^{\zeta'+\eta'}_{\alpha'+\beta'}\in\mathfrak{p}^{\top}$
  is apparently true in the former case.
  As ${\cal S}^{\eta'}_{\beta'}\neq{\cal S}^{\bf 0}_{\hspace{.5pt}\bf 0}$,
  it is sufficient to confine the attention to $su(2^p)$ with $p>1$.
  Given the Lie algebra $su(2^p)$ being nonabelian, $p>1$,
  there exists at least one spinor ${\cal S}^{\xi}_{\gamma}$
  commuting with neither of ${\cal S}^{\zeta'}_{\alpha'}$
  and ${\cal S}^{\eta'}_{\beta'}$, which is allowed by the identities
  $\xi\cdot\alpha'+\zeta'\cdot\gamma=\xi\cdot\beta'+\eta'\cdot\gamma=1$.
  Suppose that ${\cal S}^{\xi}_{\gamma}\in\mathfrak{p}^{\top}$;
  a same result can be obtained for
  ${\cal S}^{\xi}_{\gamma}\in\mathfrak{t}^{\top}$.
  The nonvanishing commutator
  $[{\cal S}^{\zeta'}_{\alpha'},{\cal S}^{\xi}_{\gamma}]\neq 0$
  produces the bi-additive ${\cal S}^{\zeta'+\xi}_{\alpha'+\gamma}$
  absorbed in $\mathfrak{t}^{\top}$ by the rule of the decomposition condition
  $[\mathfrak{p}^{\top},\mathfrak{p}^{\top}]\subset\mathfrak{t}^{\top}$,
  while the commutator $[{\cal S}^{\eta'}_{\beta'},{\cal S}^{\xi}_{\gamma}]\neq 0$
  creates the bi-additive ${\cal S}^{\eta'+\xi}_{\beta'+\gamma}$
  belonging to $\mathfrak{p}^{\top}$
  by the rule
  $[\mathfrak{t}^{\top},\mathfrak{p}^{\top}]\subset\mathfrak{p}^{\top}$.
  Then, following the rule
  $[\mathfrak{t}^{\top},\mathfrak{p}^{\top}]\subset\mathfrak{p}^{\top}$
  again,
  the subspace $\mathfrak{p}^{\top}$ covers the bi-additive
  ${\cal S}^{\zeta'+\eta'}_{\alpha'+\beta'}
  ={\cal S}^{\zeta'+\xi+\eta'+\xi}_{\alpha'+\gamma+\beta'+\gamma}$
  of the two spinors
  ${\cal S}^{\zeta'+\xi}_{\alpha'+\gamma}\in\mathfrak{t}^{\top}$
  and
  ${\cal S}^{\eta'+\xi}_{\beta'+\gamma}\in\mathfrak{p}^{\top}$.
  Thus, the 2nd inclusion
  $\{\mathfrak{t}^{\top},\mathfrak{p}^{\top}\}\subset\mathfrak{p}^{\top}$
  is asserted.

  The final relation
  is to be affirmed by showing the inclusion
  ${\cal S}^{\eta_1+\eta_2}_{\beta_1+\beta_2}\in\mathfrak{t}^{\top}$
  for all
  ${\cal S}^{\eta_1}_{\beta_1}$ and ${\cal S}^{\eta_2}_{\beta_2}\in\mathfrak{t}^{\top}$.
  Similarly, given $[{\cal S}^{\eta_1}_{\beta_1},{\cal S}^{\eta_2}_{\beta_2}]\neq 0$,
  the inclusion ${\cal S}^{\eta_1+\eta_2}_{\beta_1+\beta_2}\in\mathfrak{t}^{\top}$
  simply reflects the fact of $\mathfrak{t}^{\top}$ being a subalgebra.
  To consider the case as the two spinors commute,
  let an additional spinor ${\cal S}^{\xi'}_{\gamma'}\in\mathfrak{p}$
  be taken arbitrarily.
  Only the instance is assumed that
  $[{\cal S}^{\eta_1}_{\beta_1},{\cal S}^{\xi'}_{\gamma'}]= 0$ and
  $[{\cal S}^{\eta_2}_{\beta_2},{\cal S}^{\xi'}_{\gamma'}]\neq 0$;
  a same outcome is acquirable in the other instances.
  Apparently, the bi-additive ${\cal S}^{\eta_1+\xi'}_{\beta_1+\gamma'}$ of 
  ${\cal S}^{\eta_1}_{\beta_1}\in\mathfrak{t}^{\top}$ and ${\cal S}^{\xi'}_{\gamma'}\in\mathfrak{p}^{\top}$
  is embraced in $\mathfrak{p}^{\top}$ by the above anti-commutator inclusion
  $\{\mathfrak{t}^{\top},\mathfrak{p}^{\top}\}\subset\mathfrak{p}^{\top}$.
  While guided by the rule 
  $[\mathfrak{t}^{\top},\mathfrak{p}^{\top}]\subset\mathfrak{p}^{\top}$,
  the nonvanishing commutator
  $[{\cal S}^{\eta_2}_{\beta_2},{\cal S}^{\xi'}_{\gamma'}]\neq 0$
  yields the bi-additive ${\cal S}^{\eta_2+\xi'}_{\beta_2+\gamma'}$
  also belonging to $\mathfrak{p}^{\top}$. 
  Because of the commutators 
  $[{\cal S}^{\eta_1}_{\beta_1},{\cal S}^{\eta_2}_{\beta_2}]
  =[{\cal S}^{\eta_1}_{\beta_1},{\cal S}^{\xi'}_{\gamma'}]=0$
  and
  $[{\cal S}^{\eta_2}_{\beta_2},{\cal S}^{\xi'}_{\gamma'}]\neq 0$,
  the two bi-additives
  ${\cal S}^{\eta_1+\xi'}_{\beta_1+\gamma'}$
  and
  ${\cal S}^{\eta_2+\xi'}_{\beta_2+\gamma'}$
  do not commute.
  Then, based on the inclusions
  ${\cal S}^{\eta_1+\xi'}_{\beta_1+\gamma'}\in\mathfrak{p}^{\top}$
  and
  ${\cal S}^{\eta_2+\xi'}_{\beta_2+\gamma'}\in\mathfrak{p}^{\top}$
  together with the rule 
  $[\mathfrak{p}^{\top},\mathfrak{p}^{\top}]\subset\mathfrak{t}^{\top}$,
  it is concluded that
  the subalgebra $\mathfrak{t}^{\top}$ contains the bi-additive
  ${\cal S}^{\eta_1+\eta_2}_{\beta_1+\beta_2}$.
  The proof ends.
 \end{proof}
 \vspace{6pt}
  The two associated subspaces
  $\mathfrak{p}^{\top}=su(N)-\mathfrak{t}^{\top}$
  and $\mathfrak{p}^{\bot}=su(N)-\mathfrak{t}^{\bot}$
  have separate types of maximal abelian subalgebras.
 \vspace{6pt}
 \begin{lemma}\label{lem3kindsmaxabelinp}
  In the Lie algebra $su(N)$, $2^{p-1}<N\leq 2^p$,
  the maximal abelian subalgebra of the subspace $\mathfrak{p}^{\bot}$ of a bottom Cartan decomposition
  $su(N)=\mathfrak{t}^{\bot}\oplus\mathfrak{p}^{\bot}$
  is a Cartan subalgebra $\mathfrak{C}\subset{su(N)}$ or
  a proper maximal bi-subalgebra $\mathfrak{B}$ of $\mathfrak{C}$,
  whereas the maximal abelian subalgebra of the subspace $\mathfrak{p}^{\top}$ of a top Cartan decomposition
  $su(N)=\mathfrak{t}^{\top}\oplus\mathfrak{p}^{\top}$
  is the complement $\mathfrak{B}^c=\mathfrak{C}-\mathfrak{B}$.
 \end{lemma}
 \vspace{3pt}
 \begin{proof}
  Let $\mathfrak{a}^{\top}$ (or $\mathfrak{a}^{\bot}$)
  be a maximal abelian subalgebra of the subspace $\mathfrak{p}^{\top}$
  (or $\mathfrak{p}^{\bot}$)
  and be a subset of a Cartan subalgebra $\mathfrak{C}^{\top}$
  (or $\mathfrak{C}^{\bot}$).
  Notice that the subalgebra $\mathfrak{a}^{\top}$
  (or $\mathfrak{a}^{\bot}$)
  is non-null if so is $\mathfrak{p}^{\top}$
  (or $\mathfrak{p}^{\bot}$).
  In addition, the complement $\mathfrak{C}^{\top}-\mathfrak{a}^{\top}$
  (or $\mathfrak{C}^{\bot}-\mathfrak{a}^{\bot}$)
  is a subspace of $\mathfrak{t}^{\top}$ (or $\mathfrak{t}^{\bot}$);
  it falsifies the assumption of $\mathfrak{a}^{\top}$ (or $\mathfrak{a}^{\bot}$)
  being a maximal abelian subalgebra of $\mathfrak{p}^{\top}$ (or $\mathfrak{p}^{\bot}$) otherwise.

  The assertion that
  $\mathfrak{a}^{\bot}$
  is a maximal bi-subalgebra of $\mathfrak{C}^{\bot}$
  can be validated via the inclusions
  ${\cal S}^{\zeta_1+\zeta_2}_{\alpha_1+\alpha_2}$
  and
  ${\cal S}^{\hat{\zeta}_1+\hat{\zeta}_2}_{\hat{\alpha}_1+\hat{\alpha}_2}
  \in\mathfrak{a}^{\bot}$
  for all
  ${\cal S}^{\zeta_1}_{\alpha_1},{\cal S}^{\zeta_2}_{\alpha_2}\in\mathfrak{a}^{\bot}\subset\mathfrak{p}^{\bot}$
  and
  ${\cal S}^{\hat{\zeta}_1}_{\hat{\alpha}_1},{\cal S}^{\hat{\zeta}_2}_{\hat{\alpha}_2}
  \in\mathfrak{C}^{\bot}-\mathfrak{a}^{\bot}\subset\mathfrak{t}^{\bot}$.
  By Lemma~\ref{lemtpanticomm}, both the bi-additives
  ${\cal S}^{\zeta_1+\zeta_2}_{\alpha_1+\alpha_2}$
  and
  ${\cal S}^{\hat{\zeta}_1+\hat{\zeta}_2}_{\hat{\alpha}_1+\hat{\alpha}_2}$
  are contained in $\mathfrak{p}^{\bot}$.
  Then, thanks to the vanishing commutators
  $[{\cal S}^{\zeta_1+\zeta_2}_{\alpha_1+\alpha_2},\mathfrak{a}^{\bot}]
  =[{\cal S}^{\hat{\zeta}_1+\hat{\zeta}_2}_{\hat{\alpha}_1+\hat{\alpha}_2},\mathfrak{a}^{\bot}]=0$,
  the subalgebra $\mathfrak{a}^{\bot}$ earns these two
  bi-additives.
  The feature is implied that the subalgebra
  $\mathfrak{a}^{\bot}$
  is a maximal bi-subalgebra of $\mathfrak{C}^{\bot}$,
  which is either the Cartan subalgebra $\mathfrak{C}^{\bot}$
  or a proper maximal bi-subalgebra $\mathfrak{B}^{\bot}$ of $\mathfrak{C}^{\bot}$.

  The rest is to confirm that
  a maximal abelian subalgebra $\mathfrak{a}^{\top}\subset\mathfrak{C}^{\top}$
  of $\mathfrak{p}^{\top}$ is the complement
  of a proper maximal bi-subalgebra of the Cartan subalgebra $\mathfrak{C}^{\top}$,
  or equivalently  the subspace
  $\mathfrak{J}=\mathfrak{C}^{\top}-\mathfrak{a}^{\top}\subset\mathfrak{t}^{\top}$
  is a proper maximal bi-subalgebra of $\mathfrak{C}^{\top}$.
  This is reachable by verifying the inclusions
  ${\cal S}^{\eta_1+\eta_2}_{\beta_1+\beta_2}$
  and ${\cal S}^{\hat{\eta}_1+\hat{\eta}_2}_{\hat{\beta}_1+\hat{\beta}_2}\in\mathfrak{J}$
  for all 
  ${\cal S}^{\eta_1}_{\beta_1},{\cal S}^{\eta_2}_{\beta_2}\in\mathfrak{J}
  \subset\mathfrak{t}^{\top}$
  and
  ${\cal S}^{\hat{\eta}_1}_{\hat{\beta}_1},{\cal S}^{\hat{\eta}_2}_{\hat{\beta}_2}
  \in\mathfrak{a}^{\top}\subset\mathfrak{p}^{\top}$.
  Since ${\cal S}^{\eta_i}_{\beta_i}$ and
  ${\cal S}^{\hat{\eta}_i}_{\hat{\beta}_i}\in\mathfrak{C}^{\top}$,
  $i=1,2$,
  the bi-additives ${\cal S}^{\eta_1+\eta_2}_{\beta_1+\beta_2}$
  and ${\cal S}^{\hat{\eta}_1+\hat{\eta}_2}_{\hat{\beta}_1+\hat{\beta}_2}$
  are in $\mathfrak{C}^{\top}=\mathfrak{J}\cup\mathfrak{a}^{\top}$ too.
  It further arrives at the desired that both ${\cal S}^{\eta_1+\eta_2}_{\beta_1+\beta_2}$ and
  ${\cal S}^{\hat{\eta}_1+\hat{\eta}_2}_{\hat{\beta}_1+\hat{\beta}_2}$
  belong to $\mathfrak{J}$,
  because 
  ${\cal S}^{\eta_1+\eta_2}_{\beta_1+\beta_2}$ and
  ${\cal S}^{\hat{\eta}_1+\hat{\eta}_2}_{\hat{\beta}_1+\hat{\beta}_2}\in\mathfrak{t}^{\top}$
  according to Lemma~\ref{lemtpanticomm}
  and non-null $\mathfrak{a}^{\top}\subset\mathfrak{p}^{\top}$.
  Thus, the subspace
  $\mathfrak{J}=\mathfrak{B}^{\top}\subset\mathfrak{t}^{\top}$
  is a proper maximal bi-subalgebra of $\mathfrak{C}^{\top}$.

  Recall Lemma~14 in~\cite{SuTsai1} that
  the unions
  $\mathfrak{B}^{\top}\cup{W}$
  and
  $\mathfrak{B}^{\top}\cup{\hat{W}}$
  respectively form a Cartan subalgebra in $su(N)$
  for the two conditioned subspaces $W$ and $\hat{W}$ of $\mathfrak{B}^{\top}$.
  Although the proof of the lemma has completed,
  it is worthwhile to point out that the subalgebra $\mathfrak{t}^{\top}$
  is a superset of one of the two unions.
  Let this property be proved by 
  affirming the distribution that one of the two subspaces $W$ and $\hat{W}$
  belongs to $\mathfrak{t}^{\top}$ and the other to $\mathfrak{p}^{\top}$.
  Since respectively being a coset of $\mathfrak{B}^{\top}$
  under the bi-addition by Lemma~13 in~\cite{SuTsai1},
  the two subspaces can take the forms
  $W=\{{\cal S}^{\eta+\xi}_{\beta+\gamma}:\hspace{2pt}{\cal S}^{\eta}_{\beta}\in\mathfrak{B}^{\top}\}$
  and
  $\hat{W}=\{{\cal S}^{\hat{\eta}+\hat{\xi}}_{\hat{\beta}+\hat{\gamma}}:
  \hspace{2pt}{\cal S}^{\hat{\eta}}_{\hat{\beta}}\in\mathfrak{B}^{\top}\}$
  with some ${\cal S}^{\xi}_{\gamma}\in{W}$ and ${\cal S}^{\hat{\xi}}_{\hat{\gamma}}\in\hat{W}$.
  When ${\cal S}^{\xi}_{\gamma}\in\mathfrak{t}^{\top}$
  (or ${\cal S}^{\xi}_{\gamma}\in\mathfrak{p}^{\top}$),
  each bi-additive ${\cal S}^{\eta+\xi}_{\beta+\gamma}\in{W}$
  falls in $\mathfrak{t}^{\top}$ (or $\mathfrak{p}^{\top}$)
  owing to the inclusion
  ${\cal S}^{\eta}_{\beta}\in\mathfrak{B}^{\top}\subset\mathfrak{t}^{\top}$
  and the anti-commutation relation
  $\{\mathfrak{t}^{\top},\mathfrak{t}^{\top}\}\subset\mathfrak{t}^{\top}$
  (or $\{\mathfrak{t}^{\top},\mathfrak{p}^{\top}\}\subset\mathfrak{p}^{\top}$)
  of Lemma~\ref{lemtpanticomm}.
  That is,
  $W\subset\mathfrak{t}^{\top}$
  (or $W\subset\mathfrak{p}^{\top}$).
  Then, by the consideration of
  $W\subset\mathfrak{t}^{\top}$
  (or $W\subset\mathfrak{p}^{\top}$),
  the subspace $\hat{W}$ can be proved to be in
  $\mathfrak{p}^{\top}$
  (or $\mathfrak{t}^{\top}$).
  According to Lemma~9 in~\cite{SuTsai1},
  the bi-additive ${\cal S}^{\xi+\hat{\xi}}_{\gamma+\hat{\gamma}}$
  of ${\cal S}^{\xi}_{\gamma}\in{W}$ and ${\cal S}^{\hat{\xi}}_{\hat{\gamma}}\in\hat{W}$
  is contained in $\mathfrak{a}^{\top}=\mathfrak{C}^{\top}-\mathfrak{B}^{\top}$
  and thus in $\mathfrak{p}^{\top}$.
  With the noncommuting of ${\cal S}^{\xi}_{\gamma}$
  and ${\cal S}^{\xi+\hat{\xi}}_{\gamma+\hat{\gamma}}$
  derived from the commutator
  $[{\cal S}^{\xi}_{\gamma},{\cal S}^{\hat{\xi}}_{\hat{\gamma}}]\neq 0$
  based on Lemma~10 in~\cite{SuTsai1},
  the bi-additive
  ${\cal S}^{\hat{\xi}}_{\hat{\gamma}}={\cal S}^{\xi+\xi+\hat{\xi}}_{\gamma+\gamma+\hat{\gamma}}$
  of the two commuting spinors is embraced in
  $\mathfrak{p}^{\top}$ (or $\mathfrak{t}^{\top}$)
  if
  ${\cal S}^{\xi}_{\gamma}\in\mathfrak{t}^{\top}$
  (or ${\cal S}^{\xi}_{\gamma}\in\mathfrak{p}^{\top}$)
  through the rule of the decomposition condition
  $[\mathfrak{t}^{\top},\mathfrak{p}^{\top}]\subset\mathfrak{p}^{\top}$
  (or $[\mathfrak{p}^{\top},\mathfrak{p}^{\top}]\subset\mathfrak{t}^{\top}$).
  Hence, when
  $W\subset\mathfrak{t}^{\top}$
  (or $W\subset\mathfrak{p}^{\top}$)
  so that the spinor
  ${\cal S}^{\xi}_{\gamma}\in{W}$
  is in $\mathfrak{t}^{\top}$
  (or $\mathfrak{p}^{\top}$)
  and thus the inclusion
  ${\cal S}^{\hat{\xi}}_{\hat{\gamma}}\in\mathfrak{p}^{\top}$
  (or $\mathfrak{t}^{\top}$)
  holds,
  the subspace
  $\mathfrak{p}^{\top}$
  (or $\mathfrak{t}^{\top}$)
  encompasses all the bi-additives
  ${\cal S}^{\hat{\eta}+\hat{\xi}}_{\hat{\beta}+\hat{\gamma}}\in\hat{W}$
  by virtue of the inclusion
  ${\cal S}^{\hat{\eta}}_{\hat{\beta}}\in\mathfrak{B}^{\top}\subset\mathfrak{t}^{\top}$
  and the relation
  $\{\mathfrak{t}^{\top},\mathfrak{p}^{\top}\}\subset\mathfrak{p}^{\top}$
  (or $\{\mathfrak{t}^{\top},\mathfrak{t}^{\top}\}\subset\mathfrak{t}^{\top}$)
  of Lemma~\ref{lemtpanticomm}.
  This implies that
  $\hat{W}\subset\mathfrak{p}^{\top}$
  (or $\mathfrak{t}^{\top}$)
  if $W\subset\mathfrak{t}^{\top}$
  (or $W\subset\mathfrak{p}^{\top}$).
  Thus, either the distribution holds
  $W\subset\mathfrak{t}^{\top}$ and $\hat{W}\subset\mathfrak{p}^{\top}$
  or the other
  $W\subset\mathfrak{p}^{\top}$
  and
  $\hat{W}\subset\mathfrak{t}^{\top}$. 
 \end{proof}
 \vspace{6pt}
 As will be demonstrated in a later episode~\cite{SuTsai3}, a Cartan decomposition $su(N)=\mathfrak{t}\oplus\mathfrak{p}$
 of the Lie algebra $su(N)$ is considered as the
 {\em $\mathfrak{t}$-$\mathfrak{p}$ decomposition of the $1$st level},
 which is denoted as $\mathfrak{t}\oplus\mathfrak{p}=\mathfrak{t}_{[1]}\oplus\mathfrak{p}_{[1]}$
 for $\mathfrak{t}=\mathfrak{t}_{[1]}$ and
 $\mathfrak{p}=\mathfrak{p}_{[1]}$.
 The subalgebra $\mathfrak{t}_{[1]}=\mathfrak{t}_{[2]}\oplus\mathfrak{p}_{[2]}$
 allows a further Cartan decomposition which is a
 {\em $\mathfrak{t}$-$\mathfrak{p}$ decomposition of the $2$nd level}.
 Generating refined Cartan decompositions in $su(N)$ is a
 recursive procedure that a {\em $\mathfrak{t}$-$\mathfrak{p}$ decomposition of the $l$-th level}
 $\mathfrak{t}_{[l-1]}=\mathfrak{t}_{[l]}\oplus\mathfrak{p}_{[l]}$
 is the subalgebra of an $(l-1)$-th level
 $\mathfrak{t}_{[l-1]}\oplus\mathfrak{p}_{[l-1]}$, here $1\leq l\leq 2p$
 and $su(N)=\mathfrak{t}_{[0]}$ with $\mathfrak{p}_{[0]}=\{0\}$.
 It will be shown that these $\mathfrak{t}$-$\mathfrak{p}$ decompositions of higher levels
 follow the commutation and anti-commutation relations
 similar to the decomposition condition and anti-commutation
 relations as of Lemma~\ref{lemtpanticomm}.
 Furthermore, the maximal abelian subalgebra
 of the subspace $\mathfrak{p}_{[l]}$ of an $l$-th-level
 decomposition $\mathfrak{t}_{[l]}\oplus\mathfrak{p}_{[l]}$
 is either a bi-subalgebra or a coset of a bi-subalgebra,
 {\em cf.} Lemma~\ref{lem3kindsmaxabelinp}.

  Based on the above lemmas,
  the relation between Cartan decompositions and quotient-algebra partitions over $su(N)$
  can be clarified as follows.
 \vspace{6pt}
 \begin{thm}\label{thmsuNCDandQAP}
  The Lie algebra $su(N)$, $2^{p-1}<N\leq 2^p$,
  admits a Cartan decomposition $su(N)=\mathfrak{t}\oplus\mathfrak{p}$ iff, under the tri-addition,
  the subalgebra $\mathfrak{t}$ is a proper maximal subgroup of a quotient-algebra partition over $su(N)$ and
  the subspace $\mathfrak{p}$ is the coset of $\mathfrak{t}$ in this partition.
 \end{thm}
 \vspace{3pt}
 \begin{proof}
  The implication will be affirmed first that,
  given a Cartan decomposition $su(N)=\mathfrak{t}\oplus\mathfrak{p}$,
  there exists a quotient-algebra partition over $su(N)$ such that
  the subalgebra $\mathfrak{t}$ is a proper maximal subgroup of
  the partition under the tri-addition
  and the subspace $\mathfrak{p}$ is the coset of $\mathfrak{t}$
  under the same operation.
  As described in Lemma~\ref{lem3kindsmaxabelinp}, when
  a bottom decomposition
  $\mathfrak{t}\oplus\mathfrak{p}=\mathfrak{t}^{\bot}\oplus\mathfrak{p}^{\bot}$
  is concerned, 
  the maximal abelian subalgebra  subalgebra
  $\mathfrak{p}=\mathfrak{p}^{\bot}$
  is a Cartan subalgebra $\mathfrak{C}'$ or a $1$st maximal bi-subalgebra $\mathfrak{B}'$
  of $\mathfrak{C}'$ and, when
  $\mathfrak{t}\oplus\mathfrak{p}=\mathfrak{t}^{\top}\oplus\mathfrak{p}^{\top}$
  is a top decomposition,
  the maximal abelian subalgebra of $\mathfrak{p}=\mathfrak{p}^{\top}$
  is the complement $\mathfrak{B}^c=\mathfrak{C}-\mathfrak{B}$
  of a $1$st maximal bi-subalgebra $\mathfrak{B}$ in a Cartan subalgebra $\mathfrak{C}$.
  Note that the subalgebra $\mathfrak{t}^{\top}$ encloses
  a Cartan subalgebra $\widetilde{\mathfrak{C}}$ being a superset of $\mathfrak{B}$,
  {\em cf.} the proof of Lemma~\ref{lem3kindsmaxabelinp}.
  Let $\bar{\mathcal{B}}$ be identical to
  $\mathfrak{C}'\subset\mathfrak{p}^{\bot}$ or to $\mathfrak{B}'\subset\mathfrak{p}^{\bot}$
  for the bottom decomposition
  $\mathfrak{t}^{\bot}\oplus\mathfrak{p}^{\bot}$
  or be identical to
  $\widetilde{\mathfrak{C}}\subset\mathfrak{t}^{\top}$
  for the top decomposition
  $\mathfrak{t}^{\top}\oplus\mathfrak{p}^{\top}$.
  It will be manifest that
  the quotient-algebra partition
  $\{\mathcal{P}_{\mathcal{Q}}(\mathfrak{B}^{[r]})\}$,
  generated by a bi-subalgebra $\mathfrak{B}^{[r]}$ of
  $\bar{\mathcal{B}}$,
  is a partition required for the implication.
  Here the subalgebra $\mathfrak{B}^{[r]}$ is at most as large as $\bar{\mathcal{B}}$
  and is an $r$-th maximal bi-subalgebra of
  $\mathfrak{C}'$ when $\mathfrak{t}\oplus\mathfrak{p}=\mathfrak{t}^{\bot}\oplus\mathfrak{p}^{\bot}$
  or of $\widetilde{\mathfrak{C}}$
  when
  $\mathfrak{t}\oplus\mathfrak{p}=\mathfrak{t}^{\top}\oplus\mathfrak{p}^{\top}$.
  Affirming the implication will be achieved by establishing the following three assertions.
  At first, both $\mathfrak{t}$ and $\mathfrak{p}$ are respectively composed of the conditioned subspaces of
  $\{\mathcal{P}_{\mathcal{Q}}(\mathfrak{B}^{[r]})\}$.
  Next,
  $\mathfrak{t}$ and $\mathfrak{p}$ are disjoint in this partition;
  that is,
  the intersection $\mathfrak{t}\cap\mathfrak{p}$
  contains no any conditioned subspace of
  $\{\mathcal{P}_{\mathcal{Q}}(\mathfrak{B}^{[r]})\}$.
  Lastly, the conditioned subspaces of $\mathfrak{t}\oplus\mathfrak{p}$
  satisfy the set of distribution rules, for all
  $\mathfrak{B}_1,\mathfrak{B}_2\in\mathcal{G}(\mathfrak{C})$,
  $\epsilon,\sigma\in{Z_2}$ and $i,j\in{Z^r_2}$,
 \begin{align}\label{eqrulegenW}
  &W^{\epsilon+\sigma}(\mathfrak{B}_1\sqcap\mathfrak{B}_2,\mathfrak{B}^{[r]};i+j)\subset\mathfrak{t}\hspace{1pt},
  \hspace{1pt}\text{ if } \hspace{2pt}W^{\epsilon}(\mathfrak{B}_1,\mathfrak{B}^{[r]};i)\subset\mathfrak{t}
  \hspace{3pt}\text{ and }\hspace{2pt}W^{\sigma}(\mathfrak{B}_2,\mathfrak{B}^{[r]};j)\subset\mathfrak{t};\notag\\
  &W^{\epsilon+\sigma}(\mathfrak{B}_1\sqcap\mathfrak{B}_2,\mathfrak{B}^{[r]};i+j)\subset\mathfrak{p},
  \text{ if } \hspace{2pt}W^{\epsilon}(\mathfrak{B}_1,\mathfrak{B}^{[r]};i)\subset\mathfrak{t}
  \hspace{3pt}\text{ and }\hspace{2pt}W^{\sigma}(\mathfrak{B}_2,\mathfrak{B}^{[r]};j)\subset\mathfrak{p};\notag\\
  &W^{\epsilon+\sigma}(\mathfrak{B}_1\sqcap\mathfrak{B}_2,\mathfrak{B}^{[r]};i+j)\subset\mathfrak{t}\hspace{1pt},
  \hspace{0pt}\text{ if } \hspace{2pt}W^{\epsilon}(\mathfrak{B}_1,\mathfrak{B}^{[r]};i)\subset\mathfrak{p}
  \hspace{2pt}\text{ and }\hspace{2pt}W^{\sigma}(\mathfrak{B}_2,\mathfrak{B}^{[r]};j)\subset\mathfrak{p}.
 \end{align}

  The 1st assertion 
  is to be confirmed by showing that,
  for an arbitrary spinor ${\cal S}^{\zeta}_{\alpha}\in\mathfrak{t}$
  (or $\in\mathfrak{p}$),
  the subalgebra $\mathfrak{t}$ (or $\mathfrak{p}$) is a superset of
  the conditioned subspace
  ${W}^{\epsilon}(\mathfrak{B},\mathfrak{B}^{[r]};i)\in\{\mathcal{P}_{\mathcal{Q}}(\mathfrak{B}^{[r]})\}$
  containing 
  ${\cal S}^{\zeta}_{\alpha}$.
  Here only the case of a top decomposition
  $\mathfrak{t}\oplus\mathfrak{p}=\mathfrak{t}^{\top}\oplus\mathfrak{p}^{\top}$
  is examined;
  the confirmation for a bottom decomposition
  $\mathfrak{t}\oplus\mathfrak{p}=\mathfrak{t}^{\bot}\oplus\mathfrak{p}^{\bot}$
  can be obtained via a similar argument.
  Known from Corollary~\ref{coroanticommclose},
  the bi-additive ${\cal S}^{\zeta+\eta}_{\alpha+\beta}$
  of ${\cal S}^{\zeta}_{\alpha}\in\mathfrak{t}^{\top}$
  (or ${\cal S}^{\zeta}_{\alpha}\in\mathfrak{p}^{\top}$) and another arbitrary spinor
  ${\cal S}^{\eta}_{\beta}\in{W}^\epsilon(\mathfrak{B},\mathfrak{B}^{[r]};i)$
  must belong to the bi-subalgebra
  $\mathfrak{B}^{[r]}={W}^1(\mathfrak{C},\mathfrak{B}^{[r]};\mathbf{0})$.
  Because of the inclusion
  $\mathfrak{B}^{[r]}\subset\widetilde{\mathfrak{C}}\subset\mathfrak{t}^{\top}$,
  the subalgebra $\mathfrak{t}^{\top}$
  owns the bi-additive ${\cal S}^{\zeta+\eta}_{\alpha+\beta}$.
  Besides, the bi-additive ${\cal S}^{\eta}_{\beta}$
  of the two commuting spinors ${\cal S}^{\zeta}_{\alpha}\in\mathfrak{t}^{\top}$
  (or ${\cal S}^{\zeta}_{\alpha}\in\mathfrak{p}^{\top}$)
  and
  ${\cal S}^{\zeta+\eta}_{\alpha+\beta}\in\mathfrak{t}^{\top}$
  is in $\mathfrak{t}^{\top}$ (or $\mathfrak{p}^{\top}$) according to the anti-commutation
  relation $\{\mathfrak{t}^{\top},\mathfrak{t}^{\top}\}\subset\mathfrak{t}^{\top}$
  (or $\{\mathfrak{t}^{\top},\mathfrak{p}^{\top}\}\subset\mathfrak{p}^{\top}$)
  of Lemma~\ref{lemtpanticomm}.
  Therefore, each spinor ${\cal S}^{\eta}_{\beta}\in{W}^{\epsilon}(\mathfrak{B},\mathfrak{B}^{[r]};i)$
  is also in $\mathfrak{t}^{\top}$ (or
  $\mathfrak{p}^{\top}$) if ${\cal S}^{\zeta}_{\alpha}\in\mathfrak{t}^{\top}$
  (or if ${\cal S}^{\zeta}_{\alpha}\in\mathfrak{p}^{\top}$).
  Since every spinor ${\cal S}^{\zeta}_{\alpha}\in\mathfrak{t}$
  (or $\in\mathfrak{p}$) ought to dwell in a conditioned subspaces of
  $\{\mathcal{P}_{\mathcal{Q}}(\mathfrak{B}^{[r]})\}$,
  the first assertion is validated.

  Secondly,
  let the disjoint of the subalgebra $\mathfrak{t}$ and the subspace $\mathfrak{p}$
  be proved by contradiction.
  Assume there exists a conditioned subspace
  ${W}^\epsilon(\mathfrak{B},\mathfrak{B}^{[r]};i)\subset\mathfrak{t}\cap\mathfrak{p}$.
  As the subspace ${W}^\epsilon(\mathfrak{B},\mathfrak{B}^{[r]};i)$ 
  is non-null,
  since both  $\mathfrak{t}$ and  $\mathfrak{p}$
  cover the whole subspace ${W}^\epsilon(\mathfrak{B},\mathfrak{B}^{[r]};i)$
  due to the 1st assertion,
  the intersection $\mathfrak{t}\cap\mathfrak{p}$
  contains at least one spinor, which is fallacious.
  While ${W}^\epsilon(\mathfrak{B},\mathfrak{B}^{[r]};i)=\{0\}$ is null,
  the commutation relation holds
  $[{W}^1(\mathfrak{C},\mathfrak{B}^{[r]};\mathbf{0}),{W}^{\epsilon}(\mathfrak{B},\mathfrak{B}^{[r]};i)]
  \subset{W}^{1+\epsilon}(\mathfrak{B},\mathfrak{B}^{[r]};i)$,
  with the quaternion condition of closure of Eq.~\ref{eqgenWcommrankr},
  for $\mathfrak{B}^{[r]}={W}^1(\mathfrak{C},\mathfrak{B}^{[r]};\mathbf{0})$
  and the non-null conditioned subspace
  ${W}^{1+\epsilon}(\mathfrak{B},\mathfrak{B}^{[r]};i)$,
  {\em cf.} Lemma~\ref{lemcondsub}.
  Guided by the commutation relations
  $[\mathfrak{t},\mathfrak{t}]\subset\mathfrak{t}$
  and
  $[\mathfrak{t},\mathfrak{p}]\subset\mathfrak{p}$
  (or $[\mathfrak{t},\mathfrak{p}]\subset\mathfrak{p}$
   and
   $[\mathfrak{p},\mathfrak{p}]\subset\mathfrak{t}$),
   the non-null subspace
   ${W}^{1+\epsilon}(\mathfrak{B},\mathfrak{B}^{[r]};i)$
   is 
   included in $\mathfrak{t}\cap\mathfrak{p}$
   as $\mathfrak{B}^{[r]}={W}^1(\mathfrak{C},\mathfrak{B}^{[r]};\mathbf{0})\subset\mathfrak{t}$
  (or as
  ${W}^1(\mathfrak{C},\mathfrak{B}^{[r]};\mathbf{0})\subset\mathfrak{p}$).
  This contradicts the fact that $\mathfrak{t}\cap\mathfrak{p}$
  contains no spinors.
  The disjoint is asserted.

  Thanks to the quaternion condition of
  Eq.~\ref{eqgenWcommrankr} and the decomposition condition,
  confirming the 3rd assertion, for
  the conditioned subspaces of $\mathfrak{t}\oplus\mathfrak{p}$
  to obey the rules of Eq.~\ref{eqrulegenW},
  is straightforward.
  Furthermore, as a consequence of  Eq.~\ref{eqrulegenW} and Corollary~\ref{coroQAPGrankr},
  the subalgebra $\mathfrak{t}$ is a
  proper maximal subgroup of $\{\mathcal{P}_{\mathcal{Q}}(\mathfrak{B}^{[r]})\}$
  under the tri-addition and $\mathfrak{p}=\{\mathcal{P}_{\mathcal{Q}}(\mathfrak{B}^{[r]})\}-\mathfrak{t}$
  is the coset of $\mathfrak{t}$ under the same operation.

  In contrast to the above, it is plain to affirm the other implication
  that   the composition
  $\mathfrak{t}\oplus\mathfrak{p}$ forms a Cartan decomposition
  if 
  $\mathfrak{t}$ is a proper maximal subgroup
  and the complement $\mathfrak{p}=\{\mathcal{P}_{\mathcal{Q}}(\mathfrak{B}^{[r]})\}-\mathfrak{t}$
  is the coset of $\mathfrak{t}$ in a quotient-algebra partition
  $\{\mathcal{P}_{\mathcal{Q}}(\mathfrak{B}^{[r]})\}$ under the tri-addition.
  From the set of rules of Eq.~\ref{eqrulegenW}
  respected by $\mathfrak{t}$ 
  and $\mathfrak{p}$, 
  the decomposition condition immediately follows
  $[\mathfrak{t},\mathfrak{t}]\subset\mathfrak{t}$,
  $[\mathfrak{t},\mathfrak{p}]\subset\mathfrak{p}$,
  $[\mathfrak{p},\mathfrak{p}]\subset\mathfrak{t}$
  and ${\rm Tr}\{\mathfrak{t}\hspace{0.5pt}\mathfrak{p}\}=0$. 
  The proof completes.
 \end{proof}
 \vspace{6pt}
  Therefore, a choice of a Cartan decomposition of $su(N)$
  is equivalent to a determination of a proper maximal subgroup in the partition
  $\{\mathcal{P}_{\mathcal{Q}}(\mathfrak{B}^{[r]})\}$. 
  In the later episode, a consequence similar to Theorem~\ref{thmsuNCDandQAP}
  can be obtained for the $\mathfrak{t}$-$\mathfrak{p}$
  decompositions of higher levels:
  for an $l$-th-level $\mathfrak{t}$-$\mathfrak{p}$ decomposition
  $\mathfrak{t}_{[l]}\oplus\mathfrak{p}_{[l]}$,
  $1<l\leq 2p$, the subalgebra
  $\mathfrak{t}_{[l]}$ is a subgroup of a quotient-algebra
  partition under the tri-addition and $\mathfrak{p}_{[l]}$
  is a coset of $\mathfrak{t}_{[l]}$ under the same operation.

  To generate all Cartan decompositions of the Lie algebra $su(N)$,
  it is necessary to search all proper maximal subgroups of
  all quotient-algebra partitions
  over $su(N)$.
  Only two kinds of maximal subgroups will be constructed in
  the quotient-algebra partition $\{\mathcal{P}_{\mathcal{Q}}(\mathfrak{B}^{[r]})\}$
  generated by an $r$-th maximal bi-subalgebra $\mathfrak{B}^{[r]}$
  of a Cartan subalgebra $\mathfrak{C}\subset{su(N)}$.
  The kinds are decided by two selections of
  degrade conditioned subspaces of the doublet $(\mathfrak{C},\mathfrak{B}^{[r]})$
  in $\mathfrak{t}$, which embraces either the whole set of
  these conditioned subspaces or a half of them.
 \vspace{6pt}
 \begin{lemma}\label{lemdedWint}
  For a Cartan decomposition
  $su(N)=\mathfrak{t}\oplus\mathfrak{p}$ determined in the quotient-algebra partition
  $\{\mathcal{P}_{\mathcal{Q}}(\mathfrak{B}^{[r]})\}$
  generated by an $r$-th maximal bi-subalgebra $\mathfrak{B}^{[r]}$
  of a Cartan subalgebra $\mathfrak{C}\subset{su(N)}$,
  the set of all degrade conditioned subspaces of the doublet
  $(\mathfrak{C},\mathfrak{B}^{[r]})$ in the subalgebra
  $\mathfrak{t}$ forms either of the two kinds of subgroups of
  $\mathfrak{t}$ under the tri-addition,
  denoted as ${{\cal V}}^{\top}(\mathfrak{C})$ and ${{\cal V}}^{\bot}(\mathfrak{C})$
  and expressed as follows,
  \begin{align}\label{eqdedWint}
  &{{\cal V}}^{\top}(\mathfrak{C})=\{{W}^0(\mathfrak{C},\mathfrak{B}^{[r]};s),{W}^1(\mathfrak{C},\mathfrak{B}^{[r]};s):\forall\hspace{2pt}s\in{R}\}\text{ and }\notag\\
  &{{\cal V}}^{\bot}(\mathfrak{C})=\{{W}^0(\mathfrak{C},\mathfrak{B}^{[r]};s),{W}^1(\mathfrak{C},\mathfrak{B}^{[r]};t):\forall\hspace{2pt}s\in{R}\text{ and }t\in{R}^c\},
 \end{align}
  here ${R}$ being a maximal subgroup of $Z^r_2$ and ${R}^c=Z^r_2-{R}$.
 \end{lemma}
 \vspace{3pt}
 \begin{proof}
  This is a direct result of the rules of Eq.~\ref{eqrulegenW}
  when setting $\mathfrak{B}=\mathfrak{B}'=\mathfrak{C}$.
 \end{proof}
 \vspace{6pt}
  As will be illustrated later, the introduction of the
  subgroups ${{\cal V}}^{\top}(\mathfrak{C})$ and ${{\cal V}}^{\bot}(\mathfrak{C})$
  is assistant to construct various kinds of subalgebras $\mathfrak{t}$ through the tri-addition.
  Notice that the subgroup ${{\cal V}}^{\bot}(\mathfrak{C})$ consists of a half of
  the $2^{r+1}$ degrade conditioned subspaces of
  $(\mathfrak{C},\mathfrak{B}^{[r]})$ and
  ${{\cal V}}^{\top}(\mathfrak{C})$ comprises all of them
  as $R=Z^r_2$ or only a half as $R\neq{Z^r_2}$.
  Then, the Cartan subalgebra
  $\mathfrak{C}$ and its maximal bi-subalgebras appear in either $\mathfrak{t}$
  or $\mathfrak{p}$.
 \vspace{6pt}
 \begin{lemma}\label{lem3kindmaxabel}
  For a Cartan decomposition
  $\mathfrak{t}\oplus\mathfrak{p}$ determined in the
  quotient-algebra partition of rank $r$ $\{\mathcal{P}_{\mathcal{Q}}(\mathfrak{B}^{[r]})\}$
  generated by an $r$-th maximal bi-subalgebra $\mathfrak{B}^{[r]}$
  of a Cartan subalgebra $\mathfrak{C}$,
  only the distribution occurs that either the Cartan subalgebra $\mathfrak{C}$ is recovered in
  $\mathfrak{t}$ (or $\mathfrak{p}$) or
  a $1$st maximal bi-subalgebra $\mathfrak{B}^{\dag}\subset\mathfrak{C}$
  forms in $\mathfrak{t}$ (or $\mathfrak{p}$) and its complement
  $\mathfrak{B}^{\dag c}=\mathfrak{C}-\mathfrak{B}^{\dag}$ in $\mathfrak{p}$ (or $\mathfrak{t}$).
 \end{lemma}
 \vspace{2pt}
 \begin{proof}
  Recall that the subspaces ${W}^0(\mathfrak{C},\mathfrak{B}^{[r]};j)=\{0\}$ are nil
  for all $j\in{Z^r_2}$.
  Via Eq.~\ref{eqdedWint} for $R=Z^r_2$,
  the Cartan subalgebra $\mathfrak{C}=\bigcup_{\hspace{.5pt}l\in{Z^r_2}}{W}^1(\mathfrak{C},\mathfrak{B}^{[r]};l)$
  is recovered in $\mathfrak{p}$ with $\mathfrak{t}\supset{{\cal V}}^{\bot}(\mathfrak{C})$
  or recovered in $\mathfrak{t}$ with
  $\mathfrak{t}\supset{{\cal V}}^{\top}(\mathfrak{C})$.
  On the other hand as $R\neq{Z^r_2}$ and by Eq.~\ref{eqdedWint} again, a $1$st maximal bi-subalgebra
  $\mathfrak{B}^{\dag}=\bigcup_{\hspace{.5pt}s\in{R}}{W}^1(\mathfrak{C},\mathfrak{B}^{[r]};s)$
  forms in $\mathfrak{p}$ and its complement
  $\mathfrak{B}^{\dag c}=\mathfrak{C}-\mathfrak{B}^{\dag}
  =\bigcup_{\hspace{.5pt}t\in{R}^c}{W}^1(\mathfrak{C},\mathfrak{B}^{[r]};t)$ in $\mathfrak{t}$
  for $\mathfrak{t}\supset{{\cal V}}^{\bot}(\mathfrak{C})$
  or $\mathfrak{B}^{\dag}\subset\mathfrak{t}$ and $\mathfrak{B}^{\dag c}\subset\mathfrak{p}$
  for $\mathfrak{t}\supset{{\cal V}}^{\top}(\mathfrak{C})$;
  in this circumstance neither 
  $\mathfrak{t}$ nor $\mathfrak{p}$ can fully cover a Cartan
  subalgebra.
 \end{proof}
 \vspace{6pt}\noindent
  Thus, the subsets ${\cal V}^{\top}$ and ${\cal V}^{\bot}$ of
  Eq.\ref{eqdedWint} respectively belong to the subalgebra $\mathfrak{t}^{\top}$
  of a top Cartan decomposition $\mathfrak{t}^{\top}\oplus\mathfrak{p}^{\top}$
  and the subalgebra $\mathfrak{t}^{\bot}$ of a bottom
  decomposition $\mathfrak{t}^{\bot}\oplus\mathfrak{p}^{\bot}$, {\em cf.} Lemma~\ref{lem3kindsmaxabelinp}.
  The fact of recovering a Cartan subalgebra in $\mathfrak{t}$ (or $\mathfrak{p}$)
  or a $1$st maximal bi-subalgebra in $\mathfrak{t}$ (or $\mathfrak{p}$) and its complement in
  $\mathfrak{p}$ (or $\mathfrak{t}$) decides the three types of Cartan
  decompositions as briefed in the following. 

  There exist three types of Cartan
  decompositions for the Lie algebra $su(N)$ that are known as the types {\bf AI}, {\bf AII} and {\bf AIII} \cite{Helgason}
  and here denoted as $su(N)=\hat{\mathfrak{t}}_{\hspace{1pt}\rm I}\oplus\hat{\mathfrak{p}}_{\hspace{1pt}\rm I}$,
  $\hat{\mathfrak{t}}_{\hspace{1pt}{\rm II}}\oplus\hat{\mathfrak{p}}_{\hspace{1pt}{\rm II}}$
  and $\hat{\mathfrak{t}}_{\hspace{1pt}{\rm III}}\oplus\hat{\mathfrak{p}}_{\hspace{1pt}{\rm III}}$ respectively.
  The decompositions designated in \cite{Helgason} are specifically considered {\em intrinsic}.
  The intrinsic type {\bf AI} consists of the subalgebra
  $\hat{\mathfrak{t}}_{\hspace{1pt}\rm I}=so(N)$ and the vector subspace
  $\hat{\mathfrak{p}}_{\hspace{1pt}\rm I}=\{i(\ket{k}\bra{l}+\ket{l}\bra{k}):1\leq k,l\leq N\}$
  in the following exposition.
  The subalgebra $\hat{\mathfrak{t}}_{\hspace{1pt}\rm I}$ is spanned by a total number 
  $(N/2)(N-1)$ of generators and
  the set of all diagonal generators
  $\hat{\mathfrak{a}}_{\hspace{1pt}{\rm I}}=\{a_k\ket{k}\bra{k}:a_k\in\mathbb{C}\text{ and }\sum^{N}_{k=1}|a_k|^2=0,1\leq k\leq N\}$
  of $su(N)$ is a maximal abelian subalgebra in $\hat{\mathfrak{p}}_{\hspace{1pt}\rm I}$.
  Allowed in even dimensions only, the intrinsic type {\bf AII} comprises
  the subalgebra $\hat{\mathfrak{t}}_{\hspace{1pt}{\rm II}}=sp(N/2)$ 
  and the vector subspace
  $\hat{\mathfrak{p}}_{\hspace{1pt}{\rm II}}=
  \{(\begin{array}{cc}
  {\cal Z}_1&{\cal Z}_2\\
  {\cal Z}^{*}_2&-{\cal Z}^{*}_1
  \end{array}):{\cal Z}_1 \in {su}(N/2)\text{ and }{\cal Z}_2\in{so}(N/2,\mathbb{C})\}$.
  The subalgebra $\hat{\mathfrak{t}}_{\hspace{1pt}{\rm II}}$ has
  $(N/2)(N+1)$ generators and
  the set $\hat{\mathfrak{a}}_{\hspace{1pt}{\rm II}}=
  \{a_k(\ket{k}\bra{k}+\ket{k+N/2}\bra{k+N/2}):a_k\in\mathbb{C}\text{ and }\sum^{N/2}_{k=1}|a_k|^2=0,1\leq k\leq N/2\}$
  serves as a maximal ableian subalgebra in $\hat{\mathfrak{p}}_{\hspace{1pt}{\rm II}}$.
  The third, the intrinsic type {\bf AIII}
  is composed of the subalgebra
  $\hat{\mathfrak{t}}_{\hspace{1pt}{\rm III}}=c\otimes su(m)\otimes su(n)$, $m+n=N$, and
  the vector subspace
  $\hat{\mathfrak{p}}_{\hspace{1pt}{\rm III}}=
  \{(\begin{array}{cc}
   0&{\cal Z}\\
   -{\cal Z}^{\dagger}&0
  \end{array}):{\cal Z} \in M_{m\times n}(\mathbb{C})\}$;
  here $c$ is the center of $\hat{\mathfrak{t}}_{\hspace{1pt}{\rm III}}$,
  {\em i.e.}, $[c,\hat{\mathfrak{t}}_{\hspace{1pt}{\rm III}}]=0$.
  The number of generators of $\hat{\mathfrak{t}}_{\hspace{1pt}{\rm III}}$ totals to $m^2+n^2-1$ and
  the set $\hat{\mathfrak{a}}_{\hspace{1pt}{\rm III}}=\{i(\ket{k}\bra{k+m}-\ket{k+m}\bra{k}):1\leq k\leq n\}$
  forms a maximal abelian subalgebra in $\hat{\mathfrak{p}}_{\hspace{1pt}{\rm III}}$.
  As the dimension $N=2^p$,
  the subalgebras corresponding to these three intrinsic types
  are explicitly listed,
 \begin{align}\label{eqintrsubalg}
 \hat{\mathfrak{t}}_{\hspace{1pt}{\rm I}}= so(2^p)
 &=\{{\cal S}^{\zeta}_{\alpha}:\forall\hspace{2pt} \zeta,\alpha\in{Z^p_2},\zeta\cdot\alpha=1\},\notag\\
 \hat{\mathfrak{t}}_{\hspace{1pt}{\rm II}}= sp(2^{p-1})&=
 \{{\cal S}^{\zeta}_{\alpha}:\hspace{1pt}
 \forall\hspace{2pt} \zeta,\alpha\in Z^p_2,\epsilon_i,a_i\in{Z_2}\hspace{2pt},1\leq i\leq p,\zeta=\epsilon_1\epsilon_2\cdots\epsilon_p,\notag\\
 &\hspace{20pt}\hspace{2pt}\alpha=a_1 a_2\cdots a_p, 
 \text{ and }
 \hspace{0pt}\zeta\cdot\alpha=1+\epsilon_1+a_1\},\hspace{30pt}\text{ and}\notag\\
 \hat{\mathfrak{t}}_{\hspace{1pt}{\rm III}}
 =c\otimes su(2^{p-1})\otimes su(2^{p-1})&=
 \hspace{0pt}\{{\cal S}^{\zeta}_{\alpha}:\hspace{1pt}\forall\hspace{2pt}\zeta,\alpha\in Z^p_2,\hspace{1pt}a_i\in{Z_2},\hspace{2pt}1\leq i\leq p,\alpha=a_1 a_2\cdots a_p,\notag\\
 & \hspace{20pt}\text{and }a_1=0\},
 \end{align}
  where $c=\{{\cal S}^{\zeta_0=10\cdots 0}_{\mathbf{0}}\}$ and
  the string $\zeta_0$ has only one single nonzero bit
  in the leftmost digit.
  Following this designation,
  the maximal abelian subalgebras of
  the subspaces $\hat{\mathfrak{p}}_{\hspace{1pt}{\rm I}}$, $\hat{\mathfrak{p}}_{\hspace{1pt}{\rm II}}$
  and $\hat{\mathfrak{p}}_{\hspace{1pt}{\rm III}}$ as $N=2^p$
  are the intrinsic Cartan subalgebra
  $\hat{\mathfrak{a}}_{\hspace{1pt}{\rm I}}=\mathfrak{C}_{[\mathbf{0}]}=\{{\cal S}^{\nu_0}_{\mathbf{0}}:\forall\hspace{2pt}\nu_0\in{Z^p_2}\}\subset{su(2^p)}$,
  the $1$st maximal bi-subalgebra
  $\hat{\mathfrak{a}}_{\hspace{1pt}{\rm II}}=\mathfrak{B}^{[1]}_{intr}=\{{\cal S}^{\nu_{\mathbf{0}}}_{\mathbf{0}}:
  \forall\hspace{2pt}\nu_{\mathbf{0}}=\rho_1\rho_2\cdots\rho_p\in{Z^p_2}\text{ and }\rho_1=0\}$
  of $\mathfrak{C}_{[\mathbf{0}]}$
  and the rank-zero conditioned subspace
  $\hat{\mathfrak{a}}_{\hspace{1pt}{\rm III}}={W}^0(\mathfrak{B}^{[1]}_{intr})=
  \{{\cal S}^{\nu'_{\mathbf{0}}}_{\zeta_0=10\cdots 0}:
  \forall\hspace{2pt}\nu'_{\mathbf{0}}=\rho'_1\rho'_2\cdots\rho'_p\in{Z^p_2}\text{ and }\rho'_1=1\}$
  of $\mathfrak{B}^{[1]}_{intr}$ respectively.
  [[ It is worth noting that
  the subalgebra $\hat{\mathfrak{t}}_{\hspace{1pt}{\rm III}}$ of Eq.~\ref{eqintrsubalg}
  is a proper maximal bi-subalgebra of the bi-subalgebra $su(2^p)$, {\em cf.} Definition~3 in~\cite{SuTsai3}. ]]

  The two kinds of maximal subgroups of a partition 
  $\{\mathcal{P}_{\mathcal{Q}}(\mathfrak{B}^{[r]})\}$
  have the following forms.
 \vspace{6pt}
 \begin{lemma}\label{lem2kindmaxsubg}
  Being a group under the tri-addition,
  the quotient-algebra partition of rank $r$ $\{\mathcal{P}_{\mathcal{Q}}(\mathfrak{B}^{[r]})\}$
  generated by an $r$-th maximal bi-subalgebra
  $\mathfrak{B}^{[r]}$ of a Cartan subalgebra $\mathfrak{C}\subset su(2^p)$
  bears two kinds of maximal subgroups
  $\mathfrak{t}^{\top}$ and $\mathfrak{t}^{\bot}$ reading as
 \begin{align}\label{eqmaxsubtwokinds}
  &
  \mathfrak{t}^{\top}
  =\{{W}^0(\mathfrak{B}',\mathfrak{B}^{[r]};s),{W}^1(\mathfrak{B}',\mathfrak{B}^{[r]};s),{W}^0(\mathfrak{B}'',\mathfrak{B}^{[r]};t),{W}^1(\mathfrak{B}'',\mathfrak{B}^{[r]};t):\notag\\
  &\hspace{27pt}\forall\hspace{2pt}\mathfrak{B}'\in{\cal T},\mathfrak{B}''\in{\cal T}^c,
  \hspace{2pt}s\in{R}\text{ and }t\in{R}^c\}
  \text{ and }\notag\\
  &
  \mathfrak{t}^{\bot}
  =\{{W}^0(\mathfrak{B}',\mathfrak{B}^{[r]};s),{W}^1(\mathfrak{B}',\mathfrak{B}^{[r]};t),{W}^0(\mathfrak{B}'',\mathfrak{B}^{[r]};t),{W}^1(\mathfrak{B}'',\mathfrak{B}^{[r]};s):\notag\\
  &\hspace{27pt}\forall\hspace{2pt}\mathfrak{B}'\in{\cal T},\mathfrak{B}''\in{\cal T}^c,
  \hspace{2pt}s\in{R}\text{ and }t\in{R}^c\}
 \end{align}
  where ${\cal T}$ is a maximal subgroup of $\mathcal{G}(\mathfrak{C})$,
  $R$ is a maximal subgroup of $Z^r_2$,
  ${\cal T}^c=\mathcal{G}(\mathfrak{C})-{\cal T}$, $R^c=Z^r_2-R$,
  $\mathfrak{t}^{\top}$ is the subalgebra of a top Cartan decomposition
  $\mathfrak{t}^{\top}\oplus\mathfrak{p}^{\top}$
  and $\mathfrak{t}^{\bot}$ is the subalgebra of a bottom decomposition
  $\mathfrak{t}^{\bot}\oplus\mathfrak{p}^{\bot}$.
 \end{lemma}
 \vspace{3pt}
 \begin{proof}
  The set of conditioned subspaces forming a maximal subgroup
  $\mathfrak{t}\subset\{\mathcal{P}_{\mathcal{Q}}(\mathfrak{B}^{[r]})\}$
  is decided according to the occasion either the subgroup ${{\cal V}}^{\top}(\mathfrak{C})$
  or ${{\cal V}}^{\bot}(\mathfrak{C})$ is in $\mathfrak{t}$, {\em cf.} Eq.~\ref{eqdedWint} of Lemma~\ref{lemdedWint}.
  Based on the expression of the subgroup
  ${{\cal V}}^{\top}(\mathfrak{C})
   =\{{W}^{\epsilon}(\mathfrak{C},\mathfrak{B}^{[r]};s):\forall\hspace{2pt}\epsilon\in{Z_2},s\in{R}\}$
  (or ${{\cal V}}^{\bot}(\mathfrak{C})
  =\{{W}^0(\mathfrak{C},\mathfrak{B}^{[r]};s),{W}^1(\mathfrak{C},\mathfrak{B}^{[r]};t):\forall\hspace{2pt}s\in{R},t\in{R}^c\}$)
  of Eq.~\ref{eqdedWint},
  the set of the $2^{r+1}$ conditioned subspaces
  $\{{W}^{\epsilon}(\mathfrak{B},\mathfrak{B}^{[r]};j):\forall\hspace{2pt}\epsilon\in{Z_2},j\in{Z^r_2}\}$
  for every maximal bi-subalgebra $\mathfrak{B}\in\mathcal{G}(\mathfrak{C})$
  is divided into two subsets
  $\mathcal{J}_{\mathfrak{B}}
  =\{{W}^\epsilon(\mathfrak{B},\mathfrak{B}^{[r]};s):\forall\hspace{2pt}\epsilon\in{Z_2},s\in{R}\}$
  and
  $\hat{\mathcal{J}}_{\mathfrak{B}}
  =\{{W}^\epsilon(\mathfrak{B},\mathfrak{B}^{[r]};t):\forall\hspace{2pt}\epsilon\in{Z_2},t\in{R^c}\}$
  (or $\mathcal{K}_{\mathfrak{B}}
     =\{{W}^{0}(\mathfrak{B},\mathfrak{B}^{[r]};s),{W}^{1}(\mathfrak{B},\mathfrak{B}^{[r]};t):
  \forall\hspace{2pt}s\in{R},t\in{R^c}\}$
  and
  $\hat{\mathcal{K}}_{\mathfrak{B}}
    =\{{W}^{0}(\mathfrak{B},\mathfrak{B}^{[r]};t),{W}^{1}(\mathfrak{B},\mathfrak{B}^{[r]};s):
  \forall\hspace{2pt}s\in{R},t\in{R^c}\}$),
  here $\mathcal{J}_{\mathfrak{C}}={{\cal V}}^{\top}(\mathfrak{C})$
  (or $\mathcal{K}_{\mathfrak{C}}={{\cal V}}^{\bot}(\mathfrak{C})$), $R$ being a maximal subgroup of $Z^r_2$ and $R^c=Z^r_2-R$.
  Further imposing the decomposition condition reaches the inclusions
  of the two subsets belonging to $\mathfrak{t}$ or $\mathfrak{p}$. 

  As $\mathfrak{t}\supset{{\cal V}}^{\top}(\mathfrak{C})$,
  the distribution is validated by Eq.~\ref{eqrulegenW} that either
  $\mathcal{J}_{\mathfrak{B}}\subset\mathfrak{t}$
  and
  $\hat{\mathcal{J}}_{\mathfrak{B}}\subset\mathfrak{p}=\{\mathcal{P}_{\mathcal{Q}}(\mathfrak{B}^{[r]})\}-\mathfrak{t}$
  or $\hat{\mathcal{J}}_{\mathfrak{B}}\subset\mathfrak{t}$
  and
  $\mathcal{J}_{\mathfrak{B}}\subset\mathfrak{p}$.
  Since $\mathfrak{t}$ being maximal in $\{\mathcal{P}_{\mathcal{Q}}(\mathfrak{B}^{[r]})\}$,
  the two subspaces 
  take the unions
  $\mathfrak{t}=\mathfrak{t}^{\top}
  =\bigcup_{\mathfrak{B}'\in{\cal T},\mathfrak{B}''\in{\cal T}^c}\mathcal{J}_{\mathfrak{B}'}\cup \hat{\mathcal{J}}_{\mathfrak{B}''}$
  and
  $\mathfrak{p}=\mathfrak{p}^{\top}=\bigcup_{\mathfrak{B}'\in{\cal T},\mathfrak{B}''\in{\cal T}^c}\hat{\mathcal{J}}_{\mathfrak{B}'}\cup\mathcal{J}_{\mathfrak{B}''}$,
  where the set of the associated maximal subalgebras $\{\mathfrak{B}'\}$
  makes a maximal subgroup ${\cal T}$ of $\mathcal{G}(\mathfrak{C})$
  and 
  $\{\mathfrak{B}''\}={\cal T}^c=\mathcal{G}(\mathfrak{C})-{\cal T}$.
  Following the same argument except replacing
  $\mathcal{J}_{\mathfrak{B}}$ with $\mathcal{K}_{\mathfrak{B}}$
  and  $\hat{\mathcal{J}}_{\mathfrak{B}}$ with $\hat{\mathcal{K}}_{\mathfrak{B}}$ for every $\mathfrak{B}\in\mathcal{G}(\mathfrak{C})-\mathfrak{C}$,
  there derive the unions 
  $\mathfrak{t}=\mathfrak{t}^{\bot}=\bigcup_{\mathfrak{B}'\in{\cal T},\mathfrak{B}''\in{\cal T}^c}\mathcal{K}_{\mathfrak{B}'}\cup \hat{\mathcal{K}}_{\mathfrak{B}''}$
  and 
  $\mathfrak{p}=\mathfrak{p}^{\bot}=\bigcup_{\mathfrak{B}'\in{\cal T},\mathfrak{B}''\in{\cal T}^c}\hat{\mathcal{K}}_{\mathfrak{B}'}\cup\mathcal{K}_{\mathfrak{B}''}$
  when $\mathfrak{t}\supset{{\cal V}}^{\bot}(\mathfrak{C})$.
  The fact that $\mathfrak{t}^{\top}$ (or $\mathfrak{t}^{\bot}$)
  is the subalgebra of a top (bottom) Cartan decomposition $\mathfrak{t}^{\top}\oplus\mathfrak{p}^{\top}$
  (or $\mathfrak{t}^{\bot}\oplus\mathfrak{p}^{\bot}$) is validated
  by Corollary~\ref{coroanticommclose} together with
  Lemma~\ref{lemtpanticomm}.
 \end{proof}
 \vspace{6pt}\noindent
  The two associated subspaces
  $\mathfrak{p}^{\top}=\{\mathcal{P}_{\mathcal{Q}}(\mathfrak{B}^{[r]})\}-\mathfrak{t}^{\top}$
  and $\mathfrak{p}^{\bot}=\{\mathcal{P}_{\mathcal{Q}}(\mathfrak{B}^{[r]})\}-\mathfrak{t}^{\bot}$
  have separate types of maximal abelian subalgebras.
 \vspace{6pt}
 \begin{lemma}\label{lemmaxabelintp}
  Given Cartan decompositions
  $\mathfrak{t}^{\top}\oplus\mathfrak{p}^{\top}$ and $\mathfrak{t}^{\bot}\oplus\mathfrak{p}^{\bot}$
  determined in the quotient-algebra partition of rank $r$
  $\{\mathcal{P}_{\mathcal{Q}}(\mathfrak{B}^{[r]})\}$ given by an $r$-th maximal bi-subalgebra $\mathfrak{B}^{[r]}$
  of a Cartan subalgebra $\mathfrak{C}\subset su(2^p)$,
  the maximal abelian subalgebra of the subspace $\mathfrak{p}^{\bot}$
  is a Cartan subalgebra, denoted as $\widetilde{\mathfrak{C}}$, or
  a $1$st maximal bi-subalgebra $\widetilde{\mathfrak{B}}\subset\widetilde{\mathfrak{C}}$,
  and the maximal abelian subalgebra of $\mathfrak{p}^{\top}$
  is the complement $\widetilde{\mathfrak{B}}^c=\widetilde{\mathfrak{C}}-\widetilde{\mathfrak{B}}$.
 \end{lemma}
 \vspace{2pt}
 \begin{proof}
  Based on Theorem~\ref{thmsuNCDandQAP},
  for a Cartan decomposition $\mathfrak{t}\oplus\mathfrak{p}$,
  there exists a quotient-algebra partition $\{\mathcal{P}_{\mathcal{Q}}(\mathfrak{B}^{[r]})\}$
  over $su(N)$
  such that $\mathfrak{t}$ is a proper maximal subgroup of
  $\{\mathcal{P}_{\mathcal{Q}}(\mathfrak{B}^{[r]})\}$
  and $\mathfrak{p}=\{\mathcal{P}_{\mathcal{Q}}(\mathfrak{B}^{[r]})\}-\mathfrak{t}$
  is the complement.
  This lemma will reaffirm the
  type of the maximal abelian subalgebra $\mathfrak{a}^{\top}$
  (or $\mathfrak{a}^{\bot}$) of the subspace $\mathfrak{p}^{\top}$
  (or $\mathfrak{p}^{\bot}$)
  of a Cartan decomposition $\mathfrak{t}^{\top}\oplus\mathfrak{p}^{\top}$
  (or $\mathfrak{t}^{\bot}\oplus\mathfrak{p}^{\bot}$)
  according to the structure of $\{\mathcal{P}_{\mathcal{Q}}(\mathfrak{B}^{[r]})\}$,
  {\em cf.} Lemma~\ref{lem3kindsmaxabelinp}.
  Furthermore, the form of the subalgebra $\mathfrak{a}^{\top}$
  (or $\mathfrak{a}^{\bot}$)
  will be explicitly given, which is a union of the conditioned subspaces of
  $\mathfrak{p}^{\top}=\{\mathcal{P}_{\mathcal{Q}}(\mathfrak{B}^{[r]})\}-\mathfrak{t}^{\top}$
  (or $\mathfrak{p}^{\bot}=\{\mathcal{P}_{\mathcal{Q}}(\mathfrak{B}^{[r]})\}-\mathfrak{t}^{\bot}$).
  Owing to Eq.~\ref{eq3typsform} that will be shown in next lemma,
  the subspaces $\mathfrak{p}^{\top}$ and $\mathfrak{p}^{\bot}$ have the following forms,
  \begin{align}\label{eqpforms}
   \mathfrak{p}^{\bot}_{\mathbf{1}}
  &=\{{W}^1(\mathfrak{B},\mathfrak{B}^{[r]};l):
  \forall\hspace{2pt}\mathfrak{B}\in\mathcal{G}(\mathfrak{C})\text{ and }l\in{Z^r_2}\},\notag\\
  \overline{\mathfrak{p}}^{\bot}_{\mathbf{1}}
  &=\{{W}^0(\mathfrak{B}'',\mathfrak{B}^{[r]};l),{W}^1(\mathfrak{B}',\mathfrak{B}^{[r]};l):
  \forall\hspace{2pt}\mathfrak{B}'\in{\cal T}_0,\mathfrak{B}''\in{\cal T}^c_0\text{ and }l\in{Z^r_2}\},\notag\\
  \breve{\mathfrak{p}}^{\bot}_{\omega}
  &=\{{W}^0(\mathfrak{B}',\mathfrak{B}^{[r]};t),
    {W}^1(\mathfrak{B}', \mathfrak{B}^{[r]};s),
    {W}^0(\mathfrak{B}'',\mathfrak{B}^{[r]};s),
    {W}^1(\mathfrak{B}'',\mathfrak{B}^{[r]};t):\notag\\
   &\hspace{20pt}\forall\hspace{2pt}\mathfrak{B}'\in{\cal T}_0,\mathfrak{B}''\in{\cal T}^c_0,s\in{R_0},t\in{R^c_0}
   \text{ and }\mathfrak{B}^{\dag}\in{\cal T}_0\},\notag\\
  \widetilde{\mathfrak{p}}^{\bot}_{\omega}
  &=\{{W}^0(\mathfrak{B},\mathfrak{B}^{[r]};t),{W}^1(\mathfrak{B},\mathfrak{B}^{[r]};s):
  \forall\hspace{2pt}\mathfrak{B}\in\mathcal{G}(\mathfrak{C}),s\in{R_0}\text{ and }t\in{R^c_0}\},\hspace{2pt}\omega=\mathbf{1},\mathbf{2},\notag\\
  &\hspace{3pt}\text{here }
  \breve{\mathfrak{p}}^{\bot}_{\omega}
  =\breve{\mathfrak{p}}^{\bot}_{\mathbf{1}}\text{ and }
  \widetilde{\mathfrak{p}}^{\bot}_{\omega}=\widetilde{\mathfrak{p}}^{\bot}_{\mathbf{1}}
  \text{ if }{W}^0(\mathfrak{B}^{\dag},\mathfrak{B}^{[r]};\mathbf{0})=\{0\}
  \text{ or }\notag\\
  \hspace{0pt}&\hspace{26pt}\breve{\mathfrak{p}}^{\bot}_{\omega}=\breve{\mathfrak{p}}^{\bot}_{\mathbf{2}}\text{ and }
  \widetilde{\mathfrak{p}}^{\bot}_{\omega}=\widetilde{\mathfrak{p}}^{\bot}_{\mathbf{2}}
  \text{ if }{W}^0(\mathfrak{B}^{\dag},\mathfrak{B}^{[r]};\mathbf{0})\neq\{0\},\notag\\
   \mathfrak{p}^{\top}_{\mathbf{3}}
  &=\{{W}^{\epsilon}(\mathfrak{B}'',\mathfrak{B}^{[r]};l):
  \forall\hspace{2pt}\epsilon\in{Z_2},\mathfrak{B}''\in{\cal T}^c_0\text{ and }l\in{Z^r_2}\},\notag\\
  \overline{\mathfrak{p}}^{\top}_{\mathbf{3}}
  &=\{{W}^{\epsilon}(\mathfrak{B},\mathfrak{B}^{[r]};t):
  \forall\hspace{2pt}\epsilon\in{Z_2},\mathfrak{B}\in\mathcal{G}(\mathfrak{C})\text{ and }t\in{R^c_0}\},\text{ and}\\
  \breve{\mathfrak{p}}^{\top}_{\mathbf{3}}
   &=\{{W}^{\epsilon}(\mathfrak{B}',\mathfrak{B}^{[r]};t),
    {W}^{\sigma}(\mathfrak{B}'',\mathfrak{B}^{[r]};s):\hspace{0pt}\forall\hspace{2pt}
    \epsilon,\sigma\in{Z_2},\mathfrak{B}'\in{\cal T}_0,\mathfrak{B}''\in{\cal T}^c_0,s\in{R_0}\text{ and }t\in{R^c_0}\},\notag
  \end{align}
  where $R_0$ and $T_0$ are proper maximal subgroups of $Z^r_2$ and $\mathcal{G}(\mathfrak{C})$
  respectively, ${\cal T}^c_0=\mathcal{G}(\mathfrak{C})-{\cal T}_0$,
  $R^c_0=Z^r_2-R_0$ and
  $\mathfrak{B}^{\dag}=\bigcup_{\hspace{.5pt}s\in{R_0}}{W}^1(\mathfrak{C},\mathfrak{B}^{[r]};s)$
  being a $1$st maximal bi-subalgebra of $\mathfrak{C}$ is a superset of
  $\mathfrak{B}^{[r]}={W}^1(\mathfrak{C},\mathfrak{B}^{[r]};\mathbf{0})$.
  As to be asserted in next lemma,
  the subscripts $\mathbf{1}$, $\mathbf{2}$
  and $\mathbf{3}$ indicate the Cartan decompositions
  of types {\bf AI}, {\bf AII} and {\bf AIII} respectively
  associated with the subspaces.
  When ${\cal T}=\mathcal{G}(\mathfrak{C})$ and $R=Z^r_2$,
  there obtain the subspaces $\mathfrak{p}^{\bot}=\mathfrak{p}^{\bot}_{\mathbf{1}}$
  and
  $\mathfrak{p}^{\bot}=\{0\}$,
  here $\mathfrak{t}^{\bot}=\{\mathcal{P}_{\mathcal{Q}}(\mathfrak{B}^{[r]})\}$
  being a trivial maximal subgroup.
  Setting ${\cal T}={\cal T}_0$ and $R=Z^r_2$
  results in the subspaces $\mathfrak{p}^{\bot}=\overline{\mathfrak{p}}^{\bot}_{\mathbf{1}}$
  and $\mathfrak{p}^{\top}=\mathfrak{p}^{\top}_{\mathbf{3}}$.
  The subspaces
  $\mathfrak{p}^{\bot}=\widetilde{\mathfrak{p}}^{\bot}_{\omega}$
  and $\mathfrak{p}^{\top}=\overline{\mathfrak{p}}^{\top}_{\mathbf{3}}$
  appear when ${\cal T}=\mathcal{G}(\mathfrak{C})$ and $R=R_0$.
  Finally the subgroups ${\cal T}={\cal T}_0$ and $R=R_0$ bring forth
  the subspaces
  $\mathfrak{p}^{\bot}=\breve{\mathfrak{p}}^{\bot}_{\omega}$
  and
  $\mathfrak{p}^{\top}=\breve{\mathfrak{p}}^{\top}_{\mathbf{3}}$.
  The subscript of $\breve{\mathfrak{p}}^{\bot}_{\omega}$
  (or $\widetilde{\mathfrak{p}}^{\bot}_{\omega}$)
  refers to the type {\bf AI} as $\omega=\mathbf{1}$ or {\bf AII} as $\omega=\mathbf{2}$.
  Notice that the choice $\mathfrak{B}^{\dag}\in{\cal T}_0$ of $\mathfrak{p}^{\bot}_{\omega}$
  is simply for convenience 
  and is equivalent to the alternative $\mathfrak{B}^{\dag}\in{\cal T}^c_0$
  producing the same subspace.
  Due to the fact of the union
  $\mathfrak{C}=\bigcup_{\hspace{.5pt}l\in{Z^r_2}}{W}^1(\mathfrak{C},\mathfrak{B}^{[r]};l)$
  being recovered in $\mathfrak{p}^{\bot}_{\mathbf{1}}$, $\overline{\mathfrak{p}}^{\bot}_{\mathbf{1}}$ or
  $\mathfrak{t}^{\top}_{\mathbf{3}}$ by Eq.~\ref{eqpforms},
  the Cartan subalgebra of the form
 \begin{align}\label{eqmaxabelformAI}
  \mathfrak{a}^{\bot}=\mathfrak{C}^{\bot}=\mathfrak{C}
  =\mathfrak{a}_{\mathfrak{p}^{\bot}_{\mathbf{1}}}
  =\mathfrak{a}_{\overline{\mathfrak{p}}^{\bot}_{\mathbf{1}}}
  =\bigcup_{\hspace{.5pt}l\in{Z^r_2}}{W}^1(\mathfrak{C},\mathfrak{B}^{[r]};l)
 \end{align}
  is a maximal abelian subalgebra of
  $\mathfrak{p}^{\bot}_{\mathbf{1}}$ and $\overline{\mathfrak{p}}^{\bot}_{\mathbf{1}}$.
  The complement
 \begin{align}\label{eqmaxabelformAIII}
  \mathfrak{a}^{\top}={W}^{\epsilon}(\mathfrak{B}'')
  =\mathfrak{a}_{\mathfrak{p}_{\mathbf{3}}^{\top}}
  =\bigcup_{\hspace{.5pt}l\in{Z^r_2}}{W}^{\epsilon}(\mathfrak{B}'',\mathfrak{B}^{[r]};l)
 \end{align}
  of a $1$st maximal bi-subalgebra $\mathfrak{B}^{\top}=\mathfrak{B}''$ of the Cartan subalgebra
  $\mathfrak{C}^{\top}=\mathfrak{B}''\cup{W}^{\epsilon}(\mathfrak{B}'')$
  is a counterpart in $\mathfrak{p}_{\mathbf{3}}^{\top}$, {\em cf.} Lemma~14 in~\cite{SuTsai1}.

  Since $\mathfrak{B}^{\dag}$ is a superset of
  $\mathfrak{B}^{[r]}={W}^1(\mathfrak{C},\mathfrak{B}^{[r]};\mathbf{0})$,
  the subspaces ${W}^{\epsilon_0}(\mathfrak{B}^{\dag},\mathfrak{B}^{[r]};t)=\{0\}$ with $t\in{R^c_0}$
  and ${W}^{1+\epsilon_0}(\mathfrak{B}^{\dag},\mathfrak{B}^{[r]};s)=\{0\}$ with $s\in{R_0}$ are null
  as ${W}^{\epsilon_0}(\mathfrak{B}^{\dag},\mathfrak{B}^{[r]};\mathbf{0})\neq\{0\}$
  ({\em i.e.},
  ${W}^{1+\epsilon_0}(\mathfrak{B}^{\dag},\mathfrak{B}^{[r]};\mathbf{0})=\{0\}$),
  $\epsilon_0\in{Z_2}$,
  by Lemma~\ref{lemcondsub} and the commutation relation
  of Eq.~\ref{eqgenWcommrankr}.
  This yields the unions
  ${W}^0(\mathfrak{B}^{\dag})
  =\bigcup_{\hspace{.5pt}s\in{R_0}}{W}^0(\mathfrak{B}^{\dag},\mathfrak{B}^{[r]};s)$
  and
  ${W}^1(\mathfrak{B}^{\dag})
  =\bigcup_{\hspace{.5pt}t\in{R^c_0}}{W}^1(\mathfrak{B}^{\dag}, \mathfrak{B}^{[r]};t)$
  if ${W}^0(\mathfrak{B}^{\dag},\mathfrak{B}^{[r]};\mathbf{0})\neq\{0\}$
  or
  ${W}^0(\mathfrak{B}^{\dag})
  =\bigcup_{\hspace{.5pt}t\in{R^c_0}}{W}^0(\mathfrak{B}^{\dag},\mathfrak{B}^{[r]};t)$
  and
  ${W}^1(\mathfrak{B}^{\dag})
  =\bigcup_{\hspace{.5pt}s\in{R_0}}{W}^1(\mathfrak{B}^{\dag}, \mathfrak{B}^{[r]} ;s)$
  if ${W}^0(\mathfrak{B}^{\dag},\mathfrak{B}^{[r]};\mathbf{0})=\{0\}$,
  each of which is a conditioned subspace of $\mathfrak{B}^{\dag}$.
  The subspace
  $\breve{\mathfrak{p}}^{\bot}_{\omega}$
  ($\widetilde{\mathfrak{p}}^{\bot}_{\omega}$) thus has two options
  $\breve{\mathfrak{p}}^{\bot}_{\mathbf{1}}$
  ($\widetilde{\mathfrak{p}}^{\bot}_{\mathbf{1}}$)
  as ${W}^0(\mathfrak{B}^{\dag},\mathfrak{B}^{[r]};\mathbf{0})=\{0\}$
  or
  $\breve{\mathfrak{p}}^{\bot}_{\mathbf{2}}$
  ($\widetilde{\mathfrak{p}}^{\bot}_{\mathbf{2}}$)
  as
  ${W}^0(\mathfrak{B}^{\dag},\mathfrak{B}^{[r]};\mathbf{0})\neq\{0\}$.
  Due to the fact that both the subspaces ${W}^0(\mathfrak{B}^{\dag})$ and ${W}^1(\mathfrak{B}^{\dag})$
  are included either in $\breve{\mathfrak{p}}^{\bot}_{\mathbf{1}}$
  ($\widetilde{\mathfrak{p}}^{\bot}_{\mathbf{1}}$) or
  in $\breve{\mathfrak{t}}^{\bot}_{\mathbf{2}}$
  ($\widetilde{\mathfrak{t}}^{\bot}_{\mathbf{2}}$),
  the Cartan subalgebra
 \begin{align}\label{eqmaxabelformAIbreve}
  \mathfrak{a}^{\bot}
  =\mathfrak{a}_{\breve{\mathfrak{p}}^{\bot}_{\mathbf{1}}}
  =\mathfrak{a}_{\widetilde{\mathfrak{p}}^{\bot}_{\mathbf{1}}}
  =\mathfrak{C}^{\bot}=\mathfrak{B}^{\dag}\cup{W}^\epsilon(\mathfrak{B}^{\dag}),\hspace{2pt}\epsilon\in{Z_2},
 \end{align}
  is a maximal abelian subalgebra of $\breve{\mathfrak{p}}^{\bot}_{\mathbf{1}}$
  and $\widetilde{\mathfrak{p}}^{\bot}_{\mathbf{1}}$,
  {\em cf.} Lemma~14 in~\cite{SuTsai1}.
  While the $1$st maximal bi-subalgebra
 \begin{align}\label{eqmaxabelformAIIbreve}
  \mathfrak{a}^{\bot}
  =\mathfrak{a}_{\breve{\mathfrak{p}}^{\bot}_{\mathbf{2}}}
  =\mathfrak{a}_{\widetilde{\mathfrak{p}}^{\bot}_{\mathbf{2}}}
  =\mathfrak{B}^{\dag},
 \end{align}
  is a maximal abelian subalgebra of $\breve{\mathfrak{p}}^{\bot}_{\mathbf{2}}$
  and $\widetilde{\mathfrak{p}}^{\bot}_{\mathbf{2}}$.
  Moreover, the subalgebra $\mathfrak{B}^{\dag}$ is
  a subset of $\overline{\mathfrak{t}}^{\top}_{\mathbf{3}}$
  ($\breve{\mathfrak{t}}^{\top}_{\mathbf{3}}$)
  and
  either ${W}^0(\mathfrak{B}^{\dag})$ or ${W}^1(\mathfrak{B}^{\dag})$
  belongs to $\overline{\mathfrak{p}}^{\top}_{\mathbf{3}}$
  ($\breve{\mathfrak{p}}^{\top}_{\mathbf{3}}$).
 Therefore, the union
  $\mathfrak{B}^{\dag}\cup{W}^0(\mathfrak{B}^{\dag})$
  or $\mathfrak{B}^{\dag}\cup{W}^1(\mathfrak{B}^{\dag})$,
  which is a Cartan subalgebra by Lemma~14 in~\cite{SuTsai1},
  can be constructed in $\overline{\mathfrak{t}}^{\top}_{\mathbf{3}}$
  ($\breve{\mathfrak{t}}^{\top}_{\mathbf{3}}$),
  and the complement
 \begin{align}\label{eqmaxabelformAIIIbreve}
  \mathfrak{a}^{\top}
  =\mathfrak{a}_{\overline{\mathfrak{p}}^{\top}_{\mathbf{3}}}
  =\mathfrak{a}_{\breve{\mathfrak{p}}^{\top}_{\mathbf{3}}}
  ={W}^\epsilon(\mathfrak{B}^{\dag}),\hspace{2pt}\epsilon\in{Z_2},
 \end{align}
  of $\mathfrak{B}^{\dag}$ in
  $\mathfrak{C}^{\top}=\mathfrak{B}^{\dag}\cup{W}^\epsilon(\mathfrak{B}^{\dag})$
  is a maximal abelian subalgebra of
  $\overline{\mathfrak{p}}^{\top}_{\mathbf{3}}$ 
  and $\breve{\mathfrak{p}}^{\top}_{\mathbf{3}}$.
  It is noted that the formations of maximal abelian subalgebras in
  $\mathfrak{p}^{\bot}$ ($\mathfrak{p}^{\top}$) are not unique.
 \end{proof}
 \vspace{6pt}\noindent
  The decompositions $\mathfrak{t}^{\top}\oplus\mathfrak{p}^{\top}$
  and $\mathfrak{t}^{\bot}\oplus\mathfrak{p}^{\bot}$ lead to
  the three types of Cartan decompositions.
 \vspace{6pt}
 \begin{lemma}\label{lemmaxsubCD}
  Both determined in the quotient algebra
  partition of rank $r$ $\{\mathcal{P}_{\mathcal{Q}}(\mathfrak{B}^{[r]})\}$
  given by an $r$-th maximal bi-subalgebra $\mathfrak{B}^{[r]}$ of a Cartan subalgebra $\mathfrak{C}$,
  the decomposition
  $\mathfrak{t}^{\bot}\oplus\mathfrak{p}^{\bot}$ is a Cartan decomposition of type {\bf AI} or {\bf AII}
  and $\mathfrak{t}^{\top}\oplus\mathfrak{p}^{\top}$ is a type {\bf AIII}.
 \end{lemma}
 \vspace{3pt}
 \begin{proof}
  The proof is essentially based on the general forms of the
  proper maximal subgroups $\mathfrak{t}^{\top}$ and $\mathfrak{t}^{\bot}$ of the
  partition $\{\mathcal{P}_{\mathcal{Q}}(\mathfrak{B}^{[r]})\}$
  under the tri-addition, {\em cf.} Theorem~\ref{thmsuNCDandQAP}.
  According to the designation of the maximal subgroups $R\subseteq{Z^r_2}$
  and ${\cal T}\subseteq\mathcal{G}(\mathfrak{C})$ of Eq.~\ref{eqmaxsubtwokinds}
  to be the whole groups $Z^r_2$ and $\mathcal{G}(\mathfrak{C})$ or a proper one 
  $R_0\subset{Z^r_2}$ and ${\cal T}_0\subset\mathcal{G}(\mathfrak{C})$ respectively,
  the subalgebras $\mathfrak{t}^{\top}$ and $\mathfrak{t}^{\bot}$
  take the forms,
  \begin{align}\label{eq3typsform}
   \mathfrak{t}^{\bot}_{\mathbf{1}}
   &
   =
   \{{W}^0(\mathfrak{B},\mathfrak{B}^{[r]};l):\forall\hspace{2pt}\mathfrak{B}\in\mathcal{G}(\mathfrak{C})\text{ and }l\in{Z^r_2}\},\notag\\
   \overline{\mathfrak{t}}^{\bot}_{\mathbf{1}}
   &
   =\{{W}^0(\mathfrak{B}',\mathfrak{B}^{[r]};l),
    {W}^1(\mathfrak{B}'',\mathfrak{B}^{[r]};l):
   \hspace{0pt}\forall\hspace{2pt}\mathfrak{B}'\in{\cal T}_0,\mathfrak{B}''\in{\cal T}^c_0\text{ and }l\in{Z^r_2}\},\notag\\
   \breve{\mathfrak{t}}^{\bot}_{\omega}
   &
   =\{{W}^0(\mathfrak{B}',\mathfrak{B}^{[r]};s),
    {W}^1(\mathfrak{B}',\mathfrak{B}^{[r]};t),
    {W}^0(\mathfrak{B}'',\mathfrak{B}^{[r]};t),
    {W}^1(\mathfrak{B}'',\mathfrak{B}^{[r]};s):\notag\\
   &
   \hspace{20pt}\forall\hspace{2pt}\mathfrak{B}'\in{\cal T}_0,\mathfrak{B}''\in{\cal T}^c_0,s\in{R_0},t\in{R^c_0}
   \text{ and }\mathfrak{B}^{\dag}\in{\cal T}_0\},\notag\\
   \widetilde{\mathfrak{t}}^{\bot}_{\omega}
   &
   =\{{W}^0(\mathfrak{B},\mathfrak{B}^{[r]};s),
    {W}^1(\mathfrak{B},\mathfrak{B}^{[r]};t):
   \hspace{0pt}\forall\hspace{2pt}\mathfrak{B}\in\mathcal{G}(\mathfrak{C}),s\in{R_0}\text{ and }t\in{R^c_0}\},\hspace{2pt}\omega=\mathbf{1},\mathbf{2},\notag\\
   &\hspace{3pt}\text{here }\breve{\mathfrak{t}}^{\bot}_{\omega}=\breve{\mathfrak{t}}^{\bot}_{\mathbf{1}}
   \text{ and }\widetilde{\mathfrak{t}}^{\bot}_{\omega}=\widetilde{\mathfrak{t}}^{\bot}_{\mathbf{1}}
   \text{ if }{W}^0(\mathfrak{B}^{\dag},\mathfrak{B}^{[r]};\mathbf{0})=\{0\}\text{ or}\notag\\
   &\hspace{25pt}\breve{\mathfrak{t}}^{\bot}_{\omega}=\breve{\mathfrak{t}}^{\bot}_{\mathbf{2}}
   \text{ and }\widetilde{\mathfrak{t}}^{\bot}_{\omega}=\widetilde{\mathfrak{t}}^{\bot}_{\mathbf{2}}
   \text{ if }{W}^0(\mathfrak{B}^{\dag},\mathfrak{B}^{[r]};\mathbf{0})\neq\{0\},\notag\\
   \mathfrak{t}^{\top}_{\mathbf{3}}
   &
   =
   \{{W}^{\epsilon}(\mathfrak{B}',\mathfrak{B}^{[r]};l):\forall\hspace{2pt}\epsilon\in{Z_2},\mathfrak{B}'\in{\cal T}_0\text{ and }l\in{Z^r_2}\},\notag\\
   \overline{\mathfrak{t}}^{\top}_{\mathbf{3}}
   &
   =
   \{{W}^{\epsilon}(\mathfrak{B},\mathfrak{B}^{[r]};s):\forall\hspace{2pt}\epsilon\in{Z_2},\mathfrak{B}\in\mathcal{G}(\mathfrak{C})\text{ and }s\in{R_0}\},\text{ and}\\
   \breve{\mathfrak{t}}^{\top}_{\mathbf{3}}
   &
   =
   \{{W}^{\epsilon}(\mathfrak{B}',\mathfrak{B}^{[r]};s),
    {W}^{\sigma}(\mathfrak{B}'',\mathfrak{B}^{[r]};t):\forall\hspace{2pt}
    \epsilon,\sigma\in{Z_2},\mathfrak{B}'\in{\cal T}_0,\mathfrak{B}''\in{\cal T}^c_0,s\in{R_0}\text{ and }t\in{R^c_0}\},
    \notag
  \end{align}
  where $\mathfrak{B}^{\dag}=\bigcup_{\hspace{.5pt}s\in{R_0}}{W}^1(\mathfrak{C},\mathfrak{B}^{[r]};s)$
  is a $1$st maximal bi-subalgebra of $\mathfrak{C}$,
  ${\cal T}_0$ is a proper maximal subgroup of $\mathcal{G}(\mathfrak{C})$,
  $R_0$ is a proper maximal subgroup of $Z^r_2$, ${\cal T}^c_0=\mathcal{G}(\mathfrak{C})-{\cal T}_0$
  and $R^c_0=Z^r_2-R_0$.
  Through calculating numbers of generators 
  and constructing the versions in {\em the intrinsic quotient-algebra partition},
  the subalgebras
  $\mathfrak{t}^{\bot}_{\mathbf{1}}$
  ($\overline{\mathfrak{t}}^{\bot}_{\mathbf{1}}$,
   $\breve{\mathfrak{t}}^{\bot}_{\mathbf{1}}$ or
   $\widetilde{\mathfrak{t}}^{\bot}_{\mathbf{1}}$),
  $\breve{\mathfrak{t}}^{\bot}_{\mathbf{2}}$ ($\widetilde{\mathfrak{t}}^{\bot}_{\mathbf{2}}$)
  and
  $ \mathfrak{t}^{\top}_{\mathbf{3}}$
  ($\overline{\mathfrak{t}}^{\top}_{\mathbf{3}}$ or $\breve{\mathfrak{t}}^{\top}_{\mathbf{3}}$)
  will be affirmed to correspond to the Cartan decompositions of types {\bf AI}, {\bf AII} and {\bf AIII} respectively.

  The number of generators of each subalgebra of Eq.~\ref{eq3typsform} is obtained by
  counting the contributions from its non-null degrade and regular conditioned subspaces.
  For every maximal bi-subalgebra
  $\mathfrak{B}\in\mathcal{G}(\mathfrak{C})$,
  either each of the $2^r$ conditioned subspaces
  $\{{W}^{\epsilon}(\mathfrak{B},\mathfrak{B}^{[r]};i):\forall\hspace{2pt}i\in{Z^r_2}\}$
  is a regular subspace as
  $\mathfrak{B}\nsupseteq\mathfrak{B}^{[r]}$
  or a half of them are the null subspace and the other half are a degrade non-null
  subspace as $\mathfrak{B}\supset\mathfrak{B}^{[r]}$.
  A non-null degrade subspace has $2^{p-r}$ and a regular one has $2^{p-r-1}$ generators.
  Since comprising $2^{r-1}(2^r-1)$
  non-null degrade and $2^r(2^p-2^r)$ regular subspaces,
  the subalgebra $\mathfrak{t}^{\bot}_{\mathbf{1}}$
  ($\overline{\mathfrak{t}}^{\bot}_{\mathbf{1}}$, $\breve{\mathfrak{t}}^{\bot}_{\mathbf{1}}$
  or $\tilde{\mathfrak{t}}^{\bot}_{\mathbf{1}}$)
  has the number $2^{p-1}(2^p-1)$ of generators, which is
  identical to that of the subalgebra $\hat{\mathfrak{t}}_{{\hspace{1pt}\rm I}}$
  of the Cartan  decomposition of type {\bf AI} of Eq.~\ref{eqintrsubalg}.
  Coinciding with that of $\hat{\mathfrak{t}}_{{\hspace{1pt}\rm II}}$
  of the type-{\bf AII} decomposition of Eq.~\ref{eqintrsubalg},
  the subalgebra $\breve{\mathfrak{t}}^{\bot}_{\mathbf{2}}$ ($\widetilde{\mathfrak{t}}^{\bot}_{\mathbf{2}}$)
  consists of in total $2^{p-1}(2^p+1)$ generators contributed by
  its $2^{r-1}(2^r+1)$ non-null degrade and $2^r(2^p+2^r)$ regular subspaces.
  The subalgebra  $\overline{\mathfrak{t}}^{\top}_{\mathbf{3}}$ ($\breve{\mathfrak{t}}^{\top}_{\mathbf{3}}$)
  has $2^{2r-1}$ non-null degrade and $2^r(2^p-2^r)$ regular subspaces.
  The subalgebra $\mathfrak{t}^{\top}_{\mathbf{3}}$ has
  $2^{2r}$ non-null degrade and $2^r(2^p-2^{r+1})$ regular subspaces
  (or $2^{2r-1}$ non-null degrade and $2^r(2^p-2^r)$ regular subspaces)
  as the subgroup ${\cal T}_0$ of $\mathfrak{t}^{\top}_{\mathbf{3}}$ embraces all the
  $2^r$ (or a half of $2^{r-1}$) maximal bi-subalgebras
  being a superset of $\mathfrak{B}^{[r]}$.
  The three subalgebras $\mathfrak{t}^{\top}_{\mathbf{3}}$, $\overline{\mathfrak{t}}^{\top}_{\mathbf{3}}$
  and $\breve{\mathfrak{t}}^{\top}_{\mathbf{3}}$
  have the same number $2^{2p-1}$ of generators
  identical to that of the subalgebra $\hat{\mathfrak{t}}_{{\hspace{1pt}\rm III}}$
  of the type-{\bf AIII} decomposition of Eq.~\ref{eqintrsubalg}.

  Now let the {\em intrinsic} decompositions of the three types be constructed in
  the {\em intrinsic quotient-algebra partition}.
  For any other decomposition of the same type is related to the intrinsic one simply by a conjugate transformation.
  The intrinsic quotient-algebra partition of rank $r$
  $\{\mathcal{P}_{\mathcal{Q}}(\mathfrak{B}^{[r]}_{intr})\}$ is generated by
  the $r$-th maximal bi-subalgebra
  $\mathfrak{B}^{[r]}_{intr}
  =\{{\cal S}^{\nu_{\mathbf{0}}}_{{\bf 0}}:
  \forall\hspace{2pt} \nu_{\mathbf{0}}\in{Z^p_2},\rho_i\in{Z_2},1\leq i\leq p,
  \hspace{2pt}\nu_{\mathbf{0}}
  =\rho_1\rho_2\cdots\rho_p\text{ and }\rho_{i\leq r}=0\}$
  of the intrinsic Cartan subalgebra
  $\mathfrak{C}_{[\mathbf{0}]}=\{{\cal S}^{\nu_0}_{{\bf 0}}:\forall\hspace{2pt} \nu_0\in{Z^p_2}\}\subset su(2^p)$.
  The conditioned subspaces in $\{\mathcal{P}_{\mathcal{Q}}(\mathfrak{B}^{[r]}_{intr})\}$
  have the form,
  $\forall\hspace{2pt}\epsilon\in{Z_2}$, $l\in{Z^r_2}$ and $\alpha\in{Z^p_2}$,
 \begin{align}\label{eqgenWinQAPrankr}
  &{W}^\epsilon(\mathfrak{B}_{\alpha},\mathfrak{B}^{[r]}_{intr};l)\notag\\
  &\hspace{-3pt}=\{{\cal S}^{\zeta}_{\alpha}:
  \forall\hspace{2pt} \zeta\in{Z^p_2},\hspace{2pt}\rho_i\in{Z_2},\hspace{2pt}r<i\leq p,\hspace{2pt}
  \zeta=l\circ\rho_{r+1}\rho_{r+2}\cdots\rho_p\text{ and }\zeta\cdot\alpha=1+\epsilon\},
 \end{align}
  or explicitly read as
 \begin{align}
  &{W}^0(\mathfrak{B}_{\alpha=\mathbf{0}},\mathfrak{B}^{[r]}_{intr};l)=\{0\},\notag\\
  &{W}^1(\mathfrak{B}_{\alpha=\mathbf{0}},\mathfrak{B}^{[r]}_{intr};l)=\mathfrak{B}^{[r,l]}_{intr}\notag\\
  &\hspace{-3pt}=\{{\cal S}^{\nu_l}_{\mathbf{0}}:
  \forall\hspace{2pt} \nu_l\in{Z^p_2},\hspace{2pt}\rho_i\in{Z_2},\hspace{2pt}r<i\leq p,\hspace{2pt}
  \nu_l=l\circ\rho_{r+1}\rho_{r+2}\cdots\rho_p\},\notag\\
  &{W}^0(\mathfrak{B}_{\alpha\neq\mathbf{0}},\mathfrak{B}^{[r]}_{intr};l)=\hat{W}(\mathfrak{B}_{\alpha\neq\mathbf{0}},\mathfrak{B}^{[r]}_{intr};l)\notag \\
  &\hspace{-3pt}=\{{\cal S}^{\hat{\zeta}}_{\alpha}:
  \forall\hspace{2pt} \hat{\zeta}\in{Z^p_2},\hspace{2pt}\hat{\tau}_i\in{Z_2},\hspace{2pt}r<i\leq p,\hspace{2pt}
  \hat{\zeta}=l\circ\hat{\tau}_{r+1}\hat{\tau}_{r+2}\cdots\hat{\tau}_p\text{ and }\hat{\zeta}\cdot\alpha=1\},\text{ and}\notag \\
  &{W}^1(\mathfrak{B}_{\alpha\neq\mathbf{0}},\mathfrak{B}^{[r]}_{intr};l)={W}(\mathfrak{B}_{\alpha},\mathfrak{B}^{[r]}_{intr};l)\notag\\
  &\hspace{-3pt}=\{{\cal S}^{\zeta}_{\alpha}:
  \forall\hspace{2pt} \zeta\in{Z^p_2},\hspace{2pt}\tau_i\in{Z_2},\hspace{2pt}r<i\leq p,\hspace{2pt}
  \zeta=l\circ\tau_{r+1}\tau_{r+2}\cdots\tau_p\text{ and }\zeta\cdot\alpha=0\},
 \end{align}
  where
  $\mathfrak{B}_{\alpha}=\{{\cal S}^{\nu}_{\mathbf{0}}:\forall\hspace{2pt} \nu\in{Z^p_2},\hspace{2pt}\nu\cdot\alpha=0\}
  \in\mathcal{G}(\mathfrak{C}_{[\mathbf{0}]})$
  is a maximal bi-subalgebra of $\mathfrak{B}_{\mathbf{0}}=\mathfrak{C}_{[\mathbf{0}]}$.
  The symbol ``$\circ$" denotes the concatenation of two strings, namely
  the first $r$ digits of the string $\nu_l$ coinciding with those of the index string $l$ and
  similarly for the strings $\zeta$ and $\hat{\zeta}$.
  This index specification, differing from the choice 
  in Lemma~\ref{lemQArankforsuN}, is 
  convenient despite
  multiple alternatives of the subspace labelling, {\em cf.} Lemma~\ref{lemcosetparinC}.
  Besides being associated to the subspace index $l$, the generators of
  the conditioned subspace ${W}^\epsilon(\mathfrak{B}_{\alpha},\mathfrak{B}^{[r]}_{intr};l)$
  share the same binary-partitioning string $\alpha$ and self
  parity $1+\epsilon$.

  Since the subalgebra of the form
 \begin{align}\label{eqintrAIt}
  \mathfrak{t}^{\bot}_{\mathbf{1}}=\hat{\mathfrak{t}}_{\mathbf{1}}
  =\{{W}^0(\mathfrak{B}_{\beta},\mathfrak{B}^{[r]}_{intr};l):\forall\hspace{2pt}\mathfrak{B}_{\beta}\in\mathcal{G}(\mathfrak{C}_{[\mathbf{0}]}),
  \hspace{1pt}l\in{Z^r_2}\}
 \end{align}
  derived according to Eq.~\ref{eq3typsform} is identical to the algebra $\hat{\mathfrak{t}}_{\hspace{1pt}{\rm I}}=so(2^p)$ of
  Eq.~\ref{eqintrsubalg},
  the subalgebra $\hat{\mathfrak{t}}_{\mathbf{1}}$
  and its complement
  $\mathfrak{p}^{\bot}_{\mathbf{1}}=\hat{\mathfrak{p}}_{\mathbf{1}}
  =\{\mathcal{P}_{\mathcal{Q}}(\mathfrak{B}^{[r]}_{intr})\}-\hat{\mathfrak{t}}_{\mathbf{1}}$
  form the intrinsic Cartan decomposition of type {\bf AI}.
  By choosing the maximal subgroups
  $\hat{{\cal T}}_0=\{\mathfrak{B}_{\beta}:\forall\hspace{2pt}\beta=b_1b_2\cdots b_p\in{Z^p_2}\text{ and }b_1=0\}\subset\mathcal{G}(\mathfrak{C}_{[\mathbf{0}]})$
  and
  $\hat{R}_0=\{s:\forall\hspace{2pt}s=\sigma_1\sigma_2\cdots\sigma_r\in{Z^r_2},\hspace{2pt}\sigma_1=0\}\subset{Z^r_2}$,
  there create the $1$st maximal bi-subalgebra
  $\hat{\mathfrak{B}}^{\dag}=
  \bigcup_{\hspace{.5pt}s\in{\hat{R}_0}}{W}^1(\mathfrak{C}_{[\mathbf{0}]},\mathfrak{B}^{[r]}_{intr};s)\subset\mathfrak{C}_{[\mathbf{0}]}$
  and the conditioned subspaces ${W}^0(\hat{\mathfrak{B}}^{\dag})
  =\bigcup_{\hspace{.5pt}s\in{\hat{R}_0}}(\hat{\mathfrak{B}}^{\dag},\mathfrak{B}^{[r]}_{intr};s)$
  and
  ${W}^1(\hat{\mathfrak{B}}^{\dag})
  =\bigcup_{\hspace{.5pt}t\in{\hat{R}^c_0}}(\hat{\mathfrak{B}}^{\dag},\mathfrak{B}^{[r]}_{intr};t)$
  of $\hat{\mathfrak{B}}^{\dag}$, here ${W}^0(\hat{\mathfrak{B}}^{\dag},\mathfrak{B}^{[r]}_{intr};\mathbf{0})\neq\{0\}$.
  As taking $R_0=\hat{R}_0$ and ${\cal T}_0=\hat{{\cal T}}_0$ of Eq.~\ref{eq3typsform},
  the subalgebra
 \begin{align}\label{eqintrAIIt}
  \breve{\mathfrak{t}}^{\bot}_{\mathbf{2}}=\hat{\mathfrak{t}}_{\mathbf{2}}
  =\{&{W}^0(\mathfrak{B}_{\beta},\mathfrak{B}^{[r]}_{intr};s),{W}^1(\mathfrak{B}_{\beta},\mathfrak{B}^{[r]}_{intr};t),
  {W}^0(\mathfrak{B}_{\beta'},\mathfrak{B}^{[r]}_{intr};t),{W}^1(\mathfrak{B}_{\beta'}, \mathfrak{B}^{[r]}_{intr};s):
  \notag\\
  &\forall\hspace{2pt}\mathfrak{B}_{\beta}\in\hat{{\cal T}}_0,\hspace{2pt}
  \mathfrak{B}_{\beta'}\in\hat{{\cal T}}^c_0=\mathcal{G}(\mathfrak{C}_{[\mathbf{0}]})-\hat{{\cal T}}_0,
  \hspace{2pt}s\in{\hat{R}_0}\text{ and }t\in{\hat{R}^c_0}={Z^r_2}-{\hat{R}_0}\}
 \end{align}
  of Eq.~\ref{eq3typsform} is identified to be the algebra
  $\hat{\mathfrak{t}}_{\hspace{1pt}{\rm II}}=sp(2^{p-1})$ of
  Eq.~\ref{eqintrsubalg}.
  Thus the composition $\hat{\mathfrak{t}}_{\mathbf{2}}\oplus\hat{\mathfrak{p}}_{\mathbf{2}}$
  is the intrinsic Cartan decomposition of type {\bf AII},
  here $\breve{\mathfrak{p}}^{\bot}_{\mathbf{2}}=\hat{\mathfrak{p}}_{\mathbf{2}}
  =\{\mathcal{P}_{\mathcal{Q}}(\mathfrak{B}^{[r]}_{intr})\}-\hat{\mathfrak{t}}_{\mathbf{2}}$.
  Lastly, the subalgebra
 \begin{align}\label{eqintrAIIIt}
  \mathfrak{t}^{\top}_{\mathbf{3}}=\hat{\mathfrak{t}}_{\mathbf{3}}=
  \{{W}^{\epsilon}(\mathfrak{B}_{\beta},\mathfrak{B}^{[r]}_{intr};i):
  \forall\hspace{2pt}\mathfrak{B}_{\beta}\in\hat{{\cal T}}_0,\hspace{2pt}\epsilon\in{Z_2}\text{ and }i\in{Z^r_2}\}
 \end{align}
  is identical to
  $\hat{\mathfrak{t}}_{\hspace{1pt}{\rm III}}=c\otimes su(2^{p-1})\otimes su(2^{p-1})$ of
  Eq.~\ref{eqintrsubalg}
  and the decomposition
  $\hat{\mathfrak{t}}_{\mathbf{3}}\oplus\hat{\mathfrak{p}}_{\mathbf{3}}$
  is the intrinsic decomposition of type {\bf AIII},
  here $\mathfrak{p}^{\top}_{\mathbf{3}}=\hat{\mathfrak{p}}_{\mathbf{3}}
  =\{\mathcal{P}_{\mathcal{Q}}(\mathfrak{B}^{[r]}_{intr})\}-\hat{\mathfrak{t}}_{\mathbf{3}}$.
  A type-{\bf A$\Omega$} decomposition can be mapped to its 
  intrinsic of the same type by the conjugate transformation
  delivered in Appendix~\ref{appTransCDs}.
  Depicted in Appendix~\ref{appExpCD} are
  examples of decompositions for $su(8)$. 
 \end{proof}
 \vspace{6pt}
 The arguments of Lemmas from~\ref{lem2kindmaxsubg} to~\ref{lemmaxsubCD}
 can concluded into a terse law of type decision.
 \vspace{6pt}
 \begin{thm}\label{thmtypeCDabel}
 A Cartan decomposition $\mathfrak{t}\oplus\mathfrak{p}$ 
 is a Cartan decomposition of type {\bf AI} if a Cartan subalgebra $\mathfrak{C}$
 is recovered in the subspace $\mathfrak{p}$,
 a decomposition of type {\bf AII}
 if the maximal abelian subalgebra in $\mathfrak{p}$
 is a proper maximal bi-subalgebra $\mathfrak{B}$ of $\mathfrak{C}$,
 or a type {\bf AIII} if the maximal abelian
 subalgebra in $\mathfrak{p}$ is a complement $\mathfrak{B}^{c}=\mathfrak{C}-\mathfrak{B}$.
 \end{thm}
 \vspace{6pt}
  As a consequence of appropriate distributions of conditioned subspaces,
  there determine all the three types of Cartan decompositions
  in a quotient-algebra partition.
 \vspace{6pt}
 \begin{cor}\label{coroCDtype}
  The quotient-algebra partition of rank $r$ $\{\mathcal{P}_{\mathcal{Q}}(\mathfrak{B}^{[r]})\}$
  generated by an $r$-th maximal bi-subalgebra
  $\mathfrak{B}^{[r]}$
  of a Cartan subalgebra $\mathfrak{C}\subset{su(N)}$
  admits Cartan decompositions of types {\bf AI}, {\bf AII} and {\bf AIII} as $1\leq r<p$
  and the types {\bf AI} and {\bf AIII} as $r=0$.
 \end{cor}
 \vspace{2pt}
 \begin{proof}
  This corollary is evident because every subalgebra of Eq.~\ref{eq3typsform} of Lemma~\ref{lemmaxsubCD}
  is acquirable as $1\leq r<p$
  and only the subalgebras $\mathfrak{t}^{\bot}_{\mathbf{1}}$,
  $\overline{\mathfrak{t}}^{\bot}_{\mathbf{1}}$ and
  $\mathfrak{t}^{\top}_{\mathbf{3}}$ as $r=0$.
 \end{proof}
 \vspace{6pt} \noindent

  To construct a Cartan decomposition
  $su(N)=\mathfrak{t}\oplus\mathfrak{p}$ in a quotient or
  a co-quotient algebra $\{{\cal Q}(\mathcal{A})\}$~\cite{Su,SuTsai1},
  the center subalgebra $\mathcal{A}$ has to be absorbed in $\mathfrak{p}$
  such that if one conditioned subspace of a conjugate pair is in
  $\mathfrak{t}$, the other subspace of the pair shall belong
  to $\mathfrak{p}$.
  Rather than only the type {\bf AI} in a quotient algebra of rank zero,
  there allow two types of Cartan decompositions
  in a quotient algebra of rank $\hspace{0pt} r>0$.
 \vspace{6pt} \noindent
 \begin{cor}\label{coroCDQA}
  The quotient algebra of rank $r$\hspace{2pt} $\{{\cal Q}(\mathfrak{B}^{[r]};2^{p+r}-1)\}$
  given by an $r$-th maximal bi-subalgebra $\mathfrak{B}^{[r]}$ of a Cartan subalgebra $\mathfrak{C}$
  admits Cartan decompositions of types {\bf AI} and {\bf AII} as $1\leq r<p$
  and solely the type {\bf AI} as $r=0$.
 \end{cor}
 \vspace{3pt}
 \begin{proof}
  This lemma follows the inclusion that 
  the center subalgebra $\mathfrak{B}^{[r]}$
  is a subset of the subspace $\mathfrak{p}^{\bot}_{\mathbf{1}}$
  ($\overline{\mathfrak{p}}^{\bot}_{\mathbf{1}}$, $\breve{\mathfrak{p}}^{\bot}_{\omega}$
  or $\widetilde{\mathfrak{p}}^{\bot}_{\omega}$)
  by Eq.~\ref{eqpforms}
  or 
  of the subalgebra $\mathfrak{t}^{\top}_{\mathbf{3}}$
  ($\overline{\mathfrak{t}}^{\top}_{\mathbf{3}}$ or $\breve{\mathfrak{t}}^{\top}_{\mathbf{3}}$)
  by Eq.~\ref{eq3typsform}.
 \end{proof}
 \vspace{6pt} \noindent
  While decompositions of all the three types are attainable in
  co-quotient algebras. 
 \vspace{6pt}
 \begin{cor}\label{coroCDcoQAdeg}
  With a non-null degrade conditioned subspace
  ${W}^{\epsilon}(\mathfrak{B},\mathfrak{B}^{[r]};l)\in\{{\cal Q}(\mathfrak{B}^{[r]})\}$ else from $\mathfrak{B}^{[r]}$
  as the center subalgebra,
  the co-quotient algebra of rank $r$
  $\{{\cal Q}({W}^{\epsilon}(\mathfrak{B},\mathfrak{B}^{[r]};l); 
  2^{p+r}-2^{2r-2})\}$
  given by ${W}^{\epsilon}(\mathfrak{B},\mathfrak{B}^{[r]};l)$
  admits Cartan decompositions of types {\bf AI}, {\bf AII} and {\bf AIII} as $1<r<p$
  and the types {\bf AI} and {\bf AIII} as $r=1$,
  here $\mathfrak{B}\in\mathcal{G}(\mathfrak{C})$,
  $\mathfrak{B}\supset\mathfrak{B}^{[r]}$, $\epsilon\in{Z_2}$ and
  $l\in{Z^r_2}$.
 \end{cor}
 \vspace{3pt}
 \begin{proof}
  Since the center subalgebra
  ${W}^{\epsilon}(\mathfrak{B},\mathfrak{B}^{[r]};l)$
  appears in every subalgebra of Eq.~\ref{eqpforms}
  as $1< r<p$, all the three types of Cartan decompositions can be determined in the co-quotient algebra.
  Yet as $r=1$, the subspace
  $\breve{\mathfrak{p}}^{\bot}_{\mathbf{2}}$
  ($\widetilde{\mathfrak{p}}^{\bot}_{\mathbf{2}}$)
  of Eq.~\ref{eqpforms} contains no non-null degrade conditioned
  subspaces except $\mathfrak{B}^{[1]}$, which excludes the type {\bf AII} from this structure.
  Notice that such a co-quotient algebra starts with rank one because
  the only non-null degrade conditioned subspace in a quotient-algebra partition
  of rank zero is a Cartan subalgebra.
 \end{proof}
 \vspace{6pt}\noindent

 \vspace{6pt}
 \begin{cor}\label{coroCDcoQAreg}
  With a regular conditioned subspace
  ${W}^{\sigma}(\mathfrak{B}',\mathfrak{B}^{[r]};s)\in\{{\cal Q}(\mathfrak{B}^{[r]})\}$
  as the center subalgebra,
  the co-quotient algebra of rank $r$
  $\{{\cal Q}({W}^{\sigma}(\mathfrak{B}',\mathfrak{B}^{[r]};s);2^{p+r}-1)\}$
  given by ${W}^{\sigma}(\mathfrak{B}',\mathfrak{B}^{[r]};s)$
  admits the Cartan decompositions of types {\bf AI}, {\bf AII} and {\bf AIII} as $1\leq r<p$
  and the types {\bf AI} and {\bf AIII} as $r=0$,
  here $\mathfrak{B}'\in\mathcal{G}(\mathfrak{C})$, $\mathfrak{B}'\nsupseteq\mathfrak{B}^{[r]}$,
  $\sigma\in{Z_2}$ and $s\in{Z^r_2}$.
 \end{cor}
 \vspace{3pt}
 \begin{proof}
  Since every subspace of Eq.~\ref{eqpforms}
  covers the center subalgebra
  ${W}^{\sigma}(\mathfrak{B}',\mathfrak{B}^{[r]};s)$
  as $1\leq r<p$,
  there permit all the three types of Cartan decompositions in such a co-quotient algebra.
  But the type {\bf AII} is forbidden as $r=0$ 
  due to 
  $\mathfrak{p}^{\bot}_{\mathbf{1}}=\widetilde{\mathfrak{p}}^{\bot}_{\omega}$
  and
  $\overline{\mathfrak{p}}^{\bot}_{\mathbf{1}}=\breve{\mathfrak{p}}^{\bot}_{\omega}$
  by Eq.~\ref{eqpforms}.
 \end{proof}
 \vspace{6pt}\noindent
  Differing from Corollaries~\ref{coroCDcoQAdeg} and~\ref{coroCDcoQAreg},
  Corollary~\ref{coroCDQA} is applied to algebras of even dimensions only.

  A classical method of searching Cartan decompositions of a Lie
  algebra $\mathfrak{g}$ is based on the hunt of
  the {\em involutions} of $\mathfrak{g}$~\cite{Helgason}.
  An involution $T:\mathfrak{g}\rightarrow\mathfrak{g}$ is a linear automorphism
  satisfying the condition $T^2=I$ and preserving every inner product
  $<X, Y>=<T(X), T(Y)>$ defined on $\mathfrak{g}$ for all $X,Y\in\mathfrak{g}$.
  Having the eigenvalue $\pm 1$,
  an involution $T$ spilts
  $su(2^p)=\mathfrak{t}\oplus\mathfrak{p}$ into two eigenspaces
  $\mathfrak{t}=T(\mathfrak{t})$ of the eigenvalue $+1$ and $\mathfrak{p}=-T(\mathfrak{p})$
  of the eigenvalue $-1$.
  It is easy to show that the composition $\mathfrak{t}\oplus\mathfrak{p}$
  obeys the conditions $[\mathfrak{t},\mathfrak{t}]\subset\mathfrak{t}$,
  $[\mathfrak{t},\mathfrak{p}]\subset\mathfrak{p}$ and
  $[\mathfrak{p},\mathfrak{p}]\subset\mathfrak{t}$.
  Furthermore, owing to the preservation of inner products under $T$,
  the orthogonality condition is acquired
  ${\rm Tr}(\mathfrak{t}\hspace{.5pt}\mathfrak{p})={\rm Tr}(T(\mathfrak{t})T(\mathfrak{p}))
  =-{\rm Tr}(\mathfrak{t}\hspace{.5pt}\mathfrak{p})=0$.
  Thus, an involution of $\mathfrak{g}$ leads to a Cartan decomposition.
  Conversely, a Cartan decomposition of $\mathfrak{g}$ corresponds to
  an involution of $\mathfrak{g}$.

  There exist three types of Cartan decompositions in $su(2^p)$
  by Theorem~\ref{thmtypeCDabel}.
  Let $T_{\Omega}$  denote an involution of a decomposition of type {\bf A$\Omega$}
  $\mathfrak{t}_{\Omega}\oplus\mathfrak{p}_{\Omega}$,
  here the type {\bf A$\Omega$} being referring to type {\bf AI}, {\bf AII} or {\bf AIII}.
  The construction of $T_{\Omega}$ is realized by acquiring the involution
  $\hat{T}_{\Omega}$ of the intrinsic decomposition of the same type
  $\hat{\mathfrak{t}}_{\Omega}\oplus\hat{\mathfrak{p}}_{\Omega}$
  and then map $\hat{T}_{\Omega}$ to $T_{\Omega}=Q_{\Omega}\hat{T}_{\Omega}Q^{\dag}_{\Omega}$
  through the transformation $Q_{\Omega}$ obeying
  $\mathfrak{t}_{\Omega}=Q^{\dag}_{\Omega}\hat{\mathfrak{t}}_{\Omega}Q_{\Omega}$
  and
  $\mathfrak{p}_{\Omega}=Q^{\dag}_{\Omega}\hat{\mathfrak{p}}_{\Omega}Q_{\Omega}$,
  {\em cf.} Appendix~\ref{appTransCDs}.
  Thus, the involutions of all Cartan decompositions can be constructed
  as long as the involutions of the intrinsic decompositions of the three types are given.
  \vspace{6pt}
  \begin{cor}
   In the Lie algebra $su(2^p)$,
   the involutions of the intrinsic Cartan decompositions of
   type {\bf AI} $\hat{\mathfrak{t}}_{{\rm I}}\oplus\hat{\mathfrak{p}}_{{\rm I}}$,
   type {\bf AII} $\hat{\mathfrak{t}}_{{\rm II}}\oplus\hat{\mathfrak{p}}_{{\rm II}}$
   and type {\bf AIII}
   $\hat{\mathfrak{t}}_{{\rm III}}\oplus\hat{\mathfrak{p}}_{{\rm III}}$
   respectively take the forms
   $\hat{T}_{{\rm I}}({\cal S}^{\zeta}_{\alpha})=(-1)^{1+\zeta\cdot\alpha}{\cal S}^{\zeta}_{\alpha}$,
   $\hat{T}_{{\rm II}}({\cal S}^{\eta}_{\beta})
  ={\cal S}^{\tau_0}_{\tau_0}{\cal S}^{\eta}_{\beta}{\cal S}^{\tau_0}_{\tau_0}$
  and
   $\hat{T}_{{\rm III}}({\cal S}^{\xi}_{\gamma})
  ={\cal S}^{\tau_0}_{\mathbf{0}}{\cal S}^{\xi}_{\gamma}{\cal S}^{\tau_0}_{\mathbf{0}}$
  for all
  ${\cal S}^{\zeta}_{\alpha}\in\hat{\mathfrak{t}}_{{\rm I}}\oplus\hat{\mathfrak{p}}_{{\rm I}}$,
  ${\cal S}^{\eta}_{\beta}\in\hat{\mathfrak{t}}_{{\rm II}}\oplus\hat{\mathfrak{p}}_{{\rm II}}$
  and
  ${\cal S}^{\xi}_{\gamma}\in\hat{\mathfrak{t}}_{{\rm III}}\oplus\hat{\mathfrak{p}}_{{\rm III}}$,
  where the string $\tau_0=10\cdots 0\in{Z^p_2}$ has only one single nonzero bit
  in the leftmost digit.
  \end{cor}
  \vspace{6pt}

 By slightly extending the proofs of
 Lemmas~\ref{lemmaxabelintp} and~\ref{lemmaxsubCD},
 it is not hard to read an implication of identical sets of decompositions.
 That is,
 the set of type-{\bf AI} decompositions determined from quotient-algebra
 partitions of rank zero generated by the set of all Cartan subalgebras
 $\{\mathfrak{C}\}\subset su(2^p)$ is identical to
 that determined from partitions of rank $r$ generated by the set of
 all $r$-th maximal bi-subalgebras $\{\mathfrak{B}^{[r]}\}\subset\{\mathfrak{C}\}\subset su(2^p)$,
 $1\leq r<p$.
 Similarly, an identical set of type-{\bf AII} decompositions is developed
 from partitions of rank one generated by the set of all
 $1$st maximal bi-subalgebra $\{\mathfrak{B}^{[1]}\}$ of $\{\mathfrak{C}\}$
 and from those of rank $r'$ generated by the set of all $r'$-th
 maximal bi-subalgebras $\{\mathfrak{B}^{[r']}\}$ in $\{\mathfrak{C}\}\subset su(2^p)$,
 $1<r'<p$.
 Also, the quotient-algebra partitions of rank zero and rank $r$ respectively generated by
 $\{\mathfrak{C}\}$ and $\{\mathfrak{B}^{[r']}\}$ of $\{\mathfrak{C}\}$
 contribute an identical set of the type {\bf AIII}. 
 To proceed with the assertion, let a set of Cartan
 decompositions be denoted with the specifications of the type and the
 bi-subalgebras generating the partitions within which the decompositions are decided. 
 \vspace{6pt}
 \begin{defn}\label{defCDset}
 The decomposition set $\mathfrak{D}_{\Omega}\{\mathfrak{B}^{[r]}\}$ is a
 collection of Cartan decompositions of type {\bf A$\Omega$} determined within 
 quotient-algebra partitions generated by the set of all $r$-th maximal
 bi-subalgebras $\{\mathfrak{B}^{[r]}\}$ in Cartan subalgebras $\{\mathfrak{C}\}\subset su(2^p)$, $0\leq r<p$;
 the type {\bf A$\Omega$} is referring to the type {\bf AI}, {\bf AII} or {\bf AIII}.
 \end{defn}
 \vspace{6pt} \noindent
 Locating its Cartan subalgebras is essential to compassing the
 possible decompositions of a Lie algebra.
 It is reminded that, as introduced in~\cite{Su} and proved in~\cite{SuTsai1},
 the set of all Cartan subalgebras of $su(N)$ can be grown through the procedure of
 the {\em subalgebra extension} based the structures of rank-zero quotient algebras.
 Then the production of abelian bi-subalgebras becomes systematic~\cite{SuTsai1} as long as the complete set of
 Cartan subalgebras is provided.
 Decompositions of a unitary Lie algebra culminate with the following relation.
 \vspace{6pt}
 \begin{thm}\label{thmidentCD}
 The decomposition sets of the three types generated by bi-subalgebras of different ranks
 are related according to the identities
 $\mathfrak{D}_{{\rm I}}\{\mathfrak{C}\}=\mathfrak{D}_{{\rm I}}\{\mathfrak{B}^{[r]}\}$,
 $\mathfrak{D}_{{\rm II}}\{\mathfrak{B}^{[1]}\}=\mathfrak{D}_{{\rm II}}\{\mathfrak{B}^{[r']}\}$
 and
 $\mathfrak{D}_{{\rm III}}\{\mathfrak{C}\}=\mathfrak{D}_{{\rm III}}\{\mathfrak{B}^{[r'']}\}$ 
 for $1\leq r,r''<p$ and $1<r'<p$.
\end{thm}
 \vspace{3pt}
 \begin{proof}
 To verify the first identity
 $\mathfrak{D}_{{\rm I}}\{\mathfrak{C}\}=\mathfrak{D}_{{\rm I}}\{\mathfrak{B}^{[r]}\}$,
 suppose that
 $\mathfrak{t}_{\hspace{1pt}{\rm I}}\oplus\mathfrak{p}_{\hspace{1pt}{\rm I}}\in\mathfrak{D}_{{\rm I}}\{\mathfrak{C}\}$
 is an arbitrary type-{\bf AI} decomposition determined from the
 quotient-algebra of rank zero given by a Cartan subalgebra
 $\mathfrak{C}_{\mathbf{1}}\subset\mathfrak{p}_{\hspace{1pt}{\rm I}}$ of $su(2^p)$.
 This decomposition
 can also be determined from a quotient-algebra partition of rank $r$
 given by an $r$-th maximal bi-subalgebra $\mathfrak{B}^{[r]}_{\mathbf{1}}\subset\mathfrak{C}_{\mathbf{1}}$
 as per Lemma~\ref{lemmaxsubCD}, which leads to
 the inclusion 
 $\mathfrak{D}_{{\rm I}}\{\mathfrak{C}\}\subset\mathfrak{D}_{{\rm I}}\{\mathfrak{B}^{[r]}\}$.
 On the other hand, a decomposition
 $\mathfrak{t}^{\dag}_{\hspace{1pt}{\rm I}}\oplus\mathfrak{p}^{\dag}_{\hspace{1pt}{\rm I}}\in\mathfrak{D}_{{\rm I}}\{\mathfrak{B}^{[r]}\}$
 determined from a quotient-algebra partition of rank $r$ given by $\mathfrak{B}^{\dag[r]}_{\mathbf{1}}$
 belongs to $\mathfrak{D}_{{\rm I}}\{\mathfrak{C}\}$ too and thus
 $\mathfrak{D}_{{\rm I}}\{\mathfrak{B}^{[r]}\}\subset\mathfrak{D}_{{\rm I}}\{\mathfrak{C}\}$,
 because the maximal abelian subalgebra of the subspace
 $\mathfrak{p}^{\dag}_{\hspace{1pt}{\rm I}}$ is a Cartan subalgebra $\mathfrak{C}^{\dag}_{\mathbf{1}}\supset\mathfrak{B}^{\dag[r]}_{\mathbf{1}}$
 by Lemma~\ref{lemmaxsubCD}.
 It asserts the identity
 $\mathfrak{D}_{{\rm I}}\{\mathfrak{C}\}=\mathfrak{D}_{{\rm I}}\{\mathfrak{B}^{[r]}\}$.

 Let
 $\mathfrak{t}_{\hspace{1pt}{\rm II}}\oplus\mathfrak{p}_{\hspace{1pt}{\rm II}}\in\mathfrak{D}_{{\rm II}}\{\mathfrak{B}^{[1]}\}$
 be an arbitrary type-{\bf AII} decomposition determined from
 the quotient-algebra partition of rank one given by
 a maximal bi-subalgebra $\mathfrak{B}^{[1]}_{\mathbf{2}}\subset\mathfrak{p}_{\hspace{1pt}{\rm II}}$
 of a Cartan subalgebra $\mathfrak{C}_{\mathbf{2}}$ 
 with
 the coset $\mathfrak{B}^{[1,1]}_{\mathbf{2}}=\mathfrak{C}_{\mathbf{2}}-\mathfrak{B}^{[1]}_{\mathbf{2}}\subset\mathfrak{t}_{\hspace{1pt}{\rm II}}$.
 The same decomposition
 can be determined from a quotient algebra of rank $r'$ given by an $r'$-th maximal bi-subalgebra
 $\mathfrak{B}^{[r']}_{\mathbf{2}}\subset\mathfrak{B}^{[1]}_{\mathbf{2}}$ according to Lemma~\ref{lemmaxsubCD},
 where
 $\mathfrak{B}^{[1]}_{\mathbf{2}}=\bigcup_{\hspace{.5pt}l'\in R}\mathfrak{B}^{[r',l']}_{\mathbf{2}}$,
 $\mathfrak{B}^{[1,1]}_{\mathbf{2}}=\bigcup_{\hspace{.5pt}s'\in R^c}\mathfrak{B}^{[r',s']}_{\mathbf{2}}$,
 $R$ is a maximal subgroup of $Z^{r'}_2$ and $R^c=Z^{r'}_2-R$.
 The inclusion is thus reached that
 $\mathfrak{D}_{{\rm II}}\{\mathfrak{B}^{[1]}\}\subset\mathfrak{D}_{{\rm II}}\{\mathfrak{B}^{[r']}\}$.
 On the other hand, for a decomposition
 $\mathfrak{t}^{\dag}_{\hspace{1pt}{\rm II}}\oplus\mathfrak{p}^{\dag}_{\hspace{1pt}{\rm II}}\in\mathfrak{D}_{{\rm II}}\{\mathfrak{B}^{[r']}\}$
 determined from a quotient-algebra partition of rank $r'$ given by
 $\mathfrak{B}^{\dag[r']}_{\mathbf{2}}$, since the
 maximal abelian subalgebra in the subspace $\mathfrak{p}^{\dag}_{\hspace{1pt}{\rm II}}$ is a
 $1$st maximal bi-subalgebra $\mathfrak{B}^{\dag}_{\mathbf{2}}\supset\mathfrak{B}^{\dag[r']}_{\mathbf{2}}$
 by Lemma~\ref{lemmaxsubCD}, the decomposition
 $\mathfrak{t}^{\dag}_{\hspace{1pt}{\rm II}}\oplus\mathfrak{p}^{\dag}_{\hspace{1pt}{\rm II}}$
 is obtainable as well in the quotient-algebra partition of rank one given
 by $\mathfrak{B}^{\dag}_{\mathbf{2}}$.
 It deduces that
 $\mathfrak{t}^{\dag}_{\hspace{1pt}{\rm II}}\oplus\mathfrak{p}^{\dag}_{\hspace{1pt}{\rm II}}\in\mathfrak{D}_{{\rm II}}\{\mathfrak{B}^{[1]}\}$
 and then $\mathfrak{D}_{{\rm II}}\{\mathfrak{B}^{[r]}\}\subset\mathfrak{D}_{{\rm II}}\{\mathfrak{B}^{[1]}\}$.
 Therefore, the $2$nd relation
 $\mathfrak{D}_{{\rm II}}\{\mathfrak{B}^{[1]}\}=\mathfrak{D}_{{\rm II}}\{\mathfrak{B}^{[r']}\}$ is affirmed.

 Now assume
 $\mathfrak{t}_{\hspace{1pt}{\rm III}}\oplus\mathfrak{p}_{\hspace{1pt}{\rm III}}\in
 \mathfrak{D}_{{\rm III}}\{\mathfrak{C}\}$
 to be an arbitrary type-{\bf AIII} decomposition determined from the quotient-algebra
 partition of rank zero given by the Cartan subalgebra
 $\mathfrak{C}_{\mathbf{3}}\subset\mathfrak{p}_{\hspace{1pt}{\rm III}}$
 There has the inclusion
 $\mathfrak{D}_{{\rm III}}\{\mathfrak{C}_{\mathbf{3}}\}\subset\mathfrak{D}_{{\rm III}}\{\mathfrak{B}^{[r'']}\}$,
 for the same decomposition is achievable
 in a quotient-algebra partition given by an $r''$-th maximal bi-subalgebra
 $\mathfrak{B}^{[r'']}_{\mathbf{3}}\subset\mathfrak{C}_{\mathbf{3}}$
 according to Lemma~\ref{lemmaxsubCD},
 where $\mathfrak{C}_{\mathbf{3}}=\bigcup_{\hspace{.5pt}l''\in \hat{R}}\mathfrak{B}^{[r'',l'']}_{\mathbf{3}}$ is the union
 of the cosets $\mathfrak{B}^{[r'',l'']}_{\mathbf{3}}$ of
 $\mathfrak{B}^{[r'']}_{\mathbf{3}}=\mathfrak{B}^{[r'',\mathbf{0}]}_{\mathbf{3}}$ in $\mathfrak{C}_{\mathbf{3}}$,
 $l''\in{Z^{r''}_2}$.
 On the other hand, a decomposition
 $\mathfrak{t}^{\dag}_{\hspace{1pt}{\rm III}}\oplus\mathfrak{p}^{\dag}_{\hspace{1pt}{\rm III}}\in\mathfrak{D}_{{\rm III}}\{\mathfrak{B}^{[r'']}\}$
 determined from a quotient-algebra partition given by a $r''$-th maximal bi-subalgebra
 ${\mathfrak{B}^{\dag}_{\mathbf{3}}}^{[r'']}={\mathfrak{B}^{\dag}_{\mathbf{3}}}^{[r'',\mathbf{0}]}$
 is also allowed in the quotient-algebra given by a Cartan
 subalgebra $\mathfrak{C}^{\dag}_{\mathbf{3}}=\bigcup_{\hspace{.5pt}s''\in{Z^{r''}_2}}{\mathfrak{B}^{\dag}_{\mathbf{3}}}^{[r'',s'']}$
 according to Lemma~\ref{lemmaxsubCD}.
 It derives that $\mathfrak{t}^{\dag}_{\hspace{1pt}{\rm III}}\oplus\mathfrak{p}^{\dag}_{\hspace{1pt}{\rm III}}\in\mathfrak{D}_{{\rm II}}\{\mathfrak{B}^{[1]}\}$
 and then $\mathfrak{D}_{{\rm II}}\{\mathfrak{C}\}\subset\mathfrak{D}_{{\rm II}}\{\mathfrak{B}^{[r'']}\}$.
 The last identity $\mathfrak{D}_{{\rm III}}\{\mathfrak{C}\}=\mathfrak{D}_{{\rm III}}\{\mathfrak{B}^{[r'']}\}$
 is hence asserted.
 \end{proof}
 \vspace{6pt}
 These identities convey the truth that, independent of the rank, the identical set
 of Cartan decompositions of a certain type is yielded
 provided all quotient-algebra partitions generated by the complete set of bi-subalgebras
 of a specified rank are examined. 
 Accordingly, the sets
 $\mathfrak{D}_{{\rm I}}\{\mathfrak{C}\}=\mathfrak{D}_{{\rm I}}\{\mathfrak{B}^{[r]}\}$,
 $\mathfrak{D}_{{\rm II}}\{\mathfrak{B}^{[r]}\}$ and
 $\mathfrak{D}_{{\rm III}}\{\mathfrak{C}\}=\mathfrak{D}_{{\rm III}}\{\mathfrak{B}^{[r]}\}$
 are considered
 the complete sets of Cartan decompositions of types {\bf AI}, {\bf AII} and {\bf AIII}
 of $su(2^p)$ respectively.
 In practice, it is favorable to determine the three complete sets of decompositions
 within quotient-algebra partitions generated by the sets of all
 Cartan subalgebras $\{\mathfrak{C}\}$ and
 of all the $1$st maximal bi-subalgebras $\{\mathfrak{B}^{[1]}\}$ in $\{\mathfrak{C}\}$.

 Owing to the existence of 
 quotient-algebra partitions in arbitrary dimensions,
 decompositions of all the three types with dimensions not necessarily of powers of $2$
 are accessible as well in these structures resorting to the removing process~\cite{Su}.
 Such examples will be prepared in the sequel work~\cite{SuTsai3}. 

\nonumsection{References}

\appendix{~~Transformation Connecting Quotient-Algebra Partitions\label{appTransOAPs}}
  The transformation mapping a
  quotient-algebra partition of rank $r$ to the {\em canonical} or the {\em intrinsic} 
  of the same rank will be constructed
  in this appendix, $0\leq r<p$.
  Remind that the canonical
  quotient-algebra partition of rank $r$ $\{{\cal P}_{{\cal Q}}(\mathfrak{B}^{[r]}_{can})\}$
  as defined in Lemma~\ref{lemQArankforsuN} is generated by the $r$-th maximal bi-subalgebra
  $\mathfrak{B}^{[r]}_{can}=\{{\cal S}^{\nu_{\mathbf{0}}}_{\mathbf{0}}:
  \forall\hspace{2pt} \nu_{\mathbf{0}}\in{Z^p_2},\hspace{2pt}\rho_i\in{Z_2},\hspace{2pt}1<i\leq p,\hspace{2pt}
  \nu_{\mathbf{0}}=\rho_{1}\rho_{2}\cdots\rho_p\text{ and }\rho_{i>r}=0\}$
  of the intrinsic Cartan subalgebra
  $\mathfrak{C}_{[\mathbf{0}]}=\{{\cal S}^{\nu_0}_{\mathbf{0}}:\nu_0\in Z^p_2\}\subset su(2^p)$.
  While the intrinsic quotient-algebra partition of rank $r$ $\{{\cal P}_{{\cal Q}}(\mathfrak{B}^{[r]}_{intr})\}$
  results from the $r$-th maximal bi-subalgebra
  $\mathfrak{B}^{[r]}_{intr}
  =\{{\cal S}^{\mu_{\mathbf{0}}}_{{\bf 0}}:
  \forall\hspace{2pt} \mu_{\mathbf{0}}\in{Z^p_2},\sigma_i\in{Z_2},1\leq i\leq p,
  \hspace{2pt}\mu_{\mathbf{0}}
  =\sigma_1\sigma_2\cdots\sigma_p\text{ and }\sigma_{i\leq r}=0\}$
  of $\mathfrak{C}_{[\mathbf{0}]}$, {\em cf.} Lemma~\ref{lemmaxsubCD}.
  The following 
  establishes the action $Q^{(r)}$ transforming
  the quotient-algebra partition of rank $r$
  $\{{\cal P}_{{\cal Q}}(\widetilde{\mathfrak{B}}^{[r]})\}$,
  here $\widetilde{\mathfrak{B}}^{[r]}$ being either
  $\mathfrak{B}^{[r]}_{can}$ or $\mathfrak{B}^{[r]}_{intr}$,
  to that of the same rank
  $\{{\cal P}_{{\cal Q}}(\mathfrak{B}^{[r]})\}$ 
  given by an $r$-th maximal bi-subalgebra $\mathfrak{B}^{[r]}$
  of a Cartan subalgebra $\mathfrak{C}\subset{su(2^p)}$,
  specifically 
  $\{{\cal P}_{{\cal Q}}(\widetilde{\mathfrak{B}}^{[r]})\}
  =Q^{(r)}\{{\cal P}_{{\cal Q}}(\mathfrak{B}^{[r]})\}Q^{(r)\dag}$.

  Similarly let the required action be constructed in compositions of
  {\em basic transformations}~\cite{SuTsai1}
 \begin{align}\label{eqbasic-trans}
  h^{\zeta}_{\alpha}=
  \frac{1}{\sqrt{2}}({\cal S}^{\nu}_{\mathbf 0}+i\cdot(-i)^{\zeta\cdot\alpha}{\cal S}^{\zeta}_{\alpha})
  \hspace{3pt}\text{ with }\hspace{2pt}
  \hspace{3pt}\nu\cdot\alpha=0,
 \end{align}
  where the parity $\zeta\cdot\alpha$ takes either $0$ or $1$.
  For the mathematical simplicity, it is set $\nu={\mathbf 0}$.
  Every such transformation is a spinor-to-spinor mapping according to
  the commutation relation of two spinors:
  the operator $h^{\zeta}_{\alpha}$ sends a spinor generator  ${\cal S}^{\eta}_{\beta}$ to
  itself if $[{\cal S}^{\zeta}_{\alpha},{\cal S}^{\eta}_{\beta}]=0$ or to
  ${\cal S}^{\zeta+\eta}_{\alpha+\beta}$ if  $[{\cal S}^{\zeta}_{\alpha},{\cal S}^{\eta}_{\beta}]\neq 0$.
  Based on Theorem~\ref{thmgenWcommrankr},
  two quotient-algebra partitions are identical if they
  are generated by a same $r$-th bi-subalgebra of a Cartan subalgebra.
  Thus the transformation $Q^{(r)}$ is simply an action
  mapping the bi-subalgebra $\mathfrak{B}^{[r]}$ to
  $\widetilde{\mathfrak{B}}^{[r]}$.
  Suppose $\widetilde{\mathfrak{B}}^{[r]}$ and $\mathfrak{B}^{[r]}$ are
  spanned respectively by the two independent sets of spinors
  $\{{\cal S}^{\nu_1}_{\mathbf{0}},{\cal S}^{\nu_2}_{\mathbf{0}},\cdots,{\cal S}^{\nu_{p-r}}_{\mathbf{0}}\}$
  and
  $\{{\cal S}^{\zeta_1}_{\alpha_1},{\cal S}^{\zeta_2}_{\alpha_2},\cdots,{\cal S}^{\zeta_{p-r}}_{\alpha_{p-r}}\}$.
  Similar to Eq.~6.4 in~\cite{SuTsai1}, each spinor ${\cal S}^{\zeta_i}_{\alpha_i}$
  converts into a diagonal 
  ${\cal S}^{\mu_i}_{\mathbf{0}}$
  via the {\em diagonalization operator of rank $r$}
 \begin{align}\label{eqdialrankr}
  R^{(r)}=h^{\eta_1}_{\alpha_1}h^{\eta_2}_{\alpha_2}\cdots h^{\eta_{p-r}}_{\alpha_{p-r}},
 \end{align}
  where $\eta_i\cdot\alpha_j+\zeta_j\cdot\alpha_i=\delta_{ij}$
  for $1\leq i,j\leq p-r$.
  Then the transformed spinor 
  ${\cal S}^{\mu_i}_{\mathbf{0}}=R^{(r)}{\cal S}^{\zeta_i}_{\alpha_i}{R^{(r)}}^{\dag}$
  can be mapped to
  ${\cal S}^{\nu_i}_{\mathbf{0}}$ through an iterative procedure, {\em cf.} Lemma~22 and Eq.~7.6 in~\cite{SuTsai1}.
  The action $e_{\beta_1}=h^{\mathbf{0}}_{\beta_1}h^{\xi_1}_{\beta_1}$
  initiates the iteration by sending ${\cal S}^{\mu_1}_{\mathbf{0}}$
  to ${\cal S}^{\nu_1}_{\mathbf{0}}$
  and ${\cal S}^{\mu_j}_{\mathbf{0}}$
  to ${\cal S}^{\mu^{(1)}_j}_{\mathbf{0}}$,
  here $2\leq j\leq p-r$, $\nu_1\cdot\beta_1=1$, $\nu_1=\mu_1+\xi_1$ and
  $\mu^{(1)}_j=\mu_j$ or $\mu^{(1)}_j=\mu_j+\xi_1$.
  After the application of the $2$nd action
  $e_{\beta_2}=h^{\mathbf{0}}_{\beta_2}h^{\xi_2}_{\beta_2}$,
  the spinor ${\cal S}^{\nu_1}_{\mathbf{0}}$ remains unchanged,
  ${\cal S}^{\mu^{(1)}_2}_{\mathbf{0}}$ is mapped to
  ${\cal S}^{\nu_2}_{\mathbf{0}}$ and ${\cal S}^{\mu^{(1)}_j}_{\mathbf{0}}$
  to ${\cal S}^{\mu^{(2)}_j}_{\mathbf{0}}$,
  for $3\leq j\leq p-r$, $\nu_1\cdot\beta_2=0$, $\nu_2\cdot\beta_2=1$,
  $\nu_2=\mu^{(1)}_2+\xi_2$, and
  $\mu^{(2)}_j=\mu^{(1)}_j$ or $\mu^{(2)}_j=\mu^{(1)}_j+\xi_2$.

  To the $k$-th iteration step,
  the operator
  $e_{\beta_k}=h^{\mathbf{0}}_{\beta_k}h^{\xi_k}_{\beta_k}$
  keeps the first $k-1$ spinors ${\cal S}^{\nu_i}_{\mathbf{0}}$
  unchanged, but maps the spinor ${\cal S}^{\mu^{(k-1)}_k}_{\mathbf{0}}$
  to ${\cal S}^{\nu_k}_{\mathbf{0}}$ and
  ${\cal S}^{\mu^{(k-1)}_j}_{\mathbf{0}}$ to ${\cal S}^{\mu^{(k)}_j}_{\mathbf{0}}$,
  here $1\leq i<k<j\leq p-r$, $\nu_i\cdot\beta_k=0$,
  $\nu_k\cdot\beta_k=1$,
  $\nu_k=\mu^{(k-1)}_k+\xi_k$, and $\mu^{(k)}_j=\mu^{(k-1)}_j$ or $\mu^{(k)}_j=\mu^{(k-1)}_j+\xi_k$.
  Each spinor ${\cal S}^{\mu_i}_{\mathbf{0}}$ is 
  delivered to ${\cal S}^{\nu_i}_{\mathbf{0}}$ by the {\em exchanged operator of rank $r$}
  when the iteration ends
 \begin{align}\label{eqexchagrankr}
  E^{(r)}=\prod^{p-r}_{k=1}e_{\beta_k}
  =e_{\beta_{p-r}}e_{\beta_{p-r-2}}\cdots e_{\beta_1},
 \end{align}
  {\em i.e.}, ${\cal S}^{\nu_i}_{\mathbf{0}}=E^{(r)}{\cal S}^{\mu_i}_{\mathbf{0}}{E^{(r)}}^{\dag}$,
  where the $k$-th action $e_{\beta_k}=h^{\mathbf{0}}_{\beta_k}h^{\xi_k}_{\beta_k}$
  satisfies the conditions $\nu_k\cdot\beta_i=0$, $\nu_k\cdot\beta_k=1$,
  $\nu_k=\mu^{(k-1)}_k+\xi_k$,
  and $\mu^{(k)}_j=\mu^{(k-1)}_j$ or $\mu^{(k)}_j=\mu^{(k-1)}_j+\xi_k$ for $1\leq i< k<j\leq p-r$.
  Accordingly, the required transformation is a composition of the
  diagonaliztion operator of Eq.~\ref{eqdialrankr} and the exchange operator
  of Eq.~\ref{eqexchagrankr}
 \begin{align}\label{eqtransQAP}
  Q^{(r)}=E^{(r)}R^{(r)}.
 \end{align}

 \appendix{~~Transformation Connecting Cartan Decompositions\label{appTransCDs}}
  This appendix is to formulate
  the transformation $Q_{\Omega}$ mapping a Cartan decomposition of type {\bf A$\Omega$}
  $\mathfrak{t}_{\Omega}\oplus\mathfrak{p}_{\Omega}$
  to the {\em intrinsic} of the same type $\hat{\mathfrak{t}}_{\Omega}\oplus\hat{\mathfrak{p}}_{\Omega}$
  such that
  $Q_{\Omega}\mathfrak{t}_{\Omega}Q^\dag_{\Omega}=\hat{\mathfrak{t}}_{\Omega}$
  and
  $Q_{\Omega}\mathfrak{p}_{\Omega}Q^\dag_{\Omega}=\hat{\mathfrak{p}}_{\Omega}$,
  here the subscript {\bf A$\Omega$} being referring to
  type {\bf AI}, {\bf AII} or {\bf AIII}.
  A desired transformation can be a composition of
  two actions $Q_{\Omega}=Q_{in,\Omega}Q^{(r)}$.
  The operator $Q^{(r)}$ as of Eq.~\ref{eqtransQAP} maps the decomposition
  $\mathfrak{t}_{\Omega}\oplus\mathfrak{p}_{\Omega}$
  to a same type $\mathfrak{t}^{\dag}_{\Omega}\oplus\mathfrak{p}^{\dag}_{\Omega}$ 
  determined in the intrinsic quotient-algebra partition
  $\{\mathcal{P}_{\mathcal{Q}}(\mathfrak{B}^{[r]}_{intr})\}$.
  The attention thus has to focus only on the 2nd action $Q_{in,\Omega}$
  that further sends $\mathfrak{t}^{\dag}_{\Omega}\oplus\mathfrak{p}^{\dag}_{\Omega}$
  to the intrinsic
  $\hat{\mathfrak{t}}_{\Omega}\oplus\hat{\mathfrak{p}}_{\Omega}$.
  Based on Eqs.~\ref{eq3typsform} and~\ref{eqgenWinQAPrankr}
  in the proof of Lemma~\ref{lemmaxsubCD},
  the general forms for the subalgebras $\mathfrak{t}^{\dag}_{\Omega}$
  of decompositions
  $\mathfrak{t}^{\dag}_{\Omega}\oplus\mathfrak{p}^{\dag}_{\Omega}$ in
  $\{\mathcal{P}_{\mathcal{Q}}(\mathfrak{B}^{[r]}_{intr})\}$
  are listed as follows,
  \begin{align}\label{eqintrsubalg3typsgform}
   \hat{\mathfrak{t}}^{\bot}_{\mathbf{1}}
   &
   =
   \{{W}^0(\mathfrak{B}_{\beta},\mathfrak{B}^{[r]}_{intr};l):\forall\hspace{2pt}\beta\in{Z^p_2}\text{ and }l\in{Z^r_2}\},\notag\\
   \overline{\mathfrak{t}}^{\bot}_{\mathbf{1}}
   &
   =\{{W}^0(\mathfrak{B}_{\beta'},\mathfrak{B}^{[r]}_{intr};l),
    {W}^1(\mathfrak{B}_{\beta''},\mathfrak{B}^{[r]}_{intr};l):
   \hspace{0pt}\forall\hspace{2pt}\beta'\in{T}_0,\beta''\in{T}^c_0\text{ and }l\in{Z^r_2}\},\notag\\
   \breve{\mathfrak{t}}^{\bot}_{\omega}
   &
   =\{{W}^0(\mathfrak{B}_{\beta'},\mathfrak{B}^{[r]}_{intr};s),
    {W}^1(\mathfrak{B}_{\beta'},\mathfrak{B}^{[r]}_{intr};t),
    {W}^0(\mathfrak{B}_{\beta''},\mathfrak{B}^{[r]}_{intr};t),
    {W}^1(\mathfrak{B}_{\beta''},\mathfrak{B}^{[r]}_{intr};s):\notag\\
   &
   \hspace{20pt}\forall\hspace{2pt}\beta'\in{T}_0,\beta''\in{T}^c_0,s\in{R_0},t\in{R^c_0}
   \text{ and }\beta^{\dag}\in{T}_0\},\notag\\
   \widetilde{\mathfrak{t}}^{\bot}_{\omega}
   &
   =\{{W}^0(\mathfrak{B}_{\beta},\mathfrak{B}^{[r]}_{intr};s),
    {W}^1(\mathfrak{B}_{\beta},\mathfrak{B}^{[r]}_{intr};t):
   \hspace{0pt}\forall\hspace{2pt}\beta\in{Z^p_2},s\in{R_0}\text{ and }t\in{R^c_0}\},\hspace{2pt}\omega=\mathbf{1},\mathbf{2},\notag\\
   &\hspace{5pt}\text{here }\breve{\mathfrak{t}}^{\bot}_{\omega}=\breve{\mathfrak{t}}^{\bot}_{\mathbf{1}}
   \text{ and }\widetilde{\mathfrak{t}}^{\bot}_{\omega}=\widetilde{\mathfrak{t}}^{\bot}_{\mathbf{1}}
   \text{ if }{W}^0(\mathfrak{B}_{\beta^{\dag}},\mathfrak{B}^{[r]};\mathbf{0})=\{0\}\text{ or}\notag\\
   &\hspace{27pt}\breve{\mathfrak{t}}^{\bot}_{\omega}=\breve{\mathfrak{t}}^{\bot}_{\mathbf{2}}
   \text{ and }\widetilde{\mathfrak{t}}^{\bot}_{\omega}=\widetilde{\mathfrak{t}}^{\bot}_{\mathbf{2}}
   \text{ if }{W}^0(\mathfrak{B}_{\beta^\dag},\mathfrak{B}^{[r]};\mathbf{0})\neq\{0\},\notag\\
   \widetilde{\mathfrak{t}}^{\top}_{\mathbf{3}}
   &
   =
   \{{W}^{\epsilon}(\mathfrak{B}_{\beta'},\mathfrak{B}^{[r]}_{intr};l):\forall\hspace{2pt}\epsilon\in{Z_2},\beta'\in{T}_0\text{ and }l\in{Z^r_2}\},\notag\\
   \overline{\mathfrak{t}}^{\top}_{\mathbf{3}}
   &
   =
   \{{W}^{\epsilon}(\mathfrak{B}_{\beta},\mathfrak{B}^{[r]}_{intr};s):\forall\hspace{2pt}\epsilon\in{Z_2},\beta\in{Z^p_2}\text{ and }s\in{R_0}\},\text{ and}\\
   \breve{\mathfrak{t}}^{\top}_{\mathbf{3}}
   &
   =
   \{{W}^{\epsilon}(\mathfrak{B}_{\beta'},\mathfrak{B}^{[r]}_{intr};s),
    {W}^{\sigma}(\mathfrak{B}_{\beta''},\mathfrak{B}^{[r]}_{intr};t):\forall\hspace{2pt}
    \epsilon,\sigma\in{Z_2},\beta'\in{T}_0,\beta''\in{T}^c_0,s\in{R_0}\text{ and }t\in{R^c_0}\},
    \notag
  \end{align}
   where
   $\mathfrak{B}_{\beta}
   =\{{\cal S}^{\zeta}_{\mathbf{0}}:\forall\hspace{2pt}\zeta\cdot\beta=0\}\in\mathcal{G}(\mathfrak{C}_{[\mathbf{0}]})$
   is a maximal bi-subalgebra of $\mathfrak{C}_{[\mathbf{0}]}$,
   $\beta^\dag=10\cdots 0$ is a $p$-digit string of all zero bits except the leftmost,
   $T_0$ (or $R_0$) is a proper maximal subgroup of $Z^p_2$
   (or $Z^r_2$) and $T^c_0=Z^p_2-T_0$ (or $R^c_0=Z^r_2-R_0$).
   Among them,
   the subspace
   $\hat{\mathfrak{t}}^{\bot}_{\mathbf{1}}=\hat{\mathfrak{t}}_{\mathbf{1}}=so(2^p)$
   is the subalgebra
   of the intrinsic decomposition of type {\bf AI}
   $\hat{\mathfrak{t}}_{\mathbf{1}}\oplus\hat{\mathfrak{p}}_{\mathbf{1}}$.
   Assigning
   $R_0=\hat{R}_0=\{s:\forall\hspace{2pt}s=\sigma_1\sigma_2\cdots\sigma_r\in{Z^r_2},\hspace{2pt}\sigma_1=0\}$ 
   and
   $T_0=\hat{T}_0=\{\hat{\beta}:\forall\hspace{2pt}\hat{\beta}=b_1b_2\cdots b_p\in{Z^p_2}\text{ and }b_1=0\}$,
   $\breve{\mathfrak{t}}^{\bot}_{\mathbf{2}}=\hat{\mathfrak{t}}_{\mathbf{2}}=sp(2^{p-1})
   =\{{W}^0(\mathfrak{B}_{\beta},\mathfrak{B}^{[r]}_{intr};s),{W}^1(\mathfrak{B}_{\beta},\mathfrak{B}^{[r]}_{intr};t),
   {W}^0(\mathfrak{B}_{\beta'}, \mathfrak{B}^{[r]}_{intr};t),{W}^1(\mathfrak{B}_{\beta'}, \mathfrak{B}^{[r]}_{intr};s):
   \forall\hspace{2pt}\hat{\beta}\in\hat{T}_0,\hspace{2pt}
   \hat{\beta}'\in\hat{T}^c_0=Z^p_2-\hat{T}_0,
   \hspace{2pt}s\in{\hat{R}_0}\text{ and }t\in{\hat{R}^c_0}={Z^r_2}-{\hat{R}_0}\}$
   is correspondent with the subalgebra of
   the intrinsic type-{\bf AII} decomposition $\hat{\mathfrak{t}}_{\mathbf{2}}\oplus\hat{\mathfrak{p}}_{\mathbf{2}}$.
   The subalgebra of the intrinsic type-{\bf AIII} 
   has the form
   $\hat{\mathfrak{t}}_{\mathbf{3}}=c\otimes{su(2^{p-1})}\otimes{su(2^{p-1})}
   =\{{W}^{\epsilon}(\mathfrak{B}_{\hat{\beta}},\mathfrak{B}^{[r]}_{intr};l):
   \forall\hspace{2pt}\epsilon\in{Z_2},\hat{\beta}\in\hat{T}_0\text{ and }l\in{Z^r_2}\}$
   as choosing $T_0=\hat{T}_0$
   in $\widetilde{\mathfrak{t}}^{\top}_{\mathbf{3}}$; 
   the center $c=\{{\cal S}^{\beta_c}_{\mathbf{0}}\}$,
   $\beta_c=\beta^\dag$,
   commutes with $\hat{\mathfrak{t}}_{\mathbf{3}}$.

   The subalgebra $\overline{\mathfrak{t}}^{\bot}_{\mathbf{1}}$ of Eq.~\ref{eqintrsubalg3typsgform}
   can be transformed to
   $\hat{\mathfrak{t}}_{\mathbf{1}}$
   by the operator
  \begin{align}\label{eqtranstyp1bar}
   \overline{Q}_{in,\mathbf{1}}=h^{\xi}_{\mathbf{0}},
  \end{align}
   with $\xi\cdot\beta'=1+\xi\cdot\beta''=0$
   for $\xi\in{Z^p_2}$, $\beta'\in{T}_0$ and $\beta''\in{T^c_0}$.
   This action leaves the subspace ${W}^0(\mathfrak{B}_{\beta'},\mathfrak{B}^{[r]}_{intr};l)$
   intact but maps the subspace
   ${W}^1(\mathfrak{B}_{\beta''},\mathfrak{B}^{[r]}_{intr};l)$
   of parity one to ${W}^0(\mathfrak{B}_{\beta''}, \mathfrak{B}^{[r]}_{intr};l)$ of parity zero.
   By virtue of the action 
  \begin{align}\label{eqtranstyp1bre}
  \breve{Q}_{in,\mathbf{1}}=\overline{Q}_{in,\mathbf{1}}h^{\mathbf{0}}_{\gamma},
  \end{align}
   the subalgebra $\breve{\mathfrak{t}}^{\bot}_{\mathbf{1}}$
   of Eq.~\ref{eqintrsubalg3typsgform} is sent to $\overline{\mathfrak{t}}^{\bot}_{\mathbf{1}}$
   by $h^{\mathbf{0}}_{\gamma}$ and then to $\hat{\mathfrak{t}}_{\mathbf{1}}$
   by $\overline{Q}_{in,\mathbf{1}}$, 
   here $\gamma=\gamma_0\circ\gamma_1$
   being a concatenated string of $\gamma_0\in{Z^{p-r}_2}$ and $\gamma_1\in{Z^r_2}$
   such that $s\cdot\gamma_1=1+t\cdot\gamma_1=0$
   for $s\in{R_0}$ and $t\in{R^c_0}$.
   Being transformed to $\breve{\mathfrak{t}}^{\bot}_{\mathbf{1}}$ by
   $h^{\xi'}_{\mathbf{0}}$ firstly,
   the subalgebra
   $\widetilde{\mathfrak{t}}^{\bot}_{\mathbf{1}}$
   turns into the intrinsic via the composition
  \begin{align}\label{eqtranstyp1qta}
   \widetilde{Q}_{in,\mathbf{1}}=\breve{Q}_{in,\mathbf{1}}h^{\xi'}_{\mathbf{0}}.
  \end{align}
   The operator $h^{\xi'}_{\mathbf{0}}$ meets the conditions
   $\xi'\cdot\beta=0$ as $\beta\in{T_0}$
   or $\xi'\cdot\beta=1$ as $\beta\in{T^c_0}$
   as well as $\xi'_1\in{R_0}$
   in the concatenation $\xi'=\xi'_0\circ\xi'_1$
   of $\xi'_0\in{Z^{p-r}_2}$ and $\xi'_1\in{Z^r_2}$.

   The subalgebra
   $\widetilde{\mathfrak{t}}^{\top}_{\mathbf{3}}$ of Eq.~\ref{eqintrsubalg3typsgform}
   is mapped to the intrinsic $\hat{\mathfrak{t}}_{\mathbf{3}}$,
   through the transformation
  \begin{align}\label{eqtranstyp3qta}
   \widetilde{Q}_{in,\mathbf{3}}=h^{\mu}_{\tau}h^{\nu}_{\tau}
  \end{align}
   with $(\mu+\nu)\cdot\tau=1$ and
   $\tau\in{T^c_0}\cap{\hat{T}^c_0}\subset\mathcal{G}(\mathfrak{C}_{[\mathbf{0}]})$.
   A conditioned subspace
   ${W}^\epsilon(\mathfrak{B}_{\beta'},\mathfrak{B}^{[r]}_{intr};l)\in\widetilde{\mathfrak{t}}^{\top}_{\mathbf{3}}$
   remains invariant if $\beta'\in{T}_0\cap\hat{T}_0$ under
   $\widetilde{Q}_{in,\mathbf{3}}$,  
   but changes into
   ${W}^\epsilon(\mathfrak{B}_{\beta''},\mathfrak{B}^{[r]}_{intr};l)$
   with $\beta''\in{T^c_0}\cap{\hat{T}_0}$ if $\beta'\in{T}_0\cap\hat{T}^c_0$.
   The composite action 
  \begin{align}\label{eqtranstyp3bre}
   \breve{Q}_{in,\mathbf{3}}=\widetilde{Q}_{in,\mathbf{3}}h^{\mathbf{0}}_{\tau'}
  \end{align}
   first maps the subalgebra $\breve{\mathfrak{t}}^{\top}_{\mathbf{3}}$
   to $\widetilde{\mathfrak{t}}^{\top}_{\mathbf{3}}$
   and then to the intrinsic, 
   here $\tau'\in{T^c_0}$ and $s\cdot\tau'_1=1+t\cdot\tau'_1=0$
   for $\tau'_1\in{Z^r_2}$ and $\tau'=\tau'_0\circ\tau'_1$.
   Under the application of the operator
 \begin{align}\label{eqtranstyp3bar}
  \overline{Q}_{in,\mathbf{3}}=\breve{Q}_{in,\mathbf{3}}h^{\mu'}_{\mathbf{0}},
 \end{align}
  the subalgebra $\overline{\mathfrak{t}}^{\top}_{\mathbf{3}}$
  of Eq.~\ref{eqintrsubalg3typsgform} is transformed to
  $\breve{\mathfrak{t}}^{\top}_{\mathbf{3}}$
  by $h^{\mu'}_{\mathbf{0}}$ and then unto $\hat{\mathfrak{t}}_{\mathbf{3}}$
  by $\breve{Q}_{in,\mathbf{3}}$.
  The concatenated string $\mu'=\mu'_0\circ\mu'_1$,
  $\mu'_0\in{Z^{p-r}_2}$ and $\mu'_1\in{Z^r_2}$, satisfies the requirement 
  $\mu'_1\in{R_0}$, $\mu'\cdot\beta=0$ as $\beta\in{T_0}$
  or $\mu'\cdot\beta=1$ as $\beta\in{T^c_0}$. 

  For the type {\bf AII}, 
  the transformation of the form
 \begin{align}\label{eqtranstyp2}
  \breve{Q}_{in,\mathbf{2}}=h^{\nu}_{\mathbf{0}}h^{\mathbf{0}}_{\kappa}
 \end{align}
  maps $\breve{\mathfrak{t}}^{\bot}_{\mathbf{2}}$ to the intrinsic
  $\hat{\mathfrak{t}}_{\mathbf{2}}$.
  The component $h^{\nu}_{\mathbf{0}}$
  realizes the identity
  $\nu\cdot\beta'=\nu\cdot\beta''=0$
  for $\beta'\in{T_0\cap{\hat{T}_0}}$ and $\beta''\in{T^c_0\cap{\hat{T}^c_0}}$
  or $\nu\cdot\beta'=\nu\cdot\beta''=1$
  for $\beta'\in{T_0\cap{\hat{T}^c_0}}$ and
  $\beta''\in{T^c_0\cap{\hat{T}_0}}$.
  While the other component $h^{\mathbf{0}}_{\kappa}$
  with the concatenation $\kappa=\kappa_0\circ\kappa_1$ of $\kappa_0\in{Z^{p-r}_2}$
  and $\kappa_1\in{Z^r_2}$
  follows the condition
  $s\cdot\kappa_1=t\cdot\kappa_1=0$ as
  $s\in{R_0\cap{\hat{R}_0}}$
  and
  $t\in{R^c_0\cap{\hat{R}^c_0}}$
  or 
  $s\cdot\kappa_1=t\cdot\kappa_1=1$
  as $s\in{R_0\cap{\hat{R}^c_0}}$
  and
  $t\in{R^c_0\cap{\hat{R}_0}}$.
  The action $\breve{Q}_{in,\mathbf{2}}$ converts the conditioned subspaces
  ${W}^\epsilon(\mathfrak{B}_{\beta'},\mathfrak{B}^{[r]}_{intr};s)$
  (or ${W}^\epsilon(\mathfrak{B}_{\beta''},\mathfrak{B}^{[r]}_{intr};t)$)
  of $\beta'\in{T_0}$ and $s\in{R_0}$ (or $\beta''\in{T^c_0}$ and $t\in{R^c_0}$)
  to those
  ${W}^\epsilon(\mathfrak{B}_{\hat{\beta}'},\mathfrak{B}^{[r]}_{intr};\hat{s})$
  (or ${W}^\epsilon(\mathfrak{B}_{\hat{\beta}''},\mathfrak{B}^{[r]}_{intr};\hat{t})$)
  of $\hat{\beta}'\in{\hat{T}_0}$ and $\hat{s}\in{\hat{R}_0}$
  (or $\hat{\beta}''\in{\hat{T}^c_0}$ and $\hat{t}\in{\hat{R}^c_0}$).
  At last, the subalgebra $\widetilde{\mathfrak{t}}^{\bot}_{\mathbf{2}}$ 
  is delivered to the intrinsic via the operator  
 \begin{align}\label{eqtranstyp2qta}
  \widetilde{Q}_{in,\mathbf{2}}=\breve{Q}_{in,\mathbf{2}}\widetilde{Q}_{in,\mathbf{1}},
 \end{align}
  noting that $\widetilde{Q}_{in,\mathbf{1}}$ sends
  $\widetilde{\mathfrak{t}}^{\bot}_{\mathbf{2}}$ to
  $\breve{\mathfrak{t}}^{\bot}_{\mathbf{2}}$.

  In conclusion, the transformation mapping a Cartan decomposition
  of type {\bf A$\Omega$} to the intrinsic of the same type
  is written in the composition
 \begin{align}\label{eqtransCDform}
  Q_{\Omega}=Q_{in,\Omega}Q^{(r)},
 \end{align}
  where $Q^{(r)}$ is the action of Eq.~\ref{eqtransQAP}
  and $Q_{in,{\rm I}}=\overline{Q}_{in,\mathbf{1}}$, $\breve{Q}_{in,\mathbf{1}}$
  or $\widetilde{Q}_{in,\mathbf{1}}$ for type {\bf AI},
  $Q_{in,{\rm II}}=\breve{Q}_{in,\mathbf{2}}$ or $\widetilde{Q}_{in,\mathbf{2}}$ for type {\bf AII}
  and
  $Q_{in,{\rm III}}=\widetilde{Q}_{in,\mathbf{3}}$,
  $\breve{Q}_{in,\mathbf{3}}$ or $\overline{Q}_{in,\mathbf{3}}$
  for type {\bf AIII}
  as given in from Eqs.~\ref{eqtranstyp1bar} to~\ref{eqtranstyp2}.

 \newpage
 \appendix{~~Tables of Quotient Algebras\label{appQAFigs}}

 \begin{figure}[htbp]
 \begin{center}
 \[
\begin{array}{c}
 \mathfrak{C}_{[\mathbf{0}]}\\
 \\
 {\cal S}^{000}_{000},\hspace{2pt}{\cal S}^{001}_{000},\hspace{2pt}{\cal S}^{010}_{000},\hspace{2pt}{\cal S}^{011}_{000},
 \hspace{2pt}{\cal S}^{100}_{000},\hspace{2pt}{\cal S}^{101}_{000},\hspace{2pt}{\cal S}^{110}_{000},\hspace{2pt}{\cal S}^{111}_{000}
 \end{array}\]
 \[\begin{array}{ccccc}
 {W}(\mathfrak{B}_{001})
 &{\cal S}^{000}_{001},\hspace{2pt}{\cal S}^{010}_{001},\hspace{2pt}{\cal S}^{100}_{001},\hspace{2pt}{\cal S}^{110}_{001}
 &\hspace{20pt}
 &{\cal S}^{001}_{001},\hspace{2pt}{\cal S}^{011}_{001},\hspace{2pt}{\cal S}^{101}_{001},\hspace{2pt}{\cal S}^{111}_{001}
 &\hat{W}(\mathfrak{B}_{001})\\
 &&&&\\
 W(\mathfrak{B}_{010})
 &{\cal S}^{000}_{010},\hspace{2pt}{\cal S}^{001}_{010},\hspace{2pt}{\cal S}^{100}_{010},\hspace{2pt}{\cal S}^{101}_{010}
 &
 &{\cal S}^{010}_{010},\hspace{2pt}{\cal S}^{011}_{010},\hspace{2pt}{\cal S}^{110}_{010},\hspace{2pt}{\cal S}^{111}_{010}
 &\hat{W}(\mathfrak{B}_{010})\\
 &&&&\\
 W(\mathfrak{B}_{011})
 &{\cal S}^{000}_{011},\hspace{2pt}{\cal S}^{011}_{011},\hspace{2pt}{\cal S}^{100}_{011},\hspace{2pt}{\cal S}^{111}_{011}
 &
 &{\cal S}^{001}_{011},\hspace{2pt}{\cal S}^{010}_{011},\hspace{2pt}{\cal S}^{101}_{011},\hspace{2pt}{\cal S}^{110}_{011}
 &\hat{W}(\mathfrak{B}_{011})\\
 &&&&\\
 W(\mathfrak{B}_{100})
 &{\cal S}^{000}_{100},\hspace{2pt}{\cal S}^{001}_{100},\hspace{2pt}{\cal S}^{010}_{100},\hspace{2pt}{\cal S}^{011}_{100}
 &
 &{\cal S}^{100}_{100},\hspace{2pt}{\cal S}^{101}_{100},\hspace{2pt}{\cal S}^{110}_{100},\hspace{2pt}{\cal S}^{111}_{100}
 &\hat{W}(\mathfrak{B}_{100})\\
 &&&&\\
\end{array} \]
 \vspace{15pt}
 \[ \hspace{10pt}\begin{array}{ccc}
 \mathfrak{B}_{001}=
 \{\hspace{2pt}{\cal S}^{000}_{000},{\cal S}^{010}_{000},{\cal S}^{100}_{000},{\cal S}^{110}_{000}\hspace{2pt}\},
 &\hspace{10pt}
 &\mathfrak{B}_{100}=
 \{\hspace{2pt}{\cal S}^{000}_{000},{\cal S}^{001}_{000},{\cal S}^{010}_{000},{\cal S}^{011}_{000}\hspace{2pt}\},
 \\
 \\
 \mathfrak{B}_{010}=
 \{\hspace{2pt}{\cal S}^{000}_{000},{\cal S}^{001}_{000},{\cal S}^{100}_{000},{\cal S}^{101}_{000}\hspace{2pt}\},
 &
 &\mathfrak{B}_{101}=
 \{\hspace{2pt}{\cal S}^{000}_{000},{\cal S}^{010}_{000},{\cal S}^{101}_{000},{\cal S}^{111}_{000}\hspace{2pt}\},
 \\
 \\
 \mathfrak{B}_{011}=
 \{\hspace{2pt}{\cal S}^{000}_{000},{\cal S}^{011}_{000},{\cal S}^{100}_{000},{\cal S}^{111}_{000}\hspace{2pt}\},
 &
 &\mathfrak{B}_{110}=
 \{\hspace{2pt}{\cal S}^{000}_{000},{\cal S}^{001}_{000},{\cal S}^{110}_{000},{\cal S}^{111}_{000}\hspace{2pt}\},\\
 &&\\
 &&\mathfrak{B}_{111}=
 \{\hspace{2pt}{\cal S}^{000}_{000},{\cal S}^{011}_{000},{\cal S}^{101}_{000},{\cal S}^{110}_{000}\hspace{2pt}\}.
 \end{array} \]
 \end{center}
 \vspace{6pt}
 \fcaption{The intrinsic quotient algebra of rank zero given by the intrinsic Cartan subalgebra
 $\mathfrak{C}_{[{\bf 0}]}\subset su(8)$ and the complete set of corresponding maximal bi-subalgebras.\label{figsu8intrQArank0}}
 \end{figure}

 \begin{figure}[htbp]
 \begin{center}
  \[\begin{array}{c}
   \mathfrak{B}^{[1]}_{intr}\\
   \\
   {\cal S}^{000}_{000},{\cal S}^{001}_{000},{\cal S}^{010}_{000},{\cal S}^{011}_{000}
   \end{array}\]
  \[\begin{array}{ccccc}
  \hspace{0pt}\mathfrak{B}^{[1,1]}_{intr}=\mathfrak{C}_{[\mathbf{0}]}-\mathfrak{B}^{[1]}_{intr}
  &{\cal S}^{100}_{000},{\cal S}^{101}_{000},{\cal S}^{110}_{000},{\cal S}^{111}_{000}&&\{0\}&\\
  \\
  {W}(\mathfrak{B}_{100},\mathfrak{B}^{[1]}_{intr};0)
  &{\cal S}^{000}_{100},{\cal S}^{001}_{100},{\cal S}^{010}_{100},{\cal S}^{011}_{100}&&\{0\}&\hat{W}(\mathfrak{B}_{100},\mathfrak{B}^{[1]}_{intr};0)\\
  &&&&\\
  {W}(\mathfrak{B}_{100},\mathfrak{B}^{[1]}_{intr};1)&\{0\}&&{\cal S}^{100}_{100},{\cal S}^{101}_{100},{\cal S}^{110}_{100},{\cal S}^{111}_{100}&\hspace{0pt}\hat{W}(\mathfrak{B}_{100},\mathfrak{B}^{[1]}_{intr};1)\\
  \\
  W(\mathfrak{B}_{001},\mathfrak{B}^{[1]}_{intr};0)&{\cal S}^{000}_{001},{\cal S}^{010}_{001}
  &&{\cal S}^{001}_{001},{\cal S}^{011}_{001}&\hat{W}(\mathfrak{B}_{001},\mathfrak{B}^{[1]}_{intr};0)\\
  \\
\end{array}\]
 \end{center}
 \vspace{6pt}
 \fcaption{The intrinsic quotient algebra of rank one given by the
 intrinsic center subalgebra
 $\mathfrak{B}^{[1]}_{intr}=\mathfrak{B}^{[1,0]}_{intr}=\mathfrak{B}_{100}\subset\mathfrak{C}_{[\mathbf{0}]}\subset su(8)$.\label{figsu8QArank1intr}}
 \end{figure}

 \begin{figure}[htbp]
 \begin{center}
  \[ \begin{array}{c}
   \mathfrak{B}^{[1,1]}_{intr}\\
   \\
   {\cal S}^{100}_{000},{\cal S}^{101}_{000},{\cal S}^{110}_{000},{\cal S}^{111}_{000}
   \end{array}\]
  \[\begin{array}{ccccc}
  \mathfrak{B}^{[1]}_{intr}
  &{\cal S}^{000}_{000},{\cal S}^{001}_{000},{\cal S}^{010}_{000},{\cal S}^{011}_{000}&&\{0\}&\\
  \\
  W(\mathfrak{B}_{100},\mathfrak{B}^{[1]}_{intr};0)&{\cal S}^{000}_{100},{\cal S}^{001}_{100},{\cal S}^{010}_{100},{\cal S}^{011}_{100}
  &&{\cal S}^{100}_{100},{\cal S}^{101}_{100},{\cal S}^{110}_{100},{\cal S}^{111}_{100}&\hspace{0pt}\hat{W}(\mathfrak{B}_{100},\mathfrak{B}^{[1]}_{intr};1)\\
  \\
  {W}(\mathfrak{B}_{100},\mathfrak{B}^{[1]}_{intr};1)
  &\{0\}&&\{0\}&\hat{W}(\mathfrak{B}_{100},\mathfrak{B}^{[1]}_{intr};0)\\
  \\
  W(\mathfrak{B}_{001},\mathfrak{B}^{[1]}_{intr};0)&{\cal S}^{000}_{001},{\cal S}^{010}_{001}
  &&{\cal S}^{101}_{001},{\cal S}^{111}_{001}&\hat{W}(\mathfrak{B}_{001},\mathfrak{B}^{[1]}_{intr};1)\\
  \\
\end{array} \]
 \end{center}
 \vspace{6pt}
 \fcaption{The co-quotient algebra of rank one given by the coset
 $\mathfrak{B}^{[1,1]}_{intr}=\mathfrak{C}_{[{\bf 0}]}-\mathfrak{B}^{[1]}_{intr}$.\label{figsu8coQArank1inC0}}
 \end{figure}

 \newpage
 \begin{figure}[htbp]
 \begin{center}
  \[\begin{array}{c}
 \mathfrak{B}^{[2]}_{intr}\\
 \\
 {\cal S}^{000}_{000},\hspace{2pt}{\cal S}^{001}_{000}
 \end{array}\]
 \[\begin{array}{ccccccc}
 \mathfrak{B}^{[2,\hspace{2pt}01]}_{intr}&{\cal S}^{010}_{000},{\cal S}^{011}_{000}&&&&\{0\}&\\
 \mathfrak{B}^{[2,\hspace{1pt}10]}_{intr}&{\cal S}^{100}_{000},{\cal S}^{101}_{000}&&&&\{0\}&\\
 \mathfrak{B}^{[2,\hspace{1pt}11]}_{intr}&{\cal S}^{110}_{000},{\cal S}^{111}_{000}&&&&\{0\}&\\
 W(\mathfrak{B}_{010},\mathfrak{B}^{[2]}_{intr};\hspace{1pt}00)&{\cal S}^{000}_{010},{\cal S}^{001}_{010}&&&&\{0\}&\hat{W}(\mathfrak{B}_{010},\mathfrak{B}^{[2]}_{intr};\hspace{1pt}00)\\
 W(\mathfrak{B}_{010},\mathfrak{B}^{[2]}_{intr};\hspace{1pt}01)&\{0\}&&&&{\cal S}^{010}_{010},{\cal S}^{011}_{010}&\hat{W}(\mathfrak{B}_{010},\mathfrak{B}^{[2]}_{intr};\hspace{1pt}01)\\
 W(\mathfrak{B}_{010},\mathfrak{B}^{[2]}_{intr};\hspace{1pt}10)&{\cal S}^{100}_{010},{\cal S}^{101}_{010}&&&&\{0\}&\hat{W}(\mathfrak{B}_{010},\mathfrak{B}^{[2]}_{intr};\hspace{1pt}10)\\
 W(\mathfrak{B}_{010},\mathfrak{B}^{[2]}_{intr};\hspace{1pt}11)&\{0\}&&&&{\cal S}^{110}_{010},{\cal S}^{111}_{010}&\hat{W}(\mathfrak{B}_{010},\mathfrak{B}^{[2]}_{intr};\hspace{1pt}11)\\
 W(\mathfrak{B}_{100},\mathfrak{B}^{[2]}_{intr};\hspace{1pt}00)&{\cal S}^{000}_{100},{\cal S}^{001}_{100}&&&&\{0\}&\hat{W}(\mathfrak{B}_{100},\mathfrak{B}^{[2]}_{intr};\hspace{1pt}00)\\
\end{array} \]
 \end{center}
 \vspace{6pt}
 \fcaption{The intrinsic quotient algebra of rank $2$ given by the
 intrinsic center subalgebra
 $\mathfrak{B}^{[2]}_{intr}=\mathfrak{B}_{010}\cap\mathfrak{B}_{100}\cap\mathfrak{B}_{110}\subset\mathfrak{C}_{[{\bf 0}]}\subset su(8)$,
 referring to Fig.~\ref{figsu8intrQArank0}. \label{figsu8QArank2intr}}
 \end{figure}

 \begin{figure}[htbp]
 \begin{center}
  \[\begin{array}{c}
 \mathfrak{B}^{[2,\hspace{2pt}01]}_{intr}\\
 \\
 {\cal S}^{010}_{000},{\cal S}^{011}_{000}
 \end{array}\]
 \[\begin{array}{ccccccc}
 \mathfrak{B}^{[2]}_{intr}&{\cal S}^{000}_{000},\hspace{2pt}{\cal S}^{001}_{000}&&&&\{0\}&\\
 \mathfrak{B}^{[2,\hspace{1pt}10]}_{intr}&{\cal S}^{100}_{000},{\cal S}^{101}_{000}&&&&\{0\}&\\
 \mathfrak{B}^{[2,\hspace{1pt}11]}_{intr}&{\cal S}^{110}_{000},{\cal S}^{111}_{000}&&&&\{0\}&\\
 W(\mathfrak{B}_{010},\mathfrak{B}^{[2]}_{intr};\hspace{1pt}00)&{\cal S}^{000}_{010},{\cal S}^{001}_{010}&&&&{\cal S}^{010}_{010},{\cal S}^{011}_{010}&\hat{W}(\mathfrak{B}_{010},\mathfrak{B}^{[2]}_{intr};\hspace{1pt}01)\\
 W(\mathfrak{B}_{010},\mathfrak{B}^{[2]}_{intr};\hspace{1pt}01)&\{0\}&&&&\{0\}&\hat{W}(\mathfrak{B}_{010},\mathfrak{B}^{[2]}_{intr};\hspace{1pt}00)\\
 W(\mathfrak{B}_{010},\mathfrak{B}^{[2]}_{intr};\hspace{1pt}11)&\{0\}&&&&\{0\}&\hat{W}(\mathfrak{B}_{010},\mathfrak{B}^{[2]}_{intr};\hspace{1pt}10)\\
 W(\mathfrak{B}_{010},\mathfrak{B}^{[2]}_{intr};\hspace{1pt}10)&{\cal S}^{100}_{010},{\cal S}^{101}_{010}&&&&{\cal S}^{110}_{010},{\cal S}^{111}_{010}&\hat{W}(\mathfrak{B}_{010},\mathfrak{B}^{[2]}_{intr};\hspace{1pt}11)\\
 W(\mathfrak{B}_{100},\mathfrak{B}^{[2]}_{intr};\hspace{1pt}01)&{\cal S}^{010}_{100},{\cal S}^{011}_{100}&&&&\{0\}&\hat{W}(\mathfrak{B}_{100},\mathfrak{B}^{[2]}_{intr};\hspace{1pt}00)\\
\end{array} \]
 \end{center}
 \vspace{6pt}
 \fcaption{A co-quotient algebra of rank $2$ given by a coset $\mathfrak{B}^{[2,\hspace{2pt}01]}_{intr}$ in the partition
 of $\mathfrak{C}_{[\mathbf{0}]}\subset su(8)$ generated by
 $\mathfrak{B}^{[2]}_{intr}=\mathfrak{B}^{[2,\hspace{2pt}00]}_{intr}\subset\mathfrak{C}_{[\mathbf{0}]}$. \label{figsu8coQArank2inC0}}
 \end{figure}

 \begin{figure}[htbp]
 \begin{center}
  \[\begin{array}{c}
   \mathfrak{B}^{[1]}_{can}\\
   \\
   {\cal S}^{000}_{000},{\cal S}^{010}_{000},{\cal S}^{100}_{000},{\cal S}^{110}_{000}
   \end{array}\]
  \[\begin{array}{ccccc}
  \hspace{0pt}\mathfrak{B}^{[1,1]}_{can}
  &{\cal S}^{001}_{000},{\cal S}^{011}_{000},{\cal S}^{101}_{000},{\cal S}^{111}_{000}&&\{0\}&\\
  \\
  W(\mathfrak{B}_{001},\mathfrak{B}^{[1]}_{can};0)&{\cal S}^{000}_{001},{\cal S}^{010}_{001},{\cal S}^{100}_{001},{\cal S}^{110}_{001}
  &&\{0\} &\hat{W}(\mathfrak{B}_{001},\mathfrak{B}^{[1]}_{can};0)\\
  \\
  W(\mathfrak{B}_{001},\mathfrak{B}^{[1]}_{can};1)&\{0\}
  &&{\cal S}^{101}_{001},{\cal S}^{111}_{001},{\cal S}^{001}_{001},{\cal S}^{011}_{001}&\hat{W}(\mathfrak{B}_{001},\mathfrak{B}^{[1]}_{can};1)\\
  \\
  W(\mathfrak{B}_{010},\mathfrak{B}^{[1]}_{can};0)&{\cal S}^{000}_{010},{\cal S}^{100}_{010}
  &&{\cal S}^{010}_{010},{\cal S}^{110}_{010}&\hat{W}(\mathfrak{B}_{010},\mathfrak{B}^{[1]}_{can};0)\\
  \\
\end{array}\]
 \end{center}
 \vspace{6pt}
 \fcaption{The canonical quotient algebra of rank one given by the canonical center subalgebra
 $\mathfrak{B}^{[1]}_{can}=\mathfrak{B}^{[1,0]}_{can}=\mathfrak{B}_{001}\subset\mathfrak{C}_{[\mathbf{0}]}\subset su(8)$.\label{figsu8canQArank1}}
 \end{figure}

 \begin{figure}[htbp]
 \begin{center}
  \[\begin{array}{c}
   \mathfrak{B}^{[1,1]}_{can}\\
   \\
   {\cal S}^{001}_{000},{\cal S}^{011}_{000},{\cal S}^{101}_{000},{\cal S}^{111}_{000}
   \end{array}\]
  \[\begin{array}{ccccc}
  \hspace{0pt}\mathfrak{B}^{[1]}_{can}
  &{\cal S}^{000}_{000},{\cal S}^{010}_{000},{\cal S}^{100}_{000},{\cal S}^{110}_{000}&&\{0\}&\\
  \\
  W(\mathfrak{B}_{001},\mathfrak{B}^{[1]}_{can};0)&{\cal S}^{000}_{001},{\cal S}^{010}_{001},{\cal S}^{100}_{001},{\cal S}^{110}_{001}
  &&{\cal S}^{101}_{001},{\cal S}^{111}_{001},{\cal S}^{001}_{001},{\cal S}^{011}_{001}&\hat{W}(\mathfrak{B}_{001},\mathfrak{B}^{[1]}_{can};1)\\
  \\
  W(\mathfrak{B}_{001},\mathfrak{B}^{[1]}_{can};1)&\{0\}
  && \{0\}  &\hat{W}(\mathfrak{B}_{001},\mathfrak{B}^{[1]}_{can};0)\\
  \\
  W(\mathfrak{B}_{010},\mathfrak{B}^{[1]}_{can};0)&{\cal S}^{000}_{010},{\cal S}^{100}_{010}
  &&{\cal S}^{011}_{010},{\cal S}^{111}_{010}&\hat{W}(\mathfrak{B}_{010},\mathfrak{B}^{[1]}_{can};1)\\
  \\
\end{array}\]
 \end{center}
 \vspace{6pt}
 \fcaption{The co-quotient algebra of rank one given by the coset
 $\mathfrak{B}^{[1,1]}_{can}=\mathfrak{C}_{[\mathbf{0}]}-\mathfrak{B}^{[1]}_{can}$.\label{figsu8cancoQArank1}}
 \end{figure}

 \begin{figure}[htbp]
 \begin{center}
  \[\begin{array}{c}
   \mathfrak{B}^{[2]}_{can}\\
   \\
   {\cal S}^{000}_{000},\hspace{2pt}{\cal S}^{100}_{000}
   \end{array}\]
  \[\begin{array}{ccccc}
  \hspace{0pt}\mathfrak{B}^{[2,\hspace{1pt}01]}_{can}
  &{\cal S}^{001}_{000},{\cal S}^{101}_{000}&\hspace{30pt}&\{0\}&
  \\
  \hspace{0pt}\mathfrak{B}^{[2,\hspace{1pt}10]}_{can}
  &{\cal S}^{010}_{000},{\cal S}^{110}_{000}&&\{0\}&
  \\
  \hspace{0pt}\mathfrak{B}^{[2,\hspace{1pt}11]}_{can}
  &{\cal S}^{011}_{000},{\cal S}^{111}_{000}&&\{0\}&
  \\
  W(\mathfrak{B}_{001},\mathfrak{B}^{[2]}_{can};00)
  &{\cal S}^{000}_{001},{\cal S}^{100}_{001}
  &
  &\{0\}
  &\hat{W}(\mathfrak{B}_{001},\mathfrak{B}^{[2]}_{can};00)
  \\
  W(\mathfrak{B}_{001},\mathfrak{B}^{[2]}_{can};01)
  &\{0\}
  &
  &{\cal S}^{101}_{001},{\cal S}^{001}_{001}
  &\hat{W}(\mathfrak{B}_{001},\mathfrak{B}^{[2]}_{can};01)
  \\
  W(\mathfrak{B}_{001},\mathfrak{B}^{[2]}_{can};10)
  &{\cal S}^{010}_{001},{\cal S}^{110}_{001}
  &
  &\{0\}
  &\hat{W}(\mathfrak{B}_{001},\mathfrak{B}^{[2]}_{can};10)
  \\
  W(\mathfrak{B}_{001},\mathfrak{B}^{[2]}_{can};11)
  &\{0\}
  &
  &{\cal S}^{111}_{001},{\cal S}^{011}_{001}
  &\hat{W}(\mathfrak{B}_{001},\mathfrak{B}^{[2]}_{can};11)
  \\
  W(\mathfrak{B}_{010},\mathfrak{B}^{[2]}_{can};00)
  &{\cal S}^{000}_{010},{\cal S}^{100}_{010}
  &
  &\{0\}
  &\hat{W}(\mathfrak{B}_{010},\mathfrak{B}^{[2]}_{can};00)
  \\
\end{array}\]
 \end{center}
 \vspace{6pt}
 \fcaption{The canonical quotient algebra of rank two given by the canonical center subalgebra
 $\mathfrak{B}^{[2]}_{can}=\mathfrak{B}^{[2,\hspace{1pt}00]}_{can}
 =\mathfrak{B}_{001}\cap\mathfrak{B}_{010}\cap\mathfrak{B}_{011}\subset\mathfrak{C}_{[\mathbf{0}]}\subset su(8)$,
 referring to Fig.~\ref{figsu8intrQArank0}.\label{figsu8canQArank2}}
 \end{figure}

 \begin{figure}[htbp]
 \begin{center}
  \[\begin{array}{c}
   \mathfrak{B}^{[2,\hspace{1pt}10]}_{can}  \\
   \\
   {\cal S}^{010}_{000},\hspace{2pt}{\cal S}^{110}_{000}
   \end{array}\]
  \[\begin{array}{ccccc}
  \hspace{-7pt}\mathfrak{B}^{[2]}_{can}
  &{\cal S}^{000}_{000},\hspace{0pt}{\cal S}^{100}_{000}&&\{0\}&
  \\
  \hspace{0pt}\mathfrak{B}^{[2,\hspace{1pt}01]}_{can}
  &{\cal S}^{001}_{000},{\cal S}^{101}_{000}&\hspace{30pt}&\{0\}&
  \\
  \hspace{0pt}\mathfrak{B}^{[2,\hspace{1pt}11]}_{can}
  &{\cal S}^{011}_{000},{\cal S}^{111}_{000}&&\{0\}&
  \\
  W(\mathfrak{B}_{001},\mathfrak{B}^{[2]}_{can};00)
  &{\cal S}^{000}_{001},{\cal S}^{100}_{001}
  &
  &\{0\}
  &\hat{W}(\mathfrak{B}_{001},\mathfrak{B}^{[2]}_{can};10)
  \\
  W(\mathfrak{B}_{001},\mathfrak{B}^{[2]}_{can};11)
  &\{0\}
  &
  &{\cal S}^{101}_{001},{\cal S}^{001}_{001}
  &\hat{W}(\mathfrak{B}_{001},\mathfrak{B}^{[2]}_{can};01)
  \\
  W(\mathfrak{B}_{001},\mathfrak{B}^{[2]}_{can};10)
  &{\cal S}^{010}_{001},{\cal S}^{110}_{001}
  &
  &\{0\}
  &\hat{W}(\mathfrak{B}_{001},\mathfrak{B}^{[2]}_{can};00)
  \\
  W(\mathfrak{B}_{001},\mathfrak{B}^{[2]}_{can};01)
  &\{0\}
  &
  &{\cal S}^{011}_{001},{\cal S}^{111}_{001}
  &\hat{W}(\mathfrak{B}_{001},\mathfrak{B}^{[2]}_{can};11)
  \\
  W(\mathfrak{B}_{010},\mathfrak{B}^{[2]}_{can};00)
  &{\cal S}^{000}_{010},{\cal S}^{100}_{010}
  &
  &{\cal S}^{010}_{010},{\cal S}^{110}_{010}
  &\hat{W}(\mathfrak{B}_{010},\mathfrak{B}^{[2]}_{can};10)
  \\
\end{array}\]
 \end{center}
 \vspace{6pt}
 \fcaption{A co-quotient algebra of rank two given by a coset
 $\mathfrak{B}^{[2,\hspace{1pt}10]}_{can}$ in the partition of
 $\mathfrak{C}_{[\mathbf{0}]}\subset su(8)$
 generated by $\mathfrak{B}^{[2]}_{can}=\mathfrak{B}^{[2,\hspace{1pt}00]}_{can}\subset\mathfrak{C}_{[\mathbf{0}]}$.\label{figsu8cancoQArank2}}
 \end{figure}

 \begin{figure}[htbp]
 \begin{center}
 \[\begin{array}{c}
 \mathfrak{C}_{\rm III}\\
 \\
 {\cal S}^{000}_{000},\hspace{2pt}{\cal S}^{001}_{000},\hspace{2pt}{\cal S}^{010}_{000},\hspace{2pt}{\cal S}^{011}_{000},
 \hspace{2pt}{\cal S}^{100}_{100},\hspace{2pt}{\cal S}^{101}_{100},\hspace{2pt}{\cal S}^{110}_{100},\hspace{2pt}{\cal S}^{111}_{100}
 \end{array}\]
 \[\begin{array}{ccccc}
 W(\hat{\mathfrak{B}}_{100,100})
 &{\cal S}^{000}_{100},\hspace{2pt}{\cal S}^{001}_{100},\hspace{2pt} {\cal S}^{010}_{100},\hspace{2pt}{\cal S}^{011}_{100}
 &
 &{\cal S}^{100}_{000},\hspace{2pt}{\cal S}^{101}_{000},\hspace{2pt} {\cal S}^{110}_{000},\hspace{2pt}{\cal S}^{111}_{000}
 &\hat{W}(\hat{\mathfrak{B}}_{100,100})\\
 \\
 W(\hat{\mathfrak{B}}_{001,100})
 &{\cal S}^{000}_{001},\hspace{2pt}{\cal S}^{010}_{001},\hspace{2pt}{\cal S}^{100}_{101},\hspace{2pt}{\cal S}^{110}_{101}
 &\hspace{20pt}
 &{\cal S}^{001}_{001},\hspace{2pt}{\cal S}^{011}_{001},\hspace{2pt} {\cal S}^{101}_{101},\hspace{2pt}{\cal S}^{111}_{101}
 &\hat{W}(\hat{\mathfrak{B}}_{001,100})\\
 \\
 W(\hat{\mathfrak{B}}_{101,100})
 &{\cal S}^{101}_{001},\hspace{2pt}{\cal S}^{111}_{001},\hspace{2pt} {\cal S}^{000}_{101},\hspace{2pt}{\cal S}^{010}_{101}
 &
 &{\cal S}^{100}_{001},\hspace{2pt}{\cal S}^{110}_{001},\hspace{2pt}{\cal S}^{001}_{101},\hspace{2pt}{\cal S}^{011}_{101}
 &\hat{W}(\hat{\mathfrak{B}}_{101,100})\\
 \\
 W(\hat{\mathfrak{B}}_{010,100})
 &{\cal S}^{000}_{010},\hspace{2pt}{\cal S}^{001}_{010},\hspace{2pt} {\cal S}^{100}_{110},\hspace{2pt}{\cal S}^{101}_{110}
 &
 &{\cal S}^{010}_{010},\hspace{2pt}{\cal S}^{011}_{010},\hspace{2pt} {\cal S}^{110}_{110},\hspace{2pt}{\cal S}^{111}_{110}
 &\hat{W}(\hat{\mathfrak{B}}_{010,100})\\
 \\
\end{array} \]
 \vspace{15pt}
 \[\hspace{9pt}\begin{array}{ccc}
 \hspace{0pt}\hat{\mathfrak{B}}_{001,100}=
 \{\hspace{2pt}{\cal S}^{000}_{000},{\cal S}^{010}_{000},{\cal S}^{100}_{100},{\cal S}^{110}_{100}\hspace{2pt}\},
 &\hspace{10pt}
 &\hat{\mathfrak{B}}_{100,100}=
 \{\hspace{2pt}{\cal S}^{000}_{000},{\cal S}^{001}_{000},{\cal S}^{010}_{000},{\cal S}^{011}_{000}\hspace{2pt}\},
 \\
 \\
 \hat{\mathfrak{B}}_{010,100}=
 \{\hspace{2pt}{\cal S}^{000}_{000},{\cal S}^{001}_{000},{\cal S}^{100}_{100},{\cal S}^{101}_{100}\hspace{2pt}\},
 &
 &\hat{\mathfrak{B}}_{101,100}=
 \{\hspace{2pt}{\cal S}^{000}_{000},{\cal S}^{010}_{000},{\cal S}^{101}_{100},{\cal S}^{111}_{100}\hspace{2pt}\},
 \\
 \\
 \hat{\mathfrak{B}}_{011,100}=
 \{\hspace{2pt}{\cal S}^{000}_{000},{\cal S}^{011}_{000},{\cal S}^{100}_{100},{\cal S}^{111}_{100}\hspace{2pt}\},
 &
 &\hat{\mathfrak{B}}_{110,100}=
 \{\hspace{2pt}{\cal S}^{000}_{000},{\cal S}^{001}_{000},{\cal S}^{110}_{100},{\cal S}^{111}_{100}\hspace{2pt}\},
 \\
 \\
 &&\hat{\mathfrak{B}}_{111,100}=
 \{\hspace{2pt}{\cal S}^{000}_{000},{\cal S}^{011}_{000},{\cal S}^{101}_{100},{\cal S}^{110}_{100}\hspace{2pt}\}.
 \end{array} \]
 \end{center}
 \vspace{6pt}
 \fcaption{A quotient algebra of rank zero given by a Cartan subalgebra $\mathfrak{C}_{\rm III}\subset su(8)$
 and the complete set of corresponding maximal bi-subalgebras.\label{figsu8QArank0inCIII}}
 \end{figure}

 \begin{figure}[htbp]
  \begin{center}
   \[\begin{array}{c}
   \hat{\mathfrak{B}}^{[1]}_{intr}\\
   \\
   {\cal S}^{000}_{000},{\cal S}^{001}_{000},{\cal S}^{010}_{000},{\cal S}^{011}_{000}
   \end{array}\]
  \[\begin{array}{cccc}
  \hat{\mathfrak{B}}^{[1,1]}_{intr}=\mathfrak{C}_{\rm III}-\hat{\mathfrak{B}}^{[1]}_{intr}
  &{\cal S}^{100}_{100},{\cal S}^{101}_{100},{\cal S}^{110}_{100},{\cal S}^{111}_{100}
  &\{0\}&\\
  \\
  {W}(\hat{\mathfrak{B}}_{100,100},\hat{\mathfrak{B}}^{[1]}_{intr};0)
  &{\cal S}^{000}_{100},{\cal S}^{001}_{100},{\cal S}^{010}_{100},{\cal S}^{011}_{100}
  &\{0\}&\hat{W}(\hat{\mathfrak{B}}_{100,100},\hat{\mathfrak{B}}^{[1]}_{intr};0)\\
  \\
  {W}(\hat{\mathfrak{B}}_{100,100},\hat{\mathfrak{B}}^{[1]}_{intr};1)
  &\{0\}
  &{\cal S}^{100}_{000},{\cal S}^{101}_{000},{\cal S}^{110}_{000},{\cal S}^{111}_{000}&\hspace{0pt}\hat{W}(\hat{\mathfrak{B}}_{100,100},\hat{\mathfrak{B}}^{[1]}_{intr};1)\\
  \\
  {W}(\hat{\mathfrak{B}}_{001,100},\hat{\mathfrak{B}}^{[1]}_{intr};0)&{\cal S}^{000}_{001},{\cal S}^{010}_{001}
  &{\cal S}^{001}_{001},{\cal S}^{011}_{001}&\hat{W}(\hat{\mathfrak{B}}_{001,100},\hat{\mathfrak{B}}^{[1]}_{intr};0)\\
  \\
\end{array}\]
  \end{center}
  \vspace{6pt}
 \fcaption{The quotient algebra of rank one given by a maximal bi-subalgebra
 $\hat{\mathfrak{B}}^{[1]}_{intr}=\hat{\mathfrak{B}}^{[1,0]}_{intr}=\hat{\mathfrak{B}}_{100,100}\subset\mathfrak{C}_{\rm III}\subset su(8)$.\label{figsu8QArank1inCIII}}
 \end{figure}

 \begin{figure}[htbp]
 \begin{center}
 \[\begin{array}{c}
   \hat{\mathfrak{B}}^{[1,1]}_{intr}\\
   \\
   {\cal S}^{100}_{100},{\cal S}^{101}_{100},{\cal S}^{110}_{100},{\cal S}^{111}_{100}
   \end{array}\]
  \[\begin{array}{cccc}
  \hat{\mathfrak{B}}^{[1]}_{intr}
  &{\cal S}^{000}_{000},{\cal S}^{001}_{000},{\cal S}^{010}_{000},{\cal S}^{011}_{000}
  &\{0\}&\\
  \\
  {W}(\hat{\mathfrak{B}}_{100,100},\hat{\mathfrak{B}}^{[1]}_{intr};0)
  &{\cal S}^{000}_{100},{\cal S}^{001}_{100},{\cal S}^{010}_{100},{\cal S}^{011}_{100}
  &{\cal S}^{100}_{000},{\cal S}^{101}_{000},{\cal S}^{110}_{000},{\cal S}^{111}_{000}&\hat{W}(\hat{\mathfrak{B}}_{100,100},\hat{\mathfrak{B}}^{[1]}_{intr};1)\\
  \\
  {W}(\hat{\mathfrak{B}}_{100,100},\hat{\mathfrak{B}}^{[1]}_{intr};1)
  &\{0\}
  &\{0\}&\hspace{0pt}\hat{W}(\hat{\mathfrak{B}}_{100,100},\hat{\mathfrak{B}}^{[1]}_{intr};0)\\
  \\
  {W}(\hat{\mathfrak{B}}_{001,100},\hat{\mathfrak{B}}^{[1]}_{intr};0)&{\cal S}^{000}_{001},{\cal S}^{010}_{001}
  &{\cal S}^{101}_{101},{\cal S}^{111}_{101}&\hat{W}(\hat{\mathfrak{B}}_{001,100},\hat{\mathfrak{B}}^{[1]}_{intr};1)\\
  \\
\end{array}\]
 \end{center}
 \vspace{6pt}
 \fcaption{The intrinsic co-quotient algebra of rank one given by the
 intrinsic center subalgebra
 $\hat{\mathfrak{B}}^{[1,1]}_{intr}=\mathfrak{C}_{\rm III}-\hat{\mathfrak{B}}^{[1]}_{intr}\subset su(8)$.\label{figsu8coQArank1intr}}
 \end{figure}

 \begin{figure}[htbp]
 \begin{center}
  \[\begin{array}{c}
  \hat{\mathfrak{B}}^{[2]}_{intr}\\
  \\
  {\cal S}^{000}_{000},\hspace{2pt}{\cal S}^{001}_{000}
  \end{array}\]
 \[\begin{array}{ccccccc}
 \hat{\mathfrak{B}}^{[2,\hspace{2pt}01]}_{intr}&{\cal S}^{010}_{000},{\cal S}^{011}_{000}&&&&\{0\}&\\
 \hat{\mathfrak{B}}^{[2,\hspace{1pt}10]}_{intr}&{\cal S}^{100}_{100},{\cal S}^{101}_{100}&&&&\{0\}&\\
 \hat{\mathfrak{B}}^{[2,\hspace{1pt}11]}_{intr}&{\cal S}^{110}_{100},{\cal S}^{111}_{100}&&&&\{0\}&\\
 {W}(\hat{\mathfrak{B}}_{100,100},\hat{\mathfrak{B}}^{[2]}_{intr};\hspace{1pt}00)&{\cal S}^{000}_{100},{\cal S}^{001}_{100}&&&&\{0\}&\hat{W}(\hat{\mathfrak{B}}_{100,100},\hat{\mathfrak{B}}^{[2]}_{intr};\hspace{1pt}00)\\
 {W}(\hat{\mathfrak{B}}_{100,100},\hat{\mathfrak{B}}^{[2]}_{intr};\hspace{1pt}01)&{\cal S}^{010}_{100},{\cal S}^{011}_{100}&&&&\{0\}&\hat{W}(\hat{\mathfrak{B}}_{100,100},\hat{\mathfrak{B}}^{[2]}_{intr};\hspace{1pt}01)\\
 {W}(\hat{\mathfrak{B}}_{100,100},\hat{\mathfrak{B}}^{[2]}_{intr};\hspace{1pt}10)&\{0\}&&&&{\cal S}^{100}_{000},{\cal S}^{101}_{000}&\hat{W}(\hat{\mathfrak{B}}_{100,100},\hat{\mathfrak{B}}^{[2]}_{intr};\hspace{1pt}10)\\
 {W}(\hat{\mathfrak{B}}_{100,100},\hat{\mathfrak{B}}^{[2]}_{intr};\hspace{1pt}11)&\{0\}&&&&{\cal S}^{110}_{000},{\cal S}^{111}_{000}&\hat{W}(\hat{\mathfrak{B}}_{100,100},\hat{\mathfrak{B}}^{[2]}_{intr};\hspace{1pt}11)\\
 {W}(\hat{\mathfrak{B}}_{010,100},\hat{\mathfrak{B}}^{[2]}_{intr};\hspace{1pt}00)&{\cal S}^{000}_{010},{\cal S}^{001}_{010}&&&&\{0\}&\hat{W}(\hat{\mathfrak{B}}_{010,100},\hat{\mathfrak{B}}^{[2]}_{intr};\hspace{1pt}00)\\
\end{array} \]
 \end{center}
 \vspace{6pt}
 \fcaption{A quotient algebra of rank $2$ given by a $2$nd maximal bi-subalgebra
 $\hat{\mathfrak{B}}^{[2]}_{intr}=\hat{\mathfrak{B}}^{[2,\hspace{2pt}00]}_{intr}=\hat{\mathfrak{B}}_{010,100}\cap\hat{\mathfrak{B}}_{110,100}\cap\hat{\mathfrak{B}}_{010,100}
 \subset\mathfrak{C}_{\rm III}\subset su(8)$, referring to Fig.~\ref{figsu8QArank0inCIII}. \label{figsu8QArank2inCIII}}
 \end{figure}

 \begin{figure}[htbp]
 \begin{center}
  \[\begin{array}{c}
 \hat{\mathfrak{B}}^{[2,\hspace{1pt}10]}_{intr}\\
 \\
 {\cal S}^{100}_{100},{\cal S}^{101}_{100}
 \end{array}\]
 \[\begin{array}{ccccccc}
 \hat{\mathfrak{B}}^{[2,\hspace{2pt}00]}_{intr}&{\cal S}^{000}_{000},\hspace{2pt}{\cal S}^{001}_{000}&&&&\{0\}&\\
 \hat{\mathfrak{B}}^{[2,\hspace{2pt}01]}_{intr}&{\cal S}^{010}_{000},{\cal S}^{011}_{000}&&&&\{0\}&\\
 \hat{\mathfrak{B}}^{[2,\hspace{1pt}11]}_{intr}&{\cal S}^{110}_{100},{\cal S}^{111}_{100}&&&&\{0\}&\\
 {W}(\hat{\mathfrak{B}}_{100,100},\hat{\mathfrak{B}}^{[2]}_{intr};\hspace{1pt}00)&{\cal S}^{000}_{100},{\cal S}^{001}_{100}&&&&{\cal S}^{100}_{000},{\cal S}^{101}_{000}&\hat{W}(\hat{\mathfrak{B}}_{100,100},\hat{\mathfrak{B}}^{[2]}_{intr};\hspace{1pt}10)\\
 {W}(\hat{\mathfrak{B}}_{100,100},\hat{\mathfrak{B}}^{[2]}_{intr};\hspace{1pt}11)&\{0\}&&&&\{0\}&\hat{W}(\hat{\mathfrak{B}}_{100,100},\hat{\mathfrak{B}}^{[2]}_{intr};\hspace{1pt}01)\\
 {W}(\hat{\mathfrak{B}}_{100,100},\hat{\mathfrak{B}}^{[2]}_{intr};\hspace{1pt}10)&\{0\}&&&&\{0\}&\hat{W}(\hat{\mathfrak{B}}_{100,100},\hat{\mathfrak{B}}^{[2]}_{intr};\hspace{1pt}00)\\
 {W}(\hat{\mathfrak{B}}_{100,100},\hat{\mathfrak{B}}^{[2]}_{intr};\hspace{1pt}01)&{\cal S}^{010}_{100},{\cal S}^{011}_{100}&&&&{\cal S}^{110}_{000},{\cal S}^{111}_{000}&\hat{W}(\hat{\mathfrak{B}}_{100,100},\hat{\mathfrak{B}}^{[2]}_{intr};\hspace{1pt}11)\\
 {W}(\hat{\mathfrak{B}}_{010,100},\hat{\mathfrak{B}}^{[2]}_{intr};\hspace{1pt}10)&{\cal S}^{100}_{110},{\cal S}^{101}_{110}&&&&\{0\}&\hat{W}(\hat{\mathfrak{B}}_{010,100},\hat{\mathfrak{B}}^{[2]}_{intr};\hspace{1pt}00)\\
\end{array} \]
 \end{center}
 \vspace{6pt}
 \fcaption{The intrinsic co-quotient algebra of rank $2$ given by the coset
 $\hat{\mathfrak{B}}^{[2,\hspace{1pt}10]}_{intr}$ in the partition of $\mathfrak{C}_{\rm III}\subset su(8)$
 generated by $\hat{\mathfrak{B}}^{[2]}_{intr}=\hat{\mathfrak{B}}^{[2,\hspace{2pt}00]}_{intr}$. \label{figsu8coQArank2intr}}
 \end{figure}

 \begin{figure}[htbp]
 \begin{center}
 \[\begin{array}{c}
 \mathfrak{C}\\
 \\
 \hspace{20pt}{\cal S}^{000}_{000},\hspace{2pt}{\cal S}^{001}_{000},\hspace{2pt}{\cal S}^{010}_{010},\hspace{2pt}{\cal S}^{011}_{010},
 \hspace{2pt}{\cal S}^{100}_{100},\hspace{2pt}{\cal S}^{101}_{100},\hspace{2pt}{\cal S}^{110}_{110},\hspace{2pt}{\cal S}^{111}_{110}
 \end{array}\]
 \[\begin{array}{ccccc}
 W(\mathfrak{B}_{1})
 &{\cal S}^{100}_{000},\hspace{2pt}{\cal S}^{101}_{000},\hspace{2pt} {\cal S}^{110}_{010},\hspace{2pt}{\cal S}^{111}_{010}
 &
 &{\cal S}^{000}_{100},\hspace{2pt}{\cal S}^{001}_{100},\hspace{2pt} {\cal S}^{010}_{110},\hspace{2pt}{\cal S}^{011}_{110}
 &\hat{W}(\mathfrak{B}_{1})\\
 \\
 W(\mathfrak{B}_{2})
 &{\cal S}^{010}_{000},\hspace{2pt}{\cal S}^{011}_{000},\hspace{2pt}{\cal S}^{110}_{100},\hspace{2pt}{\cal S}^{111}_{100}
 &\hspace{20pt}
 &{\cal S}^{000}_{010},\hspace{2pt}{\cal S}^{001}_{010},\hspace{2pt} {\cal S}^{100}_{110},\hspace{2pt}{\cal S}^{101}_{110}
 &\hat{W}(\mathfrak{B}_{2})\\
 \\
 W(\mathfrak{B}_{3})
 &{\cal S}^{110}_{000},\hspace{2pt}{\cal S}^{111}_{000},\hspace{2pt}{\cal S}^{000}_{110},\hspace{2pt}{\cal S}^{001}_{110}
 &
 &{\cal S}^{100}_{010},\hspace{2pt}{\cal S}^{101}_{010},\hspace{2pt} {\cal S}^{010}_{100},\hspace{2pt}{\cal S}^{011}_{100}
 &\hat{W}(\mathfrak{B}_{3})\\
 \\
 W(\mathfrak{B}_{4})
 &{\cal S}^{000}_{001},\hspace{2pt}{\cal S}^{010}_{011},\hspace{2pt} {\cal S}^{100}_{101},\hspace{2pt}{\cal S}^{110}_{111}
 &
 &{\cal S}^{001}_{001},\hspace{2pt}{\cal S}^{011}_{011},\hspace{2pt} {\cal S}^{101}_{101},\hspace{2pt}{\cal S}^{111}_{111}
 &\hat{W}(\mathfrak{B}_{4})\\
 \\
\end{array} \]
 \vspace{15pt}
 \[ \hspace{10pt}\begin{array}{ccc}
 \mathfrak{B}_{1}=
 \{\hspace{2pt}{\cal S}^{000}_{000},{\cal S}^{001}_{000},{\cal S}^{010}_{010},{\cal S}^{011}_{010}\hspace{2pt}\},
 &\hspace{10pt}
 &\mathfrak{B}_{4}=
 \{\hspace{2pt}{\cal S}^{000}_{000},{\cal S}^{010}_{010},{\cal S}^{100}_{100},{\cal S}^{110}_{110}\hspace{2pt}\},
 \\
 \\
 \mathfrak{B}_{2}=
 \{\hspace{2pt}{\cal S}^{000}_{000},{\cal S}^{001}_{000},{\cal S}^{100}_{100},{\cal S}^{101}_{100}\hspace{2pt}\},
 &
 &
 \mathfrak{B}_{5}=
 \{\hspace{2pt}{\cal S}^{000}_{000},{\cal S}^{010}_{010},{\cal S}^{101}_{100},{\cal S}^{111}_{110}\hspace{2pt}\},
 \\
 \\
 \mathfrak{B}_{3}=
 \{\hspace{2pt}{\cal S}^{000}_{000},{\cal S}^{001}_{000},{\cal S}^{110}_{110},{\cal S}^{101}_{110}\hspace{2pt}\},
 &
 &
 \mathfrak{B}_{6}=
 \{\hspace{2pt}{\cal S}^{000}_{000},{\cal S}^{011}_{010},{\cal S}^{100}_{100},{\cal S}^{111}_{110}\hspace{2pt}\},
 \\
 \\
 &&
 \mathfrak{B}_{7}=
 \{\hspace{2pt}{\cal S}^{000}_{000},{\cal S}^{011}_{010},{\cal S}^{101}_{100},{\cal S}^{110}_{110}\hspace{2pt}\}.
 \end{array}\]
 \end{center}
 \vspace{6pt}
 \fcaption{A quotient algebra of rank zero given by a Cartan subalgebra $\mathfrak{C}\subset su(8)$
 and the complete set of corresponding maximal bi-subalgebras.\label{figsu8QArank0inC}}
 \end{figure}

 \begin{figure}[htbp]
 \begin{center}
  \[\begin{array}{c}
   \mathfrak{B}^{[1]}\\
   \\
   {\cal S}^{000}_{000},{\cal S}^{001}_{000},{\cal S}^{010}_{010},{\cal S}^{011}_{010}
   \end{array}\]
  \[\begin{array}{ccccc}
  \mathfrak{B}^{[1,1]}=\mathfrak{C}-\mathfrak{B}^{[1]}
  &{\cal S}^{100}_{100},{\cal S}^{101}_{100},{\cal S}^{110}_{110},{\cal S}^{111}_{110}&&\{0\}&\\
  \\
  {W}(\mathfrak{B}_1,\mathfrak{B}^{[1]};0)
  &{\cal S}^{100}_{000},{\cal S}^{101}_{000},{\cal S}^{110}_{010}\,{\cal S}^{111}_{010}&&\{0\}&\hat{W}(\mathfrak{B}_1,\mathfrak{B}^{[1]};0)\\
  \\
  {W}(\mathfrak{B}_1,\mathfrak{B}^{[1]};1)&\{0\}&&{\cal S}^{000}_{100},{\cal S}^{001}_{100},{\cal S}^{010}_{110},{\cal S}^{011}_{110}&\hspace{0pt}\hat{W}(\mathfrak{B}_1,\mathfrak{B}^{[1]};1)\\
  \\
  {W}(\mathfrak{B}_2,\mathfrak{B}^{[1]};0)&{\cal S}^{010}_{000},{\cal S}^{011}_{000}
  &&{\cal S}^{000}_{010},{\cal S}^{001}_{010}&\hat{W}(\mathfrak{B}_2,\mathfrak{B}^{[1]};0)\\
  \\
\end{array} \]
 \end{center}
 \vspace{6pt}
 \fcaption{A quotient algebra of rank one given by a maximal bi-subalgebra
 $\mathfrak{B}^{[1]}=\mathfrak{B}^{[1,0]}=\mathfrak{B}_1\subset\mathfrak{C}\subset su(8)$.\label{figsu8QArank1inC}}
 \end{figure}

 \begin{figure}[htbp]
 \begin{center}
 \[\begin{array}{c}
   \mathfrak{B}^{[1,1]}\\
   \\
   {\cal S}^{100}_{100},{\cal S}^{101}_{100},{\cal S}^{110}_{110},{\cal S}^{111}_{110}
   \end{array}\]
  \[\begin{array}{ccccc}
  \mathfrak{B}^{[1]}
  &{\cal S}^{000}_{000},{\cal S}^{001}_{000},{\cal S}^{010}_{010},{\cal S}^{011}_{010}
  &&\{0\}&\\
  \\
  {W}(\mathfrak{B}_1,\mathfrak{B}^{[1]};0)
  &{\cal S}^{100}_{000},{\cal S}^{101}_{000},{\cal S}^{110}_{010}\,{\cal S}^{111}_{010}
  &&
  {\cal S}^{000}_{100},{\cal S}^{001}_{100},{\cal S}^{010}_{110},{\cal S}^{011}_{110}&\hat{W}(\mathfrak{B}_1,\mathfrak{B}^{[1]};1)\\
  \\
  {W}(\mathfrak{B}_1,\mathfrak{B}^{[1]};1)&\{0\}&&\{0\}&\hspace{0pt}\hat{W}(\mathfrak{B}_1,\mathfrak{B}^{[1]};0)\\
  \\
  {W}(\mathfrak{B}_2,\mathfrak{B}^{[1]};0)&{\cal S}^{010}_{000},{\cal S}^{011}_{000}
  &&{\cal S}^{100}_{110},{\cal S}^{101}_{110}&\hat{W}(\mathfrak{B}_2,\mathfrak{B}^{[1]};1)\\
  \\
\end{array}\]
 \end{center}
 \vspace{6pt}
 \fcaption{A co-quotient algebra of rank one given by the coset
 $\mathfrak{B}^{[1,1]}=\mathfrak{C}-\mathfrak{B}^{[1]}$.\label{figsu8coQArank1inC}}
 \end{figure}

 \begin{figure}[htbp]
 \begin{center}
 \[\begin{array}{c}
 \mathfrak{B}^{[2]}\\
 \\
 {\cal S}^{000}_{000},\hspace{2pt}{\cal S}^{010}_{010}
 \end{array}\]
 \[\begin{array}{ccccccc}
 \mathfrak{B}^{[2,\hspace{2pt}01]}&{\cal S}^{001}_{000},{\cal S}^{011}_{010}&&&&\{0\}&\\
 \mathfrak{B}^{[2,\hspace{1pt}10]}&{\cal S}^{100}_{100},{\cal S}^{110}_{110}&&&&\{0\}&\\
 \mathfrak{B}^{[2,\hspace{1pt}11]}&{\cal S}^{101}_{100},{\cal S}^{111}_{110}&&&&\{0\}&\\
 {W}(\mathfrak{B}_1,\mathfrak{B}^{[2]};\hspace{1pt}00)&{\cal S}^{100}_{000},{\cal S}^{110}_{010}&&&&\{0\}&\hat{W}(\mathfrak{B}_1,\mathfrak{B}^{[2]};\hspace{1pt}00)\\
 {W}(\mathfrak{B}_1,\mathfrak{B}^{[2]};\hspace{1pt}01)&{\cal S}^{101}_{000},{\cal S}^{111}_{010}&&&&\{0\}&\hat{W}(\mathfrak{B}_1,\mathfrak{B}^{[2]};\hspace{1pt}01)\\
 {W}(\mathfrak{B}_1,\mathfrak{B}^{[2]};\hspace{1pt}10)&\{0\}&&&&{\cal S}^{000}_{100},{\cal S}^{010}_{110}&\hat{W}(\mathfrak{B}_1,\mathfrak{B}^{[2]};\hspace{1pt}10)\\
 {W}(\mathfrak{B}_1,\mathfrak{B}^{[2]};\hspace{1pt}11)&\{0\}&&&&{\cal S}^{001}_{100},{\cal S}^{011}_{001}&\hat{W}(\mathfrak{B}_1,\mathfrak{B}^{[2]};\hspace{1pt}11)\\
 {W}(\mathfrak{B}_4,\mathfrak{B}^{[2]};\hspace{1pt}00)&{\cal S}^{000}_{001},{\cal S}^{010}_{011}&&&&\{0\}&\hat{W}(\mathfrak{B}_4,\mathfrak{B}^{[2]};\hspace{1pt}00)\\
\end{array} \]
 \end{center}
 \vspace{6pt}
 \fcaption{A quotient algebra of rank $2$ given by a $2$nd maximal bi-subalgebra
 $\mathfrak{B}^{[2]}=\mathfrak{B}^{[2,\hspace{2pt}00]}=\mathfrak{B}_1\cap\mathfrak{B}_4\cap\mathfrak{B}_5\subset\mathfrak{C}\subset su(8)$,
 referring to Fig.~\ref{figsu8QArank0inC}. \label{figsu8QArank2inC}}
 \end{figure}

 \begin{figure}[htbp]
 \begin{center}
 \[\begin{array}{c}
 \mathfrak{B}^{[2,\hspace{1pt}11]}\\
 \\
 {\cal S}^{101}_{100},{\cal S}^{111}_{110}
 \end{array}\]
 \[\begin{array}{ccccccc}
 \mathfrak{B}^{[2,\hspace{2pt}00]}&{\cal S}^{000}_{000},\hspace{2pt}{\cal S}^{010}_{010}&&&&\{0\}&\\
 \mathfrak{B}^{[2,\hspace{2pt}01]}&{\cal S}^{001}_{000},{\cal S}^{011}_{010}&&&&\{0\}&\\
 \mathfrak{B}^{[2,\hspace{1pt}10]}&{\cal S}^{100}_{100},{\cal S}^{110}_{110}&&&&\{0\}&\\
 {W}(\mathfrak{B}_1,\mathfrak{B}^{[2]};\hspace{1pt}00)&{\cal S}^{100}_{000},{\cal S}^{110}_{010}&&&&{\cal S}^{001}_{100},{\cal S}^{011}_{001}&\hat{W}(\mathfrak{B}_1,\mathfrak{B}^{[2]};\hspace{1pt}11)\\
 {W}(\mathfrak{B}_1,\mathfrak{B}^{[2]};\hspace{1pt}10)&\{0\}&&&&\{0\}&\hat{W}(\mathfrak{B}_1,\mathfrak{B}^{[2]};\hspace{1pt}01)\\
 {W}(\mathfrak{B}_1,\mathfrak{B}^{[2]};\hspace{1pt}01)&{\cal S}^{101}_{000},{\cal S}^{111}_{010}&&&&{\cal S}^{000}_{100},{\cal S}^{010}_{110}&\hat{W}(\mathfrak{B}_1,\mathfrak{B}^{[2]};\hspace{1pt}10)\\
 {W}(\mathfrak{B}_1,\mathfrak{B}^{[2]};\hspace{1pt}11)&\{0\}&&&&\{0\}&\hat{W}(\mathfrak{B}_1,\mathfrak{B}^{[2]};\hspace{1pt}00)\\
 {W}(\mathfrak{B}_4,\mathfrak{B}^{[2]};\hspace{1pt}11)&\{0\}&&&&\{0\}&\hat{W}(\mathfrak{B}_4,\mathfrak{B}^{[2]};\hspace{1pt}00)\\
\end{array} \]
 \end{center}
 \vspace{6pt}
 \fcaption{A co-quotient algebra of rank $2$ given by a coset
 $\mathfrak{B}^{[2,11]}$ in the partition of $\mathfrak{C}\subset su(8)$ generated by
 $\mathfrak{B}^{[2]}=\mathfrak{B}^{[2,\hspace{2pt}00]}\subset\mathfrak{C}$. \label{figsu8coQArank2inC}}
 \end{figure}

 \begin{figure}[htbp]
 \begin{center}
 \[\begin{array}{c}
 \mathfrak{C}_{[\mathbf{0}]}\\
 \\
 \{\hspace{2pt}I,\hspace{2pt}d_{15},\hspace{2pt}d_{25},\hspace{2pt}d_{35},\hspace{2pt}d_{45}\hspace{2pt}\}
 \end{array}\]
 \[\begin{array}{ccccc}
 W(\mathfrak{B}_{001})
 &\lambda_{12},\hspace{2pt}\lambda_{34}
 &
 &\hat{\lambda}_{12},\hspace{2pt}\hat{\lambda}_{34}
 &\hat{W}(\mathfrak{B}_{001})\\
 \\
 W(\mathfrak{B}_{010})
 &\lambda_{13},\hspace{2pt}\lambda_{24}
 &\hspace{20pt}
 &\hat{\lambda}_{13},\hspace{2pt}\hat{\lambda}_{24}
 &\hat{W}(\mathfrak{B}_{010})\\
 \\
 W(\mathfrak{B}_{011})
 &\lambda_{14},\hspace{2pt}\lambda_{23}
 &
 &\hat{\lambda}_{14},\hspace{2pt}\hat{\lambda}_{23}
 &\hat{W}(\mathfrak{B}_{011})\\
 \\
 W(\mathfrak{B}_{100})
 &\lambda_{15}
 &
 &\hat{\lambda}_{15}
 &\hat{W}(\mathfrak{B}_{100})\\
 \\
 W(\mathfrak{B}_{101})
 &\lambda_{25}
 &
 &\hat{\lambda}_{25}
 &\hat{W}(\mathfrak{B}_{101})\\
 \\
 W(\mathfrak{B}_{110})
 &\lambda_{35}
 &
 &\hat{\lambda}_{35}
 &\hat{W}(\mathfrak{B}_{110})\\
 \\
 W(\mathfrak{B}_{111})
 &\lambda_{45}
 &
 &\hat{\lambda}_{45}
 &\hat{W}(\mathfrak{B}_{111})
 \end{array}%
\]
 \end{center}
 \vspace{6pt}
 \fcaption{The intrinsic quotient algebra of rank zero given by the intrinsic Cartan subalgebra
 $\mathfrak{C}_{[\mathbf{0}]}\subset su(5)$ and the complete set of corresponding maximal bi-subalgebras;
 here $I$ is the identity matrix in $su(5)$.\label{figsu5QArank0intr}}
 \end{figure}

 \begin{figure}[htbp]
 \begin{center}
 \[\begin{array}{c}
 \mathfrak{C}\\
 \\
 \{\hspace{2pt}d_{15}+d_{25},\hspace{2pt}d_{35}+d_{45},\hspace{2pt}\lambda_{12},\hspace{2pt}\lambda_{34}\hspace{2pt}\}
 \end{array}\]
 \[\begin{array}{ccccc}
 W(\mathfrak{B}_{1})
 &d_{12},\hspace{2pt}d_{34}
 &
 &\hat{\lambda}_{12},\hspace{2pt}\hat{\lambda}_{34}
 &\hat{W}(\mathfrak{B}_{1})\\
 \\
 W(\mathfrak{B}_{2})
 &\lambda_{13}+\lambda_{24},\hspace{2pt}\lambda_{14}+\lambda_{23}
 &\hspace{20pt}
 &\hat{\lambda}_{13}+\hat{\lambda}_{24},\hspace{2pt}\hat{\lambda}_{14}+\hat{\lambda}_{23}
 &\hat{W}(\mathfrak{B}_{2})\\
 \\
 W(\mathfrak{B}_{3})
 &\lambda_{13}-\lambda_{24},\hspace{2pt}\lambda_{14}-\lambda_{23}
 &
 &\hat{\lambda}_{13}-\hat{\lambda}_{24},\hspace{2pt}\hat{\lambda}_{14}-\hat{\lambda}_{23}
 &\hat{W}(\mathfrak{B}_{3})\\
 \\
 W(\mathfrak{B}_{4})
 &\lambda_{15}+\lambda_{25}
 &
 &\hat{\lambda}_{15}+\hat{\lambda}_{25}
 &\hat{W}(\mathfrak{B}_{4})\\
 \\
 W(\mathfrak{B}_{5})
 &\lambda_{15}-\lambda_{25}
 &
 &\hat{\lambda}_{15}-\hat{\lambda}_{25}
 &\hat{W}(\mathfrak{B}_{5})\\
 \\
 W(\mathfrak{B}_{6})
 &\lambda_{35}+\lambda_{45}
 &
 &\hat{\lambda}_{35}+\hat{\lambda}_{45}
 &\hat{W}(\mathfrak{B}_{6})\\
 \\
 W(\mathfrak{B}_{7})
 &\lambda_{35}-\lambda_{45}
 &
 &\hat{\lambda}_{35}-\hat{\lambda}_{45}
 &\hat{W}(\mathfrak{B}_{7})
 \end{array} \]
 \vspace{15pt}
 \[\hspace{0pt}\begin{array}{c}
  \mathfrak{B}_{1}=\{\hspace{2pt}d_{15}+d_{25},\hspace{2pt}d_{35}+d_{45}\hspace{2pt}\},
 \\
 \\
 \hspace{0pt}\mathfrak{B}_{2}=\mathfrak{B}_{3}=\{\hspace{2pt}\lambda_{13}+\lambda_{24}\hspace{2pt}\},
 \hspace{2pt}\mathfrak{B}_{4}=\mathfrak{B}_{5}=\{\hspace{2pt}\lambda_{34}\hspace{2pt}\},
 \hspace{2pt}\mathfrak{B}_{6}=\mathfrak{B}_{7}=\{\hspace{2pt}\lambda_{12}\hspace{2pt}\}.
 \end{array}\]
 \end{center}
 \vspace{6pt}
 \fcaption{A quotient algebra of rank zero given by a Cartan subalgebra $\mathfrak{C}\subset su(5)$
 and the complete set of corresponding maximal bi-subalgebras.\label{figsu5QArank1inC}}
 \end{figure}

 \clearpage
 \begin{figure}[htbp]
 \begin{center}
 \[\hspace{-15pt}\begin{array}{c}
 \hspace{30pt}\mathfrak{C}_{[\mathbf{0}]}\\
 \\
 \hspace{30pt}I\otimes{\cal S}^{0}_{\hspace{.70pt}0},\hspace{2pt}I\otimes{\cal S}^{1}_{\hspace{.9pt}0},\hspace{2pt}\mu_3\otimes{\cal S}^{0}_{\hspace{.70pt}0},
 \hspace{2pt}\mu_3\otimes{\cal S}^{1}_{\hspace{.9pt}0},\hspace{2pt}\mu_8\otimes{\cal S}^{0}_{\hspace{.70pt}0},
 \hspace{2pt}\mu_8\otimes{\cal S}^{1}_{\hspace{.9pt}0}
 \end{array}\]
 \[\hspace{-20pt}\begin{array}{ccccccc}
 &&{W}(\mathfrak{B}_{001})
 &I\otimes{\cal S}^{\hspace{.1pt}0}_{1},\hspace{2pt}\mu_3\otimes{\cal S}^{\hspace{.1pt}0}_{1},\hspace{2pt}
 \mu_8\otimes{\cal S}^{\hspace{.1pt}0}_{1}
 &\hspace{24pt}
 &I\otimes{\cal S}^{1}_{\hspace{.7pt}1},\hspace{2pt}\mu_3\otimes{\cal S}^{1}_{\hspace{.7pt}1},\hspace{2pt}
 \mu_8\otimes{\cal S}^{\hspace{-.5pt}1}_{1}
 &\hat{W}(\mathfrak{B}_{001})\\
 &&&&&&\\
 &&{W}(\mathfrak{B}_{010})
 &\mu_1\otimes{\cal S}^{0}_{\hspace{.70pt}0},\hspace{2pt}\mu_1\otimes{\cal S}^{1}_{\hspace{.9pt}0}
 &
 &\mu_2\otimes{\cal S}^{0}_{\hspace{.70pt}0},\hspace{2pt}\mu_2\otimes{\cal S}^{1}_{\hspace{.9pt}0}
 &\hat{W}(\mathfrak{B}_{010})\\
 &&&&&&\\
 &&{W}(\mathfrak{B}_{011})
 &\mu_1\otimes{\cal S}^{\hspace{.1pt}0}_{1},\hspace{2pt}\mu_2\otimes{\cal S}^{\hspace{-.5pt}1}_{1}
 &
 &\mu_1\otimes{\cal S}^{\hspace{-.5pt}1}_{1},\hspace{2pt}\mu_2\otimes{\cal S}^{\hspace{.1pt}0}_{1}
 &\hat{W}(\mathfrak{B}_{011})\\
 &&&&&&\\
 &&{W}(\mathfrak{B}_{100})
 &\mu_4\otimes{\cal S}^{0}_{\hspace{.70pt}0},\hspace{2pt}\mu_4\otimes{\cal S}^{1}_{\hspace{.9pt}0}
 &
 &\mu_5\otimes{\cal S}^{0}_{\hspace{.70pt}0},\hspace{2pt}\mu_5\otimes{\cal S}^{1}_{\hspace{.9pt}0}
 &\hat{W}(\mathfrak{B}_{100})\\
 &&&&&&\\
\end{array}%
 \]
 \end{center}
 \vspace{6pt}
 \fcaption{The quotient algebra of rank zero given by the intrinsic Cartan subalgebra $\mathfrak{C}_{[\mathbf{0}]}\subset su(6)$
 and the complete set of corresponding maximal bi-subalgebras; here $\mu_i$ denote the Gell-Mann
 matrices, $1\leq i\leq 8$, and $I$ is the $3\times 3$ identity matrix.\label{figsu6QArank0intr}}
 \end{figure}

 \begin{figure}[htbp]
 \begin{center}
 \[\hspace{-12pt}\begin{array}{c}
 \hspace{0pt}\mathfrak{B}^{[1]}_{can}\\
 \\
 \hspace{0pt}I\otimes{\cal S}^{0}_{\hspace{.70pt}0},\hspace{2pt}\mu_3\otimes{\cal S}^{0}_{\hspace{.70pt}0},
 \hspace{2pt}\mu_8\otimes{\cal S}^{0}_{\hspace{.70pt}0}
 \end{array}\]
 \[\hspace{-12pt}\begin{array}{ccccc}
 \mathfrak{B}^{[1,1]}_{can}=\mathfrak{C}_{[\mathbf{0}]}-\mathfrak{B}^{[1]}_{can}
 &I\otimes{\cal S}^{1}_{\hspace{.9pt}0},\mu_3\otimes{\cal S}^{1}_{\hspace{.9pt}0},\hspace{2pt}
 \mu_8\otimes{\cal S}^{1}_{\hspace{.9pt}0}
 &\hspace{0pt}
 &\{0\}
 &\\
 \\
 {W}(\mathfrak{B}_{001},\mathfrak{B}^{[1]}_{can};0)
 &I\otimes{\cal S}^{\hspace{.1pt}0}_{1},\hspace{2pt}\mu_3\otimes{\cal S}^{\hspace{.1pt}0}_{1},\hspace{2pt}
 \mu_8\otimes{\cal S}^{\hspace{.1pt}0}_{1}
 &\hspace{0pt}
 &\{0\}
 &\hat{W}(\mathfrak{B}_{001},\mathfrak{B}^{[1]}_{can};0)\\
 \\
 {W}(\mathfrak{B}_{001},\mathfrak{B}^{[1]}_{can};1)
 &\{0\}
 &\hspace{0pt}
 &I\otimes{\cal S}^{1}_{\hspace{.7pt}1},\hspace{2pt}\mu_3\otimes{\cal S}^{1}_{\hspace{.7pt}1},\hspace{2pt}
 \mu_8\otimes{\cal S}^{\hspace{-.5pt}1}_{1}
 &\hat{W}(\mathfrak{B}_{001},\mathfrak{B}^{[1]}_{can};1)\\
 \\
 {W}(\mathfrak{B}_{010},\mathfrak{B}^{[1]}_{can};0)
 &\mu_1\otimes{\cal S}^{0}_{\hspace{.70pt}0}
 &
 &\mu_2\otimes{\cal S}^{0}_{\hspace{.70pt}0}
 &\hat{W}(\mathfrak{B}_{010},\mathfrak{B}^{[1]}_{can};0)\\
 \\
\end{array} \]
 \end{center}
 \vspace{6pt}
 \fcaption{A quotient algebra of rank one given by a maximal bi-subalgebra
 $\mathfrak{B}^{[1]}_{can}=\mathfrak{B}^{[1,0]}=\mathfrak{B}_{001}\subset\mathfrak{C}_{[\mathbf{0}]}\subset su(6)$.\label{figsu6canQArank1inC}}
 \end{figure}

 \begin{figure}[htbp]
 \begin{center}
 \[\begin{array}{c}
 \hspace{30pt}\mathfrak{B}^{[1,1]}_{can}\\
 \\
 \hspace{30pt}I\otimes{\cal S}^{1}_{\hspace{.9pt}0},\mu_3\otimes{\cal S}^{1}_{\hspace{.9pt}0},\hspace{2pt}
 \mu_8\otimes{\cal S}^{1}_{\hspace{.9pt}0}
 \end{array}\]
 \[\begin{array}{ccccc}
 \mathfrak{B}^{[1]}_{can}
 &I\otimes{\cal S}^{0}_{\hspace{.70pt}0},\hspace{2pt}\mu_3\otimes{\cal S}^{0}_{\hspace{.70pt}0},
 \hspace{2pt}\mu_8\otimes{\cal S}^{0}_{\hspace{.70pt}0}
 &\hspace{0pt}
 &\{0\}
 &\\
 \\
 {W}(\mathfrak{B}_{001},\mathfrak{B}^{[1]}_{can};0)
 &I\otimes{\cal S}^{\hspace{.1pt}0}_{1},\hspace{2pt}\mu_3\otimes{\cal S}^{\hspace{.1pt}0}_{1},\hspace{2pt}
 \mu_8\otimes{\cal S}^{\hspace{.1pt}0}_{1}
 &\hspace{0pt}
 &I\otimes{\cal S}^{1}_{\hspace{.7pt}1},\hspace{2pt}\mu_3\otimes{\cal S}^{1}_{\hspace{.7pt}1},\hspace{2pt}
 \mu_8\otimes{\cal S}^{\hspace{-.5pt}1}_{1}
 &\hat{W}(\mathfrak{B}_{001},\mathfrak{B}^{[1]}_{can};1)\\
 \\
 {W}(\mathfrak{B}_{001},\mathfrak{B}^{[1]}_{can};1)
 &\{0\}
 &\hspace{0pt}
 &\{0\}
 &\hat{W}(\mathfrak{B}_{001},\mathfrak{B}^{[1]}_{can};0)\\
 \\
 {W}(\mathfrak{B}_{010},\mathfrak{B}^{[1]}_{can};1)
 &\mu_1\otimes{\cal S}^{1}_{\hspace{.9pt}0}
 &
 &\mu_2\otimes{\cal S}^{0}_{\hspace{.70pt}0}
 &\hat{W}(\mathfrak{B}_{010},\mathfrak{B}^{[1]}_{can};0)\\
 \\
\end{array} \]
 \end{center}
 \vspace{6pt}
 \fcaption{A co-quotient algebra of rank one given by the complement
 $\mathfrak{B}^{[1,1]}_{can}=\mathfrak{C}_{[\mathbf{0}]}-\mathfrak{B}^{[1]}_{can}$.\label{figsu6coQArank1inC}}
 \end{figure}

 \begin{figure}[htbp]
 \begin{center}
 \[\begin{array}{c}
 \mathfrak{C}_{[\mathbf{0}]}\\
 \\
 \{\hspace{2pt}d_{17},\hspace{2pt}d_{27},\hspace{2pt}d_{37},\hspace{2pt}d_{47},\hspace{2pt}d_{57},\hspace{2pt}d_{67}\hspace{2pt}\}
 \end{array}\]
 \[\begin{array}{ccccc}
 W(\mathfrak{B}_{001})
 &\lambda_{12},\hspace{2pt}\lambda_{34},\hspace{2pt}\lambda_{56}
 &
 &\hat{\lambda}_{12},\hspace{2pt}\hat{\lambda}_{34},\hspace{2pt}\hat{\lambda}_{56}
 &\hat{W}(\mathfrak{B}_{001})\\
 \\
 W(\mathfrak{B}_{010})
 &\lambda_{13},\hspace{2pt}\lambda_{24},\hspace{2pt}\lambda_{57}
 &\hspace{20pt}
 &\hat{\lambda}_{13},\hspace{2pt}\hat{\lambda}_{24},\hspace{2pt}\hat{\lambda}_{57}
 &\hat{W}(\mathfrak{B}_{010})\\
 \\
 W(\mathfrak{B}_{011})
 &\lambda_{14},\hspace{2pt}\lambda_{23},\hspace{2pt}\lambda_{67}
 &
 &\hat{\lambda}_{14},\hspace{2pt}\hat{\lambda}_{23},\hspace{2pt}\hat{\lambda}_{67}
 &\hat{W}(\mathfrak{B}_{011})\\
 \\
 W(\mathfrak{B}_{100})
 &\lambda_{15},\hspace{2pt}\lambda_{26},\hspace{2pt}\lambda_{37}
 &
 &\hat{\lambda}_{15},\hspace{2pt}\hat{\lambda}_{26},\hspace{2pt}\hat{\lambda}_{37}
 &\hat{W}(\mathfrak{B}_{100})\\
 \\
 W(\mathfrak{B}_{101})
 &\lambda_{16},\hspace{2pt}\lambda_{25},\hspace{2pt}\lambda_{47}
 &
 &\hat{\lambda}_{16},\hspace{2pt}\hat{\lambda}_{25},\hspace{2pt}\hat{\lambda}_{47}
 &\hat{W}(\mathfrak{B}_{101})\\
 \\
 W(\mathfrak{B}_{110})
 &\lambda_{17},\hspace{2pt}\lambda_{35},\hspace{2pt}\lambda_{46}
 &
 &\hat{\lambda}_{17},\hspace{2pt}\hat{\lambda}_{35},\hspace{2pt}\hat{\lambda}_{46}
 &\hat{W}(\mathfrak{B}_{110})\\
 \\
 W(\mathfrak{B}_{111})
 &\lambda_{27},\hspace{2pt}\lambda_{36},\hspace{2pt}\lambda_{45}
 &
 &\hat{\lambda}_{27},\hspace{2pt}\hat{\lambda}_{36},\hspace{2pt}\hat{\lambda}_{45}
 &\hat{W}(\mathfrak{B}_{111})
 \end{array}%
\]
 \end{center}
 \vspace{6pt}
 \fcaption{The intrinsic quotient algebra of rank zero given by the intrinsic Cartan subalgebra
 $\mathfrak{C}_{[\mathbf{0}]}\subset su(7)$ and the complete set of corresponding maximal bi-subalgebras.\label{figsu7QArank0intr}}
 \end{figure}

 \begin{figure}[htbp]
 \begin{center}
 \[\begin{array}{c}
 \mathfrak{C}\\
 \\
 \{\hspace{2pt}d_{17}+d_{27},\hspace{2pt}d_{37}+d_{47},\hspace{2pt}d_{57}+d_{67}\hspace{2pt},\lambda_{12},\hspace{2pt}\lambda_{34},\hspace{2pt}\lambda_{56}\hspace{2pt}\}
 \end{array}\]
 \[\begin{array}{ccccc}
 W(\mathfrak{B}_{1})
 &d_{12},\hspace{2pt}d_{34},\hspace{2pt}d_{56}
 &
 &\hat{\lambda}_{12},\hspace{2pt}\hat{\lambda}_{34},\hspace{2pt}\hat{\lambda}_{56}
 &\hat{W}(\mathfrak{B}_{1})\\
 \\
 W(\mathfrak{B}_{2})
 &\lambda_{13}+\lambda_{24},\hspace{2pt}\lambda_{14}+\lambda_{23},\hspace{2pt}\lambda_{57}+\lambda_{67}
 &\hspace{20pt}
 &\hat{\lambda}_{13}+\hat{\lambda}_{24},\hspace{2pt}\hat{\lambda}_{14}+\hat{\lambda}_{23},\hspace{2pt}\hat{\lambda}_{57}+\hat{\lambda}_{67}
 &\hat{W}(\mathfrak{B}_{2})\\
 \\
 W(\mathfrak{B}_{3})
 &\lambda_{13}-\lambda_{24},\hspace{2pt}\lambda_{14}-\lambda_{23},\hspace{2pt}\lambda_{57}-\lambda_{67}
 &\hspace{20pt}
 &\hat{\lambda}_{13}-\hat{\lambda}_{24},\hspace{2pt}\hat{\lambda}_{14}-\hat{\lambda}_{23},\hspace{2pt}\hat{\lambda}_{57}-\hat{\lambda}_{67}
 &\hat{W}(\mathfrak{B}_{3})\\
 \\
 W(\mathfrak{B}_{4})
 &\lambda_{15}+\lambda_{26},\hspace{2pt}\lambda_{16}+\lambda_{25},\hspace{2pt}\lambda_{37}+\lambda_{47}
 &
 &\hat{\lambda}_{15}-\hat{\lambda}_{26},\hspace{2pt}\hat{\lambda}_{16}-\lambda_{25},\hspace{2pt}\hat{\lambda}_{37}-\hat{\lambda}_{47}
 &\hat{W}(\mathfrak{B}_{4})\\
 \\
\end{array}%
\]
 \end{center}
 \vspace{6pt}
 \fcaption{A quotient algebra of rank zero given by a Cartan subalgebra $\mathfrak{C}\subset su(7)$
 and the complete set of corresponding maximal bi-subalgebras.\label{figsu7QArank0inC}}
 \end{figure}

 \begin{figure}[htbp]
 \begin{center}
 \[%
 \]
 \vspace{6pt}
 \fcaption{The intrinsic quotient algebra of rank zero given by the intrinsic Cartan subalgebra
 $\mathfrak{C}_{[{\bf 0}]}\subset su(16)$ followed by the complete set of corresponding maximal bi-subalgebras.\label{figsu16intrQArank0}}
 \end{figure}

 \begin{figure}[htbp]
 \begin{center}
 \[%
 &\hat{W}(\mathfrak{B}_{0001},\mathfrak{B}^{[1]}_{can};1)
 \\
 &&&&\\
 {W}(\mathfrak{B}_{0010},\mathfrak{B}^{[1]}_{can};0)
 &
 {\cal S}^{0000}_{0010},{\cal S}^{0100}_{0010},{\cal S}^{1000}_{0010},{\cal S}^{1100}_{0010}
 &
 &{\cal S}^{0010}_{0010},{\cal S}^{0110}_{0010},{\cal S}^{1010}_{0010},{\cal S}^{1110}_{0010}
 &\hat{W}(\mathfrak{B}_{0010},\mathfrak{B}^{[1]}_{can};0)
 \\
 {W}(\mathfrak{B}_{0010},\mathfrak{B}^{[1]}_{can};1)
 &{\cal S}^{0001}_{0010},{\cal S}^{0101}_{0010},{\cal S}^{1001}_{0010},{\cal S}^{1101}_{0010}
 &
 &
 {\cal S}^{0011}_{0010},{\cal S}^{0111}_{0010},{\cal S}^{1011}_{0010},{\cal S}^{1111}_{0010}
 &\hat{W}(\mathfrak{B}_{0010},\mathfrak{B}^{[1]}_{can};1)
 \\
 {W}(\mathfrak{B}_{0011},\mathfrak{B}^{[1]}_{can};0)
 &{\cal S}^{0000}_{0011},{\cal S}^{0100}_{0011},{\cal S}^{1000}_{0011},{\cal S}^{1100}_{0011}
 &
 &
 {\cal S}^{0010}_{0011},{\cal S}^{0110}_{0011},{\cal S}^{1010}_{0011},{\cal S}^{1110}_{0011}
 &\hat{W}(\mathfrak{B}_{0011},\mathfrak{B}^{[1]}_{can};0)
 \\
 {W}(\mathfrak{B}_{0011},\mathfrak{B}^{[1]}_{can};1)
 &{\cal S}^{0011}_{0011},{\cal S}^{0111}_{0011},{\cal S}^{1011}_{0011},{\cal S}^{1111}_{0011}
 &
 &{\cal S}^{0001}_{0011},{\cal S}^{0101}_{0011},{\cal S}^{1001}_{0011},{\cal S}^{1101}_{0011}
 &\hat{W}(\mathfrak{B}_{0011},\mathfrak{B}^{[1]}_{can};1)
 \\
 {W}(\mathfrak{B}_{0100},\mathfrak{B}^{[1]}_{can};0)
 &{\cal S}^{0000}_{0100},{\cal S}^{0010}_{0100},{\cal S}^{1000}_{0100},{\cal S}^{1010}_{0100}
 &
 &
 {\cal S}^{0100}_{0100},{\cal S}^{0110}_{0100},{\cal S}^{1100}_{0100},{\cal S}^{1110}_{0100}
 &\hat{W}(\mathfrak{B}_{0100},\mathfrak{B}^{[1]}_{can};0)
 \\
 {W}(\mathfrak{B}_{0100},\mathfrak{B}^{[1]}_{can};1)
 &{\cal S}^{0001}_{0100},{\cal S}^{0011}_{0100},{\cal S}^{1001}_{0100},{\cal S}^{1011}_{0100}
 &
 &{\cal S}^{0101}_{0100},{\cal S}^{0111}_{0100},{\cal S}^{1101}_{0100},{\cal S}^{1111}_{0100}
 &\hat{W}(\mathfrak{B}_{0100},\mathfrak{B}^{[1]}_{can};1)
 \\
 {W}(\mathfrak{B}_{0101},\mathfrak{B}^{[1]}_{can};0)
 &
 {\cal S}^{0000}_{0101},{\cal S}^{0010}_{0101},{\cal S}^{1000}_{0101},{\cal S}^{1010}_{0101}
 &
 &
 {\cal S}^{0100}_{0101},{\cal S}^{0110}_{0101},{\cal S}^{1100}_{0101},{\cal S}^{1110}_{0101}
 &\hat{W}(\mathfrak{B}_{0101},\mathfrak{B}^{[1]}_{can};0)
 \\
 {W}(\mathfrak{B}_{0101},\mathfrak{B}^{[1]}_{can};1)
 &{\cal S}^{0101}_{0101},{\cal S}^{0111}_{0101},{\cal S}^{1101}_{0101},{\cal S}^{1111}_{0101}
 &
 &{\cal S}^{0001}_{0101},{\cal S}^{0011}_{0101},{\cal S}^{1001}_{0101},{\cal S}^{1011}_{0101}
 &\hat{W}(\mathfrak{B}_{0101},\mathfrak{B}^{[1]}_{can};1)
 \\
 {W}(\mathfrak{B}_{0110},\mathfrak{B}^{[1]}_{can};0)
 &
 {\cal S}^{0000}_{0110},{\cal S}^{0110}_{0110},{\cal S}^{1000}_{0110},{\cal S}^{1110}_{0110}
 &
 &
 {\cal S}^{0010}_{0110},{\cal S}^{0100}_{0110},{\cal S}^{1010}_{0110},{\cal S}^{1100}_{0110}
 &\hat{W}(\mathfrak{B}_{0110},\mathfrak{B}^{[1]}_{can};0)
 \\
 {W}(\mathfrak{B}_{0110},\mathfrak{B}^{[1]}_{can};1)
 &{\cal S}^{0001}_{0110},{\cal S}^{0111}_{0110},{\cal S}^{1001}_{0110},{\cal S}^{1111}_{0110}
 &
 &{\cal S}^{0011}_{0110},{\cal S}^{0101}_{0110},{\cal S}^{1011}_{0110},{\cal S}^{1101}_{0110}
 &\hat{W}(\mathfrak{B}_{0110},\mathfrak{B}^{[1]}_{can};1)
 \\
 {W}(\mathfrak{B}_{0111},\mathfrak{B}^{[1]}_{can};0)
 &
 {\cal S}^{0000}_{0111},{\cal S}^{0110}_{0111},{\cal S}^{1000}_{0111},{\cal S}^{1110}_{0111}
 &
 &
 {\cal S}^{0010}_{0111},{\cal S}^{0100}_{0111}, {\cal S}^{1010}_{0111},{\cal S}^{1100}_{0111}
 &\hat{W}(\mathfrak{B}_{0111},\mathfrak{B}^{[1]}_{can};0)
 \\
 {W}(\mathfrak{B}_{0111},\mathfrak{B}^{[1]}_{can};1)
 &{\cal S}^{0011}_{0111},{\cal S}^{0101}_{0111},{\cal S}^{1011}_{0111},{\cal S}^{1101}_{0111}
 &
 &{\cal S}^{0001}_{0111},{\cal S}^{0111}_{0111},{\cal S}^{1001}_{0111},{\cal S}^{1111}_{0111}
 &\hat{W}(\mathfrak{B}_{0111},\mathfrak{B}^{[1]}_{can};1)
 \\
 {W}(\mathfrak{B}_{1001},\mathfrak{B}^{[1]}_{can};0)
 &
 {\cal S}^{0000}_{1001},{\cal S}^{0010}_{1001},{\cal S}^{0100}_{1001},{\cal S}^{0110}_{1001}
 &
 &{\cal S}^{1000}_{1001},{\cal S}^{1010}_{1001},{\cal S}^{1100}_{1001},{\cal S}^{1110}_{1001}
 &\hat{W}(\mathfrak{B}_{1001},\mathfrak{B}^{[1]}_{can};0)
 \\
 {W}(\mathfrak{B}_{1001},\mathfrak{B}^{[1]}_{can};1)
 &{\cal S}^{1001}_{1001},{\cal S}^{1011}_{1001},{\cal S}^{1101}_{1001},{\cal S}^{1111}_{1001}
 &
 &{\cal S}^{0001}_{1001},{\cal S}^{0011}_{1001},{\cal S}^{0101}_{1001},{\cal S}^{0111}_{1001}
 &\hat{W}(\mathfrak{B}_{1001},\mathfrak{B}^{[1]}_{can};1)
 \\
 {W}(\mathfrak{B}_{1010},\mathfrak{B}^{[1]}_{can};0)
 &
 {\cal S}^{0000}_{1010},{\cal S}^{0100}_{1010},{\cal S}^{1010}_{1010},{\cal S}^{1111}_{1010}
 &
 &
 {\cal S}^{0010}_{1010},{\cal S}^{0110}_{1010}, {\cal S}^{1000}_{1010},{\cal S}^{1100}_{1010}
 &\hat{W}(\mathfrak{B}_{1010},\mathfrak{B}^{[1]}_{can};0)
 \\
 {W}(\mathfrak{B}_{1010},\mathfrak{B}^{[1]}_{can};1)
 &{\cal S}^{0001}_{1010},{\cal S}^{0101}_{1010},{\cal S}^{1011}_{1010},{\cal S}^{1110}_{1010}
 &
 &{\cal S}^{0011}_{1010},{\cal S}^{0111}_{1010},{\cal S}^{1001}_{1010},{\cal S}^{1101}_{1010}
 &\hat{W}(\mathfrak{B}_{1010},\mathfrak{B}^{[1]}_{can};1)
 \\
 {W}(\mathfrak{B}_{1011},\mathfrak{B}^{[1]}_{can};0)
 &
 {\cal S}^{0000}_{1011},{\cal S}^{0100}_{1011},{\cal S}^{1010}_{1011},{\cal S}^{1110}_{1011}
 &
 &
 {\cal S}^{0010}_{1011},{\cal S}^{0110}_{1011}, {\cal S}^{1000}_{1011},{\cal S}^{1100}_{1011}
 &\hat{W}(\mathfrak{B}_{1011},\mathfrak{B}^{[1]}_{can};0)
 \\
 {W}(\mathfrak{B}_{1011},\mathfrak{B}^{[1]}_{can};1)
 &{\cal S}^{0011}_{1011},{\cal S}^{0111}_{1011},{\cal S}^{1001}_{1011},{\cal S}^{1101}_{1011}
 &
 &{\cal S}^{0001}_{1011},{\cal S}^{0101}_{1011},{\cal S}^{1011}_{1011},{\cal S}^{1111}_{1011}
 &\hat{W}(\mathfrak{B}_{1011},\mathfrak{B}^{[1]}_{can};1)
 \\
 {W}(\mathfrak{B}_{1100},\mathfrak{B}^{[1]}_{can};0)
 &
 {\cal S}^{0000}_{1100},{\cal S}^{0010}_{1100},{\cal S}^{1100}_{1100},{\cal S}^{1110}_{1100}
 &
 &
 {\cal S}^{0100}_{1100},{\cal S}^{0110}_{1100}, {\cal S}^{1000}_{1100},{\cal S}^{1010}_{1100}
 &\hat{W}(\mathfrak{B}_{1100},\mathfrak{B}^{[1]}_{can};0)
 \\
 {W}(\mathfrak{B}_{1100},\mathfrak{B}^{[1]}_{can};1)
 &{\cal S}^{0001}_{1100},{\cal S}^{0011}_{1100},{\cal S}^{1101}_{1100},{\cal S}^{1111}_{1100}
 &
 &{\cal S}^{0101}_{1100},{\cal S}^{0111}_{1100},{\cal S}^{1001}_{1100},{\cal S}^{1011}_{1100}
 &\hat{W}(\mathfrak{B}_{1100},\mathfrak{B}^{[1]}_{can};1)
 \\
 {W}(\mathfrak{B}_{1101},\mathfrak{B}^{[1]}_{can};0)
 &
 {\cal S}^{0000}_{1101},{\cal S}^{0010}_{1101},{\cal S}^{1100}_{1101},{\cal S}^{1110}_{1101}
 &
 &{\cal S}^{0100}_{1101},{\cal S}^{0110}_{1101}, {\cal S}^{1000}_{1101},{\cal S}^{1010}_{1101}
 &\hat{W}(\mathfrak{B}_{1101},\mathfrak{B}^{[1]}_{can};0)
 \\
 {W}(\mathfrak{B}_{1101},\mathfrak{B}^{[1]}_{can};1)
 &{\cal S}^{0101}_{1101},{\cal S}^{0111}_{1101},{\cal S}^{1001}_{1101},{\cal S}^{1011}_{1101}
 &
 &{\cal S}^{0001}_{1101},{\cal S}^{0011}_{1101},{\cal S}^{1101}_{1101},{\cal S}^{1111}_{1101}
 &\hat{W}(\mathfrak{B}_{1101},\mathfrak{B}^{[1]}_{can};1)
 \\
 {W}(\mathfrak{B}_{1110},\mathfrak{B}^{[1]}_{can};0)
 &
 {\cal S}^{0000}_{1110},{\cal S}^{0110}_{1110},{\cal S}^{1010}_{1110},{\cal S}^{1100}_{1110}
 &
 &
 {\cal S}^{0010}_{1110},{\cal S}^{0100}_{1110}, {\cal S}^{1000}_{1110},{\cal S}^{1110}_{1110}
 &\hat{W}(\mathfrak{B}_{1110},\mathfrak{B}^{[1]}_{can};0)
 \\
 {W}(\mathfrak{B}_{1110},\mathfrak{B}^{[1]}_{can};1)
 &{\cal S}^{0001}_{1110},{\cal S}^{0111}_{1110},{\cal S}^{1011}_{1110},{\cal S}^{1101}_{1110}
 &
 &{\cal S}^{0011}_{1110},{\cal S}^{0101}_{1110},{\cal S}^{1001}_{1110},{\cal S}^{1111}_{1110}
 &\hat{W}(\mathfrak{B}_{1110},\mathfrak{B}^{[1]}_{can};1)
 \\
 {W}(\mathfrak{B}_{1111},\mathfrak{B}^{[1]}_{can};0)
 &
 {\cal S}^{0000}_{1111},{\cal S}^{0110}_{1111}, {\cal S}^{1010}_{1111},{\cal S}^{1100}_{1111}
 &
 &
 {\cal S}^{0010}_{1111},{\cal S}^{0100}_{1111},{\cal S}^{1000}_{1111},{\cal S}^{1110}_{1111}
 &\hat{W}(\mathfrak{B}_{1111},\mathfrak{B}^{[1]}_{can};0)
 \\
 {W}(\mathfrak{B}_{1111},\mathfrak{B}^{[1]}_{can};1)
 &{\cal S}^{0011}_{1111},{\cal S}^{0101}_{1111},{\cal S}^{1001}_{1111},{\cal S}^{1111}_{1111}
 &
 &{\cal S}^{0001}_{1111},{\cal S}^{0111}_{1111},{\cal S}^{1011}_{1111},{\cal S}^{1101}_{1111}
 &\hat{W}(\mathfrak{B}_{1111},\mathfrak{B}^{[1]}_{can};1)
 \end{array} \]
 \end{center}
 \vspace{6pt}
 \fcaption{The canonical quotient algebra of rank one given by the canonical center subalgebra
 $\mathfrak{B}^{[1]}_{can}=\mathfrak{B}^{[1,0]}_{can}=\mathfrak{B}_{0001}\subset\mathfrak{C}_{[\mathbf{0}]}\subset su(16)$.\label{figsu16canQArank1}}
 \end{figure}

 \begin{figure}[htbp]
 \begin{center}
 \[\begin{array}{c}
 \mathfrak{B}^{[1,1]}_{can}\\
 \\
 {\cal S}^{0001}_{0000},\hspace{2pt}{\cal S}^{0011}_{0000},\hspace{2pt}{\cal S}^{0101}_{0000},\hspace{2pt}{\cal S}^{0111}_{0000},
 \hspace{2pt}{\cal S}^{1001}_{0000},\hspace{2pt}{\cal S}^{1011}_{0000},\hspace{2pt}{\cal S}^{1101}_{0000},\hspace{2pt}{\cal S}^{1111}_{0000}
 \end{array}\]
 \[\hspace{-30pt}\begin{array}{ccccc}
 \mathfrak{B}^{[1]}_{can}&
 \hspace{3pt}
 \begin{array}{c}
 {\cal S}^{0000}_{0000},{\cal S}^{0010}_{0000},{\cal S}^{0100}_{0000},{\cal S}^{0110}_{0000},\\
 \hspace{-3pt}{\cal S}^{1000}_{0000},{\cal S}^{1010}_{0000},{\cal S}^{1100}_{0000},{\cal S}^{1110}_{0000}
 \end{array}
 &&\{0\}&\\
 \\
 {W}(\mathfrak{B}_{0001},\mathfrak{B}^{[1]}_{can};0)
 &\hspace{3pt}
 \begin{array}{c}
 {\cal S}^{0000}_{0001},{\cal S}^{0010}_{0001},{\cal S}^{0100}_{0001},{\cal S}^{0110}_{0001},
 \\
 \hspace{-3pt}{\cal S}^{1000}_{0001},{\cal S}^{1010}_{0001},{\cal S}^{1100}_{0001},{\cal S}^{1110}_{0001}
 \end{array}
 &\hspace{0pt}
 &
 \hspace{3pt}
 \begin{array}{c}
 {\cal S}^{0001}_{0001},{\cal S}^{0011}_{0001},{\cal S}^{0101}_{0001},{\cal S}^{0111}_{0001},
 \\
 \hspace{-3pt}{\cal S}^{1001}_{0001},{\cal S}^{1011}_{0001},{\cal S}^{1101}_{0001},{\cal S}^{1111}_{0001}
 \end{array}
 &\hat{W}(\mathfrak{B}_{0001},\mathfrak{B}^{[1]}_{can};1)\\
 &&&&\\
 {W}(\mathfrak{B}_{0001},\mathfrak{B}^{[1]}_{can};1)
 &\{0\}
 &
 &\{0\}
 &\hat{W}(\mathfrak{B}_{0001},\mathfrak{B}^{[1]}_{can};0)
 \\
 &&&&\\
 {W}(\mathfrak{B}_{0010},\mathfrak{B}^{[1]}_{can};0)
 &
 {\cal S}^{0000}_{0010},{\cal S}^{0100}_{0010},{\cal S}^{1000}_{0010},{\cal S}^{1100}_{0010}
 &
 &{\cal S}^{0011}_{0010},{\cal S}^{0111}_{0010},{\cal S}^{1011}_{0010},{\cal S}^{1111}_{0010}
 &\hat{W}(\mathfrak{B}_{0010},\mathfrak{B}^{[1]}_{can};1)
 \\
 {W}(\mathfrak{B}_{0010},\mathfrak{B}^{[1]}_{can};1)
 &{\cal S}^{0001}_{0010},{\cal S}^{0101}_{0010},{\cal S}^{1001}_{0010},{\cal S}^{1101}_{0010}
 &
 &
 {\cal S}^{0010}_{0010},{\cal S}^{0110}_{0010},{\cal S}^{1010}_{0010},{\cal S}^{1110}_{0010}
 &\hat{W}(\mathfrak{B}_{0010},\mathfrak{B}^{[1]}_{can};0)
 \\
 {W}(\mathfrak{B}_{0011},\mathfrak{B}^{[1]}_{can};1)
 &{\cal S}^{0011}_{0011},{\cal S}^{0111}_{0011},{\cal S}^{1011}_{0011},{\cal S}^{1111}_{0011}
 &
 &
 {\cal S}^{0010}_{0011},{\cal S}^{0110}_{0011},{\cal S}^{1010}_{0011},{\cal S}^{1110}_{0011}
 &\hat{W}(\mathfrak{B}_{0011},\mathfrak{B}^{[1]}_{can};0)
 \\
 {W}(\mathfrak{B}_{0011},\mathfrak{B}^{[1]}_{can};0)
 &{\cal S}^{0000}_{0011},{\cal S}^{0100}_{0011},{\cal S}^{1000}_{0011},{\cal S}^{1100}_{0011}
 &
 &{\cal S}^{0001}_{0011},{\cal S}^{0101}_{0011},{\cal S}^{1001}_{0011},{\cal S}^{1101}_{0011}
 &\hat{W}(\mathfrak{B}_{0011},\mathfrak{B}^{[1]}_{can};1)
 \\
 {W}(\mathfrak{B}_{0100},\mathfrak{B}^{[1]}_{can};0)
 &{\cal S}^{0000}_{0100},{\cal S}^{0010}_{0100},{\cal S}^{1000}_{0100},{\cal S}^{1010}_{0100}
 &
 &
 {\cal S}^{0101}_{0100},{\cal S}^{0111}_{0100},{\cal S}^{1101}_{0100},{\cal S}^{1111}_{0100}
 &\hat{W}(\mathfrak{B}_{0100},\mathfrak{B}^{[1]}_{can};1)
 \\
 {W}(\mathfrak{B}_{0100},\mathfrak{B}^{[1]}_{can};1)
 &{\cal S}^{0001}_{0100},{\cal S}^{0011}_{0100},{\cal S}^{1001}_{0100},{\cal S}^{1011}_{0100}
 &
 &{\cal S}^{0100}_{0100},{\cal S}^{0110}_{0100},{\cal S}^{1100}_{0100},{\cal S}^{1110}_{0100}
 &\hat{W}(\mathfrak{B}_{0100},\mathfrak{B}^{[1]}_{can};0)
 \\
 {W}(\mathfrak{B}_{0101},\mathfrak{B}^{[1]}_{can};1)
 &
 {\cal S}^{0101}_{0101},{\cal S}^{0111}_{0101},{\cal S}^{1101}_{0101},{\cal S}^{1111}_{0101}
 &
 &
 {\cal S}^{0100}_{0101},{\cal S}^{0110}_{0101},{\cal S}^{1100}_{0101},{\cal S}^{1110}_{0101}
 &\hat{W}(\mathfrak{B}_{0101},\mathfrak{B}^{[1]}_{can};0)
 \\
 {W}(\mathfrak{B}_{0101},\mathfrak{B}^{[1]}_{can};0)
 &{\cal S}^{0000}_{0101},{\cal S}^{0010}_{0101},{\cal S}^{1000}_{0101},{\cal S}^{1010}_{0101}
 &
 &{\cal S}^{0001}_{0101},{\cal S}^{0011}_{0101},{\cal S}^{1001}_{0101},{\cal S}^{1011}_{0101}
 &\hat{W}(\mathfrak{B}_{0101},\mathfrak{B}^{[1]}_{can};1)
 \\
 {W}(\mathfrak{B}_{0110},\mathfrak{B}^{[1]}_{can};1)
 &
 {\cal S}^{0001}_{0110},{\cal S}^{0111}_{0110},{\cal S}^{1001}_{0110},{\cal S}^{1111}_{0110}
 &
 &
 {\cal S}^{0010}_{0110},{\cal S}^{0100}_{0110},{\cal S}^{1010}_{0110},{\cal S}^{1100}_{0110}
 &\hat{W}(\mathfrak{B}_{0110},\mathfrak{B}^{[1]}_{can};0)
 \\
 {W}(\mathfrak{B}_{0110},\mathfrak{B}^{[1]}_{can};0)
 &{\cal S}^{0000}_{0110},{\cal S}^{0110}_{0110},{\cal S}^{1000}_{0110},{\cal S}^{1110}_{0110}
 &
 &{\cal S}^{0011}_{0110},{\cal S}^{0101}_{0110},{\cal S}^{1011}_{0110},{\cal S}^{1101}_{0110}
 &\hat{W}(\mathfrak{B}_{0110},\mathfrak{B}^{[1]}_{can};1)
 \\
 {W}(\mathfrak{B}_{0111},\mathfrak{B}^{[1]}_{can};0)
 &
 {\cal S}^{0000}_{0111},{\cal S}^{0110}_{0111},{\cal S}^{1000}_{0111},{\cal S}^{1110}_{0111}
 &
 &
 {\cal S}^{0001}_{0111},{\cal S}^{0111}_{0111},{\cal S}^{1001}_{0111},{\cal S}^{1111}_{0111}
 &\hat{W}(\mathfrak{B}_{0111},\mathfrak{B}^{[1]}_{can};1)
 \\
 {W}(\mathfrak{B}_{0111},\mathfrak{B}^{[1]}_{can};1)
 &{\cal S}^{0011}_{0111},{\cal S}^{0101}_{0111},{\cal S}^{1011}_{0111},{\cal S}^{1101}_{0111}
 &
 &{\cal S}^{0010}_{0111},{\cal S}^{0100}_{0111}, {\cal S}^{1010}_{0111},{\cal S}^{1100}_{0111}
 &\hat{W}(\mathfrak{B}_{0111},\mathfrak{B}^{[1]}_{can};0)
 \\
 {W}(\mathfrak{B}_{1000},\mathfrak{B}^{[1]}_{can};0)
 &
 {\cal S}^{0000}_{1000},{\cal S}^{0010}_{1000},{\cal S}^{0100}_{1000},{\cal S}^{0110}_{1000}
 &
 &{\cal S}^{1001}_{1000},{\cal S}^{1011}_{1000},{\cal S}^{1101}_{1000},{\cal S}^{1111}_{1000}
 &\hat{W}(\mathfrak{B}_{1000},\mathfrak{B}^{[1]}_{can};1)
 \\
 {W}(\mathfrak{B}_{1000},\mathfrak{B}^{[1]}_{can};1)
 &{\cal S}^{0001}_{1000},{\cal S}^{0011}_{1000},{\cal S}^{0101}_{1000},{\cal S}^{0111}_{1000}
 &
 &{\cal S}^{1000}_{1000},{\cal S}^{1010}_{1000},{\cal S}^{1100}_{1000},{\cal S}^{1110}_{1000}
 &\hat{W}(\mathfrak{B}_{1000},\mathfrak{B}^{[1]}_{can};0)
 \\
 {W}(\mathfrak{B}_{1001},\mathfrak{B}^{[1]}_{can};1)
 &
 {\cal S}^{1001}_{1001},{\cal S}^{1011}_{1001},{\cal S}^{1101}_{1001},{\cal S}^{1111}_{1001}
 &
 &{\cal S}^{1000}_{1001},{\cal S}^{1010}_{1001},{\cal S}^{1100}_{1001},{\cal S}^{1110}_{1001}
 &\hat{W}(\mathfrak{B}_{1001},\mathfrak{B}^{[1]}_{can};0)
 \\
 {W}(\mathfrak{B}_{1001},\mathfrak{B}^{[1]}_{can};0)
 &{\cal S}^{0000}_{1001},{\cal S}^{0010}_{1001},{\cal S}^{0100}_{1001},{\cal S}^{0110}_{1001}
 &
 &{\cal S}^{0001}_{1001},{\cal S}^{0011}_{1001},{\cal S}^{0101}_{1001},{\cal S}^{0111}_{1001}
 &\hat{W}(\mathfrak{B}_{1001},\mathfrak{B}^{[1]}_{can};1)
 \\
 {W}(\mathfrak{B}_{1010},\mathfrak{B}^{[1]}_{can};1)
 &
 {\cal S}^{0001}_{1010},{\cal S}^{0101}_{1010},{\cal S}^{1011}_{1010},{\cal S}^{1110}_{1010}
 &
 &
 {\cal S}^{0010}_{1010},{\cal S}^{0110}_{1010}, {\cal S}^{1000}_{1010},{\cal S}^{1100}_{1010}
 &\hat{W}(\mathfrak{B}_{1010},\mathfrak{B}^{[1]}_{can};0)
 \\
 {W}(\mathfrak{B}_{1010},\mathfrak{B}^{[1]}_{can};0)
 &{\cal S}^{0000}_{1010},{\cal S}^{0100}_{1010},{\cal S}^{1010}_{1010},{\cal S}^{1111}_{1010}
 &
 &{\cal S}^{0011}_{1010},{\cal S}^{0111}_{1010},{\cal S}^{1001}_{1010},{\cal S}^{1101}_{1010}
 &\hat{W}(\mathfrak{B}_{1010},\mathfrak{B}^{[1]}_{can};1)
 \\
 {W}(\mathfrak{B}_{1011},\mathfrak{B}^{[1]}_{can};0)
 &
 {\cal S}^{0000}_{1011},{\cal S}^{0100}_{1011},{\cal S}^{1010}_{1011},{\cal S}^{1110}_{1011}
 &
 &
 {\cal S}^{0001}_{1011},{\cal S}^{0101}_{1011},{\cal S}^{1011}_{1011},{\cal S}^{1111}_{1011}
 &\hat{W}(\mathfrak{B}_{1011},\mathfrak{B}^{[1]}_{can};1)
 \\
 {W}(\mathfrak{B}_{1011},\mathfrak{B}^{[1]}_{can};1)
 &{\cal S}^{0011}_{1011},{\cal S}^{0111}_{1011},{\cal S}^{1001}_{1011},{\cal S}^{1101}_{1011}
 &
 &{\cal S}^{0010}_{1011},{\cal S}^{0110}_{1011}, {\cal S}^{1000}_{1011},{\cal S}^{1100}_{1011}
 &\hat{W}(\mathfrak{B}_{1011},\mathfrak{B}^{[1]}_{can};0)
 \\
 {W}(\mathfrak{B}_{1100},\mathfrak{B}^{[1]}_{can};1)
 &
 {\cal S}^{0001}_{1100},{\cal S}^{0011}_{1100},{\cal S}^{1101}_{1100},{\cal S}^{1111}_{1100}
 &
 &
 {\cal S}^{0100}_{1100},{\cal S}^{0110}_{1100}, {\cal S}^{1000}_{1100},{\cal S}^{1010}_{1100}
 &\hat{W}(\mathfrak{B}_{1100},\mathfrak{B}^{[1]}_{can};0)
 \\
 {W}(\mathfrak{B}_{1100},\mathfrak{B}^{[1]}_{can};0)
 &{\cal S}^{0000}_{1100},{\cal S}^{0010}_{1100},{\cal S}^{1100}_{1100},{\cal S}^{1110}_{1100}
 &
 &{\cal S}^{0101}_{1100},{\cal S}^{0111}_{1100},{\cal S}^{1001}_{1100},{\cal S}^{1011}_{1100}
 &\hat{W}(\mathfrak{B}_{1100},\mathfrak{B}^{[1]}_{can};1)
 \\
 {W}(\mathfrak{B}_{1101},\mathfrak{B}^{[1]}_{can};0)
 &
 {\cal S}^{0000}_{1101},{\cal S}^{0010}_{1101},{\cal S}^{1100}_{1101},{\cal S}^{1110}_{1101}
 &
 &{\cal S}^{0001}_{1101},{\cal S}^{0011}_{1101},{\cal S}^{1101}_{1101},{\cal S}^{1111}_{1101}
 &\hat{W}(\mathfrak{B}_{1101},\mathfrak{B}^{[1]}_{can};1)
 \\
 {W}(\mathfrak{B}_{1101},\mathfrak{B}^{[1]}_{can};1)
 &{\cal S}^{0101}_{1101},{\cal S}^{0111}_{1101},{\cal S}^{1001}_{1101},{\cal S}^{1011}_{1101}
 &
 &{\cal S}^{0100}_{1101},{\cal S}^{0110}_{1101}, {\cal S}^{1000}_{1101},{\cal S}^{1010}_{1101}
 &\hat{W}(\mathfrak{B}_{1101},\mathfrak{B}^{[1]}_{can};0)
 \\
 {W}(\mathfrak{B}_{1110},\mathfrak{B}^{[1]}_{can};0)
 &
 {\cal S}^{0000}_{1110},{\cal S}^{0110}_{1110},{\cal S}^{1010}_{1110},{\cal S}^{1100}_{1110}
 &
 &
 {\cal S}^{0011}_{1110},{\cal S}^{0101}_{1110},{\cal S}^{1001}_{1110},{\cal S}^{1111}_{1110}
 &\hat{W}(\mathfrak{B}_{1110},\mathfrak{B}^{[1]}_{can};1)
 \\
 {W}(\mathfrak{B}_{1110},\mathfrak{B}^{[1]}_{can};1)
 &{\cal S}^{0001}_{1110},{\cal S}^{0111}_{1110},{\cal S}^{1011}_{1110},{\cal S}^{1101}_{1110}
 &
 &{\cal S}^{0010}_{1110},{\cal S}^{0100}_{1110}, {\cal S}^{1000}_{1110},{\cal S}^{1110}_{1110}
 &\hat{W}(\mathfrak{B}_{1110},\mathfrak{B}^{[1]}_{can};0)
 \\
 {W}(\mathfrak{B}_{1111},\mathfrak{B}^{[1]}_{can};1)
 &
 {\cal S}^{0011}_{1111},{\cal S}^{0101}_{1111},{\cal S}^{1001}_{1111},{\cal S}^{1111}_{1111}
 &
 &
 {\cal S}^{0010}_{1111},{\cal S}^{0100}_{1111},{\cal S}^{1000}_{1111},{\cal S}^{1110}_{1111}
 &\hat{W}(\mathfrak{B}_{1111},\mathfrak{B}^{[1]}_{can};0)
 \\
 {W}(\mathfrak{B}_{1111},\mathfrak{B}^{[1]}_{can};0)
 &{\cal S}^{0000}_{1111},{\cal S}^{0110}_{1111}, {\cal S}^{1010}_{1111},{\cal S}^{1100}_{1111}
 &
 &{\cal S}^{0001}_{1111},{\cal S}^{0111}_{1111},{\cal S}^{1011}_{1111},{\cal S}^{1101}_{1111}
 &\hat{W}(\mathfrak{B}_{1111},\mathfrak{B}^{[1]}_{can};1)
 \end{array} \]
 \end{center}
 \vspace{6pt}
 \fcaption{The co-quotient algebra of rank one given by the coset
 $\mathfrak{B}^{[1,0]}_{can}=\mathfrak{C}_{[\mathbf{0}]}-\mathfrak{B}^{[1]}_{can}\subset su(16)$.\label{figsu16cancoQArank1}}
 \end{figure}

 \begin{figure}[htbp]
 \begin{center}
 \[
\begin{array}{c}
 \mathfrak{B}^{[2]}_{can}\\
 \\
 {\cal S}^{0000}_{0000},\hspace{2pt}{\cal S}^{0100}_{0000},
 \hspace{2pt}{\cal S}^{1000}_{0000},\hspace{2pt}{\cal S}^{1100}_{0000}
 \end{array}\]
 \[\hspace{-30pt}\begin{array}{ccccc}
 \mathfrak{B}^{[2,\hspace{1pt}01]}_{can}
 &{\cal S}^{0001}_{0000},{\cal S}^{0101}_{0000},{\cal S}^{1001}_{0000},{\cal S}^{1101}_{0000}
 &
 &\{0\}
 &
 \\
 \mathfrak{B}^{[2,\hspace{1pt}10]}_{can}
 &{\cal S}^{0010}_{0000},{\cal S}^{0110}_{0000},{\cal S}^{1010}_{0000},{\cal S}^{1110}_{0000}
 &
 &\{0\}
 &
 \\
 \mathfrak{B}^{[2,\hspace{1pt}11]}_{can}
 &{\cal S}^{0011}_{0000},{\cal S}^{0111}_{0000}, {\cal S}^{1011}_{0000},{\cal S}^{1111}_{0000}
 &
 &\{0\}
 &
 \\
 {W}(\mathfrak{B}_{0001},\mathfrak{B}^{[2]}_{can};00)
 &\hspace{0pt}
 {\cal S}^{0000}_{0001},{\cal S}^{0100}_{0001},{\cal S}^{1000}_{0001},{\cal S}^{1100}_{0001}
 &\hspace{0pt}
 &\{0\}
 &\hat{W}(\mathfrak{B}_{0001},\mathfrak{B}^{[2]}_{can};00)
 \\
 {W}(\mathfrak{B}_{0001},\mathfrak{B}^{[2]}_{can};01)
 &\{0\}
 &
 &{\cal S}^{0001}_{0001},{\cal S}^{0101}_{0001},{\cal S}^{1001}_{0001},{\cal S}^{1101}_{0001}
 &\hat{W}(\mathfrak{B}_{0001},\mathfrak{B}^{[2]}_{can};01)
 \\
 {W}(\mathfrak{B}_{0001},\mathfrak{B}^{[2]}_{can};10)
 &{\cal S}^{0010}_{0001},{\cal S}^{0110}_{0001},{\cal S}^{1010}_{0001},{\cal S}^{1110}_{0001}
 &
 &\{0\}
 &\hat{W}(\mathfrak{B}_{0001},\mathfrak{B}^{[2]}_{can};10)
 \\
 {W}(\mathfrak{B}_{0001},\mathfrak{B}^{[2]}_{can};11)
 &\{0\}
 &
 &\hspace{0pt}
 {\cal S}^{0011}_{0001},{\cal S}^{0111}_{0001},{\cal S}^{1011}_{0001},{\cal S}^{1111}_{0001}
 &\hat{W}(\mathfrak{B}_{0001},\mathfrak{B}^{[2]}_{can};11)
 \\
 {W}(\mathfrak{B}_{0010},\mathfrak{B}^{[2]}_{can};00)
 &
 {\cal S}^{0000}_{0010},{\cal S}^{0100}_{0010},{\cal S}^{1000}_{0010},{\cal S}^{1100}_{0010}
 &
 &\{0\}
 &\hat{W}(\mathfrak{B}_{0010},\mathfrak{B}^{[2]}_{can};00)
 \\
\end{array}\]
 \end{center}
 \end{figure}
 \begin{figure}[b]
 \begin{center}
 \[\hspace{-45pt}\begin{array}{ccccc}
 {W}(\mathfrak{B}_{1000},\mathfrak{B}^{[2]}_{can};00)
 \hspace{25pt}&
 {\cal S}^{0000}_{1000},{\cal S}^{0100}_{1000}
 &\hspace{50pt}
 &{\cal S}^{1000}_{1000},{\cal S}^{1100}_{1000}
 &\hspace{25pt}\hat{W}(\mathfrak{B}_{1000},\mathfrak{B}^{[2]}_{can};00)
 \\
 {W}(\mathfrak{B}_{1000},\mathfrak{B}^{[2]}_{can};01)\hspace{25pt}
 &{\cal S}^{0001}_{1000},{\cal S}^{0101}_{1000}
 &
 &{\cal S}^{1001}_{1000},{\cal S}^{1101}_{1000}
 &\hspace{25pt}\hat{W}(\mathfrak{B}_{1000},\mathfrak{B}^{[2]}_{can};01)
 \\
 {W}(\mathfrak{B}_{1000},\mathfrak{B}^{[2]}_{can};10)\hspace{25pt}
 &{\cal S}^{0010}_{1000},{\cal S}^{0110}_{1000}
 &
 &{\cal S}^{1010}_{1000},{\cal S}^{1110}_{1000}
 &\hspace{25pt}\hat{W}(\mathfrak{B}_{1000},\mathfrak{B}^{[2]}_{can};10)
 \\
 {W}(\mathfrak{B}_{1000},\mathfrak{B}^{[2]}_{can};11)\hspace{25pt}
 &{\cal S}^{0011}_{1000},{\cal S}^{0111}_{1000}
 &
 &{\cal S}^{1011}_{1000},{\cal S}^{1111}_{1000}
 &\hspace{25pt}\hat{W}(\mathfrak{B}_{1000},\mathfrak{B}^{[2]}_{can};11)
 \\
 {W}(\mathfrak{B}_{1001},\mathfrak{B}^{[2]}_{can};00)
 \hspace{25pt}&
 {\cal S}^{0000}_{1001},{\cal S}^{0100}_{1001}
 &\hspace{50pt}
 &{\cal S}^{1000}_{1001},{\cal S}^{1100}_{1001}
 &\hspace{25pt}\hat{W}(\mathfrak{B}_{1001},\mathfrak{B}^{[2]}_{can};00)
 \\
 {W}(\mathfrak{B}_{1001},\mathfrak{B}^{[2]}_{can};01)\hspace{25pt}
 &{\cal S}^{1001}_{1001},{\cal S}^{1101}_{1001}
 &
 &{\cal S}^{0001}_{1001},{\cal S}^{0101}_{1001}
 &\hspace{25pt}\hat{W}(\mathfrak{B}_{1001},\mathfrak{B}^{[2]}_{can};01)
 \\
 {W}(\mathfrak{B}_{1001},\mathfrak{B}^{[2]}_{can};10)\hspace{25pt}
 &{\cal S}^{0010}_{1001},{\cal S}^{0110}_{1001}
 &
 &{\cal S}^{1010}_{1001},{\cal S}^{1110}_{1001}
 &\hspace{25pt}\hat{W}(\mathfrak{B}_{1001},\mathfrak{B}^{[2]}_{can};10)
 \\
 {W}(\mathfrak{B}_{1001},\mathfrak{B}^{[2]}_{can};11)\hspace{25pt}
 &{\cal S}^{1011}_{1001},{\cal S}^{1111}_{1001}
 &
 &{\cal S}^{0011}_{1001},{\cal S}^{0111}_{1001}
 &\hspace{25pt}\hat{W}(\mathfrak{B}_{1001},\mathfrak{B}^{[2]}_{can};11)
 \\
\end{array} \]
 \end{center}
 \fcaption{The canonical quotient algebra of rank two given by the canonical center subalgebra
 $\mathfrak{B}^{[2]}_{can}=\mathfrak{B}^{[2,\hspace{1pt}00]}_{can}=\mathfrak{B}_{0001}\cap\mathfrak{B}_{0010}\cap\mathfrak{B}_{0011}\subset\mathfrak{C}_{[\mathbf{0}]}\subset su(16)$,
 referring to Fig.~\ref{figsu16intrQArank0}.\label{figsu16canQArank2}}
 \end{figure}

 \begin{figure}[htbp]
 \begin{center}
 \[\begin{array}{c}
 \mathfrak{B}^{[2,\hspace{1pt}10]}_{can}\\
 \\
 {\cal S}^{0010}_{0000},\hspace{2pt}{\cal S}^{0110}_{0000},\hspace{2pt}{\cal S}^{1010}_{0000},\hspace{2pt}{\cal S}^{1110}_{0000}
 \end{array}\]
 \[\hspace{-30pt}\begin{array}{ccccc}
 \hspace{-6pt}\mathfrak{B}^{[2]}_{can}
 &{\cal S}^{0000}_{0000},{\cal S}^{0100}_{0000},{\cal S}^{1000}_{0000},{\cal S}^{1100}_{0000}
 &
 &\{0\}
 &
 \\
 \mathfrak{B}^{[2,\hspace{1pt}01]}_{can}
 &{\cal S}^{0001}_{0000},{\cal S}^{0101}_{0000},{\cal S}^{1001}_{0000},{\cal S}^{1101}_{0000}
 &
 &\{0\}
 &
 \\
 \mathfrak{B}^{[2,\hspace{1pt}11]}_{can}
 &{\cal S}^{0011}_{0000},{\cal S}^{0111}_{0000}, {\cal S}^{1011}_{0000},{\cal S}^{1111}_{0000}
 &
 &\{0\}
 &
 \\
 {W}(\mathfrak{B}_{0001},\mathfrak{B}^{[2]}_{can};00)
 &\hspace{0pt}
 {\cal S}^{0000}_{0001},{\cal S}^{0100}_{0001},{\cal S}^{1000}_{0001},{\cal S}^{1100}_{0001}
 &\hspace{0pt}
 &\{0\}
 &\hat{W}(\mathfrak{B}_{0001},\mathfrak{B}^{[2]}_{can};10)
 \\
 {W}(\mathfrak{B}_{0001},\mathfrak{B}^{[2]}_{can};11)
 &\{0\}
 &
 &{\cal S}^{0001}_{0001},{\cal S}^{0101}_{0001},{\cal S}^{1001}_{0001},{\cal S}^{1101}_{0001}
 &\hat{W}(\mathfrak{B}_{0001},\mathfrak{B}^{[2]}_{can};01)
 \\
 {W}(\mathfrak{B}_{0001},\mathfrak{B}^{[2]}_{can};10)
 &{\cal S}^{0010}_{0001},{\cal S}^{0110}_{0001},{\cal S}^{1010}_{0001},{\cal S}^{1110}_{0001}
 &
 &\{0\}
 &\hat{W}(\mathfrak{B}_{0001},\mathfrak{B}^{[2]}_{can};00)
 \\
 {W}(\mathfrak{B}_{0001},\mathfrak{B}^{[2]}_{can};01)
 &\{0\}
 &
 &\hspace{0pt}
 {\cal S}^{0011}_{0001},{\cal S}^{0111}_{0001},{\cal S}^{1011}_{0001},{\cal S}^{1111}_{0001}
 &\hat{W}(\mathfrak{B}_{0001},\mathfrak{B}^{[2]}_{can};11)
 \\
 {W}(\mathfrak{B}_{0010},\mathfrak{B}^{[2]}_{can};00)
 &
 {\cal S}^{0000}_{0010},{\cal S}^{0100}_{0010},{\cal S}^{1000}_{0010},{\cal S}^{1100}_{0010}
 &
 &{\cal S}^{0010}_{0010},{\cal S}^{0110}_{0010},{\cal S}^{1010}_{0010},{\cal S}^{1110}_{0010}
 &\hat{W}(\mathfrak{B}_{0010},\mathfrak{B}^{[2]}_{can};10)
 \\
\end{array}\]
 \end{center}
 \end{figure}
 \begin{figure}[b]
 \begin{center}
 \[\hspace{-45pt}\begin{array}{ccccc}
 {W}(\mathfrak{B}_{1000},\mathfrak{B}^{[2]}_{can};00)
 \hspace{25pt}&
 {\cal S}^{0000}_{1000},{\cal S}^{0100}_{1000}
 &\hspace{50pt}
 &{\cal S}^{1010}_{1000},{\cal S}^{1110}_{1000}
 &\hspace{25pt}\hat{W}(\mathfrak{B}_{1000},\mathfrak{B}^{[2]}_{can};10)
 \\
 {W}(\mathfrak{B}_{1000},\mathfrak{B}^{[2]}_{can};11)\hspace{25pt}
 &{\cal S}^{0011}_{1000},{\cal S}^{0111}_{1000}
 &
 &{\cal S}^{1001}_{1000},{\cal S}^{1101}_{1000}
 &\hspace{25pt}\hat{W}(\mathfrak{B}_{1000},\mathfrak{B}^{[2]}_{can};01)
 \\
 {W}(\mathfrak{B}_{1000},\mathfrak{B}^{[2]}_{can};10)\hspace{25pt}
 &{\cal S}^{0010}_{1000},{\cal S}^{0110}_{1000}
 &
 &{\cal S}^{1000}_{1000},{\cal S}^{1100}_{1000}
 &\hspace{25pt}\hat{W}(\mathfrak{B}_{1000},\mathfrak{B}^{[2]}_{can};00)
 \\
 {W}(\mathfrak{B}_{1000},\mathfrak{B}^{[2]}_{can};01)\hspace{25pt}
 &{\cal S}^{0001}_{1000},{\cal S}^{0101}_{1000}
 &
 &{\cal S}^{1011}_{1000},{\cal S}^{1111}_{1000}
 &\hspace{25pt}\hat{W}(\mathfrak{B}_{1000},\mathfrak{B}^{[2]}_{can};11)
 \\
 {W}(\mathfrak{B}_{1001},\mathfrak{B}^{[2]}_{can};10)
 \hspace{25pt}&
 {\cal S}^{0010}_{1001},{\cal S}^{0110}_{1001}
 &\hspace{50pt}
 &{\cal S}^{1000}_{1001},{\cal S}^{1100}_{1001}
 &\hspace{25pt}\hat{W}(\mathfrak{B}_{1001},\mathfrak{B}^{[2]}_{can};00)
 \\
 {W}(\mathfrak{B}_{1001},\mathfrak{B}^{[2]}_{can};01)\hspace{25pt}
 &{\cal S}^{1001}_{1001},{\cal S}^{1101}_{1001}
 &
 &{\cal S}^{0011}_{1001},{\cal S}^{0111}_{1001}
 &\hspace{25pt}\hat{W}(\mathfrak{B}_{1001},\mathfrak{B}^{[2]}_{can};11)
 \\
 {W}(\mathfrak{B}_{1001},\mathfrak{B}^{[2]}_{can};00)\hspace{25pt}
 &{\cal S}^{0000}_{1001},{\cal S}^{0100}_{1001}
 &
 &{\cal S}^{1010}_{1001},{\cal S}^{1110}_{1001}
 &\hspace{25pt}\hat{W}(\mathfrak{B}_{1001},\mathfrak{B}^{[2]}_{can};10)
 \\
 {W}(\mathfrak{B}_{1001},\mathfrak{B}^{[2]}_{can};11)\hspace{25pt}
 &{\cal S}^{1011}_{1001},{\cal S}^{1111}_{1001}
 &
 &{\cal S}^{0001}_{1001},{\cal S}^{0101}_{1001}
 &\hspace{25pt}\hat{W}(\mathfrak{B}_{1001},\mathfrak{B}^{[2]}_{can};01)
 \\
\end{array} \]
 \end{center}
 \fcaption{A co-quotient algebra of rank two given by a coset
 $\mathfrak{B}^{[2,\hspace{1pt}10]}_{can}$
 in the partition of $\mathfrak{C}_{[\mathbf{0}]}\subset su(16)$
 generated by $\mathfrak{B}^{[2]}_{can}$.\label{figsu16cancoQArank2}}
 \end{figure}

 \begin{figure}[htbp]
 \begin{center}
 \[\hspace{-20pt}\begin{array}{c}
 \mathfrak{B}^{[3]}_{can}\\
 \\
 {\cal S}^{0000}_{0000},\hspace{2pt}{\cal S}^{1000}_{0000}
 \end{array}\]
 \[\hspace{-30pt}\begin{array}{ccccc}
 \mathfrak{B}^{[3,\hspace{1pt}001]}_{can}
 &{\cal S}^{0001}_{0000},{\cal S}^{1001}_{0000}
 &\hspace{30pt}
 &\{0\}
 &
 \\
 \mathfrak{B}^{[3,\hspace{1pt}010]}_{can}
 &{\cal S}^{0010}_{0000},{\cal S}^{1010}_{0000}
 &
 &\{0\}
 &
 \\
 \mathfrak{B}^{[3,\hspace{1pt}011]}_{can}
 &{\cal S}^{0011}_{0000}, {\cal S}^{1011}_{0000}
 &
 &\{0\}
 &
 \\
 \mathfrak{B}^{[3,\hspace{1pt}100]}_{can}
 &{\cal S}^{0100}_{0000},{\cal S}^{1100}_{0000}
 &
 &\{0\}
 &
 \\
 \mathfrak{B}^{[3,\hspace{1pt}101]}_{can}
 &{\cal S}^{0101}_{0000},{\cal S}^{1101}_{0000}
 &
 &\{0\}
 &
 \\
 \mathfrak{B}^{[3,\hspace{1pt}110]}_{can}
 &{\cal S}^{0110}_{0000},{\cal S}^{1110}_{0000}
 &
 &\{0\}
 &
 \\
 \mathfrak{B}^{[3,\hspace{1pt}111]}_{can}
 &{\cal S}^{0111}_{0000},{\cal S}^{1111}_{0000}
 &
 &\{0\}
 &
 \\
 {W}(\mathfrak{B}_{0001},\mathfrak{B}^{[3]}_{can};000)
 &\hspace{0pt}
 {\cal S}^{0000}_{0001},{\cal S}^{1000}_{0001}
 &\hspace{0pt}
 &\{0\}
 &\hat{W}(\mathfrak{B}_{0001},\mathfrak{B}^{[3]}_{can};000)
 \\
\end{array}\]
 \end{center}
 \end{figure}
 \begin{figure}[htbp]
 \begin{center}
 \[\hspace{-30pt}\begin{array}{ccccc}
 {W}(\mathfrak{B}_{0100},\mathfrak{B}^{[3]}_{can};000)
 &{\cal S}^{0000}_{0100},{\cal S}^{1000}_{0100}
 &\hspace{30pt}
 &\{0\}
 &\hat{W}(\mathfrak{B}_{0100},\mathfrak{B}^{[3]}_{can};000)
 \\
 {W}(\mathfrak{B}_{0100},\mathfrak{B}^{[3]}_{can};001)
 &{\cal S}^{0001}_{0100},{\cal S}^{1001}_{0100}
 &
 &\{0\}
 &\hat{W}(\mathfrak{B}_{0100},\mathfrak{B}^{[3]}_{can};001)
 \\
 {W}(\mathfrak{B}_{0100},\mathfrak{B}^{[3]}_{can};010)
 &{\cal S}^{0010}_{0100},{\cal S}^{1010}_{0100}
 &
 &\{0\}
 &\hat{W}(\mathfrak{B}_{0100},\mathfrak{B}^{[3]}_{can};010)
 \\
 {W}(\mathfrak{B}_{0100},\mathfrak{B}^{[3]}_{can};011)
 &{\cal S}^{0011}_{0100},{\cal S}^{1011}_{0100}
 &
 &\{0\}
 &\hat{W}(\mathfrak{B}_{0100},\mathfrak{B}^{[3]}_{can};011)
 \\
 {W}(\mathfrak{B}_{0100},\mathfrak{B}^{[3]}_{can};100)
 &\{0\}
 &
 &{\cal S}^{0100}_{0100},{\cal S}^{1100}_{0100}
 &\hat{W}(\mathfrak{B}_{0100},\mathfrak{B}^{[3]}_{can};100)
 \\
 {W}(\mathfrak{B}_{0100},\mathfrak{B}^{[3]}_{can};101)
 &\{0\}
 &
 &{\cal S}^{0101}_{0100},{\cal S}^{1101}_{0100}
 &\hat{W}(\mathfrak{B}_{0100},\mathfrak{B}^{[3]}_{can};101)
 \\
 {W}(\mathfrak{B}_{0100},\mathfrak{B}^{[3]}_{can};110)
 &\{0\}
 &
 &{\cal S}^{0110}_{0100},{\cal S}^{1110}_{0100}
 &\hat{W}(\mathfrak{B}_{0100},\mathfrak{B}^{[3]}_{can};110)
 \\
 {W}(\mathfrak{B}_{0100},\mathfrak{B}^{[3]}_{can};111)
 &\{0\}
 &
 &{\cal S}^{0111}_{0100},{\cal S}^{1111}_{0100}
 &\hat{W}(\mathfrak{B}_{0100},\mathfrak{B}^{[3]}_{can};111)
 \\
\end{array}\]
 \end{center}
 \end{figure}
 \begin{figure}[b]
 \begin{center}
 \[\hspace{-50pt}\begin{array}{ccccc}
 {W}(\mathfrak{B}_{1000},\mathfrak{B}^{[3]}_{can};000)
 \hspace{25pt}&
 {\cal S}^{0000}_{1000}
 &\hspace{30pt}
 &{\cal S}^{1000}_{1000}
 &\hspace{25pt}\hat{W}(\mathfrak{B}_{1000},\mathfrak{B}^{[3]}_{can};000)
 \\
 {W}(\mathfrak{B}_{1000},\mathfrak{B}^{[3]}_{can};001)\hspace{25pt}
 &{\cal S}^{0001}_{1000}
 &
 &{\cal S}^{1001}_{1000}
 &\hspace{25pt}\hat{W}(\mathfrak{B}_{1000},\mathfrak{B}^{[3]}_{can};001)
 \\
 {W}(\mathfrak{B}_{1000},\mathfrak{B}^{[3]}_{can};010)\hspace{25pt}
 &{\cal S}^{0010}_{1000}
 &
 &{\cal S}^{1010}_{1000}
 &\hspace{25pt}\hat{W}(\mathfrak{B}_{1000},\mathfrak{B}^{[3]}_{can};010)
 \\
 {W}(\mathfrak{B}_{1000},\mathfrak{B}^{[3]}_{can};011)\hspace{25pt}
 &{\cal S}^{0011}_{1000}
 &
 &{\cal S}^{1011}_{1000}
 &\hspace{25pt}\hat{W}(\mathfrak{B}_{1000},\mathfrak{B}^{[3]}_{can};011)
 \\
 {W}(\mathfrak{B}_{1000},\mathfrak{B}^{[3]}_{can};100)
 \hspace{25pt}&
 {\cal S}^{0100}_{1000}
 &\hspace{25pt}
 &{\cal S}^{1100}_{1000}
 &\hspace{25pt}\hat{W}(\mathfrak{B}_{1000},\mathfrak{B}^{[3]}_{can};100)
 \\
 {W}(\mathfrak{B}_{1000},\mathfrak{B}^{[3]}_{can};101)\hspace{25pt}
 &{\cal S}^{0101}_{1000}
 &
 &{\cal S}^{1101}_{1000}
 &\hspace{25pt}\hat{W}(\mathfrak{B}_{1000},\mathfrak{B}^{[3]}_{can};101)
 \\
 {W}(\mathfrak{B}_{1000},\mathfrak{B}^{[3]}_{can};110)\hspace{25pt}
 &{\cal S}^{0110}_{1000}
 &
 &{\cal S}^{1110}_{1000}
 &\hspace{25pt}\hat{W}(\mathfrak{B}_{1000},\mathfrak{B}^{[3]}_{can};110)
 \\
 {W}(\mathfrak{B}_{1000},\mathfrak{B}^{[3]}_{can};111)\hspace{25pt}
 &{\cal S}^{0111}_{1000}
 &
 &{\cal S}^{1111}_{1000}
 &\hspace{25pt}\hat{W}(\mathfrak{B}_{1000},\mathfrak{B}^{[3]}_{can};111)
 \\
\end{array}%
 \]
 \end{center}
 \fcaption{The canonical quotient algebra of rank three given by the canonical center subalgebra
 $\mathfrak{B}^{[3]}_{can}=\mathfrak{B}^{[3,\hspace{1pt}000]}_{can}
 =\bigcap_{\beta\in Z^4_2,\hspace{2pt}\beta\cdot\beta_0=0}\mathfrak{B}_{\beta}\subset\mathfrak{C}_{[\mathbf{0}]}\subset su(16)$,
 here $\beta_0=1000$.\label{figsu16canQArank3}}
 \end{figure}

 \clearpage
 \begin{figure}[htbp]
 \begin{center}
 \[\hspace{-20pt}\begin{array}{c}
 \mathfrak{B}^{[3,\hspace{1pt}100]}_{can}\\
 \\
 {\cal S}^{0100}_{0000},\hspace{2pt}{\cal S}^{1100}_{0000}
 \end{array}%
 \]
 \end{center}
 \fcaption{A co-quotient algebra of rank three given by a coset
 $\mathfrak{B}^{[3,\hspace{1pt}100]}_{can}$
 in the partition of $\mathfrak{C}_{[\mathbf{0}]}\subset su(16)$
 generated by
 $\mathfrak{B}^{[3]}_{can}\subset\mathfrak{C}_{[\mathbf{0}]}$.\label{figsu16cancoQArank3}}
 \end{figure}

 \newpage
 \begin{figure}[p]
 \[\hspace{-40pt}%
      &
      \hat{W}(\mathfrak{B}_{0011})
      \\
      \\
      {W}(\mathfrak{B}_{0100})
      &
      \mu_1\otimes{\cal S}^{00}_{00},\mu_1\otimes{\cal S}^{01}_{00},\mu_1\otimes{\cal S}^{10}_{00},\mu_1\otimes{\cal S}^{11}_{00}
      &
      &
      \mu_2\otimes{\cal S}^{00}_{00},\mu_2\otimes{\cal S}^{01}_{00},\mu_2\otimes{\cal S}^{10}_{00},\mu_2\otimes{\cal S}^{11}_{00}
      &
      \hat{W}(\mathfrak{B}_{0100})
      \\
      \\
      {W}(\mathfrak{B}_{0101})
      &
      \mu_1\otimes{\cal S}^{00}_{01},\mu_1\otimes{\cal S}^{10}_{01},\mu_2\otimes{\cal S}^{01}_{01},\mu_2\otimes{\cal S}^{11}_{01}
      &
      &
      \mu_1\otimes{\cal S}^{01}_{01},\mu_1\otimes{\cal S}^{11}_{01},\mu_2\otimes{\cal S}^{00}_{01},\mu_2\otimes{\cal S}^{10}_{01}
      &
      \hat{W}(\mathfrak{B}_{0101})
      \\
      \\
      {W}(\mathfrak{B}_{0110})
      &
      \mu_1\otimes{\cal S}^{00}_{10},\mu_1\otimes{\cal S}^{01}_{10},\mu_2\otimes{\cal S}^{10}_{10},\mu_2\otimes{\cal S}^{11}_{10}
      &
      &
      \mu_1\otimes{\cal S}^{10}_{10},\mu_1\otimes{\cal S}^{11}_{10},\mu_2\otimes{\cal S}^{00}_{10},\mu_2\otimes{\cal S}^{01}_{10}
      &
      \hat{W}(\mathfrak{B}_{0110})
      \\
      \\
      {W}(\mathfrak{B}_{0111})
      &
      \mu_1\otimes{\cal S}^{00}_{11},\mu_1\otimes{\cal S}^{11}_{11},\mu_2\otimes{\cal S}^{01}_{11},\mu_2\otimes{\cal S}^{10}_{11}
      &
      &
      \mu_1\otimes{\cal S}^{01}_{11},\mu_1\otimes{\cal S}^{10}_{11},\mu_2\otimes{\cal S}^{00}_{11},\mu_2\otimes{\cal S}^{11}_{11}
      &
      \hat{W}(\mathfrak{B}_{0111})
      \\
      \\
      {W}(\mathfrak{B}_{1000})
      &
      \mu_4\otimes{\cal S}^{00}_{00},\mu_4\otimes{\cal S}^{01}_{00},\mu_4\otimes{\cal S}^{10}_{00},\mu_4\otimes{\cal S}^{11}_{00}
      &
      &
      \mu_5\otimes{\cal S}^{00}_{00},\mu_5\otimes{\cal S}^{01}_{00},\mu_5\otimes{\cal S}^{10}_{00},\mu_5\otimes{\cal S}^{11}_{00}
      &
      \hat{W}(\mathfrak{B}_{1000})
      \\
      \\
      {W}(\mathfrak{B}_{1001})
      &
      \mu_4\otimes{\cal S}^{00}_{01},\mu_4\otimes{\cal S}^{10}_{01},\mu_5\otimes{\cal S}^{01}_{01},\mu_5\otimes{\cal S}^{11}_{01}
      &
      &
      \mu_4\otimes{\cal S}^{01}_{01},\mu_4\otimes{\cal S}^{11}_{01},\mu_5\otimes{\cal S}^{00}_{01},\mu_5\otimes{\cal S}^{10}_{01}
      &
      \hat{W}(\mathfrak{B}_{1001})
      \\
      \\
      {W}(\mathfrak{B}_{1010})
      &
      \mu_4\otimes{\cal S}^{00}_{10},\mu_4\otimes{\cal S}^{01}_{10},\mu_5\otimes{\cal S}^{10}_{10},\mu_5\otimes{\cal S}^{11}_{10}
      &
      &
      \mu_4\otimes{\cal S}^{10}_{10},\mu_4\otimes{\cal S}^{11}_{10},\mu_5\otimes{\cal S}^{00}_{10},\mu_5\otimes{\cal S}^{01}_{10}
      &
      \hat{W}(\mathfrak{B}_{1010})
      \\
      \\
      {W}(\mathfrak{B}_{1011})
      &
      \mu_4\otimes{\cal S}^{00}_{11},\mu_4\otimes{\cal S}^{11}_{11},\mu_5\otimes{\cal S}^{01}_{11},\mu_5\otimes{\cal S}^{10}_{11}
      &
      &
      \mu_4\otimes{\cal S}^{01}_{11},\mu_4\otimes{\cal S}^{10}_{11},\mu_5\otimes{\cal S}^{00}_{11},\mu_5\otimes{\cal S}^{11}_{11}
      &
      \hat{W}(\mathfrak{B}_{1011})
      \\
      \\
      {W}(\mathfrak{B}_{1100})
      &
      \mu_6\otimes{\cal S}^{00}_{00},\mu_6\otimes{\cal S}^{01}_{00},\mu_6\otimes{\cal S}^{10}_{00},\mu_6\otimes{\cal S}^{11}_{00}
      &
      &
      \mu_7\otimes{\cal S}^{00}_{00},\mu_7\otimes{\cal S}^{01}_{00},\mu_7\otimes{\cal S}^{10}_{00},\mu_7\otimes{\cal S}^{11}_{00}
      &
      \hat{W}(\mathfrak{B}_{1100})
      \\
      \\
      {W}(\mathfrak{B}_{1101})
      &
      \mu_6\otimes{\cal S}^{00}_{01},\mu_6\otimes{\cal S}^{10}_{01},\mu_7\otimes{\cal S}^{01}_{01},\mu_7\otimes{\cal S}^{11}_{01}
      &
      &
      \mu_6\otimes{\cal S}^{01}_{01},\mu_6\otimes{\cal S}^{11}_{01},\mu_7\otimes{\cal S}^{00}_{01},\mu_7\otimes{\cal S}^{10}_{01}
      &
      \hat{W}(\mathfrak{B}_{1101})
      \\
      \\
      {W}(\mathfrak{B}_{1110})
      &
      \mu_6\otimes{\cal S}^{00}_{10},\mu_6\otimes{\cal S}^{01}_{10},\mu_7\otimes{\cal S}^{10}_{10},\mu_7\otimes{\cal S}^{11}_{10}
      &
      &
      \mu_6\otimes{\cal S}^{10}_{10},\mu_6\otimes{\cal S}^{11}_{10},\mu_7\otimes{\cal S}^{00}_{10},\mu_7\otimes{\cal S}^{01}_{10}
      &
      \hat{W}(\mathfrak{B}_{1110})
      \\
      \\
      {W}(\mathfrak{B}_{1111})
      &
      \mu_6\otimes{\cal S}^{00}_{11},\mu_6\otimes{\cal S}^{11}_{11},\mu_7\otimes{\cal S}^{01}_{11},\mu_7\otimes{\cal S}^{10}_{11}
      &
      &
      \mu_6\otimes{\cal S}^{01}_{11},\mu_6\otimes{\cal S}^{10}_{11},\mu_7\otimes{\cal S}^{00}_{11},\mu_7\otimes{\cal S}^{11}_{11}
      &
      \hat{W}(\mathfrak{B}_{1111})
 \end{array}\]
 \end{figure}
 \begin{figure}[h]
 \[\hspace{4pt}\begin{array}{c}
 \hspace{-137pt}\mathfrak{B}_{0001}=
 \{\hspace{2pt}{I}\otimes{\cal S}^{00}_{00},\hspace{2pt}{I}\otimes{\cal S}^{10}_{00},\hspace{2pt}\mu_3\otimes{\cal S}^{00}_{00},
   \hspace{2pt}\mu_3\otimes{\cal S}^{10}_{00},\hspace{2pt}\mu_8\otimes{\cal S}^{00}_{00},\hspace{2pt}\mu_8\otimes{\cal S}^{10}_{00}\hspace{2pt}\},
 \\
 \\
 \hspace{-137pt}\mathfrak{B}_{0010}=
 \{\hspace{2pt}{I}\otimes{\cal S}^{00}_{00},\hspace{2pt}{I}\otimes{\cal S}^{01}_{00},\hspace{2pt}\mu_3\otimes{\cal S}^{00}_{00},
   \hspace{2pt}\mu_3\otimes{\cal S}^{01}_{00},\hspace{2pt}\mu_8\otimes{\cal S}^{00}_{00},\hspace{2pt}\mu_8\otimes{\cal S}^{01}_{00}\hspace{2pt}\},
 \\
 \\
 \hspace{-137pt}\mathfrak{B}_{0011}=
 \{\hspace{2pt}{I}\otimes{\cal S}^{00}_{00},\hspace{2pt}{I}\otimes{\cal S}^{11}_{00},\hspace{2pt}\mu_3\otimes{\cal S}^{00}_{00},
   \hspace{2pt}\mu_3\otimes{\cal S}^{11}_{00},\hspace{2pt}\mu_8\otimes{\cal S}^{00}_{00},\hspace{2pt}\mu_8\otimes{\cal S}^{11}_{00}\hspace{2pt}\},
 \\
 \\
 \hspace{-62pt}\mathfrak{B}_{0100}=
 \{\hspace{2pt}{I}\otimes{\cal S}^{00}_{00},\hspace{2pt}{I}\otimes{\cal S}^{01}_{00},\hspace{2pt}{I}\otimes{\cal S}^{10}_{00},
   \hspace{2pt}{I}\otimes{\cal S}^{11}_{00},\hspace{2pt}\mu_8\otimes{\cal S}^{00}_{00},\hspace{2pt}\mu_8\otimes{\cal S}^{01}_{00}
   \hspace{2pt}\mu_8\otimes{\cal S}^{10}_{00},\hspace{2pt}\mu_8\otimes{\cal S}^{11}_{00}\hspace{2pt}\},
 \\
 \\
 \hspace{-51pt}\mathfrak{B}_{0101}=
 \{\hspace{2pt}{I}\otimes{\cal S}^{00}_{00},\hspace{2pt}{I}\otimes{\cal S}^{10}_{00},\hspace{2pt}\mu_3\otimes{\cal S}^{01}_{00},
   \hspace{2pt}\mu_3\otimes{\cal S}^{11}_{00},\hspace{2pt}\mu_8\otimes{\cal S}^{00}_{00},\hspace{2pt}\mu_8\otimes{\cal S}^{01}_{00}
   \hspace{2pt}\mu_8\otimes{\cal S}^{10}_{00},\hspace{2pt}\mu_8\otimes{\cal S}^{11}_{00}\hspace{2pt}\},
 \\
 \\
 \hspace{-51pt}\mathfrak{B}_{0110}=
 \{\hspace{2pt}{I}\otimes{\cal S}^{00}_{00},\hspace{2pt}{I}\otimes{\cal S}^{01}_{00},\hspace{2pt}\mu_3\otimes{\cal S}^{10}_{00},
   \hspace{2pt}\mu_3\otimes{\cal S}^{11}_{00},\hspace{2pt}\mu_8\otimes{\cal S}^{00}_{00},\hspace{2pt}\mu_8\otimes{\cal S}^{01}_{00}
   \hspace{2pt}\mu_8\otimes{\cal S}^{10}_{00},\hspace{2pt}\mu_8\otimes{\cal S}^{11}_{00}\hspace{2pt}\},
 \\
 \\
 \hspace{-51pt}\mathfrak{B}_{0111}=
 \{\hspace{2pt}{I}\otimes{\cal S}^{00}_{00},\hspace{2pt}{I}\otimes{\cal S}^{11}_{00},\hspace{2pt}\mu_3\otimes{\cal S}^{01}_{00},
   \hspace{2pt}\mu_3\otimes{\cal S}^{10}_{00},\hspace{2pt}\mu_8\otimes{\cal S}^{00}_{00},\hspace{2pt}\mu_8\otimes{\cal S}^{01}_{00}
   \hspace{2pt}\mu_8\otimes{\cal S}^{10}_{00},\hspace{2pt}\mu_8\otimes{\cal S}^{11}_{00}\hspace{2pt}\},
 \\
 \\
 \begin{array}{cc}
 \hspace{-20pt}\mathfrak{B}_{1000}=\mathfrak{B}_{1100}=
 \{\hspace{2pt}{I}\otimes{\cal S}^{00}_{00},\hspace{2pt}{I}\otimes{\cal S}^{11}_{00},\hspace{2pt}{I}\otimes{\cal S}^{01}_{00},
   \hspace{2pt}{I}\otimes{\cal S}^{11}_{00}\hspace{2pt}\},
 &
 \mathfrak{B}_{1001}=\mathfrak{B}_{1101}=
 \{\hspace{2pt}{I}\otimes{\cal S}^{00}_{00},\hspace{2pt}{I}\otimes{\cal S}^{10}_{00}\hspace{2pt}\},
 \end{array}
 \\
 \\
 \hspace{-101pt}\mathfrak{B}_{1010}=\mathfrak{B}_{1110}=
 \{\hspace{2pt}{I}\otimes{\cal S}^{00}_{00},\hspace{2pt}{I}\otimes{\cal S}^{01}_{00}\hspace{2pt}\},
 \hspace{5pt}
 \mathfrak{B}_{1011}=\mathfrak{B}_{1111}=
 \{\hspace{2pt}{I}\otimes{\cal S}^{00}_{00},\hspace{2pt}{I}\otimes{\cal S}^{11}_{00}\hspace{2pt}\}.
 \end{array}\]
 \vspace{6pt}
 \fcaption{The intrinsic quotient algebra of rank zero given by the intrinsic Cartan subalgebra
 $\mathfrak{C}_{[\mathbf{0}]}\subset su(12)$, followed by the complete set of corresponding maximal bi-subalgebras;
 here $\mu_i$ denote the Gell-Mann matrices and $I$ is the $3\times 3$ identity matrix. \label{figsu12QArank0intr}}
 \end{figure}

 \begin{figure}[p]
 \[\hspace{-40pt}\begin{array}{c}
   \mathfrak{B}^{[1]}_{can}\\
   \\
   \hspace{0pt}{I}\otimes{\cal S}^{00}_{00},\hspace{2pt}{I}\otimes{\cal S}^{10}_{00},\hspace{2pt}\mu_3\otimes{\cal S}^{00}_{00},
   \mu_3\otimes{\cal S}^{10}_{00},\hspace{2pt}\mu_8\otimes{\cal S}^{00}_{00},\hspace{2pt}\mu_8\otimes{\cal S}^{10}_{00}
  \end{array}\]
   \vspace{6pt}
 \[\hspace{-45pt}\begin{array}{ccccc}
      \mathfrak{B}^{[1,1]}_{can}=\mathfrak{C}_{[\mathbf{0}]}-\mathfrak{B}^{[1]}_{can}
      &
      \hspace{5pt}
      \begin{array}{c}
       \hspace{-8pt}{I}\otimes{\cal S}^{01}_{00},\hspace{2pt}{I}\otimes{\cal S}^{11}_{00},\hspace{2pt}\mu_3\otimes{\cal S}^{01}_{00},\\
       \mu_3\otimes{\cal S}^{11}_{00},\hspace{2pt}\mu_8\otimes{\cal S}^{01}_{00},\hspace{2pt}\mu_8\otimes{\cal S}^{11}_{00}
       \end{array}
      &
      &
      \{0\}
      &
      \\
      \\
      {W}(\mathfrak{B}_{0001},\mathfrak{B}^{[1]}_{can};0)
      &
      \begin{array}{c}
      \hspace{0pt}{I}\otimes{\cal S}^{00}_{01},\hspace{2pt}\mu_3\otimes{\cal S}^{00}_{01},\hspace{2pt}\mu_8\otimes{\cal S}^{00}_{01},
      \\
      \hspace{-4pt}{I}\otimes{\cal S}^{10}_{01},\hspace{2pt}\mu_3\otimes{\cal S}^{10}_{01},\hspace{2pt}\mu_8\otimes{\cal S}^{10}_{01}
      \end{array}
      &
      &
       \{0\}
      &
      \hat{W}(\mathfrak{B}_{0001},\mathfrak{B}^{[1]}_{can};0)
      \\
      \\
      {W}(\mathfrak{B}_{0001},\mathfrak{B}^{[1]}_{can};1)
      &
       \{0\}
      &
      &
      \begin{array}{c}
      \hspace{0pt}{I}\otimes{\cal S}^{01}_{01},\hspace{2pt}\mu_3\otimes{\cal S}^{01}_{01},\hspace{2pt}\mu_8\otimes{\cal S}^{01}_{01},
      \\
      \hspace{-4pt}{I}\otimes{\cal S}^{11}_{01},\hspace{2pt}\mu_3\otimes{\cal S}^{11}_{01},\hspace{2pt}\mu_8\otimes{\cal S}^{11}_{01}
      \end{array}
      &
      \hat{W}(\mathfrak{B}_{0001},\mathfrak{B}^{[1]}_{can};1)
      \\
      \\
      {W}(\mathfrak{B}_{0010},\mathfrak{B}^{[1]}_{can};0)
      &
      \hspace{0pt}{I}\otimes{\cal S}^{00}_{10},\hspace{2pt}\mu_3\otimes{\cal S}^{00}_{10},\hspace{2pt}\mu_8\otimes{\cal S}^{00}_{10}
      &
      &
      \hspace{0pt}{I}\otimes{\cal S}^{10}_{10},\hspace{2pt}\mu_3\otimes{\cal S}^{10}_{11},\hspace{2pt}\mu_8\otimes{\cal S}^{10}_{10}
      &
      \hat{W}(\mathfrak{B}_{0010},\mathfrak{B}^{[1]}_{can};0)
      \\
      {W}(\mathfrak{B}_{0010},\mathfrak{B}^{[1]}_{can};1)
      &
      \hspace{0pt}{I}\otimes{\cal S}^{01}_{10},\hspace{2pt}\mu_3\otimes{\cal S}^{01}_{10},\hspace{2pt}\mu_8\otimes{\cal S}^{01}_{10}
      &
      &
      \hspace{0pt}{I}\otimes{\cal S}^{11}_{10},\hspace{2pt}\mu_3\otimes{\cal S}^{11}_{10},\hspace{2pt}\mu_8\otimes{\cal S}^{11}_{10}
      &
      \hat{W}(\mathfrak{B}_{0010},\mathfrak{B}^{[1]}_{can};1)
      \\
      {W}(\mathfrak{B}_{0011},\mathfrak{B}^{[1]}_{can};0)
      &
       \hspace{0pt}{I}\otimes{\cal S}^{00}_{11},\hspace{2pt}\mu_3\otimes{\cal S}^{00}_{11},\hspace{2pt}\mu_8\otimes{\cal S}^{00}_{11}
      &
      &
      \hspace{0pt}{I}\otimes{\cal S}^{10}_{11},\hspace{2pt}\mu_3\otimes{\cal S}^{10}_{11},\hspace{2pt}\mu_8\otimes{\cal S}^{10}_{11}
      &
      \hat{W}(\mathfrak{B}_{0011},\mathfrak{B}^{[1]}_{can};0)
      \\
      {W}(\mathfrak{B}_{0011},\mathfrak{B}^{[1]}_{can};1)
      &
       \hspace{0pt}{I}\otimes{\cal S}^{11}_{11},\hspace{2pt}\mu_3\otimes{\cal S}^{11}_{11},\hspace{2pt}\mu_8\otimes{\cal S}^{11}_{11}
      &
      &
      \hspace{0pt}{I}\otimes{\cal S}^{01}_{11},\hspace{2pt}\mu_3\otimes{\cal S}^{01}_{01},\hspace{2pt}\mu_8\otimes{\cal S}^{01}_{11}
      &
      \hat{W}(\mathfrak{B}_{0011},\mathfrak{B}^{[1]}_{can};1)
      \\
      {W}(\mathfrak{B}_{0100},\mathfrak{B}^{[1]}_{can};0)
      &
       \hspace{0pt}\mu_1\otimes{\cal S}^{00}_{00},\hspace{2pt}\mu_1\otimes{\cal S}^{10}_{00}
      &
      &
       \hspace{0pt}\mu_2\otimes{\cal S}^{00}_{00},\hspace{2pt}\mu_2\otimes{\cal S}^{10}_{00}
      &
      \hat{W}(\mathfrak{B}_{0100},\mathfrak{B}^{[1]}_{can};0)
      \\
      {W}(\mathfrak{B}_{0100},\mathfrak{B}^{[1]}_{can};1)
      &
       \hspace{0pt}\mu_1\otimes{\cal S}^{01}_{00},\hspace{2pt}\mu_1\otimes{\cal S}^{11}_{00}
      &
      &
       \hspace{0pt}\mu_2\otimes{\cal S}^{01}_{00},\hspace{2pt}\mu_2\otimes{\cal S}^{11}_{00}
      &
      \hat{W}(\mathfrak{B}_{0100},\mathfrak{B}^{[1]}_{can};1)
      \\
      {W}(\mathfrak{B}_{0101},\mathfrak{B}^{[1]}_{can};0)
      &
       \hspace{0pt}\mu_1\otimes{\cal S}^{00}_{01},\hspace{2pt}\mu_1\otimes{\cal S}^{10}_{01}
      &
      &
       \hspace{0pt}\mu_2\otimes{\cal S}^{00}_{01},\hspace{2pt}\mu_2\otimes{\cal S}^{10}_{01}
      &
      \hat{W}(\mathfrak{B}_{0101},\mathfrak{B}^{[1]}_{can};0)
      \\
      {W}(\mathfrak{B}_{0101},\mathfrak{B}^{[1]}_{can};1)
      &
       \hspace{0pt}\mu_2\otimes{\cal S}^{01}_{01},\hspace{2pt}\mu_2\otimes{\cal S}^{11}_{01}
      &
      &
       \hspace{0pt}\mu_1\otimes{\cal S}^{01}_{01},\hspace{2pt}\mu_1\otimes{\cal S}^{11}_{01}
      &
      \hat{W}(\mathfrak{B}_{0101},\mathfrak{B}^{[1]}_{can};1)
      \\
      {W}(\mathfrak{B}_{0110},\mathfrak{B}^{[1]}_{can};0)
      &
       \hspace{0pt}\mu_1\otimes{\cal S}^{00}_{10},\hspace{2pt}\mu_2\otimes{\cal S}^{10}_{10}
      &
      &
       \hspace{0pt}\mu_2\otimes{\cal S}^{00}_{10},\hspace{2pt}\mu_1\otimes{\cal S}^{10}_{10}
      &
      \hat{W}(\mathfrak{B}_{0110},\mathfrak{B}^{[1]}_{can};0)
      \\
      {W}(\mathfrak{B}_{0110},\mathfrak{B}^{[1]}_{can};1)
      &
       \hspace{0pt}\mu_1\otimes{\cal S}^{01}_{10},\hspace{2pt}\mu_2\otimes{\cal S}^{11}_{10}
      &
      &
       \hspace{0pt}\mu_2\otimes{\cal S}^{01}_{10},\hspace{2pt}\mu_1\otimes{\cal S}^{11}_{10}
      &
      \hat{W}(\mathfrak{B}_{0110},\mathfrak{B}^{[1]}_{can};1)
      \\
      {W}(\mathfrak{B}_{0111},\mathfrak{B}^{[1]}_{can};0)
      &
       \hspace{0pt}\mu_1\otimes{\cal S}^{00}_{11},\hspace{2pt}\mu_2\otimes{\cal S}^{10}_{11}
      &
      &
       \hspace{0pt}\mu_2\otimes{\cal S}^{00}_{01},\hspace{2pt}\mu_1\otimes{\cal S}^{10}_{11}
      &
      \hat{W}(\mathfrak{B}_{0111},\mathfrak{B}^{[1]}_{can};0)
      \\
      {W}(\mathfrak{B}_{0111},\mathfrak{B}^{[1]}_{can};1)
      &
       \hspace{0pt}\mu_2\otimes{\cal S}^{01}_{11},\hspace{2pt}\mu_1\otimes{\cal S}^{11}_{11}
      &
      &
       \hspace{0pt}\mu_1\otimes{\cal S}^{01}_{11},\hspace{2pt}\mu_2\otimes{\cal S}^{11}_{11}
      &
      \hat{W}(\mathfrak{B}_{0111},\mathfrak{B}^{[1]}_{can};1)
      \\
      {W}(\mathfrak{B}_{1000},\mathfrak{B}^{[1]}_{can};0)
      &
       \hspace{0pt}\mu_4\otimes{\cal S}^{00}_{00},\hspace{2pt}\mu_4\otimes{\cal S}^{10}_{00}
      &
      &
        \hspace{0pt}\mu_5\otimes{\cal S}^{00}_{00},\hspace{2pt}\mu_5\otimes{\cal S}^{10}_{00}
      &
      \hat{W}(\mathfrak{B}_{1000},\mathfrak{B}^{[1]}_{can};0)
      \\
      {W}(\mathfrak{B}_{1000},\mathfrak{B}^{[1]}_{can};1)
      &
        \hspace{0pt}\mu_4\otimes{\cal S}^{01}_{00},\hspace{2pt}\mu_4\otimes{\cal S}^{11}_{00}
      &
      &
        \hspace{0pt}\mu_5\otimes{\cal S}^{01}_{00},\hspace{2pt}\mu_5\otimes{\cal S}^{11}_{00}
      &
      \hat{W}(\mathfrak{B}_{1000},\mathfrak{B}^{[1]}_{can};1)
      \\
      {W}(\mathfrak{B}_{1001},\mathfrak{B}^{[1]}_{can};0)
      &
       \hspace{0pt}\mu_4\otimes{\cal S}^{00}_{01},\hspace{2pt}\mu_4\otimes{\cal S}^{10}_{01}
      &
      &
        \hspace{0pt}\mu_5\otimes{\cal S}^{00}_{01},\hspace{2pt}\mu_5\otimes{\cal S}^{10}_{01}
      &
      \hat{W}(\mathfrak{B}_{1001},\mathfrak{B}^{[1]}_{can};0)
      \\
      {W}(\mathfrak{B}_{1001},\mathfrak{B}^{[1]}_{can};1)
      &
        \hspace{0pt}\mu_5\otimes{\cal S}^{01}_{01},\hspace{2pt}\mu_5\otimes{\cal S}^{11}_{01}
      &
      &
        \hspace{0pt}\mu_4\otimes{\cal S}^{01}_{01},\hspace{2pt}\mu_4\otimes{\cal S}^{11}_{01}
      &
      \hat{W}(\mathfrak{B}_{1001},\mathfrak{B}^{[1]}_{can};1)
      \\
      {W}(\mathfrak{B}_{1010},\mathfrak{B}^{[1]}_{can};0)
      &
       \hspace{0pt}\mu_4\otimes{\cal S}^{00}_{10},\hspace{2pt}\mu_5\otimes{\cal S}^{10}_{10}
      &
      &
        \hspace{0pt}\mu_5\otimes{\cal S}^{00}_{10},\hspace{2pt}\mu_4\otimes{\cal S}^{10}_{10}
      &
      \hat{W}(\mathfrak{B}_{1010},\mathfrak{B}^{[1]}_{can};0)
      \\
      {W}(\mathfrak{B}_{1010},\mathfrak{B}^{[1]}_{can};1)
      &
        \hspace{0pt}\mu_4\otimes{\cal S}^{01}_{10},\hspace{2pt}\mu_5\otimes{\cal S}^{11}_{10}
      &
      &
        \hspace{0pt}\mu_5\otimes{\cal S}^{01}_{10},\hspace{2pt}\mu_4\otimes{\cal S}^{11}_{10}
      &
      \hat{W}(\mathfrak{B}_{1010},\mathfrak{B}^{[1]}_{can};1)
      \\
      {W}(\mathfrak{B}_{1011},\mathfrak{B}^{[1]}_{can};0)
      &
       \hspace{0pt}\mu_4\otimes{\cal S}^{00}_{11},\hspace{2pt}\mu_5\otimes{\cal S}^{10}_{11}
      &
      &
        \hspace{0pt}\mu_5\otimes{\cal S}^{00}_{11},\hspace{2pt}\mu_4\otimes{\cal S}^{10}_{11}
      &
      \hat{W}(\mathfrak{B}_{1011},\mathfrak{B}^{[1]}_{can};0)
      \\
      {W}(\mathfrak{B}_{1011},\mathfrak{B}^{[1]}_{can};1)
      &
        \hspace{0pt}\mu_5\otimes{\cal S}^{01}_{11},\hspace{2pt}\mu_4\otimes{\cal S}^{11}_{11}
      &
      &
        \hspace{0pt}\mu_4\otimes{\cal S}^{01}_{11},\hspace{2pt}\mu_5\otimes{\cal S}^{11}_{11}
      &
      \hat{W}(\mathfrak{B}_{1011},\mathfrak{B}^{[1]}_{can};1)
      \\
      {W}(\mathfrak{B}_{1100},\mathfrak{B}^{[1]}_{can};0)
      &
       \hspace{0pt}\mu_6\otimes{\cal S}^{00}_{00},\hspace{2pt}\mu_6\otimes{\cal S}^{10}_{00}
      &
      &
        \hspace{0pt}\mu_7\otimes{\cal S}^{00}_{00},\hspace{2pt}\mu_7\otimes{\cal S}^{10}_{00}
      &
      \hat{W}(\mathfrak{B}_{1100},\mathfrak{B}^{[1]}_{can};0)
      \\
      {W}(\mathfrak{B}_{1100},\mathfrak{B}^{[1]}_{can};1)
      &
        \hspace{0pt}\mu_6\otimes{\cal S}^{01}_{00},\hspace{2pt}\mu_6\otimes{\cal S}^{11}_{00}
      &
      &
        \hspace{0pt}\mu_7\otimes{\cal S}^{01}_{00},\hspace{2pt}\mu_7\otimes{\cal S}^{11}_{00}
      &
      \hat{W}(\mathfrak{B}_{1100},\mathfrak{B}^{[1]}_{can};1)
      \\
      {W}(\mathfrak{B}_{1101},\mathfrak{B}^{[1]}_{can};0)
      &
       \hspace{0pt}\mu_6\otimes{\cal S}^{00}_{01},\hspace{2pt}\mu_6\otimes{\cal S}^{10}_{01}
      &
      &
        \hspace{0pt}\mu_7\otimes{\cal S}^{00}_{01},\hspace{2pt}\mu_7\otimes{\cal S}^{10}_{01}
      &
      \hat{W}(\mathfrak{B}_{1101},\mathfrak{B}^{[1]}_{can};0)
      \\
      {W}(\mathfrak{B}_{1101},\mathfrak{B}^{[1]}_{can};1)
      &
        \hspace{0pt}\mu_7\otimes{\cal S}^{01}_{01},\hspace{2pt}\mu_7\otimes{\cal S}^{11}_{01}
      &
      &
        \hspace{0pt}\mu_6\otimes{\cal S}^{01}_{01},\hspace{2pt}\mu_6\otimes{\cal S}^{11}_{01}
      &
      \hat{W}(\mathfrak{B}_{1101},\mathfrak{B}^{[1]}_{can};1)
      \\
      {W}(\mathfrak{B}_{1110},\mathfrak{B}^{[1]}_{can};0)
      &
       \hspace{0pt}\mu_6\otimes{\cal S}^{00}_{10},\hspace{2pt}\mu_7\otimes{\cal S}^{10}_{10}
      &
      &
        \hspace{0pt}\mu_7\otimes{\cal S}^{00}_{10},\hspace{2pt}\mu_6\otimes{\cal S}^{10}_{10}
      &
      \hat{W}(\mathfrak{B}_{1110},\mathfrak{B}^{[1]}_{can};0)
      \\
      {W}(\mathfrak{B}_{1110},\mathfrak{B}^{[1]}_{can};1)
      &
        \hspace{0pt}\mu_6\otimes{\cal S}^{01}_{10},\hspace{2pt}\mu_7\otimes{\cal S}^{11}_{10}
      &
      &
        \hspace{0pt}\mu_7\otimes{\cal S}^{01}_{10},\hspace{2pt}\mu_6\otimes{\cal S}^{11}_{10}
      &
      \hat{W}(\mathfrak{B}_{1110},\mathfrak{B}^{[1]}_{can};1)
      \\
      {W}(\mathfrak{B}_{1111},\mathfrak{B}^{[1]}_{can};0)
      &
       \hspace{0pt}\mu_6\otimes{\cal S}^{00}_{11},\hspace{2pt}\mu_7\otimes{\cal S}^{10}_{11}
      &
      &
        \hspace{0pt}\mu_7\otimes{\cal S}^{00}_{11},\hspace{2pt}\mu_6\otimes{\cal S}^{10}_{11}
      &
      \hat{W}(\mathfrak{B}_{1111},\mathfrak{B}^{[1]}_{can};0)
      \\
      {W}(\mathfrak{B}_{1111},\mathfrak{B}^{[1]}_{can};1)
      &
        \hspace{0pt}\mu_7\otimes{\cal S}^{01}_{11},\hspace{2pt}\mu_6\otimes{\cal S}^{11}_{11}
      &
      &
        \hspace{0pt}\mu_6\otimes{\cal S}^{01}_{11},\hspace{2pt}\mu_7\otimes{\cal S}^{11}_{11}
      &
      \hat{W}(\mathfrak{B}_{1111},\mathfrak{B}^{[1]}_{can};1)
 \end{array}\]
 \vspace{6pt}
 \fcaption{The canonical quotient algebra of rank one given by the maximal bi-subalgebra
 $\mathfrak{B}^{[1]}_{can}=\mathfrak{B}^{[1,0]}_{can}=\mathfrak{B}_{0001}\subset\mathfrak{C}_{[\mathbf{0}]}\subset su(12)$. \label{figsu12canQArank1}}
 \end{figure}

 \begin{figure}[p]
 \[\hspace{-40pt}\begin{array}{c}
   \mathfrak{B}^{[1,1]}_{can}\\
   \\
   {I}\otimes{\cal S}^{01}_{00},\hspace{2pt}{I}\otimes{\cal S}^{11}_{00},\hspace{2pt}\mu_3\otimes{\cal S}^{01}_{00},
   \hspace{2pt}\mu_3\otimes{\cal S}^{11}_{00},\hspace{2pt}\mu_8\otimes{\cal S}^{01}_{00},\hspace{2pt}\mu_8\otimes{\cal S}^{11}_{00}
   \hspace{0pt}
  \end{array}\]
   \vspace{6pt}
 \[\hspace{-45pt}\begin{array}{ccccc}
      \mathfrak{B}^{[1]}_{can}
      &
      \hspace{5pt}
      \begin{array}{c}
       \hspace{-8pt}{I}\otimes{\cal S}^{00}_{00},\hspace{2pt}{I}\otimes{\cal S}^{10}_{00},\hspace{2pt}\mu_3\otimes{\cal S}^{00}_{00},\\
       \mu_3\otimes{\cal S}^{10}_{00},\hspace{2pt}\mu_8\otimes{\cal S}^{00}_{00},\hspace{2pt}\mu_8\otimes{\cal S}^{10}_{00}
       \end{array}
      &
      &
      \{0\}
      &
      \\
      \\
      {W}(\mathfrak{B}_{0001},\mathfrak{B}^{[1]}_{can};0)
      &
      \begin{array}{c}
      \hspace{0pt}{I}\otimes{\cal S}^{00}_{01},\hspace{2pt}\mu_3\otimes{\cal S}^{00}_{01},\hspace{2pt}\mu_8\otimes{\cal S}^{00}_{01},
      \\
      \hspace{-4pt}{I}\otimes{\cal S}^{10}_{01},\hspace{2pt}\mu_3\otimes{\cal S}^{10}_{01},\hspace{2pt}\mu_8\otimes{\cal S}^{10}_{01}
      \end{array}
      &
      &
       \begin{array}{c}
      \hspace{0pt}{I}\otimes{\cal S}^{01}_{01},\hspace{2pt}\mu_3\otimes{\cal S}^{01}_{01},\hspace{2pt}\mu_8\otimes{\cal S}^{01}_{01},
      \\
      \hspace{-4pt}{I}\otimes{\cal S}^{11}_{01},\hspace{2pt}\mu_3\otimes{\cal S}^{11}_{01},\hspace{2pt}\mu_8\otimes{\cal S}^{11}_{01}
      \end{array}
      &
      \hat{W}(\mathfrak{B}_{0001},\mathfrak{B}^{[1]}_{can};1)
      \\
      \\
      {W}(\mathfrak{B}_{0001},\mathfrak{B}^{[1]}_{can};1)
      &
       \{0\}
      &
      &
       \{0\}
      &
      \hat{W}(\mathfrak{B}_{0001},\mathfrak{B}^{[1]}_{can};0)
      \\
      \\
      {W}(\mathfrak{B}_{0010},\mathfrak{B}^{[1]}_{can};0)
      &
      \hspace{0pt}{I}\otimes{\cal S}^{00}_{10},\hspace{2pt}\mu_3\otimes{\cal S}^{00}_{10},\hspace{2pt}\mu_8\otimes{\cal S}^{00}_{10}
      &
      &
      \hspace{0pt}{I}\otimes{\cal S}^{11}_{10},\hspace{2pt}\mu_3\otimes{\cal S}^{11}_{10},\hspace{2pt}\mu_8\otimes{\cal S}^{11}_{10}
      &
      \hat{W}(\mathfrak{B}_{0010},\mathfrak{B}^{[1]}_{can};1)
      \\
      {W}(\mathfrak{B}_{0010},\mathfrak{B}^{[1]}_{can};1)
      &
      \hspace{0pt}{I}\otimes{\cal S}^{01}_{10},\hspace{2pt}\mu_3\otimes{\cal S}^{01}_{10},\hspace{2pt}\mu_8\otimes{\cal S}^{01}_{10}
      &
      &
      \hspace{0pt}{I}\otimes{\cal S}^{10}_{10},\hspace{2pt}\mu_3\otimes{\cal S}^{10}_{11},\hspace{2pt}\mu_8\otimes{\cal S}^{10}_{10}
      &
      \hat{W}(\mathfrak{B}_{0010},\mathfrak{B}^{[1]}_{can};0)
      \\
      {W}(\mathfrak{B}_{0011},\mathfrak{B}^{[1]}_{can};1)
      &
      \hspace{0pt}{I}\otimes{\cal S}^{11}_{11},\hspace{2pt}\mu_3\otimes{\cal S}^{11}_{11},\hspace{2pt}\mu_8\otimes{\cal S}^{11}_{11}
      &
      &
      \hspace{0pt}{I}\otimes{\cal S}^{10}_{11},\hspace{2pt}\mu_3\otimes{\cal S}^{10}_{11},\hspace{2pt}\mu_8\otimes{\cal S}^{10}_{11}
      &
      \hat{W}(\mathfrak{B}_{0011},\mathfrak{B}^{[1]}_{can};0)
      \\
      {W}(\mathfrak{B}_{0011},\mathfrak{B}^{[1]}_{can};0)
      &
      \hspace{0pt}{I}\otimes{\cal S}^{00}_{11},\hspace{2pt}\mu_3\otimes{\cal S}^{00}_{11},\hspace{2pt}\mu_8\otimes{\cal S}^{00}_{11}
      &
      &
      \hspace{0pt}{I}\otimes{\cal S}^{01}_{11},\hspace{2pt}\mu_3\otimes{\cal S}^{01}_{01},\hspace{2pt}\mu_8\otimes{\cal S}^{01}_{11}
      &
      \hat{W}(\mathfrak{B}_{0011},\mathfrak{B}^{[1]}_{can};1)
      \\
      {W}(\mathfrak{B}_{0100},\mathfrak{B}^{[1]}_{can};0)
      &
       \hspace{0pt}\mu_1\otimes{\cal S}^{00}_{00},\hspace{2pt}\mu_1\otimes{\cal S}^{10}_{00}
      &
      &
       \hspace{0pt}\mu_2\otimes{\cal S}^{01}_{00},\hspace{2pt}\mu_2\otimes{\cal S}^{11}_{00}
      &
      \hat{W}(\mathfrak{B}_{0100},\mathfrak{B}^{[1]}_{can};1)
      \\
      {W}(\mathfrak{B}_{0100},\mathfrak{B}^{[1]}_{can};1)
      &
       \hspace{0pt}\mu_1\otimes{\cal S}^{01}_{00},\hspace{2pt}\mu_1\otimes{\cal S}^{11}_{00}
      &
      &
       \hspace{0pt}\mu_2\otimes{\cal S}^{00}_{00},\hspace{2pt}\mu_2\otimes{\cal S}^{10}_{00}
      &
      \hat{W}(\mathfrak{B}_{0100},\mathfrak{B}^{[1]}_{can};0)
      \\
      {W}(\mathfrak{B}_{0101},\mathfrak{B}^{[1]}_{can};1)
      &
      \hspace{0pt}\mu_2\otimes{\cal S}^{01}_{01},\hspace{2pt}\mu_2\otimes{\cal S}^{11}_{01}
      &
      &
       \hspace{0pt}\mu_2\otimes{\cal S}^{00}_{01},\hspace{2pt}\mu_2\otimes{\cal S}^{10}_{01}
      &
      \hat{W}(\mathfrak{B}_{0101},\mathfrak{B}^{[1]}_{can};0)
      \\
      {W}(\mathfrak{B}_{0101},\mathfrak{B}^{[1]}_{can};0)
      &
       \hspace{0pt}\mu_1\otimes{\cal S}^{00}_{01},\hspace{2pt}\mu_1\otimes{\cal S}^{10}_{01}
      &
      &
       \hspace{0pt}\mu_1\otimes{\cal S}^{01}_{01},\hspace{2pt}\mu_1\otimes{\cal S}^{11}_{01}
      &
      \hat{W}(\mathfrak{B}_{0101},\mathfrak{B}^{[1]}_{can};1)
      \\
      {W}(\mathfrak{B}_{0110},\mathfrak{B}^{[1]}_{can};1)
      &
       \hspace{0pt}\mu_1\otimes{\cal S}^{01}_{10},\hspace{2pt}\mu_2\otimes{\cal S}^{11}_{10}
      &
      &
       \hspace{0pt}\mu_2\otimes{\cal S}^{00}_{10},\hspace{2pt}\mu_1\otimes{\cal S}^{10}_{10}
      &
      \hat{W}(\mathfrak{B}_{0110},\mathfrak{B}^{[1]}_{can};0)
      \\
      {W}(\mathfrak{B}_{0110},\mathfrak{B}^{[1]}_{can};0)
      &
       \hspace{0pt}\mu_1\otimes{\cal S}^{00}_{10},\hspace{2pt}\mu_2\otimes{\cal S}^{10}_{10}
      &
      &
       \hspace{0pt}\mu_2\otimes{\cal S}^{01}_{10},\hspace{2pt}\mu_1\otimes{\cal S}^{11}_{10}
      &
      \hat{W}(\mathfrak{B}_{0110},\mathfrak{B}^{[1]}_{can};1)
      \\
      {W}(\mathfrak{B}_{0111},\mathfrak{B}^{[1]}_{can};0)
      &
       \hspace{0pt}\mu_1\otimes{\cal S}^{00}_{11},\hspace{2pt}\mu_2\otimes{\cal S}^{10}_{11}
      &
      &
       \hspace{0pt}\mu_1\otimes{\cal S}^{01}_{11},\hspace{2pt}\mu_2\otimes{\cal S}^{11}_{11}
      &
      \hat{W}(\mathfrak{B}_{0111},\mathfrak{B}^{[1]}_{can};1)
      \\
      {W}(\mathfrak{B}_{0111},\mathfrak{B}^{[1]}_{can};1)
      &
       \hspace{0pt}\mu_2\otimes{\cal S}^{01}_{11},\hspace{2pt}\mu_1\otimes{\cal S}^{11}_{11}
      &
      &
       \hspace{0pt}\mu_2\otimes{\cal S}^{00}_{01},\hspace{2pt}\mu_1\otimes{\cal S}^{10}_{11}
      &
      \hat{W}(\mathfrak{B}_{0111},\mathfrak{B}^{[1]}_{can};0)
      \\
      {W}(\mathfrak{B}_{1000},\mathfrak{B}^{[1]}_{can};0)
      &
       \hspace{0pt}\mu_4\otimes{\cal S}^{00}_{00},\hspace{2pt}\mu_4\otimes{\cal S}^{10}_{00}
      &
      &
       \hspace{0pt}\mu_5\otimes{\cal S}^{01}_{00},\hspace{2pt}\mu_5\otimes{\cal S}^{11}_{00}
      &
      \hat{W}(\mathfrak{B}_{1000},\mathfrak{B}^{[1]}_{can};1)
      \\
      {W}(\mathfrak{B}_{1000},\mathfrak{B}^{[1]}_{can};1)
      &
        \hspace{0pt}\mu_4\otimes{\cal S}^{01}_{00},\hspace{2pt}\mu_4\otimes{\cal S}^{11}_{00}
      &
      &
        \hspace{0pt}\mu_5\otimes{\cal S}^{00}_{00},\hspace{2pt}\mu_5\otimes{\cal S}^{10}_{00}
      &
      \hat{W}(\mathfrak{B}_{1000},\mathfrak{B}^{[1]}_{can};0)
      \\
      {W}(\mathfrak{B}_{1001},\mathfrak{B}^{[1]}_{can};1)
      &
       \hspace{0pt}\mu_5\otimes{\cal S}^{01}_{01},\hspace{2pt}\mu_5\otimes{\cal S}^{11}_{01}
      &
      &
        \hspace{0pt}\mu_5\otimes{\cal S}^{00}_{01},\hspace{2pt}\mu_5\otimes{\cal S}^{10}_{01}
      &
      \hat{W}(\mathfrak{B}_{1001},\mathfrak{B}^{[1]}_{can};0)
      \\
      {W}(\mathfrak{B}_{1001},\mathfrak{B}^{[1]}_{can};0)
      &
        \hspace{0pt}\mu_4\otimes{\cal S}^{00}_{01},\hspace{2pt}\mu_4\otimes{\cal S}^{10}_{01}
      &
      &
        \hspace{0pt}\mu_4\otimes{\cal S}^{01}_{01},\hspace{2pt}\mu_4\otimes{\cal S}^{11}_{01}
      &
      \hat{W}(\mathfrak{B}_{1001},\mathfrak{B}^{[1]}_{can};1)
      \\
      {W}(\mathfrak{B}_{1010},\mathfrak{B}^{[1]}_{can};1)
      &
       \hspace{0pt}\mu_4\otimes{\cal S}^{01}_{10},\hspace{2pt}\mu_5\otimes{\cal S}^{11}_{10}
      &
      &
        \hspace{0pt}\mu_5\otimes{\cal S}^{00}_{10},\hspace{2pt}\mu_4\otimes{\cal S}^{10}_{10}
      &
      \hat{W}(\mathfrak{B}_{1010},\mathfrak{B}^{[1]}_{can};0)
      \\
      {W}(\mathfrak{B}_{1010},\mathfrak{B}^{[1]}_{can};0)
      &
        \hspace{0pt}\mu_4\otimes{\cal S}^{00}_{10},\hspace{2pt}\mu_5\otimes{\cal S}^{10}_{10}
      &
      &
        \hspace{0pt}\mu_5\otimes{\cal S}^{01}_{10},\hspace{2pt}\mu_4\otimes{\cal S}^{11}_{10}
      &
      \hat{W}(\mathfrak{B}_{1010},\mathfrak{B}^{[1]}_{can};1)
      \\
      {W}(\mathfrak{B}_{1011},\mathfrak{B}^{[1]}_{can};0)
      &
       \hspace{0pt}\mu_4\otimes{\cal S}^{00}_{11},\hspace{2pt}\mu_5\otimes{\cal S}^{10}_{11}
      &
      &
       \hspace{0pt}\mu_4\otimes{\cal S}^{01}_{11},\hspace{2pt}\mu_5\otimes{\cal S}^{11}_{11}
      &
      \hat{W}(\mathfrak{B}_{1011},\mathfrak{B}^{[1]}_{can};1)
      \\
      {W}(\mathfrak{B}_{1011},\mathfrak{B}^{[1]}_{can};1)
      &
        \hspace{0pt}\mu_5\otimes{\cal S}^{01}_{11},\hspace{2pt}\mu_4\otimes{\cal S}^{11}_{11}
      &
      &
        \hspace{0pt}\mu_5\otimes{\cal S}^{00}_{11},\hspace{2pt}\mu_4\otimes{\cal S}^{10}_{11}
      &
      \hat{W}(\mathfrak{B}_{1011},\mathfrak{B}^{[1]}_{can};0)
      \\
      {W}(\mathfrak{B}_{1100},\mathfrak{B}^{[1]}_{can};1)
      &
       \hspace{0pt}\mu_6\otimes{\cal S}^{01}_{00},\hspace{2pt}\mu_6\otimes{\cal S}^{11}_{00}
      &
      &
        \hspace{0pt}\mu_7\otimes{\cal S}^{00}_{00},\hspace{2pt}\mu_7\otimes{\cal S}^{10}_{00}
      &
      \hat{W}(\mathfrak{B}_{1100},\mathfrak{B}^{[1]}_{can};0)
      \\
      {W}(\mathfrak{B}_{1100},\mathfrak{B}^{[1]}_{can};0)
      &
        \hspace{0pt}\mu_6\otimes{\cal S}^{00}_{00},\hspace{2pt}\mu_6\otimes{\cal S}^{10}_{00}
      &
      &
        \hspace{0pt}\mu_7\otimes{\cal S}^{01}_{00},\hspace{2pt}\mu_7\otimes{\cal S}^{11}_{00}
      &
      \hat{W}(\mathfrak{B}_{1100},\mathfrak{B}^{[1]}_{can};1)
      \\
      {W}(\mathfrak{B}_{1101},\mathfrak{B}^{[1]}_{can};0)
      &
       \hspace{0pt}\mu_6\otimes{\cal S}^{00}_{01},\hspace{2pt}\mu_6\otimes{\cal S}^{10}_{01}
      &
      &
       \hspace{0pt}\mu_6\otimes{\cal S}^{01}_{01},\hspace{2pt}\mu_6\otimes{\cal S}^{11}_{01}
      &
      \hat{W}(\mathfrak{B}_{1101},\mathfrak{B}^{[1]}_{can};1)
      \\
      {W}(\mathfrak{B}_{1101},\mathfrak{B}^{[1]}_{can};1)
      &
        \hspace{0pt}\mu_7\otimes{\cal S}^{01}_{01},\hspace{2pt}\mu_7\otimes{\cal S}^{11}_{01}
      &
      &
        \hspace{0pt}\mu_7\otimes{\cal S}^{00}_{01},\hspace{2pt}\mu_7\otimes{\cal S}^{10}_{01}
      &
      \hat{W}(\mathfrak{B}_{1101},\mathfrak{B}^{[1]}_{can};0)
      \\
      {W}(\mathfrak{B}_{1110},\mathfrak{B}^{[1]}_{can};0)
      &
       \hspace{0pt}\mu_6\otimes{\cal S}^{00}_{10},\hspace{2pt}\mu_7\otimes{\cal S}^{10}_{10}
      &
      &
       \hspace{0pt}\mu_7\otimes{\cal S}^{01}_{10},\hspace{2pt}\mu_6\otimes{\cal S}^{11}_{10}
      &
      \hat{W}(\mathfrak{B}_{1110},\mathfrak{B}^{[1]}_{can};1)
      \\
      {W}(\mathfrak{B}_{1110},\mathfrak{B}^{[1]}_{can};1)
      &
        \hspace{0pt}\mu_6\otimes{\cal S}^{01}_{10},\hspace{2pt}\mu_7\otimes{\cal S}^{11}_{10}
      &
      &
        \hspace{0pt}\mu_7\otimes{\cal S}^{00}_{10},\hspace{2pt}\mu_6\otimes{\cal S}^{10}_{10}
      &
      \hat{W}(\mathfrak{B}_{1110},\mathfrak{B}^{[1]}_{can};0)
      \\
      {W}(\mathfrak{B}_{1111},\mathfrak{B}^{[1]}_{can};1)
      &
        \hspace{0pt}\mu_7\otimes{\cal S}^{01}_{11},\hspace{2pt}\mu_6\otimes{\cal S}^{11}_{11}
      &
      &
        \hspace{0pt}\mu_7\otimes{\cal S}^{00}_{11},\hspace{2pt}\mu_6\otimes{\cal S}^{10}_{11}
      &
      \hat{W}(\mathfrak{B}_{1111},\mathfrak{B}^{[1]}_{can};0)
      \\
      {W}(\mathfrak{B}_{1111},\mathfrak{B}^{[1]}_{can};0)
      &
        \hspace{0pt}\mu_6\otimes{\cal S}^{00}_{11},\hspace{2pt}\mu_7\otimes{\cal S}^{10}_{11}
      &
      &
        \hspace{0pt}\mu_6\otimes{\cal S}^{01}_{11},\hspace{2pt}\mu_7\otimes{\cal S}^{11}_{11}
      &
      \hat{W}(\mathfrak{B}_{1111},\mathfrak{B}^{[1]}_{can};1)
 \end{array}\]
 \vspace{6pt}
 \fcaption{The co-quotient algebra of rank one given by the coset
 $\mathfrak{B}^{[1,1]}_{can}=\mathfrak{C}_{[\mathbf{0}]}-\mathfrak{B}^{[1]}_{can}\subset su(12)$. \label{figsu12cancoQArank1}}
 \end{figure}

 \begin{figure}
 \[
\]
 \vspace{6pt}
 \fcaption{A quotient algebra of rank 2 given by a $2$nd maximal bi-subalgebra
 $\mathfrak{B}^{[2]}_{can}=\mathfrak{B}^{[2,\hspace{2pt}00]}_{can}=\mathfrak{B}_{0001}\cap\mathfrak{B}_{0010}\cap\mathfrak{B}_{0011}\subset\mathfrak{C}_{[\mathbf{0}]}\subset su(12)$,
 referring to Fig.~\ref{figsu12QArank0intr}. \label{figsu12canQArank2}}
\end{figure}

 \begin{figure}
 \[%
\]
 \vspace{6pt}
 \fcaption{A co-quotient algebra of rank 2 given by a coset
 $\mathfrak{B}^{[2,\hspace{1pt}10]}_{can}$
 in the partition of $\mathfrak{C}_{[\mathbf{0}]}\subset su(12)$ generated by $\mathfrak{B}^{[2]}_{can}\subset\mathfrak{C}_{[\mathbf{0}]}$. \label{figsu12cancoQArank2}}
\end{figure}

 \clearpage
 \appendix{~~Tables of Cartan decompositions\label{appExpCD}}

 \begin{figure}[!ht]
 \[%
\]
 \vspace{6pt}
 \fcaption{Three examples of Cartan decompositions determined from
 the intrinsic quotient algebra of $su(8)$ in Fig.~\ref{figsu8QArank1intr}:
 (a) the intrinsic decomposition of type {\bf AII}
 $\hat{\mathfrak{t}}_{\hspace{1pt}{\rm II}}\oplus\hat{\mathfrak{p}}_{\hspace{1pt}{\rm II}}$;
 (b) a decomposition of type {\bf AI} and a type {\bf AII}. \label{figsu8tpintrrank1inC0}}
 \end{figure}

  \begin{figure}[htbp]
 \begin{center}
 \[\hspace{0pt}
 %
\]
 \vspace{6pt}
 \fcaption{Two examples of Cartan decompositions of type {\bf AII} determined from the intrinsic quotient algebra of $su(8)$ in
 Fig.~\ref{figsu8QArank2intr}:
 (a) the intrinsic decomposition $\hat{\mathfrak{t}}_{\hspace{0pt}{\rm II}}\oplus\hat{\mathfrak{p}}_{\hspace{0pt}{\rm II}}$
 and (b) a non-intrinsic one. \label{figsu8tpintrrank2}}
 \end{center}
 \end{figure}

 \begin{figure}[htbp]
 \[%
\]
 \fcaption{Two examples of Cartan decompositions determined from the quotient algebra of $su(8)$
 in Fig.~\ref{figsu8QArank1inC}:
 (a) a type {\bf AI} and (b) a type {\bf AII}. \label{figsu8tprank1}}
 \end{figure}

  \begin{figure}[htbp]
 \[\hspace{-20pt}%
\]
 \fcaption{Two examples of Cartan decompositions determined from the quotient algebra of $su(8)$ in
 Fig.~\ref{figsu8QArank2inC}: (a) a type {\bf AI} and (b) a type {\bf AII}. \label{figsu8tprank2inC}}
 \end{figure}

 \begin{figure}[htbp]
 \[\hspace{0pt}%
\]
 \vspace{6pt}
 \fcaption{Three examples of Cartan decompositions determined from
 the intrinsic co-quotient algebra of $su(8)$ in
 Fig.~\ref{figsu8coQArank1intr}:
 (a) the intrinsic decomposition of type {\bf AIII}
 $\hat{\mathfrak{t}}_{\hspace{1pt}{\rm III}}\oplus\hat{\mathfrak{p}}_{\hspace{1pt}{\rm III}}$;
 (b) a type {\bf AI} and a type {\bf AIII}. \label{figsu8tpintrcorank1inCIII}}
 \end{figure}

  \begin{figure}[htbp]
 \[\hspace{-8pt}%
\]
 \vspace{6pt}
 \fcaption{Two examples of Cartan decompositions determined from the intrinsic co-quotient algebra of $su(8)$ in
 Fig.~\ref{figsu8QArank2inCIII}:
 (a) the intrinsic decomposition of type {\bf AIII}
 $\hat{\mathfrak{t}}_{\hspace{0pt}{\rm III}}\oplus\hat{\mathfrak{p}}_{\hspace{0pt}{\rm III}}$
 and (b) a decomposition of type {\bf AII}. \label{figsu8tpintrcorank2}}
 \end{figure}

 \begin{figure}[htbp]
 \[%
\]
 \fcaption{Two examples of Cartan decompositions determined from the co-quotient
 algebra of $su(8)$ in Fig.~\ref{figsu8coQArank1inC}: (a) a type {\bf AI} and (b) a type {\bf AIII}. \label{figsu8tpcorank1}}
 \end{figure}

 \begin{figure}[htbp]
 \[\hspace{-20pt}%
\]
 \fcaption{Two examples of Cartan decompositions determined from the co-quotient algebra of $su(8)$ in
 Fig.~\ref{figsu8coQArank2inC}: (a) a type {\bf AII} and (b) a type {\bf AIII}. \label{figsu8tpcorank2inC}}
 \end{figure}

\appendix{~~Bi-Subalgebra Partition of Higher Order and Quotient-Algebra Partition of Higher Rank\label{appQAPDisplay}}
\begin{figure}[htbp]
%
\\
\end{tabular} \right\}\mathcal{W}({\mathfrak{B}}_{2^{p}-1},\mathfrak{B}^{[r]};i)\]
\end{tabular}
\\
\fcaption{A schematic diagram of a bi-subalgebra partition of
 order $p$ in (a) and a refined version of order $p+r$ in (b);
 the partition in (a) is
 decided by the abelian group $\mathcal{G}(\mathfrak{C})$ comprising all maximal bi-subalgebras of
 a Cartan subalgebra ${\mathfrak{C}}\subset su(N)$
 and
 the partition in (b) is obtained by first partitioning $\mathfrak{C}$
 into $2^r$ cosets $\mathfrak{B}^{[r,l]}$ of an $r$-th maximal bi-subalgebra
 $\mathfrak{B}^{[r]}\subset\mathfrak{C}$ and then partitioning each conjugate-pair subspace ${\cal W}(\mathfrak{B}_m)$
 into $2^r$ partitioned
 conjugate-pair subspaces ${\cal W}(\mathfrak{B}_m,\mathfrak{B}^{[r]};i)$
 via the coset rule of partition,
 here $i,l\in{Z^r_2}$,
 $\mathfrak{B}^{[r]}=\mathfrak{B}^{[r,\mathbf{0}]}$,
 $\mathfrak{B}_m\in\mathcal{G}(\mathfrak{C})$
 and $1\leq m< 2^p$.
  \label{FigBPrankrDig}}
\end{figure}

\begin{figure}[htbp]
\[\left.
\begin{tabular}{p{4.0cm}}
$~~~~~~~~~~~$
\begin{tabular}{|p{1.0cm}p{1.0cm}|}
\hline
&\\
\hline
\end{tabular}\vspace{0.25cm}
\\
$~~~~~~~~~~~$
\begin{tabular}{|p{1.0cm}p{1.0cm}|}
\hline
&\\
\hline
\end{tabular}\\
$~~~~~~~~~~~~~~~~~~~~~~${$\vdots$}\\
$~~~~~~~~~~~$
\begin{tabular}{|p{1.0cm}p{1.0cm}|}
\hline
&\\
\hline
\end{tabular}
\end{tabular} \right\}{\mathfrak{B}}^{[r, l]}\]
\vspace{0.5cm}
\[W({\mathfrak{B}}_{1},\mathfrak{B}^{[r]};i)\left.
\begin{tabular}{p{7cm}}
$~~~~~~~~~~~~~~~~~~$
\begin{tabular}{|>{\columncolor{PineGreen}}p{1.0cm}|p{1.7cm}p{1.7cm}|>{\columncolor{PineGreen}}p{1.0cm}|}
\cline{1-1}\cline{4-4}
& & & \\
\cline{1-1}\cline{4-4}
\end{tabular}\\
$~~~~~~~~~~~~~~~~~~$
\begin{tabular}{|>{\columncolor{PineGreen}}p{1.0cm}|p{1.7cm}p{1.7cm}|>{\columncolor{PineGreen}}p{1.0cm}|}
\cline{1-1}\cline{4-4}
& & & \\
\cline{1-1}\cline{4-4}
\end{tabular}\\
$~~~${$\vdots$}$~~~~~~~~~~~~~~~~~~~~~~~~${$\vdots$}$~~$
\begin{tabular}{|>{\columncolor{PineGreen}}p{1.0cm}|p{1.7cm}p{1.7cm}|>{\columncolor{PineGreen}}p{1.0cm}|}
\cline{1-1}\cline{4-4}
& & & \\
\cline{1-1}\cline{4-4}
\end{tabular}\\
\end{tabular} \right.\hat{W}({\mathfrak{B}}_{1},\mathfrak{B}^{[r]};i)\]
\vspace{0.5cm}
\[W({\mathfrak{B}}_{2},\mathfrak{B}^{[r]};i)\left.
\begin{tabular}{p{7cm}}
$~~~~~~~~~~~~~~~~~~$
\begin{tabular}{|>{\columncolor{Tan}}p{1.0cm}|p{1.7cm}p{1.7cm}|>{\columncolor{Tan}}p{1.0cm}|}
\cline{1-1}\cline{4-4}
& & & \\
\cline{1-1}\cline{4-4}
\end{tabular}\\
$~~~~~~~~~~~~~~~~~~$
\begin{tabular}{|>{\columncolor{Tan}}p{1.0cm}|p{1.7cm}p{1.7cm}|>{\columncolor{Tan}}p{1.0cm}|}
\cline{1-1}\cline{4-4}
& & & \\
\cline{1-1}\cline{4-4}
\end{tabular}\\
$~~~${$\vdots$}$~~~~~~~~~~~~~~~~~~~~~~~~${$\vdots$}$~~$
\begin{tabular}{|>{\columncolor{Tan}}p{1.0cm}|p{1.7cm}p{1.7cm}|>{\columncolor{Tan}}p{1.0cm}|}
\cline{1-1}\cline{4-4}
& & & \\
\cline{1-1}\cline{4-4}
\end{tabular}\\
\end{tabular} \right.\hat{W}({\mathfrak{B}}_{2},\mathfrak{B}^{[r]};i)\]
\vspace{0.5cm}

$~~~~~~~~~~~~~~~~~~~~~~~~~~~~~~~~~~~~~~~~~~~~~~~~~~~~~~~~~~~~~~${$\vdots$}
\[W({\mathfrak{B}}_{2^{p}-1},\mathfrak{B}^{[r]};i)\left.
\begin{tabular}{p{7cm}}
$~~~~~~~~~~~~~~~~~~$
\begin{tabular}{|>{\columncolor{CadetBlue}}p{1.0cm}|p{1.7cm}p{1.7cm}|>{\columncolor{CadetBlue}}p{1.0cm}|}
\cline{1-1}\cline{4-4}
& & & \\
\cline{1-1}\cline{4-4}
\end{tabular}\\
$~~~~~~~~~~~~~~~~~~$
\begin{tabular}{|>{\columncolor{CadetBlue}}p{1.0cm}|p{1.7cm}p{1.7cm}|>{\columncolor{CadetBlue}}p{1.0cm}|}
\cline{1-1}\cline{4-4}
& & & \\
\cline{1-1}\cline{4-4}
\end{tabular}\\
$~~~${$\vdots$}$~~~~~~~~~~~~~~~~~~~~~~~~${$\vdots$}$~~$
\begin{tabular}{|>{\columncolor{CadetBlue}}p{1.0cm}|p{1.7cm}p{1.7cm}|>{\columncolor{CadetBlue}}p{1.0cm}|}
\cline{1-1}\cline{4-4}
& & & \\
\cline{1-1}\cline{4-4}
\end{tabular}\\
\end{tabular} \right.\hat{W}({\mathfrak{B}}_{2^{p}-1},\mathfrak{B}^{[r]};i)\]
\\
 \fcaption{A schematic diagram of the quotient-algebra partition of rank $r$
  generated by an $r$-th maximal bi-subalgebra $\mathfrak{B}^{[r]}=\mathfrak{B}^{[r,\mathbf{0}]}$
  of a Cartan subalgebra $\mathfrak{C}=\bigcup_{l\in{Z^r_2}}\mathfrak{B}^{[r,l]}$;
  the pair of conditioned subspaces
  ${W}(\mathfrak{B}_m,\mathfrak{B}^{[r]};i)$ and $\hat{W}(\mathfrak{B}_m,\mathfrak{B}^{[r]};i)$
  of the doublet $(\mathfrak{B}_m,\mathfrak{B}^{[r]})$ are obtained by applying the coset rule of bisection to the
  partitioned conjugate-pair subspace
  ${\cal W}(\mathfrak{B}_m,\mathfrak{B}^{[r]};i)$ in Fig.~\ref{FigBPrankrDig}(b),
  here $\mathfrak{B}_m\in\mathcal{G}(\mathfrak{C})$ for $1\leq m<2^p$ and $i\in{Z^r_2}$.
   \label{FigQAPrankr1Dig}}
\end{figure}
\end{document}